# Geometrical aspects of Amplitudes and Correlators in $N=4$ SYM

Gabriele Dian

A Thesis presented for the degree of
Doctor of Philosophy

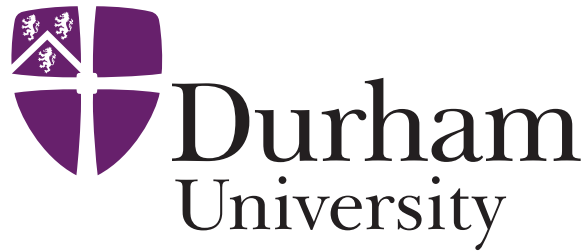

Centre for Particle Theory
Department of Mathematical Sciences
Durham University
United Kingdom

October 2022

# Geometrical aspects of Amplitudes and Correlators in N = 4 **SYM**

Gabriele Dian

Submitted for the degree of Doctor of Philosophy

October 2022

**Abstract:** This thesis describes progresses made by the author and collaborators in the positive geometry description of superamplitudes and the supercorrelators in planar $\mathcal{N} = 4$ SYM.

# Declaration

The work in this thesis is based on research carried out in the Department of Mathematical and Computer Sciences at Durham University under the supervision of Prof. Paul Heslop.

The first part of this thesis aims to introduce the reader to planar scattering amplitudes and correlators integrands in $\mathcal{N}=4$ SYM and to set the conventions used throughout the text. The second part is based on results obtained in the publications

- G. Dian and P. Heslop, Amplituhedron-like geometries, JHEP 11 (2021) 074, arXiv:2106.09372 [hep-th] ,

- G. Dian, P. Heslop, and A. Stewart, Internal boundaries of the loop amplituhedron, arXiv:2207.12464 [hep-th] .

No part of this thesis has been submitted elsewhere for any degree or qualification.

# Acknowledgements

There are many ingredients that go into a successful PhD, but an indispensable one is to have good teachers. I will like therefore to start by thanking the people from whom I learned the most.

My first thanks go to my supervisor Paul Heslop, who guided me through my studies and taught me the profession of theoretical physicist. Your kindness, respect and intellectual honesty are qualities that I deeply admire and I hope, thanks to your example, I will be able to make them my own. I promise you I'll learn more about Young tableau.

*Computing is knowledge*, once told me my undergrad thesis supervisor Raffaele Resta. I better understand what he meant now that I have experienced how fundamental explicit computations are in building our own intuition and unveiling our naïve preconceptions. But not all interesting computations can be done with pen and paper and for this the Wolfram Language has been an invaluable tool. I need therefore to thank Devendra Kapadia for providing me with the programming skills I didn't know I terribly needed. Learning something new never felt so natural thanks to the well-calibrated challenges you proposed to me daily.

During my four years at the Durham department of Mathematical Sciences, I had the fortune of finding not only stimulating researchers but also many nice human beings. In particular, I would like to thank Daniele Dorigoni, Arthur Lipstein, Patrick Dorey, Andreas Braun, Tin Sulejmanpasic, Nabil Iqbal, Iñaki García Etxebarria, Stefano Cremonesi, Douglas Smith, Magdalena Larfors, Madalena Lemos, Pankaj Vishe, Matthew Bullimore, Napat Poovuttikul (aka Nick), Federico Carta, Danny Levis, Connor Armstrong, Teresa Abel, Kieran Macfarlane. Being part of such an vibrant and friendly research environment taught me a great deal.

A special thank goes to Davide Polvara, for his spirit of playfulness which made all the steps we took together from the beginning to the end of this journey light and joyful, Dario Domingo for always being there for me when I needed him, Luigi Guerrini for all the mornings filled sharing our research and Alastair Stewart for the long time spent working together on the amplituhedron, for all the shared moments of excitement and those of disappointment.

There are many people to whom I own the immense fortune, that is the PhD, to have been paid to learn and to have concluded my studies successfully. My parents, for making me study, supporting me economically and morally. My graduate thesis supervisor and co-supervisor Roberto Valandro and Simone Giacomelli. The SAGEX network and all the people that made it possible, in particular Gabriele Travaglini, Matthias Staudacher, Paolo Benincasa, Lance Dixon, Zvi Bern, Lionel Mason, Volker Schomerus, Georgios Papathanasiou, JJ Carrasco, David Kosower, Jan Plefka, Andreas Brandhuber, Jacob Bourjaily, Johannes Blümlein and Chris White. The SAGEX ESRs, Manuel Accettulli Huber, Stefano De Angelis, Luke Corcoran, Nikolai Fadeev, Andrea Cristofoli, Ingrid Holm, Sebastian Pögel, Lorenzo Quintavalle, Marco Saragnese, Canxin Shi and Anne Spiering, with whom I shared so many moments of joy and tough challenges.

Part of my PhD has coincided with the coronavirus pandemic, when interacting with other people was not that easy. So, thanks to David Damgaard, Robert Moerman and Henrik Munch for all the stimulating talks they organized as part of *geometry and amplitudes journal club*, as well as Atul Sharma, Federico Gasparotto and Carlos Gustavo Rodriguez Fernandez, for taking part with enthusiasm. And thanks also to my old pals Christian Copetti and Francesco Cianci, always ready to talk about physics and life.

I will remember these days also as the happy days I lived in the North East. For the word "happy" would like to thank all the italian-spanish gang, too many to list, that treated me as an old friend from day one, with a special thought for those who have left us and those who have just come into the world. A special thanks to Antonella and Roberto for making me and Barbara feel like living in a *paesello* and to Andreas and Kübra for the good wine, good food, good friends and good times.

Finally, thanks again to Paul, Alastair, Dario, Davide and Francesco for reviewing the manuscript and to Douglas Smith and Tomasz Lukowski, my PhD examiners, for further insightful suggestions and corrections.

*De più poderia dir, ma me voi destrigar,*
*Perchè da vu mi bramo sentirme confutar.*

— from *Critica del Filosofo Inglese* by G. Baffo

*This thesis is dedicated
to my life partner*

Barbara

# Contents

# List of Figures

# Chapter 1

# Preface

The amplituhedron is a geometrical object which gives a beautiful intrinsic definition of the perturbative expansion of planar amplitudes in $\mathcal{N} = 4$ super Yang-Mills (SYM), allowing for entirely novel expressions for amplitudes to be found. In this framework, amplitude integrands are obtained as a differential form, called the canonical form, of the amplituhedron. The boundary structure of the amplituhedron then encodes the full singularity structure of the integrand. In the last few years tight relations between scattering amplitudes and geometrical objects called positive geometries have been uncovered, generalizing the amplituhedron framework beyond $\mathcal{N} = 4$ SYM.

The main objective of this thesis is to sharpen our understating of the amplituhedron and its canonical form as well as extend this new paradigm to correlators. This would simultaneously improve our ability to compute physical observables and possibly illuminate underlying structures in quantum field theory (QFT).

## Thesis Outline

The content of this thesis is presented in two parts titled:

- Part 1: Review of Concepts

- Part 2: Loop Amplituhedron and Squared Amplituhedron as WPGs

The first part mainly serves as an introduction to the amplituhedron. We review the notions of amplitude and correlator integrand and their supersymmetric generalization to superfunctions as well as their connection to differential forms and positive geometries. The second part instead consists of the original work published in [1, 2]. The content of these papers is adapted to be coherent as a whole and sometimes extended.

Part 1 contains the following chapters:

- Chapter 2 is an invitation to superamplitudes and supercorrelators in planar $\mathcal{N}=4$ SYM. We define the main quantity in which we are interested which is the color ordered amplitude integrand. We then introduce superamplitudes in super momentum twistor space and supercorrelators in supertwistor space. We conclude by formulating the duality between the square of the superamplitude and the supercorrelator.

- Chapter 3 introduces positive geometries, the canonical form and the fundamental mathematical tools to deal with these objects. We then define the amplituhedron, the correlahedron and the squared amplituhedron.

Part 2 contains the following chapters

- Chapter 4 shows why the loop amplituhedron is not a positive geometry and how positive geometries can be generalized to weighted positive geometries (WPG) to include it.

- Chapter 5 explores some boundaries of the loop amplituhedron related to setting all loop propagators on-shell. Then the canonical form for some special cases is computed at arbitrary loop order.

- Chapter 6 shows that the squared amplituhedron is a WPG given by the union of a set of geometries corresponding to products of amplitudes called amplituhedron-like geometries. Finally, the positroid tiling of the amplituhedron is used to prove that the canonical form of the tree-level $\overline{\text{MHV}}$ squared amplituhedron corresponds to the square of the superamplitude.

- The thesis ends with the conclusions and an outlook on future work in Chapter 7.

## 1.1 On the Amplituhedron and positive geometries

One of the best understood and studied QFTs is $\mathcal{N}=4$ SYM [3, 4]. The reason is that, despite the richness of phenomena that it can describe, the Lagrangian is completely determined by its symmetries, a signal of a hidden simplicity. Moreover, at the origin of the moduli space, the theory is also conformal invariant which unlocks all the power of CFT techniques [5]. Nevertheless, it is not straightforward to make use of the full supersymmetry to compute observables for the simple fact that it is not possible to write a manifestly $\mathcal{N} = 4$ supersymmetric off-shell Lagrangian and therefore to write supersymmetric Feynman diagrams.

This obstacle has inspired many researchers to look directly at the representation of supersymmetry on (quasi)-physical observables like amplitudes and correlators. These can be represented as manifestly supersymmetric quantities called superamplitudes and supercorrelators on various superspace constructions. Superfunctions are polynomials in auxiliary Grassmann (anticommuting) variables, which transform under spacetime and supersymmetry transformations. The coefficients of superamplitudes and supercorrelators are then ordinary amplitudes and correlators. Taking an approach typical in the study of CFT, one can then compute the perturbative expansion of the latter simply by constraining them with their symmetries and their analytic properties, like behaviour on singularities and branch points, when studied for complex momenta [6–8].

Particularly striking is the case of planar $\mathcal{N} = 4$ where amplitude integrands, that is the amplitude before integrating the loop variables[1], can be computed by a generalization of the BCFW recursion relation [9]. An unexpected outcome of this investigation was discovering that planar amplitudes in $\mathcal{N} = 4$ turn out to be invariant under an infinite dimensional group called the Yangian [10], completely hidden by the Lagrangian description of the theory. The study of this new symmetry led to discovering a new structure in algebraic geometry, associated with contour integrals over the Grassmannian, which allows to write the amplitude as a sum of manifestly Yangian invariant terms [11–13]. At the same time the remarkable observation was made by A. Hodges that NMHV amplitudes can be thought of as the volume of certain polytopes in momentum twistor space [14].

In December 2013, N. Arkani-Hamed and J. Trnka published " The Amplituhedron" and were able to make a connection between contour integrals over the Grassmannian and the polytope description of the amplitude, opening new perspectives and puzzles [15]. The idea is that any n-point superamplitude, where the sum of the helicities of the scattering particles is equal to n − 2(k + 2), can be mapped to a volume form[2] in the Grassmannian Gr(k, k + 4), that is the space of k-hyperplanes in k + 4 dimension. In this space, a geometrical object, called *the amplituhedron*, can be defined such that its boundary structure encodes all the singularities of the volume form, characterizing it completely.

Since the original paper, there has been an enormous amount of progress in understanding various properties of the amplituhedron, such as: its boundary structure [16–18], relation with Yangian symmetry [19, 20], parity [21], its tilings [22–28] and a formal proof of the tree-level conjecture [29]. In particular, in 2017 three fundamental breakthroughs led to the explosion of new results in the geometrical description of amplitudes we have been seeing in the last 4 years.

---
[1]Amplitude integrands are well defined only in particular cases, see section 2.2.
[2]A volume form or top form is a differential form of degree equal to the differentiable manifold dimension.

The first is the rigorous mathematical definition of the map between the amplituhedron and the amplitude [30]. This is called the canonical form and all geometries possessing a canonical form are then called positive geometries. The second is a characterization of the amplituhedron, proposed in [31, 32] and recently proved in [33], in terms of an integer called flipping (or winding) number, a topological invariant connected to the ordering of the kinematical data once projected down to one dimension. This picture allowed for a more practical analysis of the amplituhedron geometry that led among other important results to proving perturbative unitarity from the amplituhedron [34]. Last but not least is the positive geometry description of tree-level amplitudes in the massless biadjoint $\phi^3$ theory [35], which showed how supersymmetry and gauge interactions are not a fundamental ingredients for this type of geometrical structures to appear. This construction has been since then partially generalized to: massive particles [36], higher polynomial interaction [37–41], multi fields [42] and loops [43, 44]. Moreover, the ideas presented in these three seminal papers led to a positive geometry description of tree-level amplitudes in $\mathcal{N} = 4$ in spinors helicity variables [45], in ABJM theory [46, 47], connections to the CHY formalism [48, 49] and cosmological correlators [50].

Crucially for this thesis, in 2017 a geometrical object called the *correlahedron* was introduced in [51] and conjectured to be equivalent to the correlator of stress-energy tensor supermultiplets in planar $\mathcal{N} = 4$ SYM. The correlahedron is not itself a positive geometry, but evidence was given that it nevertheless possesses a well-defined volume form that should yield the correlator. Correlators in planar N = 4 SYM have very direct and surprising connections with scattering amplitudes. In particular, the square of the superamplitude at all loops can be obtained as a limit of the tree-level correlator [52–57]. The same limit performed geometrically on the correlahedron defines a new geometry called the squared amplituhedron, whose canonical form is thus conjectured to correspond to the square of the superamplitude.

A detailed study of the squared amplituhedron geometry has been carried out in [1] and represents one of the main topics of this thesis. While the initial aim was to give further evidence for the correlahedron conjecture, this study led to more fundamental considerations about the sum and product of superamplitudes in the positive geometry framework. The challenges with the product in the amplituhedron space can be summarized as follows:

- Given two superamplitudes with volume forms in $\text{Gr}(k_1, k_1 + 4)$ and $\text{Gr}(k_2, k_2 + 4)$, it is not obvious how to obtain the volume form of the product, which is a

- form in $\mathrm{Gr}(k_1 + k_2, k_1 + k_2 + 4)$.

- Given two amplituhedra, it is not obvious how to derive the geometry of the product.

The first problem found a concrete answer in formula (6.1.2). The second problem found a general, but unsatisfactory, answer using the technology of positroids/on-shell diagrams [58] and a partial but elegant answer for the product of parity conjugate amplitudes. In this case, the geometry is given by a direct generalization of the amplituhedron to arbitrary flipping number we called *amplituhedron-like geometries*. Using on-shell diagrams it is possible to generate a tiling of the amplituhedron and of geometry of the product. This description is unsatisfactory, both because the number of tiles grows with the complexity of the amplitude and because it reveals the boundaries of the single tile, rather than the ones of the final geometry. Nevertheless, the positroid tiling allowed us to prove that the canonical form of amplituhedron-like geometries corresponds to the product of amplitudes and to prove that the maximal squared amplituhedron canonical form gives indeed the sum of products of parity conjugate amplitudes.

This last statement brings us directly to the subtle point of the sum of canonical forms and union of positive geometries, which we started to investigate in [1] and more clearly addressed in [2]. In fact, the squared amplituhedron and, to our great surprise, the loop amplituhedron are not strictly speaking positive geometries. Nevertheless, it is possible to express them as a union of positive geometries and slightly generalize the concept of positive geometry to include them in the definition as we will briefly illustrate now.

It is easy to show that the canonical form of the disjoint union of two geometries is equal to the sum of two canonical forms. The union of two geometries instead can have two outcomes: the two geometries can be disjoint or they can overlap. In the case where their interiors overlap and the canonical form of the union does not correspond to the sum of the canonical forms and the two differ by the canonical form of the intersection. It therefore makes sense to define the union of two geometries as a region paired with a weight function counting in each point the number of times the geometries do overlap, together with a sign taking care of orientations as illustrated in figure 1.1. In [2] we made this intuition precise by generalising what we mean by a geometrical region by introducing the concept of *weighted positive geometries* (WPGs).

So what do these overlappings have to do with the squared amplituhedron and the loop amplituhedron? In the first case, we have that the squared amplituhedron is given by almost-disconnected unions of amplituhedron-like geometries, which means that while the interior of the amplituhedron-like geometry is disconnected they

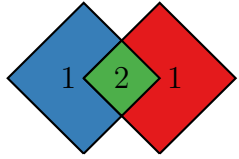

(a) Union of two squares with the same orientation.

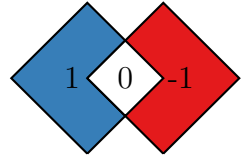

(b) Union of two squares with opposite orientation.

Figure 1.1: Examples of the sum of two overlapping WPGs.

overlap on codimension-2 or higher boundaries. As a consequence, even though the squared amplituhedron can be described by a WPGs with weight 1 in its interior, some of its boundaries are given by overlapping unions of geometries and can take various weights. Even more interesting is the case of the loop amplituhedron. The latter in fact has a connected interior, but it has nevertheless boundaries given by almost disconnected unions! Indeed a simple example of these boundaries is given by points in 3d satisfying $z > 0, z + xy > 0$, which pictured from below looks like the following:

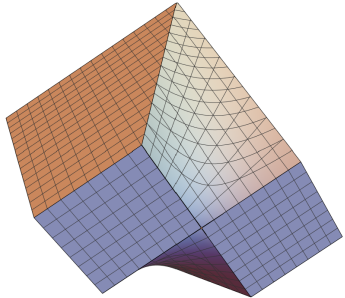

This geometry contains a boundary (nearest to the viewer in the above picture) at $z = 0$, which is given by $xy > 0$; two opposite quadrants of a plane. This 2d boundary in turn has a boundary at $y = 0$, consisting of two 1-dimensional regions $x > 0, x < 0$ with opposite orientation. Finally, this 1d region contains a 0-dimensional overlapping boundary at $x = 0$.

Overlapping boundaries emerge in the loop amplituhedron when, in some sense, loop-loop propagators are set on-shell. The geometry of these boundaries for a general amplituhedron has been derived in [2], together with the canonical form at all loops for one particular class of boundaries in the MHV amplituhedron.

In this thesis, we will review in detail the concepts outlined in this introduction and give the reader a brief but hopefully fairly pedagogical review of all the mathematical tools needed to understand the results presented in [1, 2]. We conclude with a discussion on possible future directions. Have a good reading!

# Part I

# Review of Concepts

# Chapter 2

# Planar Integrands in $\mathcal{N} = 4$ SYM

The aim of this chapter is to give a review of the basics of amplitudes and correlators in $\mathcal{N} = 4$ SYM to set the stage for the definition of the amplituhedron/correlahedron and the canonical form. We will start introducing the concepts of planarity and color ordering, which are essential to define the amplitude integrand. Then, we will specialize to $\mathcal{N} = 4$ SYM and describe amplitudes and correlators on superspace. We will introduce supertwistors and momentum supertwistor and we will use them to rewrite the supercorrelators in a form that makes manifest their superspace symmetries. We conclude the chapter by presenting the correlator/amplitude duality whose geometric avatar will be one of the main topics of this thesis.

## 2.1 Planar color ordered amplitudes

color decomposition, originally derived in the seminal works [59, 60] to study tree gluon amplitudes, has been key to the computation of scattering amplitudes and to the study of their analytic properties. Here, we are not interested in deriving how amplitudes can be decomposed into color ordered amplitudes but rather to explain the general idea and to highlight its key properties. We will focus on the special case relevant to the amplituhedron, that is amplitudes appearing in the large N limit of a U(N) or SU(N) gauge theory with only adjoint particles. Given $(T^a)_i^j$ the base elements in the adjoint representation of the relevant Lie algebra, any Feynman diagram can be written as a product of a function of the kinematics, that is momenta and polarization vectors, times a sum of products of T traces. Thanks to a series of iterated identities it is possible to show that an n-point l-loop gluon amplitude $\hat{M}_{n,l}$ can always be written in the color decomposition form, that is

$$\hat{M}_{n,l}^{a_1,\cdots,a_n} = g^{n+2(l-1)} \left( N^l \sum_{\sigma \in S_n/Z_n} \text{Tr}(T^{a_{\sigma(1)}} \cdots T^{a_{\sigma(n)}}) M_{n,l} + \mathcal{O}\left(\frac{1}{N}\right) \right), \quad (2.1.1)$$

where the indices $\{a_{\sigma(1)}, \cdots, a_{\sigma(n)}\}$ label the color of the external particles, $S_n$ is the set of all permutations of n objects and $Z_n$ is the subset of all cyclic permutations that preserve the trace. We suppressed the dependence of $\hat{M}_{n,l}$ and $M_{n,l}$ on momenta and spin to keep the expression compact. The functions $M_{n,l}$ are then called *color ordered amplitudes*. Notice that because of the symmetry of the trace under cyclic permutations, $M_{n,l}$ is cyclic invariant under the permutation of the external data, that is momenta and spin. First, we want to highlight that if we take the limit $N \to \infty$ while keeping the so-called 't Hooft coupling $\lambda = g^2 N$ finite, the amplitude reduces to the leading term of this expansion, which is given by color ordered amplitudes. Secondly, an important fact is that the color ordered amplitudes can be computed diagrammatically using some modified Feynman rules. Differently from the amplitude itself, instead of summing over all diagrams to compute the color ordered amplitude we have to sum over only cyclically ordered planar diagrams[1]. The latter are diagrams for which the external leg terminates on a circle in the order dictated by the trace factor and propagators do not cross, i.e. the graph must be planar. The prescription of summing over only planar diagrams comes from the fact that all non-planar diagrams are subleading in the $1/N$ expansion and instead all planar diagrams contribute to the leading term.

As an example, consider the pure YM Lagrangian

$$\mathcal{L} = \frac{1}{4} \mathrm{Tr} F_{\mu\nu} F^{\mu\nu}, \tag{2.1.2}$$

with $F_{\mu\nu} = \partial_\mu A_\nu - \partial_\nu A_\mu - \frac{ig}{2}[A_\mu, A_\nu]$ and $A_\mu = A_\mu^a T^a$. In the *Gervais-Neveu gauge* [61,62], the Lagrangian takes the form

$$\mathcal{L} = \mathrm{Tr}\left(-\frac{1}{2}\partial_\mu A_\nu \partial^\mu A^\nu - i\sqrt{2} g \partial^\mu A^\nu A_\nu A_\mu + \frac{g^2}{4} A^\mu A^\nu A_\nu A_\mu\right). \tag{2.1.3}$$

External legs are given by the polarization vector as in the amplitudes but without the color factor. The color ordered amplitudes can then be computed by summing over all the planar diagrams where the external legs are ordered and with Feynman rules

$$\text{\small(propagator)} = \frac{\eta_{\mu\nu}}{p^2}, \tag{2.1.4}$$

$$\text{\small(3-vertex)} = -\sqrt{2}(\eta^{\mu_1\mu_2} p_1^{\mu_3} + \eta^{\mu_2\mu_3} p_2^{\mu_1} + \eta^{\mu_3\mu_1} p_3^{\mu_2}), \tag{2.1.5}$$

---

[1] A planar graph is a graph that can be embedded in the plane, i.e. it can be drawn in such a way that no edges cross each other.

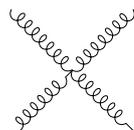

$$\phantom{xxxxxxxxx} = \eta^{\mu_1\mu_2}\eta^{\mu_3\mu_4} \; , \tag{2.1.6}$$

where $\eta^{\mu\nu}$ is the Minkowski flat space metric. The color ordered tree level 4-particle amplitude, with generic helicity content, will be given by the diagrammatic expansion

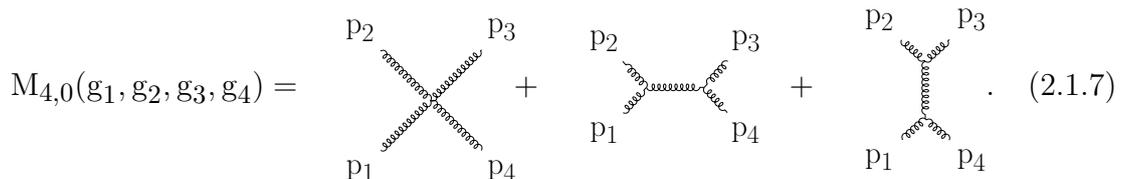

$$M_{4,0}(g_1, g_2, g_3, g_4) = \quad\quad + \quad\quad + \quad\quad . \tag{2.1.7}$$

Finally, following the review [63], the cross section for gluon scattering, or more generally for particles in the adjoint representation of the gauge group, can be directly expressed in terms of color ordered amplitudes. Consider a tree level n-point gluon amplitude $\hat{M}_n(\{p_1, h_1\}, \cdots, \{p_n, h_n\})$ where $p_i$ and $h_i$ are the momenta and helicities of outgoing scattering particles. The cross section for such process is given by

$$d\sigma^{\text{tree}} \approx \sum_{a=1}^{N_c^2-1} |\hat{M}_n|^2 \; , \tag{2.1.8}$$

where the sum is over final-state colors, is given by

$$d\sigma^{\text{tree}} \approx N_c^n \left( \sum_{\sigma \in S_n/Z_n} |M_n(\sigma(1), \sigma(2), \cdots, \sigma(n))|^2 + \mathcal{O}(1/N_c^2) \right) \; . \tag{2.1.9}$$

## 2.2 Region variables and integrands

The tree level amplitude is in general a well defined rational function of the external kinematics. The loop amplitude integrand instead, that is the expression derived at loop level using Feynman diagrams before integrating over the loops, is defined up to a shift of the loop momenta

$$l_i^\mu = l_i^\mu + a_i^\mu \; , \tag{2.2.1}$$

where $a_i$ is an arbitrary constant. While in general defining the integrand as an analytic function is an open problem, in the planar limit it can be defined unambiguously using the so-called region variables [64–66]. We will now illustrate how these variables are defined and how to relate them to the internal and external momenta.

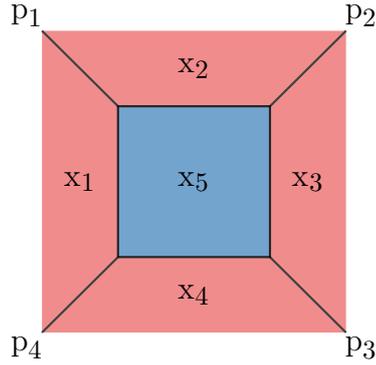

Figure 2.1: Region variables associated to the box diagram. The internal region is colored in blue while the external regions are colored in red.

When a planar graph is drawn with no crossing edges, it divides the plane into a set of regions, called faces. To any planar graph g we can associate a dual graph. The latter has a vertex for each face of g and an edge for each pair of vertices corresponding to adjacent faces. We will call the faces corresponding to loops *internal faces* and we will call *external faces* the remaining ones, that is the unbounded regions in the external part of the diagram. An example of a 4-point 1-loop diagram is given in figure 2.1.

In an n-particle planar Feynman diagram we have n-external regions. For simplicity, let's assume that the external momenta are ordered cyclically as $p_1, \cdots p_n$. We start by assigning the four-vector $x_1$ to the face between the external leg $p_n$ and $p_1$ and then continue cyclically. Each external leg with momenta $p_i$ will correspond to the boundary between the two regions $x_i$ and $x_{i+1}$. We can then define a relation between the region variable and the momenta as

$$\begin{aligned} s_i p_i &= x_{i+1} - x_i, \qquad \text{for } i < n, \\ s_n p_n &= x_1 - x_n. \end{aligned} \qquad (2.2.2)$$

where $s_i = \pm 1$ is equal to 1 for incoming particles and –1 for outgoing particles. More generally, we can define the momentum flowing in any leg as the difference between the two adjacent region variables, so the momentum q flowing in between two faces $x_i$ and $x_j$ can be written as $q = x_i - x_j$ with $i < j$. Notice that in the x variables the momentum conservation at each vertex is trivialized. In particular, for any values of the x's, the associated momenta automatically satisfies the external momentum conservation, that is

$$\sum_{i=0}^{n} s_i p_i = \sum_{i=0}^{n} x_{i+1} - x_i = -x_1 + x_{n+1} = 0 \ . \tag{2.2.3}$$

We turn our attention now to internal regions. These will be l for an l-loop graph and we will label them with the variables $x_i$ for $n < i \leq n + l$. There are different ways to label internal regions and depending on this choice different functions can be obtained. We can obtain a unique integrand by simply symmetrising over all possible labelling. Any l-loop diagram will then correspond in general to l! rational functions. For example, the double box diagram can be written using region variables as

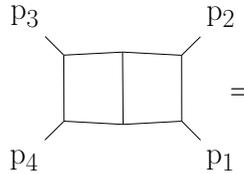

$$= \frac{1}{x_{14}^2 x_{23}^2 x_{15}^2 x_{25}^2 x_{56}^2 x_{45}^2 x_{26}^2 x_{36}^2 x_{46}^2} + \text{ sym } 5 \leftrightarrow 6 \ , \tag{2.2.4}$$

where $x_{i,j}^2 = (x_i - x_j)^2$. Finally, we would like to remark that one has to remember that on-shell conditions still apply and read

$$x_{\widehat{ii+1}}^2 = m_i^2 \ , \qquad \text{for } i < n \text{ and } \widehat{n+1} = 1 \ . \tag{2.2.5}$$

## 2.3 The Superamplitude

Supersymmetry is a transformation that mixes states of different helicity. In $\mathcal{N} = 4$ SYM, its action is described by four complex super charges $Q^{A\alpha}$, $Q^\dagger_{A\dot{\alpha}}$, where I goes from one to four and transforms under an internal SU(4) symmetry called R-symmetry, while $\alpha, \dot{\alpha}$ are spacetime spinorial indices. Amplitudes in $\mathcal{N} = 4$ SYM are annihilated by the action of $Q^{A\alpha}$, $Q^\dagger_{A\dot{\alpha}}$. A non-trivial representation of these operators is obtained by considering amplitudes as a function of the on-shell superspace which we will now describe.

In the on-shell superspace formulation, particle creation operators are dressed with Grassmann odd variables carrying an R-symmetry index and read

$$\begin{aligned}
&g^+ &&\text{positive helicity gluon} \\
&\eta^A \lambda_A &&\text{positive helicity spinors} \\
&\eta^A \eta^B \phi_{AB} &&\text{scalars} \\
&\eta^A \eta^B \eta^C \lambda_{ABC} &&\text{negative helicity spinors} \\
&\eta^1 \eta^2 \eta^3 \eta^4 g^- &&\text{negative helicity gluon} 
\end{aligned} \tag{2.3.1}$$

Amplitudes involving these dressed states will be a function of the external kinematics and the Grassmannian variables $\eta$, which together form the on-shell superspace. it is possible to write compactly all n-particle amplitudes in one object by considering the scattering of states generated by the on-shell chiral superfield

$$O = g^+ + \eta^A \lambda_A + +\eta^A \eta^B \phi_{AB} + \eta^A \eta^B \eta^C \lambda_{ABC} + \eta^1 \eta^2 \eta^3 \eta^4 g^- \ . \qquad (2.3.2)$$

In n-particle scattering amplitude each $\eta$ will carry two indices: an R-charge index, going from 1 to 4, and a particle index going from 1 to n. The vacuum expectation value of n on-shell superfields

$$\mathcal{M}(p_1, \cdots, p_n, \eta_1, \cdots, \eta_n) = \langle 0|O_1(p_1, \eta_1) \cdots O_n(p_n, \eta_n)|0\rangle \ , \qquad (2.3.3)$$

is called the superamplitude. Each coefficient of the $\eta$ polynomial corresponds to an ordinary amplitude. The particle content of the coefficient is given by its $\eta$ degree. For example, if the degree in $\eta_a$ is 0, the a$^{\text{th}}$ particle will be a positive helicity gluon. Amplitudes in $\mathcal{N} = 4$ SYM are invariant under the SU(4) R-symmetry group. Any product of four $\eta$'s of the form $\epsilon_{ijkl}\eta_a^i \eta_b^j \eta_c^k \eta_d^l$, is an R-symmetry invariant. Therefore, all $\eta$ monomials in the superamplitude have an $\eta$ degree which is a multiple of four. Amplitudes with the same $\eta$ degree are related by supersymmetry Ward identities. It makes sense then to consider polynomials of homogeneous degree separately

$$\mathcal{M}_n = \mathcal{M}_{n,0} + \mathcal{M}_{n,1} + \mathcal{M}_{n,2} + \cdots + \mathcal{M}_{n,n-4} \ , \qquad (2.3.4)$$

where $\mathcal{M}_{n,k}$ have a uniform 4(k+2) degree in the $\eta$'s and is called the N$^k$MHV sector. If we choose to use spinor helicity variables to represent the kinematic space, the one-particle on-shell superspace is made by the vectors $(|p\rangle, |p], \eta^A)$, where $|p\rangle, |p]$ is the spinor helicity variable associated to the momentum p as we describe in appendix A. The action of the supercharges on a function of the n-particle superspace reads

$$Q^{A\alpha} = \sum_{i=1}^n [p_i|^\alpha \partial_{\eta_A} \ , \qquad Q^\dagger_{A\dot\alpha} = \sum_{i=1}^n |p_i\rangle_{\dot\alpha} \, \eta_A \ . \qquad (2.3.5)$$

All functions annihilated by $Q^\dagger_{A\dot\alpha}$ must have a prefactor

$$\delta^{(8)}(q^\dagger) = \frac{1}{2^4} q^\dagger_{A\dot\alpha} q^{\dagger A\dot\alpha} = \frac{1}{2^4} \prod_{A=1}^4 \sum_{i=1}^n \langle ij\rangle \eta_{iA} \eta_{jA} \ . \qquad (2.3.6)$$

This prefactor is sometimes called for its behavior under Grassmann integration the super-momentum delta function. We conclude this section giving the expression of the simplest amplitude in $\mathcal{N} = 4$; the MHV amplitude. Its expression on superspace can be derived using the supersymmetric version of the BCFW recursion relation, originally derived by Britto, Cachazo, Feng and Witten in [67, 68], and reads

$$\mathcal{M}_{n,0}(|i\rangle, |i], \eta_i) = \frac{\delta^{(4)}(\sum |i\rangle [i|)\delta^{(8)}(\sum |i\rangle \eta_i)}{\langle 12 \rangle \cdots \langle n1 \rangle} \,, \tag{2.3.7}$$

where $\delta^{(4)}(\sum |i\rangle [i|)$ is the total momentum conservation delta function. From the superamplitude $\mathcal{M}_{n,0}$ we can recover the ordinary amplitudes by projection on any $\eta$ monomial. In particular, if we project onto the monomial $\eta_1^4 \eta_2^4$ we get

$$M_{n,0}(g_1^-, g_2^-, g_3^+, \cdots, g_n^+) = \frac{\langle 12 \rangle^4}{\langle 12 \rangle \cdots \langle n1 \rangle} \,, \tag{2.3.8}$$

which is the famous Parke-Taylor amplitude [60, 69] originally conjectured by Parke and Taylor for the n-gluon amplitude in pure YM. The two color ordered amplitudes are identical because all tree level pure gluon amplitudes in $\mathcal{N} = 4$ SYM are actually equal YM amplitudes.

For reasons that will be clear later, it is convenient to consider the superamplitude divided by the MHV tree level amplitude, which we will indicate with the symbol $\mathcal{A}_n$. More specifically, for a fixed $N^k$MHV sector we have

$$\mathcal{M}_{n,k}(x_i, \eta_i) = \mathcal{M}_{n,0}(x_i, \eta_i) \times \mathcal{A}_{n,k}(x_i, \eta_i) \,. \tag{2.3.9}$$

In particular we have $\mathcal{A}_{n,0} = 1$.

### 2.3.1 The squared superamplitude

As with the superamplitude, the product of the full superamplitude (the sum over all $N^k$MHV sectors) with itself can be written as a superfunction. We refer to this as the superamplitude squared. Given that the expansion of NMHV sectors of the superamplitude reads

$$\mathcal{A}_n = \sum_{k=0}^{n-4} \mathcal{A}_{n,k} \,, \tag{2.3.10}$$

the superamplitude squared can be written as

$$\mathcal{A}_n^2 = \sum_{k=0}^{n-4} \sum_{k'=0}^{k} \mathcal{A}_{n,k'} \mathcal{A}_{n,k-k'} \,, \tag{2.3.11}$$

where we stress the fact that these products are between functions of anti-commuting variables. Each product $\mathcal{A}_{n,k'} \mathcal{A}_{n,k-k'}$ has uniform degree in $\eta$ equal to $4(k' + (k - $

k$'$)) = 4k. Therefore, we can define the superamplitude squared helicity sectors as

$$(\mathcal{A}^2)_{n,k} = \sum_{k'=0}^{k} \mathcal{A}_{n,k'} \mathcal{A}_{n,k-k'} \,. \qquad (2.3.12)$$

At loop level the amplitude is a double sum over MHV degree and loop level:

$$\mathcal{A}_n = \sum_{l=0}^{\infty} \sum_{k=0}^{n-4} \int d\mu_l \, \mathcal{A}_{n,k,l} \,. \qquad (2.3.13)$$

Here the amplitude $\mathcal{A}_{n,k,l}$ is symmetric respect to the loop variables $\{AB_1, \cdots, AB_l\}$ and we define the integration measure as weighted by $1/l!$ compared to l copies of the 1-loop measure:

$$d\mu_l[(AB)_1, .., (AB)_l] := \frac{d\mu_1[(AB)_1]..d\mu_1[(AB)_l]}{l!} \,. \qquad (2.3.14)$$

Then when we take the square of the amplitude we obtain

$$\mathcal{A}_n^2 = \sum_{l=0}^{\infty} \sum_{k=0}^{n-k} \sum_{l'=0}^{l} \sum_{k'=0}^{k} \int d\mu_{l'} d\mu_{l-l'} \mathcal{A}_{n,k',l'} \mathcal{A}_{n,k-k',l-l'} := \sum_{l=0}^{\infty} \sum_{k=0}^{n-k} \int d\mu_l \, (\mathcal{A}^2)_{n,k,l} \,. \qquad (2.3.15)$$

Thus by the $N^k$MHV l-loop squared amplitude we mean

$$(\mathcal{A}^2)_{n,k,l} = \sum_{l'=0, k'=0}^{l,k} \binom{l}{l'} \mathcal{A}_{n,k',l'} \mathcal{A}_{n,k-k',l-l'} \,. \qquad (2.3.16)$$

The numerical factor arises from matching the measures $\int d\mu_{l'} d\mu_{l-l'} = \binom{l}{l!} \int d\mu_l$. Note however that we have not specified the distribution of the l loop variables between the two factors $\mathcal{A}_{n,k',l'}$ and $\mathcal{A}_{n,k-k',l-l'}$. The most natural choice is to have a completely symmetric distribution in which case there are exactly $\binom{l}{l!}$ inequivalent ways to do this and the squared amplitude simply sums over all these inequivalent distributions (2.3.16).

## 2.4 Twistors and momentum twistors

As we already mentioned, $\mathcal{N} = 4$ SYM at the origin of the moduli space is a superconformal quantum field theory (SCFT). This means that it is a supersymmetric theory invariant under the conformal group, which is composed by the Poincaré transformations plus the special conformal transformations and dilatation. Here we will see how to linearise the action of the superconformal group in spacetime and the dual superconformal group in momentum space.

### 2.4.1 Twistors

Famously, the action of the conformal group can be linearised using a a set of variables called *twistors*, originally introduced by R. Penrose [70] to describe flat Minkowski spacetime. One way to approach twistor is through the so-called *embedding formalism*, originally proposed by Dirac in [71]. We start by observing that the four dimensional conformal group is isomorphic to SO(2, 4) which is also the Lorentz group of 6d flat spacetime with signature $(-,-,+,+,+,+)$. So, if we find an embedding of 4d Minkowski spacetime in 6d, the non-linearity of the conformal group will emerge from the composition of the 6d Lorentz group with the projection from 6d to 4d.

Such embedding can be obtained by considering all $X^I$, with $I = 1, \cdots, 6$, such that

$$X \cdot X = 0 \ . \tag{2.4.1}$$

Moreover, we will consider the equivalence class $X \equiv tX$, that is we consider X as elements of the projective space $\mathbb{P}^5$. Readers unfamiliar with projective space could benefit from reading the first part section 3.1 before continuing. This constraint together with the equivalence class relation lower the degrees of freedom of X from 6 to 4. We can choose a reference 6-vector I and write distances in Minkowski spacetime as

$$x_{ij}^2 = \frac{(X_i - X_j)^2}{(I \cdot X_i)(I \cdot X_j)} = -\frac{1}{2} \frac{X_i \cdot X_j}{(I \cdot X_i)(I \cdot X_j)} \ . \tag{2.4.2}$$

Note that $x_{ij}^2$ is invariant under the rescaling of $X_i$ or $X_j$, so it is a well defined function in $\mathbb{P}^5$. The $-1/2$ factor can be absorbed in the definition of I and for this reason is absent in some equations in the literature. Now we want to spinorise X. This means to contract X with the 6d Clifford algebra elements, which we will indicate as $\Gamma_a^{IJ}$ with $a = 1, \cdots 6$ and $I, J = 1 \cdots, 4$. We will define therefore

$$X^{IJ} = X^a \Gamma_a^{IJ} \ , \tag{2.4.3}$$

where $X^{IJ}$ is called the twistor line[2].

A twistor line $X^{IJ}$ is anti-symmetric in the indices IJ and therefore it can be written as

$$X^{IJ} = X_a^I X_b^J \epsilon^{ab} \ , \tag{2.4.4}$$

where $X_1, X_2$ are two points in $\mathbb{P}^3$ called twistors. We have therefore, that a point in Mikowski space $x^\mu$ corresponds to a twistor line X in $\mathbb{P}^3$ given by the points W

---

[2] We use the symbol X to indicate the twistor hoping that is nature of twistor, 4-vector or dual coordinate will be clear from the context.

such that

$$X^{[IJ}W^{K]} = 0 \,. \tag{2.4.5}$$

We can spinorise also the reference vector I, for which we will use the same symbol, so that, dropping the 1/2 factor, (2.4.2) becomes

$$x_{ij}^2 = \frac{\langle X_i X_j \rangle}{\langle IX_i \rangle \langle IX_j \rangle} \,, \tag{2.4.6}$$

where the function $\langle \cdots \rangle$ indicates the determinant of the matrix made by the elements contained in the argument. In this case $\langle X_i X_j \rangle = \epsilon_{ABCD} X_i^{AB} X_j^{CD}$ and $\langle IX_j \rangle = \epsilon_{ABCD} I^{AB} X_j^{CD}$. This spinoral version of I is usually called the infinity twistor. The complexification of the conformal group corresponds to SL(4). This implies that the numerator in (2.4.6) is conformal invariant, since the determinant is an SL(4) invariant function, while the denominator is not since I does not transform under the action of the conformal group. An important fact is that the determinant is the generator of SL(4) invariant polynomials and therefore $\langle X_i X_j \rangle$ represents the building block of conformal invariant functions.

When discussing correlator integrands we will have to deal with loops variables. For correlators, loop variables have the same degrees of freedom as external variables and therefore we can represent them using twistor lines as well.

In supersymmetric theories the conformal group is enhanced to the superconformal group. We therefore are interested in generalizing twistors in Minkowski space to supertwistors [72] in $(x^\mu, \theta_\alpha^A)$ Minkowski superspace, which is the space where the chiral correlator we will define in section 2.6 is defined. Before doing this, we first take a step back and define twistors in an alternative but equivalent way from which their supersymmetrization will be more obvious.

We start by associating $x_i^\mu$ to the $2 \times 2$ matrix

$$x^{\dot{a}b} = \sigma_\mu^{\dot{a}b} x^\mu \,, \tag{2.4.7}$$

as is usually done when introducing spinor helicity variables. Differently from massless momenta, in this case the matrix $x^{\dot{a}b}$ is not degenerate, since for generic x $\det(x) = ||x||^2 \neq 0$. Any $2 \times 2$ matrix can be then written as

$$x^{\dot{a}b} = |x\rangle_s^{\dot{a}} [x|^{sb} = |x\rangle_1^{\dot{a}} [x|^{1b} + |x\rangle_2^{\dot{a}} [x|^{2b} \,. \tag{2.4.8}$$

This construction is completely analogous to the massive spinor helicity formalism [32], where one spinorises the momenta $p^\mu$ with $p^2 = m^2$. Notice that this relation is invariant under a SU(2) transformation of the index s and the rescaling $|x\rangle_s \to t |x\rangle_s$, $|x]^s \to t^{-1} |x]^s$. This means that the pair $(|x\rangle_s, |x]^s)$ is equivalent to $(M_s^{s'} |x\rangle_{s'}, (M^{-1})_s^{s'}|x]_{s'})$, where $M \in GL(2)$. To check the degrees of freedom, we can choose representatives such that $|x\rangle_s^{\dot{a}} = \mathbb{1}_s^{\dot{a}}$, from which we see that the pair of spinors have 4 d.o.f. as expected. Then, we can define a map between $x^{\dot{a}b}$ to the spinors $|\mu]_s$ as

$$[\mu_x|^{sb} = \langle x|_{\dot{a}}^{s} x^{\dot{a}b} \,. \tag{2.4.9}$$

The inverse of (2.4.9) is simply given by the fact that x has non vanishing determinant and therefore it is invertible. The related twistors are then defined as two vectors $X_1^I, X_2^J$, where their four components correspond to

$$\begin{pmatrix} X_s^1 & X_s^2 & X_s^3 & X_s^4 \end{pmatrix} = \begin{pmatrix} |x\rangle_s^1 & |x\rangle_s^2 & |\mu_x]_s^1 & |\mu_x]_s^2 \end{pmatrix} \,. \tag{2.4.10}$$

Notice that $[\mu_x|_s$ scales with the complex little group as $\langle x|_s$. Therefore, for complex momenta, the twistors are defined up to a rescaling

$$X_s^I \sim \lambda_s X_s^I \qquad \text{with } \lambda_s \neq 0 \,, \tag{2.4.11}$$

which means each twistor has 3 degrees of freedom and is an element of $\mathbb{P}^3$. Moreover, the two twistors are defined up to a GL(2) transformation and therefore the GL(2) invariant object we should consider is the twistor line $X^{IJ} = X_a^I X_b^J \epsilon^{ab}$, as in the embedding space derivation.

To extend twistors to supertwistors, consider now points in the $\mathcal{N} = 4$ chiral superspace $(x^\mu, \theta_\alpha^A)$, where $\theta$ is a Grassmannian variable associated to the $Q_\alpha^A$ supersymmetry generator in the coset construction. Using the notation $x_i^{\dot{a}b} = |i\rangle_s^{\dot{a}} |i]^{sb}$ we can define a pair of new Grassmann variables for each $\theta_i$ as

$$\chi_{s,i}^a = \langle i_s \theta_i^A \rangle \,. \tag{2.4.12}$$

Notice that consistently $\chi_{s,i}^A$ transforms under the SU(2) gauge as the twistor $X_{s,i}^I$. A supertwistor is then defined as

$$\mathcal{X}_{s,i} = \begin{pmatrix} X_{s,i} \\ \chi_{s,i} \end{pmatrix} \in \mathbb{C}^{(4|4)}/GL(1) \,, \tag{2.4.13}$$

where $\mathbb{C}^{(4|4)}/GL(1)$ indicates that the first 4 components are bosonic defined up to rescaling and the second 4 components are anti-commuting variables. The supertwistor line $\mathcal{X}_i$ is then defined as

$$\mathcal{X}_i^{IJ} = \mathcal{X}_{a,i}^I \mathcal{X}_{a,i}^J \epsilon^{ab}, \qquad \mathcal{X}_i^{IJ} \in \mathbb{C}^{2\times(4|4)}/GL(2), \qquad I, J = 1, .., 8 \,. \tag{2.4.14}$$

### 2.4.2  Momentum twistors

In section 2.2 we saw how planar amplitudes can be rewritten in terms of dual variables $x^\mu$. We can also define the equivalent of the dual variables $x^\mu$ for the Grassmann variable $\eta$ as

$$|\theta_{iA}\rangle - |\theta_{i+1,A}\rangle = |i\rangle \, \eta_{iA} \,. \tag{2.4.15}$$

Beware that $\theta$ in this case has nothing to do with the Minkowski superspacetime $\theta$ and we use the same symbol suggestively in light of the variable identification that occurs in the amplitude/correlator duality. Amplitudes written in these variables are invariant by definition under translations $x_i^\mu \to x_i^\mu + a^\mu$ and supertranslations $|\theta_{iA}\rangle + |\phi_A\rangle$. As we just did for translation, one can consider all the transformations of the superconformal group represented on the dual coordinates space, like it were the ordinary spacetime. This group is called the dual superconformal group [73, 74]. Surprisingly, one finds that superamplitudes stripped (divided) by the MHV tree-level amplitudes, which we indicated as $\mathcal{A}_{n,k}$ in (2.3.9), are dual superconformal invariant[3]. Moreover, superconformal symmetry together with dual superconformal symmetry generates an infinite dimensional algebra known as the Yangian [10]. One can then prove that the stripped amplitudes are invariant under the dual conformal group exploiting the duality between amplitudes and (super-)Wilson loops [76–80], some aspects of which we will describe in detail later in this thesis.

Dual coordinates $x_i^\mu$ transform like spacetime four-vectors under the dual conformal group with the extra constraint $p_i^2 = (x_{i+1} - x_i)^2 = 0$ for $x_i$ with $i \leq n$. We can represent each $x_i^\mu$ using twistors $x_i^{AB}$, just this time we are in momentum space. The constraint $p_i^2$ translate to the constraints $x_{ii+1}^2 = \langle x_i x_i + 1\rangle = 0$, which geometrically means that the line $x_i$ intersects the line $x_{i+1}$. We can name the intersection point as

$$z_i = X_i \cap X_{i+1} \,. \tag{2.4.16}$$

The variable $z_i$ will be the unconstrained elements of $\mathbb{P}^3$ and are called momentum twistor variables and their super symmetric extension super momentum-twistors [12, 14]. We can fix the GL(2) gauge such that

$$z_i = X_{2,i} = X_{1,i+1} \,. \tag{2.4.17}$$

The action of the dual superconformal symmetry is linear on the Xs and as a consequence also on the zs.

To extend momentum twistors to super momentum-twistors, we can consider the

---
[3]The actual superamplitude is invariant under a similar group, where special conformal transformations are shifted by a constant term [75].

Grassmannian equivalent of dual coordinates $\theta^A_{\alpha i}$, that are defined by the relation

$$\theta^A_{\alpha i} - \theta^A_{\alpha(i+1)} = |i\rangle_\alpha \, \eta^A_i \, . \tag{2.4.18}$$

Then we define the super momentum-twistors Grassmann variables as

$$\chi^A_i = \langle i \theta^A_i \rangle = \langle i \theta^A_{i+1} \rangle \, , \tag{2.4.19}$$

which is the equivalent of (2.4.12). Super momentum-twistors are then defined as

$$\mathcal{Z}_i = \begin{pmatrix} z_i \\ \chi_i \end{pmatrix} \in \mathbb{C}^{4|4} \, , \qquad i = 1, .., n \, , \tag{2.4.20}$$

where $\mathbb{C}^{4|4}$ indicates that the first 4 components are complex numbers and the last 4 components are Grassmann odd variables. The advantage of using super momentum-twistors is that the dual superconformal group generators are particularly simple and read

$$G^I_J = \sum_{i=1}^n \mathcal{Z}^I \partial_{\mathcal{Z}^J} \, . \tag{2.4.21}$$

The complexification of the group generated by $G^I_J$ for $I, J \leq 4$ corresponds to the complexified conformal group SL(4). The only invariants of the special linear group are determinants. Four momentum twistors can be stacked together to form a $4 \times 4$ matrix whose determinant we indicate with

$$\langle z_i z_j z_k z_l \rangle = \langle ijkl \rangle = \det(z_i, z_j, z_k, z_l) \, . \tag{2.4.22}$$

These brackets provide the building blocks of the $\chi$-independent part of the amplitude. In particular, we can use (2.4.6) to write the denominator of propagators as

$$x^2_{ij} = \frac{\langle (i-1)i(j-1)j \rangle}{\langle I(i-1)i \rangle \langle I(j-1)j \rangle} \, , \tag{2.4.23}$$

where $I = \begin{pmatrix} 0 & 0 \\ 0 & 0 \\ 1 & 0 \\ 0 & 1 \end{pmatrix}$. Notice that in this expression the numerator is dual conformally invariant, while the denominator is not. We can say then that a function of momentum twistor is dual conformal invariant if it can be written in an infinity twistor free form. This is of course the case for $\mathcal{A}_{n,k}$ which we said is dual superconformal invariant.

Due to dual conformal invariance, we can write any superamplitude $\mathcal{A}_{n,k}$ purely as a function of momentum super-twistor, without the need of the infinity twistor I. We have seen that dual conformal building blocks are given by the brackets $\langle z_i z_j z_k z_l \rangle$, but what about the $\chi$ dependent dual superconformal building blocks? In the next part of this thesis, we will see that they have an elegant representation in terms of bosonized super momentum-twistors. This formulation gives a new perspective for the superamplitude that will allow us to interpret the amplitude as a differential form on the Grassmannian.

Finally we discuss how to represent loop integrands. Starting from the dual variable $x_i$ for $i > n$, it is clear that this can be represented exactly as the twistors. In the literature, the twistor of the variable $x_i$ is written as $A_i B_i$. The denominator of the propagators adjacent to an internal region of the dual graph will be given by

$$\begin{aligned} x_{ij}^2 &= \frac{\langle A_i B_i (j{-}1) j \rangle}{\langle I A_i B_i \rangle \langle I (j{-}1) j \rangle} \qquad \text{for } i > n \text{ and } j \leq n \,, \\ x_{ij}^2 &= \frac{\langle A_i B_i A_j B_j \rangle}{\langle I A_i B_i \rangle \langle I A_j B_j \rangle} \qquad \text{for } i, j > n \,. \end{aligned} \qquad (2.4.24)$$

In some cases it is useful to consider loops as the direct momentum space analogue of supertwistors. We can then write the loop supertwistors as $\begin{pmatrix} A_i & B_i \\ \chi_{A_i} & \chi_{B_i} \end{pmatrix}$.

Amplitude integrands can always be written in a dual conformal form, but loop level amplitudes need to be renormalized. It turns out that in $\mathcal{N} = 4$ SYM the dual conformal symmetry is anomalous [81]. Nevertheless, the anomaly coefficient can be predicted and the amplitude can be factorized in two parts: one which is not dual conformal symmetric but can be easily predicted and an anomaly free non-trivial part [82]. So, despite being an anomalous symmetry, dual conformal symmetry, and more generally the Yangian symmetry, remains a key ingredient for amplitude computations at loop level.

## 2.5 BCFW recursion in supermomentum twistor space

A remarkable achievement of modern amplitude techniques is the BCFW recursion; an algorithm that allows for the computation of amplitude integrands in a wide class of massless planar theories. In particular, for $\mathcal{N} = 4$ SYM the recursion can be written on superspace [83], on supertwistor space [84] and on supermomentum twistor space [9] even at loop level. In the latter formulation, amplitudes have a particularly compact form and are manifestly dual conformal invariant. Here we want to describe the super-BFCW algorithm in supermomentum twistor space highlighting some of its properties.

Consider a set of constant super momentum-twistors $(\mathcal{Z}_1, \cdots, \mathcal{Z}_n)$. We now want to analyze the analytic properties of $\hat{\mathcal{A}}_{n,k}(w) = \frac{1}{w} \mathcal{A}_{n,k}(\hat{\mathcal{Z}}_1(w), \mathcal{Z}_2, \cdots, \mathcal{Z}_n)$, where

$\hat{\mathcal{Z}}_1(w) = \mathcal{Z}_1 + w\mathcal{Z}_n$, as a function of w. The first observation is that

$$\text{Res}_{w=0}(\hat{\mathcal{A}}_{n,k}) = \mathcal{A}_{n,k} \, . \qquad (2.5.1)$$

For the global residue theorem we can then write

$$\mathcal{A}_{n,k} = \sum_i \text{Res}_{w=w_i^*}\left(\hat{\mathcal{A}}_{n,k}(w)\right) \, , \qquad (2.5.2)$$

where the $w_i^*$ represent all the poles of the amplitude, excluding the pole $w = 0$. From color ordered Feynman diagrams it is clear that the poles of the amplitude can only come from either a propagator going on-shell or from $w \to +\infty$. Each residue in (2.5.2) can be then associated with a subset of diagrams contributing to the amplitude and written in terms of product of lower loop, multiplicity, or NMHV degree amplitudes exploiting the factorization properties of amplitudes when a propagator goes on-shell (see section 7.4 of [75] for reference).

The starting point of the recursion is given by the tree-level 5-point NMHV amplitude $\mathcal{A}_{5,1}(\mathcal{Z}_1, \cdots, \mathcal{Z}_5)$. The explicit form of $\mathcal{A}_{5,1}$ is completely fixed by dual superconformal symmetry and cyclicity and reads

$$\mathcal{A}_{5,1}(\mathcal{Z}_1, \cdots, \mathcal{Z}_5) = \frac{\delta^4(\chi_1\langle 2345\rangle + \text{cyclic})}{\langle 1234\rangle \langle 2345\rangle \langle 3451\rangle \langle 4512\rangle \langle 5123\rangle} \, , \qquad (2.5.3)$$

where the Grassmannian delta function is defined as

$$\delta^4(\chi_1\langle 2345\rangle + \text{cyclic}) = \prod_{A=1}^4 (\chi_1^A\langle 2345\rangle + \text{cyclic}) \, . \qquad (2.5.4)$$

For general argument $(\mathcal{Z}_i, \mathcal{Z}_j, \mathcal{Z}_k, \mathcal{Z}_l, \mathcal{Z}_m)$, the superfunction $\mathcal{A}_{5,1}$ is also known as an R-invariant and it is indicated as $[i, j, k, l, m]$.

Here, we first give an explicit formula for the recursion and then we will unpack all the terms in the formula and the notation. The loop level BCFW recursion relation [9] can be written as

$$\mathcal{A}_{n,k}^{(L)} = \mathcal{A}_{n-1,k}(\mathcal{Z}_1, \cdots, \mathcal{Z}_{n-1}) +$$
$$+ \sum_{L'=0}^{L}\sum_{k'=0}^{k-1}\sum_{j=3}^{n-2}[j{-}1, j, n{-}1, n, 1]\mathcal{A}_{n-j+1,k'}^{(L')}(\mathcal{Z}_{I_j}, \mathcal{Z}_j, \cdots, \hat{\mathcal{Z}}_{n_j})\mathcal{A}_{j+1,k-k'-1}^{(L-L')}(\mathcal{Z}_{I_k}, \mathcal{Z}_1, \cdots, \mathcal{Z}_{j-1}) +$$
$$+ \int d^4\chi_A d^4\chi_B \, [A, B, n{-}1, n, 1]\mathcal{A}_{n+2,k+1}^{(L-1)}[\mathcal{Z}_1, \cdots, \hat{\mathcal{Z}}_{n_{AB}}, \mathcal{Z}_A, \mathcal{Z}_{\hat{B}}] \, ,$$
$$(2.5.5)$$

where

$$\begin{aligned}\hat{\mathcal{Z}}_{n_j} &= (n{-}1, n) \cap (1, j{-}1, j) \, , & \mathcal{Z}_{I_j} &= (j, j{-}1) \cap (n{-}1, n, 1) \, , \\ \hat{\mathcal{Z}}_{n_{AB}} &= (n{-}1, n) \cap (A, B, 1) \, , & \mathcal{Z}_{\hat{B}} &= (n{-}1, n) \cap (A, B, 1) \, ,\end{aligned} \qquad (2.5.6)$$

and intersections in projective space can be expanded as

$$(ij) \cap (klm) = \mathcal{Z}_i \langle jklm \rangle - \mathcal{Z}_j \langle iklm \rangle \,. \tag{2.5.7}$$

This extremely powerful formula can then be used then to compute all amplitude integrands at any multiplicity and loop order. The recursion always gives the amplitude in terms of a product of R-invariants where the Grassmannian components of the loop are integrated. To make a practical use of this formula we need to understand how to compute the product of Grassmannian delta functions and how to integrate out the $\chi_{A_i}, \chi_{B_i}$ variables.

The first problem can be solved by writing the Grassmannian delta functions in the following general form

$$\delta^{m \times k}(C \cdot \chi) := \prod_{A=1}^{m} \prod_{a=1}^{k}(C_{ia}\chi_i^A) = \prod_{A=1}^{m}\left(\sum_{I}\langle C_I \rangle \prod_{i \in I}\chi_i^A\right), \qquad \text{with } I \in \binom{[n]}{k}, \tag{2.5.8}$$

where here and in the following we will use a short-hand notation I, J etc to represent an ordered set of particle numbers. We define $[n] := \{1, 2, ..., n\}$ and then $\binom{[n]}{k}$ to be the set of all ordered sets of k elements in $[n]$. Therefore, the term $\langle C_I \rangle = \det(C_{i_1 a}, \cdots, C_{i_k a})$ represents the maximal minor of the matrix C given by the columns $I = (i_1, \cdots, i_k)$. It is easy to verify that (2.5.4) can be rewritten in this new notation as

$$\delta^{4 \times 1}(C \cdot \chi), \qquad \text{with } C_{11} = \langle 2345 \rangle, \cdots, C_{51} = \langle 1234 \rangle \,. \tag{2.5.9}$$

The product of two delta functions can be then written as

$$\delta^{m \times k_1}(C_1 \cdot \chi)\delta^{m \times k_2}(C_2 \cdot \chi) = \delta^{m \times (k_1+k_2)}\left(\begin{pmatrix} C_1 \\ C_2 \end{pmatrix} \cdot \chi\right), \tag{2.5.10}$$

where $\begin{pmatrix} C_1 \\ C_2 \end{pmatrix}$ is the matrix obtained by stacking the two matrices one over the other.

Now we need to understand how to integrate the expression $\int d^4\chi_A d^4\chi_B \delta^{m \times k}(C \cdot \chi)$. The only terms that will survive the integration are the terms in (2.5.8) labelled by an I containing the indices A, B. Notice that the integral is invariant under the rescaling of the loop super momentum-twistors. Moreover, the expression is invariant under an SL(k) transformation on the rows of C. Combining these two transformations we can rewrite C in the form $\begin{pmatrix} \widetilde{C} & 0 \\ 0 & \mathbb{1}_{2\times 2} \end{pmatrix}$. We will then have

$$\int d^4\chi_A d^4\chi_B \delta^{m\times k}(C\cdot\chi) = \prod_{A=1}^{m}\left(\sum_I \langle \widetilde{C}_I\rangle \prod_{i\in I}\chi_i^A\right), \qquad \text{with } I \in \binom{[n]-\{A,B\}}{k-2}. \tag{2.5.11}$$

Geometrically, this operation corresponds to projecting C onto the space orthogonal to the two column-vectors $C_A, C_B$.

## 2.6 The supercorrelator

In this section, we will describe the correlation functions of the operators in the stress-tensor supermultiplet $\mathcal{T}(x^\mu, \theta_\alpha^A, \bar{\theta}_{\dot\alpha}^A)$ in planar $\mathcal{N}=4$ SYM. This operator is the half-BPS operator with the lowest conformal weight in the theory and its lower component in the multiplet is equal to

$$\mathcal{O}_2^{ABCD} := \text{Tr}(\phi^{AB}\phi^{CD}) - \frac{1}{4!}\epsilon^{ABCD}\text{Tr}(\phi^{EF}\phi_{EF}), \tag{2.6.1}$$

where the trace is over the color indices and generalize $\mathcal{O}_p := \text{Tr}(\phi^p) + \cdots$, see [85] for the details of the definition.

The operator $\mathcal{O}_2$ has been intensely studied in planar $\mathcal{N}=4$ SYM for its role in the AdS/CFT correspondence [86,87] and more recently for its surprising connection to flat space amplitudes, which have been reviewed this year in [88]. Notice that, unlike for the amplitude, we don't need to perform any color ordering procedure since the correlators we consider do not carry color indices. But we can of course expand the correlator in the t'Hooft coupling and then expand each term in powers of $1/N_c$ as we do for the planar limit of the amplitude. A diagrammatic notion of non-crossing propagators/planarity is recovered for Feynman diagrams in twistor space [89].

The two main properties of the perturbative expansion of the supercorrelator we will focus on are the supercorrelator/superamplitude duality [56, 57] and the hidden permutation symmetry of the correlator between external spacetime points and loop variables [90]. We will be using these properties to describe the correlahedron conjecture [51] and to discuss the results presented in [1] on the squared amplituhedron.

The chiral part of the vector multiplet is conveniently described as a $\mathcal{N}=4$ superfield

$$W^{AB}(x,\theta,0) = \phi^{AB} + 2i\sqrt{2}\theta^{\alpha[A}\lambda_\alpha^{B]}(x) + i\sqrt{2}\theta_\alpha^{[A}\theta_\beta^{B]}F^{\alpha\beta}(x) + \cdots, \tag{2.6.2}$$

where $\phi$ and $\lambda$ are the scalar and spinors in the theory as described in (2.3.1), F is the field strength and the dots stand for non-Abelian terms proportional to the coupling constant. Remember, the multiplet transforms in the adjoint representation of the gauge group and is not itself a gauge-invariant operator. The simplest non-trivial operator we can build out of it is the stress-tensor supermultiplet $\mathcal{T}$, which can be written as a function of the vector multiplet W as

$$\mathcal{T}^{\text{ABCD}} = \text{Tr}(W^{\text{AB}}W^{\text{CD}}) - \frac{1}{4!}\epsilon^{\text{ABCD}}\text{Tr}(W^{\text{EF}}W_{\text{EF}}) \,. \qquad (2.6.3)$$

Instead of studying directly $\mathcal{T}$ we will consider a projection respect to its R-symmetry indices. The motivation for doing so comes from observing that $\mathcal{T}^{\text{ABCD}}\epsilon_{\text{ABCD}} = 0$, that it is the field has a direction in which is trivial and we would like to get rid of this redundancy[4]. To do so, consider an auxiliary variable $y^{\text{AB}} = y_a^A y_b^B \epsilon^{ab}$. Projecting $\mathcal{O}_2$ respect to y, we obtain

$$\mathcal{O}_2(x^\mu, \theta, y) := y_{\text{AB}} y_{\text{CD}} \mathcal{O}_2^{\text{ABCD}} =$$
$$= y_{\text{AB}} y_{\text{CD}} (\text{Tr}(\phi^{\text{AB}}\phi^{\text{CD}}) - \frac{1}{6}\epsilon^{\text{ABCD}}\text{Tr}(\phi^{\text{EF}}\phi_{\text{EF}})) = y_{\text{AB}} y_{\text{CD}} (\text{Tr}(\phi^{\text{AB}}\phi^{\text{CD}})) \,.$$
$$(2.6.4)$$

The introduction of the variable y also has a deeper meaning in the context of the coset representation of the superconformal group and emerges naturally in the construction of the so-called *analytic superspace* [91–93]. It makes sense therefore to consider the projection of the vector supermultiplet and stress-ternsor multiplet

$$W(x, y, \theta, \bar{\theta}) = y_{\text{AB}} W^{\text{AB}}(x, \theta, \bar{\theta}) \,, \qquad T(x, y, \theta, \bar{\theta}) = \text{Tr}(W^2) \,. \qquad (2.6.5)$$

We can now define the chiral correlator as

$$G_n(x, y, \theta) = \langle T(x_1, y_1, \theta_1, 0), \cdots, T(x_n, y_n, \theta_n, 0) \rangle \,, \qquad (2.6.6)$$

where we have set all $\bar{\theta} = 0$. Since the amplitude lives in chiral superspace, setting $\bar{\theta} = 0$ is necessary to construct an object for which the correlator/amplitude duality might hold. Moreover, the full correlator in some cases can actually be reconstructed from its chiral part [94, 95], so we don't lose any generality by restricting to the chiral sector while obtaining simpler formulas.

In the same way the superamplitude can be expanded into NMHV sectors, we can expand the supercorrelator into NMHV sectors depending of the degree of the Grassmannian variables $\theta$. We call $G_{n,k}$ the $N^k$MHV sector of the correlator. Moreover, we can then expand on loop orders and define $G_{n,k}^{(l)}$ as the l-loop contribution to the $G_{n,k}$. The 2 and 3-point correlator do not receive loop corrections and their lowest component read

$$G_2(x,y) = \frac{N_c^2 - 1}{2(4\pi)^2} \left(\frac{y_{12}^2}{x_{12}^2}\right),$$
$$G_3(x,y) = \frac{N_c^2 - 1}{(4\pi)^3} \left(\frac{y_{12}^2 y_{23}^2 y_{13}^2}{x_{12}^2 x_{23}^2 x_{13}^2}\right),$$
(2.6.7)

where $y_{ii+1}^2 = y_i^{AB} y_{i+1}^{CD} \epsilon_{ABCD} = \langle y_i y_{i+1} \rangle$. The four-point function $G_4$ is the first non-protected quantity. At Born-level the connected correlator reads

$$G_{4,0}^{(0)} = \frac{2(N_c^2 - 1)}{(4\pi^2)^4} \prod_{i=1}^{4} \frac{y_{ii+1}^2}{x_{ii+1}^2}.$$
(2.6.8)

---

[4] In the SO(6) representation of R-symmetry group this corresponds to the fact that $\mathcal{O}_2^{\text{IJ}} = \text{Tr}(\Phi^I\Phi^J) - \frac{1}{6}\delta^{\text{IJ}}\text{Tr}(\phi^K\phi_K)$ is symmetric traceless tensor.

The loop integrand $G_{4,0}^{(l)}$ only receives contribution from connected diagrams, since $G_2$ is exact at Born level, and has been computed in the planar limit up to 10 loops [96–98].

The term $G_{n,0}^{(0)}$ instead, corresponds the Born-level MHV sector; it is known at all multiplicity and its connected component reads

$$G_{n,0}^{(0)} = \frac{2(N_c^2 - 1)}{(-4\pi^2)^n} \prod_{i=1}^{n} \frac{y_{ii+1}^2}{x_{ii+1}^2} + (S_n - \text{permutations}).$$
(2.6.9)

From now on we will always indicate with $G_{n,0}^{(0)}$ the connected correlator. Notice that differently from the superamplitude, but exactly like the stripped superamplitude $\mathcal{A}_n$, the term with the lowest degree in the Grassmann variables in the supercorrelator, that is $G_{n,0}$, has degree zero.

### 2.6.1 Hidden permutation symmetry of $G_{n,n-4}$

The supercorrelator integrand $G_n$ enjoys two permutation symmetries, an $S_n$ among the n external points on superspace and $S_l$ among the l loops variables on ordinary spacetime. In [90] it was proven that it is possible to extract all the maximally nilpotent correlators $G_{n,n-4}^{(l)}$ from an $S_{n+l}$ permutation invariant function $f^{(l)}(x_1, \cdots, x_{n+l})$ defined by the equation

$$f^{(l)}(x) := \frac{1}{2} \frac{G_{4,0}^{(l)}(x,y)}{G_{4,0}^{(0)}(x,y) \xi^{(4)}(x)},$$
(2.6.10)

where $\xi^{(4)} = x_{12}^2 x_{23}^2 x_{34}^2 x_{14}^2 (x_{13}^2 x_{24}^2)^2$. This property can be proved using the Lagrangian insertion method, which also gives a similar, direct relation between $G_{n,k}^{(l)}$ and $G^{(0)}_{n+l,k+l}$ described again in [90]. The function $f^{(l)}$ has the following properties: it's conformal invariant, has scaling dimension 4 in all $x_i$, simple poles in all $x_{ij}^2$ and it is of course permutation invariant. Here we will limit our discussion to the case of the maximal nilpotent correlator.

We will state the precise relation between $f^{(l)}$ and $G^{(l)}_{n,n-4}$ expressing the correlator as the Grassmann derivative of a potential $\mathcal{G}$. The upshot of this construction is that the potential will not depend on the auxiliary variables y, which in turn will make the correlator/amplitude duality particularly easy to apply in practical computations. Moreover, the potential will be the quantity which conjecturally the correlahedron represents.

The potential $\mathcal{G}$ is implicitly defined by the relation

$$G_n(x, y, \theta) = \left(\prod_{i=1}^{n} D_4\right) \mathcal{G}_n(x, \theta) , \qquad (2.6.11)$$

where

$$D_4 = y_i^{IJ} y_i^{KL} \partial_{\theta_i^{\alpha I}} \partial_{\theta_i^{\beta J}} \partial_{\theta_{i\alpha}^{K}} \partial_{\theta_{i\beta}^{L}} . \qquad (2.6.12)$$

Notice that $\mathcal{G}_{n,k}$ is defined up to terms killed by the Grassmann derivative. Then, thanks to the hidden permutation symmetry of the correlator, the maximally nilpotent potential can be expressed at any loop and multiplicity for $n > 4$ as

$$\mathcal{G}^{(l)}_{n,n-4} = \frac{2(N_c^2 - 1)}{(-4\pi^2)^{n+l}} \mathcal{I}_n f^{(n+l)} , \qquad (2.6.13)$$

where $\mathcal{I}_n$ is the unique maximal n-point superconformal invariant and reads

$$\mathcal{I}_n = \delta^{4 \times (2n-4)}(X^\perp \cdot \theta) , \qquad (2.6.14)$$

where $X^\perp$ is the $(2n-4) \times 2n$ matrix orthogonal to the $4 \times 2n$ matrix $X = (X_1, \cdots, X_n)$, which has the twistor lines as columns.

### 2.6.2 The correlator/superamplitude duality

The correlator/superamplitude duality can be compactly stated as

$$\left(\lim_{x_{ii+1}^2 \to 0} \frac{G_{n,k}}{G_{n,0}^{(0)}}\right) \bigg|_{\mathcal{Z}_i = \mathcal{X}_i \cap \mathcal{X}_{i+1}} = \mathcal{A}_{n,k}^2 , \qquad (2.6.15)$$

or, using the potential $\mathcal{G}$, as

$$\left(\int \prod_{i=1}^{n} d^4 \chi_{2,i} \lim_{x_{ii+1}^2 \to 0} (\prod_i x_{ii+1}^2) \mathcal{G}_{n,k}\right) \bigg|_{\mathcal{Z}_i = \mathcal{X}_i \cap \mathcal{X}_{i+1}} = \mathcal{A}_{n,k}^2 . \qquad (2.6.16)$$

To make sense of this formula we first have to explain why the limit is finite and in which sense the left-hand side is dependent on super momentum-twistors.

This limit in (2.6.16), where consecutive x's become light-like separated, is known as the polygonal light-like limit, since the x's can be thought of as the vertices of a polygon with light-like edges. It is known from the OPE expansion of the correlator that the latter has simple poles on the region $x_{ij}^2 = 0$, that is $\lim_{x_{ij}^2 \to 0} \left( x_{ij}^2 \mathcal{G}_n(x, y, \theta) \right)$ is finite.

In twistor space the equation $(x_i - x_j)^2 = 0$ corresponds to

$$\langle X_i X_j \rangle = \langle X_{1,i} X_{2,i} X_{1,j} X_{2,j} \rangle = 0 \,, \tag{2.6.17}$$

which means that the two lines lie on the same plane (see section 3.1), and therefore they intersect on a point which we will suggestively call $Z_i := X_i \cap X_{i+1}$. We then fix the GL(2) gauge such that $X_{2,i} = X_{1,j} = Z_i$. Having $x_{ii+1} = 0$ for all i means that all consecutive twistor lines intersect, so we have n intersection points $Z_i$. We can then generalize this relation to supertwistors defining

$$\mathcal{Z}_i := \mathcal{X}_{1,i+1} = \mathcal{X}_i \cap \mathcal{X}_{ii+1} \,. \tag{2.6.18}$$

So the operations that we have to perform to obtain $\mathcal{A}_n$ from $\mathcal{G} \prod x_{ii+1}$ are

- take the limit $x_{ii+1} \to 0$,
- fix a gauge such that $X_{2,i} = X_{1,i+1} \to Z_i$,
- integrate out the Grassmann variables $\chi_{2,i}$.

Notice that in the third step we lower the Grassmann degree of the correlator potential by $4 \times n$. Therefore, we are left with a Grassmann degree $4(n+k) - 4n = 4k$ which is consistent with the Grassmann degree of $\mathcal{A}_{n,k}^2$.

The duality (2.6.15) holds both at tree level and for the loop integrand. Loop variables do not carry any Grassmann degree of freedom and can be straightforwardly mapped to the loop dual coordinates. In twistors coordinates we then have the identification

$$X_i = A_i B_i \,, \qquad \text{with } i > n \,. \tag{2.6.19}$$

Notice that the permutation symmetry of the correlator loop variables directly map to the permutation symmetry of the amplitude loop variables.

For the maximally nilpotent correlator this fairly complex algebraic operation, thanks to (2.6.13), simply reduces to

$$\mathcal{A}_{n,n-4,l}^2 = 2\mathcal{A}_{n,n-4,0} \lim_{x_{ii+1}^2 \to 0} \xi^{(n)} f^{(n+l-4)} \,, \tag{2.6.20}$$

where the tree-level $\overline{\text{MHV}}$ amplitude $\mathcal{A}_{n,n-4,0}$ is equal to

$$\mathcal{A}_{n,n-4,0} = \frac{\delta^{4\times k}(Z^\perp \cdot \chi)}{\prod_{i=1}^{n} x_{ii+2}^2} \,, \qquad (2.6.21)$$

except for n = 4 where $\mathcal{A}_{4,0,0} = 1$ by definition and

$$\xi^{(n)} = \prod_i^n x_{ii+1}^2 x_{ii+2}^2 \,, \qquad \text{with } x_{nn+1} = x_{1n}, \; x_{nn+2} = x_{2n} \,. \qquad (2.6.22)$$

As an example let's derive the expression of the $\mathcal{A}_{5,1}^2 = 2\mathcal{A}_{5,1}$ superamplitude squared from the correlator. Following (2.6.20), we need the expression $f^{(1)}$. Its expression can be found in [98] for example and is given by

$$f^{(1)} = \frac{1}{x_{12}^2 x_{13}^2 x_{14}^2 x_{15}^2 x_{23}^2 x_{24}^2 x_{25}^2 x_{34}^2 x_{35}^2 x_{45}^2} \,. \qquad (2.6.23)$$

So we have

$$\mathcal{A}_{5,1,0}^2 = 2\delta^{4\times k}(Z^\perp \cdot \chi) \frac{1}{x_{13}^2 x_{24}^2 x_{35}^2 x_{14}^2 x_{25}^2} \,. \qquad (2.6.24)$$

By identifying $x_{ii+2}^2 = \langle ii+1 i+2 i+3 \rangle$ and $\delta^{4\times k}(Z^\perp \cdot \chi) = \delta^4(\langle 1234 \rangle \chi_5 + \text{cyclic})$ we obtain the expression for the 5-point NMHV amplitude we have given in (2.5.3).

Again from $f^{(1)}$ we extract $\mathcal{A}_{4,0,1}^2 = 2\mathcal{A}_{4,0,1}$. From (2.6.20) we get the expression

$$\mathcal{A}_{4,0,1}^2 = 2\frac{x_{13}^2 x_{24}^2}{x_{15}^2 x_{23}^2 x_{34}^2 x_{14}^2} = 2\frac{\langle 1234 \rangle^2}{\langle AB12 \rangle \langle AB23 \rangle \langle AB34 \rangle \langle AB14 \rangle} \,, \qquad (2.6.25)$$

as expected.

As a final remark, the formula (2.6.20) can be specialized to compute $\mathcal{A}_{4,0,l}$ directly using a diagrammatic representation of the terms contributing to $f^{(l)}$ called f-graphs. So, since $f^{(l)}$ is known for $l \leq 10$, we can use the f-graphs to write an explicit formula for $\mathcal{A}_{4,0,l}$ up to ten loops.

## Chapter 3

# Amplituhedron Basics

We have seen that superamplitudes can be written in a manifestly dual superconformal invariant form and can be computed at any loop order and multiplicity by the BCFW recursion relation. Surprisingly, even more rich mathematical structures are hidden in the superamplitude integrand. The superamplitude bosonization in fact connects the $\mathcal{A}_{n,k}$ superamplitude to a top differential form on the Grassmannian $Gr(k, k+4)$, that is the space of affine k-hyperplanes in $\mathbb{R}^{k+4}$, allowing for completely different kinds of questions to be asked.

This construction was originally proposed in [14] for NMHV amplitudes, where the differential form formulation not only proved to be a great tool for the analysis of the singularity structure of the amplitude but it showed that the latter corresponds to the boundaries of a particular projective polytope[1]. The superamplitude bosonization was then extended to arbitrary amplitudes in [15], where it was conjectured that the singularities of the amplitude lie on the boundaries of a geometrical object called the Amplituhedron.

In this chapter, we will review the mathematical tools needed to define and extract amplitudes from the amplituhedron. We will start by explaining how to define regions in the Grassmannian, where particular attention will be given to convex projective polytopes. Then we will talk about how the superamplitude can be expressed as a differential form on the Grassmannian and computed as the canonical form of the amplituhedron. We will finish by introducing the correlahedron and squared amplituhedron.

## 3.1 Projective Geometry

Projective geometry is an incredibly powerful tool to map complex geometric problems to elementary linear algebra problems. The basic idea of projective space is that one can represent points in a plane as lines passing through the origin in three dimensions. As depicted in the image 3.1, this correspondence can be built by fixing a plane not passing though the origin and considering the intersection between a plane and homogeneous lines. In general, two points are needed to identify a line but, since each an homogeneous line is a line passing through the origin, one point is enough to determine it. In this image, the points are labelled by the $P_i$. You will notice that, if one multiplies the coordinates of one of the $P_i$'s by a constant, the effect will be to move it along the line it represents. So for example, both P and 3P will be equivalent because they represent the same line. More formally, given a vector space V, usually called the embedding space, and a field K, the projective space $\mathbb{P}(V)$ is defined as $V - \{0\}$ under the equivalence relation $P \equiv \lambda P$ with $P \in V$ and $\lambda \in K - \{0\}$. The projective spaces we will consider in this work are the complex projective space $\mathbb{P}(\mathbb{C}^{d+1})$ with $K = \mathbb{C}$ and the real projective space $\mathbb{P}(\mathbb{R}^{d+1})$ with $K = \mathbb{R}$. We will indicate both with the symbol $\mathbb{P}^d$ since often for our purposes the two spaces will be interchangeable or when it is clear from the context which we are referring to.

---

[1] A polytope is the higher dimensional generalization of a polygon

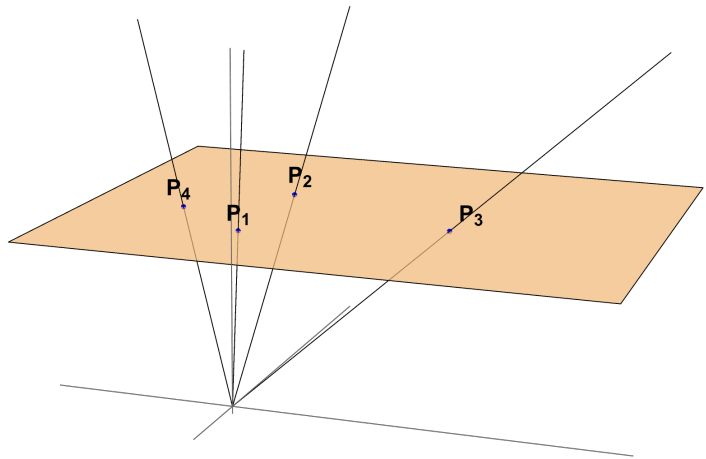

Figure 3.1: Points in $\mathbb{P}^2$.

One of the most useful formula in projective space the one that gives the expansion of a vector on a base in a vector or projective space. These will have many geometrical applications in particular the computation of the intersection of pairs of hyperplanes. Consider $d + 1$ vectors $\{w_1, \cdots, w_{d+1}\}$ in $\mathbb{R}^d$ or equivalently on $\mathbb{C}^d$. We would like first to write $w_{d+1}$ as

$$w_{d+1} = C \cdot W = c_1 w_1 + \cdots c_d w_d \qquad (3.1.1)$$

and find an explicit expression for the cs. We start from the trivial identity

$$\epsilon^{i_1 \cdots i_{d+1}} w_{i_1}^{a_1} \cdots w_{i_{d+1}}^{a_{d+1}} = 0 \, . \tag{3.1.2}$$

Then we expand this expression as

$$\epsilon^{1 i_2 \cdots i_{d+1}} w_1^{a_1} w_{i_2}^{a_2} \cdots , w_{d+1}^{a_{d+1}} + \cdots + \epsilon^{d+1 i_2 \cdots i_{d+1}} w_{d+1}^{a_1} w_{i_2}^{a_2} \cdots w_{i_{d+1}}^{a_{d+1}} = 0 \, , \tag{3.1.3}$$

from which, using the definition of the determinant, we can derive that

$$\begin{aligned} w_1 \langle w_2 \cdots w_{d+1} \rangle - w_2 \langle w_1 w_3 \cdots w_{d+1} \rangle + \cdots + \\ + (-1)^i w_i \langle w_1 \cdots w_{i-1} w_{i+1} \cdots w_{d+1} \rangle + \cdots + (-1)^d w_{d+1} \langle w_1 \cdots w_d \rangle = 0 \, . \end{aligned} \tag{3.1.4}$$

We can then solve for any of the $w_i$ to derive its expansion as a linear combination of the other ws. Notice that this formula is invariant under rescaling and therefore holds also in projective space $\mathbb{P}^{d-1}$, where it is sometimes called the *generalized Schouten identity*.

### 3.1.1 The inside of a polygon

If the determinant $\langle 1, \cdots, d \rangle$ is equal to zero it means that the d vectors in the bracket are on the same hyperplane, but what information does it give us if it is different from 0? Since the vectors can be rescaled, we can set the determinant to any value different from zero, so the only information we have is that the vectors are linearly independent. One interesting idea is to be more restrictive and consider half-lines instead of lines. This amounts to restricting the equivalence relation to $P \equiv \lambda P$ with $\lambda > 0$. This space is called oriented projective space and is sometimes indicated as $\widetilde{\mathbb{P}}^{d-1}$. In the space $\widetilde{\mathbb{P}}^2$ for example we can use the function $f(A) = \text{sign}(\langle P_1 P_2 A \rangle)$, where $P_1, P_2$ and $A$ are points in $\widetilde{\mathbb{P}}^2$, to divide it into a positive, negative and null region. Figure 3.2 gives a illustration of this function. Asking for multiple determinant functions to be positive at the same time defines more complex regions in oriented projective space.

Now we will explore the first way in which the amplituhedron can appear: the convex polygon. It is in this projective formulation that A. Hodges [14] recognized that the NMHV amplitude could be represented as the volume of a polytope. Lets start from the simplest polygon, the triangle.

Consider the set of points Y inside a triangle of vertices $\{P_1, P_2, P_3\}$ such that $\langle 123 \rangle > 0$ as illustrated in figure 3.3. Using physics to get some intuition, we can consider that each of these vertices has a mass $c_i$, and we want to calculate their

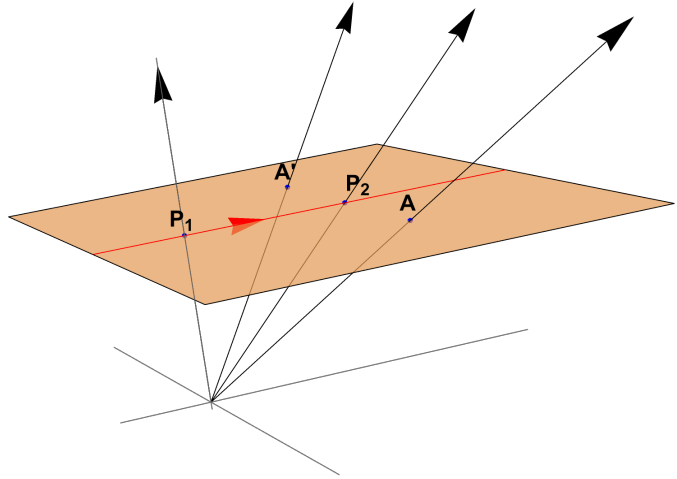

Figure 3.2: Illustration of the geometrical interpretation of the sign of the determinants $\langle P_1 P_2 A' \rangle$ and $\langle P_1 P_2 A \rangle$.

common centre of mass. The centre of mass will correspond to the weighted average:

$$Y = \frac{c_1 P_1 + c_2 P_2 + c_3 P_3}{c_1 + c_2 + c_3}, \qquad c_i > 0. \tag{3.1.5}$$

The centre of mass will always lie somewhere among the three points depending on the values of the individual masses. It is also clear that we can use two masses to span the interior of the triangle which is 2 dimensional. Since we are considering the projective triangle, we can rescale $Y, P_1, P_2, P_3$ by a positive constant and rewrite Y as

$$Y = c_1 P_1 + c_2 P_2 + c_3 P_3, \qquad c_i > 0. \tag{3.1.6}$$

where the $c_i$ can also be expressed explicitly in terms of determinants using (3.1.4), and read

$$c_1 = \frac{\langle Y23 \rangle}{\langle 123 \rangle}, \qquad c_2 = -\frac{\langle Y13 \rangle}{\langle 123 \rangle}, \qquad c_3 = \frac{\langle Y12 \rangle}{\langle 123 \rangle}. \tag{3.1.7}$$

From this expression, we can see that the triangle can also equivalently be defined as the set of all Y such that

$$\langle Y23 \rangle > 0, \qquad \langle Y31 \rangle > 0, \qquad \langle Y12 \rangle > 0, \qquad \langle 123 \rangle > 0. \tag{3.1.8}$$

This projective construction of the inside of a triangle can be generalized in a straightforward way to the case of polygons. Suppose this time that instead of three points, we have n points $\{P_1, \cdots, P_n\} \in \mathbb{P}^2$. Following the center of mass argument, we can describe the inside of n points as

$$Y = C \cdot P = c_1 P_1 + c_2 P_2 + \cdots + c_n P_n, \qquad c_i > 0, \tag{3.1.9}$$

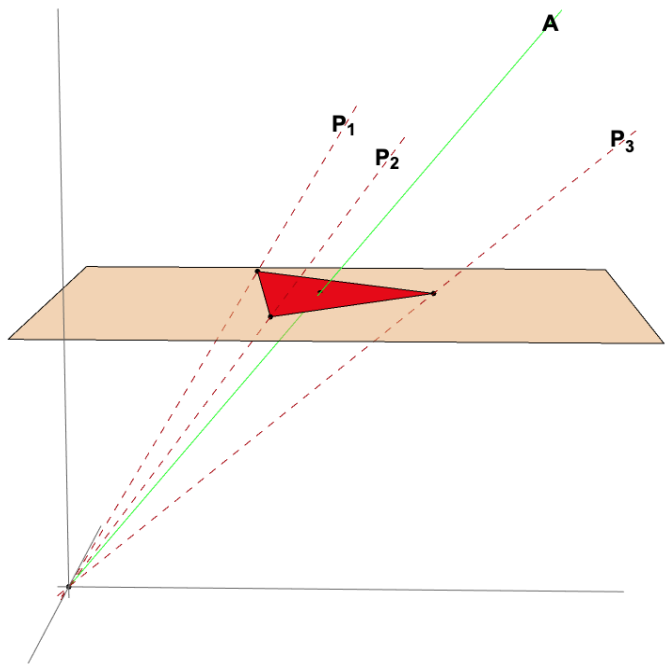

Figure 3.3: Illustration the projective triangle $P_1P_2P_3$.

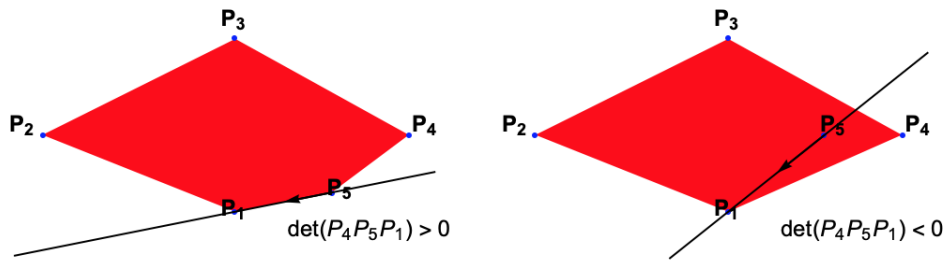

Figure 3.4: Example of a 5-point convex and non-convexity configuration.

where $C = (c_1, \cdots, c_n)$ is an n-vector and $P = (P_1, \cdots, P_n)$ is an $n \times 3$ matrix. This space is also known as the convex-hull of n points. Differently from the triangle, this time there is a new element coming into play; points can be arranged in different ways. In the case of a convex polygon, if one takes a line along one of the edges, like the line $P_1P_2$, it is clear that all the other vertices have to lie on the same side of the line, that is $\langle P_1P_2P_i \rangle > 0$ for all $i > 2$. Consider now the two configurations in image 3.4. In the right-hand figure, the points $P_1 \cdots P_5$ do not correspond to a convex polygon and their convex-hull is given by the quadrilateral $P_1, \cdots, P_4$. The non-convexity can be detected by the fact that $\langle 514 \rangle$ and $\langle 513 \rangle$ for example have different signs, therefore $B_5B_1$ cannot be the edge of a convex polytope. So, if we have n ordered points and we want to use them to represent a convex polygon, we must impose that all $\langle ijk \rangle > 0$ for $i < j < k$. If we consider the matrix $P_i^a = (P_1^a, \cdots, P_n^a)$, where i runs over the number of points and a is the dimension of the embedding space, the convexity condition is equivalent to saying that it exists an ordering such that all the ordered minors of the matrix P are positive. This definition of the interior of a polygon is highly redundant since we are using n variables to describe a 2-dimensional region. An alternative definition for the inside of an n-point convex polygon can be given as

$$\langle Y\widehat{ii+1}\rangle > 0, \qquad \forall i \leq n, \text{ with } P_{\widehat{n+1}} = -P_1,$$
$$\langle ijk\rangle > 0, \qquad \forall i < j < k. \tag{3.1.10}$$

The inequalities giving the polytope can then be derived in terms of the $\langle ijk\rangle$ brackets by parametrizing Y as $c_1 P_1 + \cdots + c_n P_n$ obtaining

$$\langle Y\widehat{ii+1}\rangle = c_1\langle 1\widehat{ii+1}\rangle + \cdots c_n\langle n\widehat{ii+1}\rangle > 0. \tag{3.1.11}$$

The equations $\langle Y\widehat{ii+1}\rangle = 0$ gives us the equations of the lines containing the edges of the polytope.

### 3.1.2 The inside of a polytope

The generalization of a polygon to higher dimensions is called a polytope. Any convex polytope can be described as the convex-hull of its vertices. The polytope with the minimal number of vertices is called a simplex. In d dimension a simplex has $d+1$ vertices. So for example in $d=1$ we have a segment, in $d=2$ we have a triangle and in $d=3$ we have a tetrahedron. The vertices of an n-point convex polytope must satisfy

$$\langle i_1 \cdots i_d\rangle \geq 0, \qquad \forall i_1 < \cdots i_d \leq n, \tag{3.1.12}$$

while its interior can be described as the convex-hull of its n vertices as

$$Y = c_1 P_1 + \cdots c_n P_n, \qquad \text{with } c_i > 0. \tag{3.1.13}$$

In projective space we can always fix one of the $c_i$ to 1, so the vector $C = (c_1, \cdots, c_n)$ has n–1 degrees of freedom. In the case of a simplex $n = d+1$, the positive coefficients of the vector c are in one to one correspondence with its interior.

It is not hard to see that the facets, that is the codimension-1 boundaries, of a polytope correspond to the equations $\langle Yi_1 \cdots i_d\rangle = 0$ for $\{i_1, \ldots, i_d\}$ such that $\langle Yi_1 \cdots i_d\rangle > 0$ for any values of the c parameters. Expanding $\langle Yi_1 \cdots i_d\rangle$ using (3.1.13) we obtain

$$\langle Yi_1 \cdots i_d\rangle = \sum_{j \neq i_1, \cdots, i_d} c_j \langle ji_1, \cdots i_d\rangle. \tag{3.1.14}$$

Since $0 < c_j < \infty$, the right hand side is positive only if all the brackets appearing in the sum are positive. We have therefore that the facets of the polygon, written as inequalities, are given by

$$
\left.\begin{array}{rcl}
\langle Y(i_1 i_1 + 1) \cdots i_{\frac{d}{2}} (i_{\frac{d}{2}} + 1) \rangle & > & 0 \\
-\langle Y i_1 i_1 + 1 \cdots i_{\frac{d}{2}} (i_{\frac{d}{2}} + 1) 1 n \rangle & > & 0
\end{array}\right\} \text{ for d odd },
$$
$$
\left.\begin{array}{rcl}
\langle Y 1 i_1 i_1 + 1 \cdots i_{\frac{d-1}{2}} (i_{\frac{d-1}{2}} + 1) \rangle & > & 0 \\
-\langle Y i_1 i_1 + 1 \cdots i_{\frac{d-1}{2}} (i_{\frac{d-1}{2}} + 1) n \rangle & > & 0
\end{array}\right\} \text{ for d even }. \qquad (3.1.15)
$$

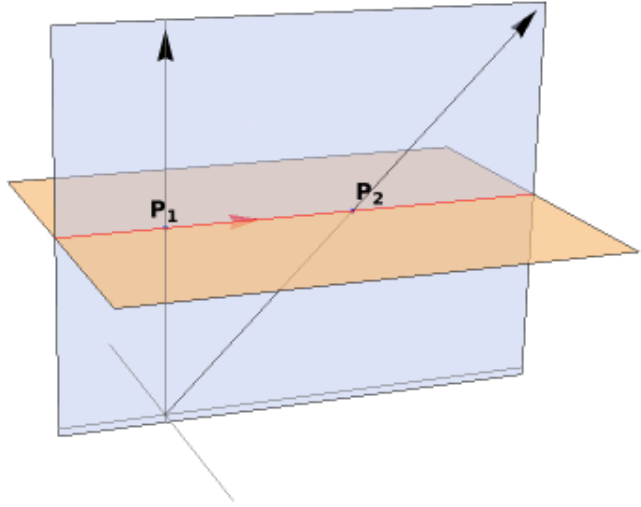

Figure 3.5: Example an oriented line in $\tilde{\mathbb{P}}^2$.

It can be shown straightforwardly that this set of constraints completely defines the polytope. However, this is not a completely trivial fact. We will see in section 3.4.1 that the amplituhedron, for example, requires additional conditions to be fully characterized.

### 3.1.3 The Grassmannian

Two linearly independent vectors $Y_1, Y_2$ span a plane. A plane in the embedding vector space represents a line in projective space, as it is illustrated in figure 3.5. If we stack $Y_1$ and $Y_2$ we can form a $2 \times d$ matrix $Y_\alpha^I = \begin{pmatrix} Y_1 \\ Y_2 \end{pmatrix}$. Notice that two matrices $Y'$ and $Y$ span the same plane, and therefore the same line in projective space, if there exists a matrix $S \in GL(2)$ such that

$$Y_\alpha^I = S_\alpha^\beta Y_\beta'^I . \qquad (3.1.16)$$

A line in projective space is therefore defined by a $2 \times d$ matrix $Y$ modulus a $GL(2)$ transformation. More generally the space of k-planes in $\mathbb{P}^{d-1}$ is called the Grassmannian $Gr(k, d)$ and is given by the space of $k \times d$ matrices mod $GL(k)$ transformations. In a similar way the space of oriented k-planes in $\tilde{\mathbb{P}}^{n-1}$ is called the oriented Grassmannian $\widetilde{Gr}(k, d)$ and is given by the space of $k \times d$ matrices mod $GL_+(k)$ transformations. Be aware that the k in k-plane always refers to the dimension of the plane in the embedding space, so for example a 1-plane corresponds to a projective point.

One way to parametrize Y is to use the $GL(k)$ symmetry to choose a representative of the form $Y = \left(\mathbb{1}_{k \times k}, A_{k \times (d-k)}\right)$. All the matrix elements of A are independent, therefore a generic $Y \in Gr(k, d)$ has $k(d-k)$ degrees of freedom. This representation of Y has a clear geometrical interpretation. Notice that writing Y in the gauge $(\mathbb{1}, A)$ is equivalent to

$$Y_\alpha = Z_\alpha + a_{\alpha, d-3} P_{d-3} + a_{\alpha, d-2} P_{d-2} + a_{\alpha, d-1} P_{d-1} + a_{\alpha, d} P_d , \qquad (3.1.17)$$

which means that $Y_\alpha$ lives in the plane $(\alpha(d-3)(d-2)(d-1)d)$. An alternative way to express this parametrization then is to say that we characterize Y by its intersection points

$$Y_\alpha = Y \cap (\alpha(d-3)(d-2)(d-1)d) . \qquad (3.1.18)$$

We can generate all sorts of parametrizations by choosing different planes. The intersection of homogeneous planes has the desirable property that in d dimensions a $k_1$-plane always intersects a $k_2$-plane in a $(k_1 + k_2 - d)$-plane. If $k_1 + k_2 - d < 0$ the two planes in general do not intersect. For example, two lines in $\mathbb{P}^2$, that is $k_1 = k_2 = 2$, always intersect in $4 - 3 = 1$ homogeneous line, that is a projective point. The intersection point of two planes $P = P_1 \cdots P_{k_1}$ and $W = W_1 \cdots W_{k_1}$ is given by the formula

$$(P \cap W) = \frac{1}{(k_1 + k_2 - d)!(k_2 - d)!} \epsilon^{i_1 \cdots i_{k_1}} P_{i_1} \cdots P_{i_{k_1 + k_2 - d - 1}} \langle P_{i_{k_1 + k_2 - d}} \cdots P_{i_{k_1}} W \rangle ,$$
$$(3.1.19)$$

which can be straightforwardly derived from (3.1.4). The intersection of two planes is symmetric. In this formulation, we are writing $(P \cap W)$ a $k_1 + k_2 - d$-plane on P but we could have as well written this formula expanding $(P \cap W)$ on W. We would like also to point out that the intersection of two hyperplanes gives a unique hyperplane in projective space but in oriented projective space instead it gives two solutions which are equal up to a sign.

Finally, let's look at yet another way to parametrize Y. The $GL(k)$ invariant data of Y is given by ratios of Y maximal minors since the latter are $SL(k)$ invariant. Maximal minors are also called Plücker coordinates. Not all Plücker coordinates are independent though and their mutual constraints are called Plücker relations.

Given $I = \{i_1, \cdots, i_k\}$ a list of k integers such that $i_j \leq d$, we will indicate with $\Delta_I(Y)$ the $Y \in Gr(k,d)$ minors involving the columns contained in I. If we fix a base $\{P_1, \cdots, P_d\}$ so that $\langle 1 \cdots d \rangle = 1$, then the minors of Y can be written as

$$\langle Y P_I \rangle = \Delta_{\bar{I}}(Y) \langle P_{\bar{I}} P_I \rangle = \text{sign}(I \cup \bar{I}) \Delta_{\bar{I}}(Y) \;, \tag{3.1.20}$$

where $\bar{I}$ is the ordered complement of I in $1,..,d$ and $P_I = P_{i_1} \ldots P_{i_k}$. As an example of Plücker relations, let's consider a line Y in $\mathbb{P}^3$. Exploiting the same trick we used to derive the generalized Schouten identities, we antisymmetrize over the six vectors $Y_1 Y_2 P_1 P_2 P_3 P_4$ obtaining

$$\epsilon_{a_1 a_2 a_3 a_4} \left( Y_1^{a_1} Y_2^{a_2} \langle 1234 \rangle + \cdots - \langle Y_1 Y_2 12 \rangle P_3^{a_1} P_4^{a_2} \right) = 0 \;. \tag{3.1.21}$$

Then, if we contract this expression with $Y_1^{a_3} Y_2^{a_4}$ we obtain

$$\langle Y12 \rangle \langle Y34 \rangle + \langle Y14 \rangle \langle Y23 \rangle + \langle Y13 \rangle \langle Y24 \rangle = 0 \;. \tag{3.1.22}$$

Examples of lines in $\mathbb{P}^3$ relevant to physics are of course the twistor lines and loop variables AB in the twistor representation. Plücker relations between the $\langle ABZ_i Z_j \rangle$ propagator are essential to consider in order to study the analytic properties of the integrand.

## 3.2 Superamplitude bosonization

One nice way to deal with the Grassmann odd nature of the superamplitude $\mathcal{A}_{n,k}$ is to attach 4k additional Grassmann odd variables $\phi_{\alpha A}$, $\alpha = 1,..,k$, $A = 1,..,4$ to each $\chi$, thus obtain commuting variables $Z_{i\alpha} := \chi_i^A \phi_{\alpha A}$ [14, 15]

$$\mathcal{Z}_i = \begin{pmatrix} z_i \\ \chi_i \end{pmatrix} \quad \rightarrow \quad Z_i(\chi_i) = \begin{pmatrix} z_i \\ \chi_i \cdot \phi_1 \\ \vdots \\ \chi_i \cdot \phi_k \end{pmatrix} \;. \tag{3.2.1}$$

We then rewrite the superamplitude in terms of these bosonised supertwistors $Z_i$. More precisely we define a map $\mathcal{B}_{k,4}$ from superamplitudes (functions of n momentum supertwistor space variables), to bosonised superamplitudes (functions of n bosonised supertwistors in $k+4$ dimensions together with a single k-plane in $k+4$ dimensions, Y)

$$\mathcal{B}_{k,4} : \mathcal{A}_{n,k}(\mathcal{Z}_i) \mapsto A_{n,k}(Z_i, Y) \;. \tag{3.2.2}$$

The map $\mathcal{B}_{k,4}$ is defined by insisting that if the bosonised Zs are written in terms of $\chi \cdot \phi$ as in (3.2.1), and Y takes the special value $Y_0$ below, then the result is the superamplitude times the product of all the $\phi$s:

$$A_{n,k}(Z_i(\chi_i), Y_0) = N(k,4) \times \prod_{\alpha=1}^{k} \prod_{A=1}^{4} \phi_{\alpha A} \times \mathcal{A}_{n,k}(\mathcal{Z}_i) ,$$

$$Y_0 = \begin{pmatrix} 0_{4\times k} \\ \mathbb{1}_{k\times k} \end{pmatrix} . \tag{3.2.3}$$

Here $N(k,m)$ is a normalisation factor to be discussed shortly. For now note that as long as $A_{n,k}(Z_i(\chi), Y_0)$ is homogeneous of degree 4k in the $\chi$s, then it will inevitably take the form of the RHS for some function of the $\chi$s, $\mathcal{A}_{n,k}(\mathcal{Z}_i)$, due to the Grassmann nature of the $\phi$s[2].

Since the bosonised $\chi$s are obtained as a product of Grassmann odd quantities, they will satisfy various non-trivial nilpotency relations between them (eg $(Z_{i\alpha})^5 = 0$) which means that (3.2.3) does not uniquely define the form of the bosonised superamplitude $A_{n,k}(Z_i, Y_0)$ if we think of it as an ordinary function of complex variables. However, the claim is that it does have a unique form with a given structure involving an emergent $SL(4+k)$ symmetry. In particular, a generic $N^k$MHV-type dual superconformal invariant can be written in a manifestly $SL(4+k)$ invariant form as the product of 4, (k + 4)-brackets[3]

$$\langle I_1 \rangle \langle I_2 \rangle \langle I_3 \rangle \langle I_4 \rangle , \tag{3.2.4}$$

where here and in the following we will use a short-hand notation I, J etc to represent an ordered set of particle numbers. We define $[n] := \{1, 2, ..., n\}$ and then $\binom{[n]}{k}$ to be the set of all ordered sets of k elements in [n]. So here $I_a \in \binom{[k+4]}{k}$. Any bosonized superamplitude can we written as a sum of terms of the form (3.2.4) times a rational function of the ordinary 4-momentum twistors.

Furthermore, 4-brackets involving twistors (2.4.22) can also be promoted to (k + 4)-brackets of bosonised supertwistors by including the (4+k) × k matrix $Y_0$, via the identity

$$\langle ijkl \rangle = \langle Y_0 ijkl \rangle , \quad \text{with} \quad Y_0 = \begin{pmatrix} 0_{4\times k} \\ \mathbb{1}_{k\times k} \end{pmatrix} . \tag{3.2.5}$$

Then there appears to be a unique way of writing a function $A_{n,k}(Z_i, Y)$ which satisfies (3.2.3), and which has manifest SL(4 + k) symmetry.

---

[2] Note that the relation is more commonly written in the form $A_{n,k}(Z_i) = N(k,4) \int d^{4k}\phi A_{n,k}(\mathcal{Z})$ which is implied by (3.2.3) but is not as strong, since $A_{n,k}(\mathcal{Z})$ could have terms of lower degree in the $\phi$s and still satisfy this integral form.

[3] In fact, one can always write it as a single bracket to the power of m, $\langle I_1 \rangle$, but it is useful to consider the more general case.

Let us consider a simple example to illustrate this. The 5 point NMHV superamplitude $\mathcal{A}_{5,1}$, whose expression is given in equation (2.5.3). The corresponding bosonised superamplitude will have the form

$$A_{5,1}(Z_i, Y_0) = [12345] := \frac{\langle 12345 \rangle^4}{\langle Y_0 1234 \rangle \langle Y_0 2345 \rangle \langle Y_0 3451 \rangle \langle Y_0 4512 \rangle \langle Y_0 5123 \rangle} \,. \tag{3.2.6}$$

The amplitude $A_{5,1}$ now manifests fully the SL(k + 4) symmetry if we allow the symmetry to act on $Y_0$ as well as the Zs. It is also straightforward to check that it satisfies (3.2.3) with $N(1,4) = 4!$. We therefore treat the Zs as projective vectors in $\mathbb{P}^{k+4}$, promote $Y_0$ from a constant to a variable $Y \in \mathrm{Gr}(k, k+4)$ and study the analytic properties of $A_{n,k}(Z, Y)$.

We will generalize this construction, as is by now standard, by considering the momentum twistor dimension and the $\chi$s R-symmetry index dimension instead to be a generic positive integer m rather than 4. Then a generic invariant is expressed on a k + m dimensional bosonized space and will read

$$\langle I_1 \rangle \langle I_2 \rangle \ldots \langle I_m \rangle \,, \tag{3.2.7}$$

where $I_a \in \binom{[k+m]}{k}$.

It is quite natural to further view the bosonised amplitude as a top form of the Grassmannian Gr(k, k+m) that Y is an element of. The dimension of Gr(k, k + m) is mk, so the amplitude will be a 4k differential form on the Grassmannian. This measure has the covariant form

$$\prod_{a=1}^{k} \langle Y d^m Y_a \rangle \,, \tag{3.2.8}$$

where $Y_a$ indicates the $a^{\text{th}}$ column of Y and $\langle Y d^m Y_a \rangle := \langle Y \overbrace{dY_a \cdots dY_a}^{m} \rangle$. Notice that the measure has weight k(m + k) in Y and thus attaching the latter to the amplitude it will have weight 0 in the Ys as well as the Zs.

This construction also extends to loops (for m = 4). As we showed in section 2.4.2, a loop is represented by a pair of bosonised supertwistors (AB) where $A, B \in \mathbb{C}^{4+k}$. Bosonised amplitudes will depend on loops through the brackets $\langle YABZ_iZ_j \rangle$. Its covariant measure reads $\langle YABd^2A \rangle \langle YABd^2B \rangle$. Loop variables always appear in the same bracket with Y. Therefore they are naturally defined on $Y^\perp$, that is the space projecting through Y, and are elements of Gr(2, 4).

Summarising, the bosonised superamplitude $A_{n,k,l}$ can be written as a rational differential form depending on $Y \in \mathrm{Gr}(k, k+4)$ and l loop variables which are lines in $Y^\perp$ so effectively $(AB)_i \in Gr(2,4)$, together with n Z's in $Gr(1, k+4)$. Remarkably the resulting differential form is the unique canonical form obtained from a simple geometrical object, the amplituhedron. In the next section, we will give a brief review of this geometrical formalism.

Finally, we discuss the normalization $N(k,m)$ appearing in the map from superspace to amplituhedron space (3.2.3). This is present simply due to the combinatorics involved in extracting the $\phi$s from the amplituhedron-type expression. It can be motivated and derived through the example of the anti-MHV $k = n - 4$ amplitude. This has a simple expression in amplituhedron space:

$$A_{n,n-4}(Z_i, Y) = \frac{\langle 1 \cdots n \rangle^4}{\prod_i \langle Y i(i+1)(i+2)(i+3) \rangle} \ . \tag{3.2.9}$$

But in order for this to give the corresponding superspace expression we need to pull out the $\phi$s, yielding a numerical factor. Explicitly then, for general m, the numerical factor $N(k,m)$ is fixed by

$$\det(\phi_i \cdot \chi_j)^m = N(k,m) \prod_{i=1}^{k} \prod_{A=1}^{m} \phi_{iA} \prod_{i=1}^{k} \prod_{A=1}^{m} \chi_i^A \ . \tag{3.2.10}$$

So for $m = 1$, for example, we don't have any R-symmetry index and every term in the (single) determinant contributes the same giving a factor of $k!$. Taking into account the re-ordering of the Grassmann variables then gives $N(k,1) = (-1)^{\lfloor \frac{k}{2} \rfloor} k!$. More generally, it has been found in [1] by explicit computation that the normalization coefficients are always consistent with the following expression

$$N(k,m) = (-1)^{\lfloor \frac{mk}{2} \rfloor}(m!)^k \prod_{j=1}^{k} \frac{(m+j)^{k-j}}{j^{k-j}} \ . \tag{3.2.11}$$

## 3.3 The canonical form

In [30] positive geometries and their canonical forms were defined. This definition is based on two important mathematical concepts: orientation forms and multivariate residues. We will start by giving a brief review of these topics in the context of algebraic geometry to then define the canonical form and give some examples.

### 3.3.1 The orientation form

A manifold $\mathcal{M}$ is oriented if it possesses a continuous top differential form $\mathcal{O}$ that is always non-vanishing in $\mathcal{M}$. The orientation is then equivalently defined by this volume form modulo positive scaling.

Here we would like to review how given an oriented manifold one can derive the induced orientation of the boundary. Suppose then $\mathcal{M}$ has a boundary $\partial \mathcal{M}$ of codimension one. The orientation $\mathcal{O}_\partial$ induced by $\mathcal{O}$ on $\partial \mathcal{M}$ will be the projection of $\mathcal{O}$ on $\partial \mathcal{M}$. We will now define what we mean by projection. To be more explicit, we can choose a basis of the dual space $\{dx_1, \cdots, dx_d\}$, where d is the dimension of $\mathcal{M}$ such that

$$\mathcal{O} = \mathcal{O}(x_1, \cdots, x_d) dx_1 \cdots dx_d , \qquad (3.3.1)$$

where the product between the differentials is understood to be a wedge product. Let the boundary be defined by $f(x) = 0$ with $f(x) > 0$ inside and $f(x) < 0$ outside the region (at least nearby). Then $\mathcal{O}_\partial$ is defined simply as

$$df \wedge \mathcal{O}_\partial = \mathcal{O}|_{\partial \mathcal{M}} . \qquad (3.3.2)$$

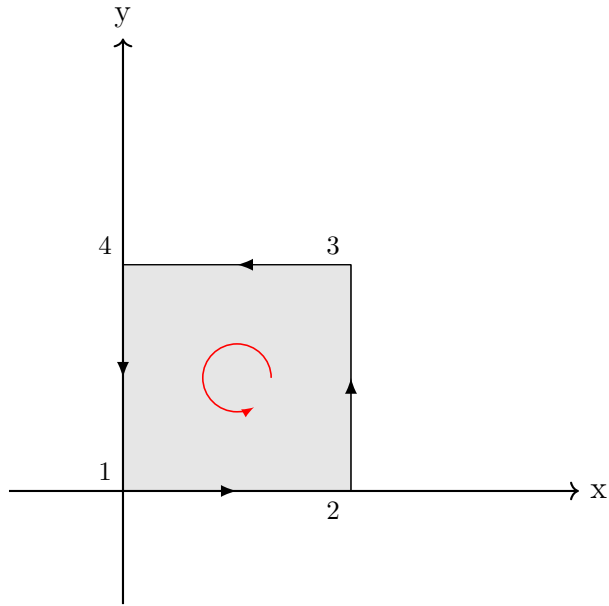

Figure 3.6: An oriented square. The square orientation is represented by a spiral, while the boundary orientation are represented by arrows.

Note that the standard convention in math literature differs by a sign, that is df represents an outward pointing differential. We make this choice so that the segment x > 0 with positive orientation form O(x) = dx has orientation $O|_{x=0} = 1$.

Let's look now at some examples and then how to compute df in practice. Consider a square in 2 dimensions as illustrated in fig: 3.6. We choose for our square an orientation form

$$\omega = dxdy . \qquad (3.3.3)$$

Segment 12 has outward-pointing 1-form df = −dy and so its orientation $\omega_{12}$ will be

given by

$$(\text{--dy})(\omega_{12}) = \omega = -\text{dy}\,\text{dx}\,, \tag{3.3.4}$$

which implies $\omega_{12} = \text{dx}$.

### 3.3.2 Multivariate residues

Multivariate residues have been intensely studied in the mathematics literature and can have different definitions. We are interested in the so-called Leary residues [99]. They have been used in recent amplitudes development to implement generalized unitarity methods [100–102] and to study the connection between amplitudes branch-cuts and integrand poles [103–106] and of course in the context to positive geometries.

Multivariate residues are most conveniently defined in the language of differential forms. Consider a d-dimensional space X and a subspace C of X, defined by the equation $f(x_1, \cdots, x_d) = 0$ where f is an irreducible polynomial. If $\omega$ is a differential k-form defined on the complement X – C, then we say that $\omega$ has a pole of order n on C if

$$\omega = \frac{\text{df}}{f^n}\Psi + \theta\,, \tag{3.3.5}$$

where $\psi$ and $\theta$ are regular and non-vanishing on S and $\theta$ is has at most a pole of order n – 1 on S. In the special case of a simple pole, that is n = 1, the residue of $\omega$ on C is defined as the restriction of $\Psi$ to the subvariety C

$$\text{Res}_C(\omega) = \Psi|_C\,. \tag{3.3.6}$$

In practical computations, we often want to write $\Psi|_C$ as a rational differential form on a patch of C. We explain how to do this when C is a rational variety in B.

### 3.3.3 Canonical form definition

A positive geometry (PG) is by definition a geometry that has a canonical form, and both the concept of positive geometry as well as its canonical form are defined recursively. A D-dimensional positive geometry is defined as the pair $(X, X_{\geq 0})$ possessing a canonical form $\Omega(X, X_{\geq 0})$ satisfying the following conditions

- X is a complex projective algebraic variety of complex dimension D, known as the embedding space. In practice for our application the algebraic variety will be a Grassmannian but the definition is given in this more general setting.

- $X_{\geq 0}$, is a closed, oriented, D-dimensional semi-algebraic subset of $X(\mathbb{R})$, the real slice of X.

- There is a unique top form $\Omega(X, X_{\geq 0})$ called the canonical form.

- Every boundary component $(C, C_{\geq 0})$ is itself a positive geometry of dimension $D-1$.

- The canonical form has no singularities inside $X_{\geq 0}$, but has simple poles on the boundary. The recursive step is then that the residue of the canonical form on each boundary component is equal to the canonical form of the boundary component itself:
$$\text{Res}_C\left(\Omega(X, X_{\geq 0})\right) = \Omega(C, C_{\geq 0}). \tag{3.3.7}$$

- The recursion is initiated by defining 0-dimensional positive geometries, for which $X_{\geq 0}$ is just a single point and $\omega(X, X_{\geq 0}) = \pm 1$ depending on the orientation.

In the following, we will follow convention and often simplify notation and refer to the positive geometry simply by $X_{\geq 0}$ instead of $(X, X_{\geq 0})$.

Note that $X_{\geq 0}$ is defined as a semi-algebraic subset of $X(\mathbb{R})$ which itself is a subset of $\mathbb{P}^n$. A semi-algebraic set is defined by a set of homogeneous real polynomial equations, $p(x) = 0$, and inequalities, $q(x) > 0$. Now inequalities $q(x) > 0$ are problematic in projective spaces since homogeneous coordinates are invariant under $x \to -x$ which may flip the sign and change the inequality. For this reason the prescription is to first define the region in $\mathbb{R}^{n+1}/\{0\}$ and then project onto $\mathbb{P}^n$ to obtain $X_{\geq 0}$.

Apart from the 0-dimensional oriented points, the simplest example of positive geometries is the oriented segment. The canonical form of the segment from a to b, that is with orientation $\mathcal{O}(x) = dx$, is

$$[a, b] = \begin{array}{c}\bullet\!\!\longrightarrow\!\!\bullet\\ a \qquad b\end{array}, \qquad \Omega([a, b]) = \frac{dx}{x-a} - \frac{dx}{x-b}. \tag{3.3.8}$$

Then the boundaries are given by $f(x) = x - b = 0$ and $f(x) = x - a = 0$ and we see from (3.3.7) that the point a has canonical form 1 and b has canonical form $-1$, which correspond to the sign of the projection of $\mathcal{O}$ on those points.

### 3.3.4 Canonical form properties

The canonical form has a series of fundamental properties that help with its computation that we would like to highlight.

**Uniqueness**

The uniqueness of the canonical form is equivalent to the statement that the algebraic variety on which the positive geometry lives has geometric genus zero ie has no non-zero holomorphic volume forms [30]. Since holomorphic forms have no poles, they could be added to any canonical form to obtain a new canonical form satisfying all the requirements and thus we would not have a unique canonical form. Conversely, under the assumption that PG have no holomorphic forms, we can proceed by induction assuming that the canonical form in d – 1 dimensions is always unique. Consider two canonical forms $\Omega_1, \Omega_2$ for the same PG. By definition, both forms have poles only on the boundary components. The residue on a boundary component is the canonical form of the boundary component, by the recursive definition of the canonical form. But since by induction we assumed that the canonical form in d – 1 dimensions is unique then we conclude that for any residue $\text{Res}(\Omega_1 - \Omega_2) = 0$, and so $\Omega_1 - \Omega_2$ has no poles and is thus a holomorphic form and so must vanish. We conclude that $\Omega_1 = \Omega_2$ and so the canonical form is unique.

**Opposite orientation**

Given two PGs $X'_\geq, X_\geq$ that differ only by the orientation, then the two canonical forms have opposite sign, that is $\Omega(X'_\geq) = -\Omega(X_\geq)$. We can prove this statement by observing that flipping the sign of the orientation also flips the sign of all the induced orientations down to 0-dimensional boundaries. Therefore, the maximal residues of the two geometries will have opposite sign. Then, using the same logic of the proof of the uniqueness of the canonical form, the statement can be proved by induction.

**Cartesian product**

Given two PGs $X_\geq, Y_\geq$, the canonical form of the Cartesian product $X_\geq \otimes Y_\geq := \{(x,y) \text{ s.t. } x \in X_\geq, y \in Y_\geq\}$ is equal to the product of the canonical form

$$\Omega(X_\geq \otimes Y_\geq) = \Omega(X_\geq)\Omega(Y_\geq) \, . \tag{3.3.9}$$

This equation holds because the codimension-1 boundaries of $X_\geq \otimes Y_\geq$ are given by the union of the $\partial X_\geq \otimes Y_\geq$ and the $X_\geq \otimes \partial Y_\geq$ codimension-1 boundaries. Also

$$\text{Res}_{\partial X_\geq \otimes Y_\geq} \left( \Omega(X_\geq)\Omega(Y_\geq) \right) = \Omega(Y_\geq) \text{Res}_{\partial X_\geq} \left( \Omega(X_\geq) \right) \, . \tag{3.3.10}$$

Using the inductive step $\Omega(\partial X_\geq \otimes Y_\geq) = \Omega(\partial X_\geq)\Omega(Y_\geq)$, it is then easy to prove that (3.3.9) holds.

This property allows computing straightforwardly the canonical form of a simplex. As we have seen in section 3.1.2, parametrizing Y as $Y = P_1 + c_2 P_2 \cdots c_{d+1} P_{d+1}$ the inside of a simplex is given by the inequalities $c_2 > 0 \wedge \cdots \wedge c_{d+1} > 0$, which, it corresponds to a Cartesian product of segments. Therefore its canonical form, which we indicate with $[1, \cdots, d+1]$, can be written in these coordinates as

$$[1 \cdots d+1] = \prod_{i=2}^{d+1} \frac{dc_i}{c_i} \ . \tag{3.3.11}$$

We can write the canonical form covariantly in terms of determinants as

$$[1 \cdots d+1] = \frac{\langle Y d^d Y \rangle \langle 1 \cdots d+1 \rangle^d}{\langle Y12 \cdots (d+1) \rangle \langle Y2 \cdots (d+1)1 \rangle \cdots \langle Y(d+1)12 \cdots d \rangle} \ . \tag{3.3.12}$$

In particular, the canonical form of a triangle will read

$$[123] = \frac{dc_2 dc_3}{c_2 c_3} = \frac{\langle Y d^2 Y \rangle \langle 123 \rangle^2}{\langle Y12 \rangle \langle Y23 \rangle \langle Y13 \rangle} \ . \tag{3.3.13}$$

**Union of geometries**

The canonical form of the union of positive geometries is a key and subtle property that will be central to the generalization of the canonical form discussed in the second part of this thesis.

The union of two completely disjoint positive geometries $X_1, X_2$ is itself a positive geometry and the canonical form is simply the sum of the two canonical forms: $\omega(X_1 \cup X_2) = \omega(X_1) + \omega(X_2)$.

A more interesting case to consider is that of two positive geometries $X_1, X_2$ which only overlap on their boundary. Firstly consider the case where they share a codimension 1 boundary. The union can only form a positive geometry if the orientations of $X_1$ and $X_2$ agree. If the orientations do agree then the canonical forms along the common boundary of $X_1$ and $X_2$ will cancel (as it must for this to be a positive geometry as this will lie in the interior of the union). A simple example is that of two triangles sharing an edge:

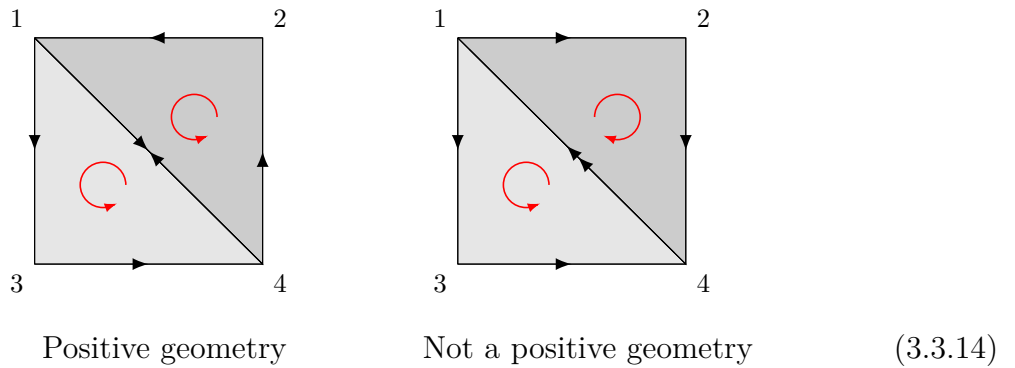

Positive geometry      Not a positive geometry      (3.3.14)

This can be then generalized to the union of an arbitrary number of PGs. Consider a PG $X_{\geq 0}$ and a set of PGs $X_{\geq 0}^{(i)}$ tiling $X_{\geq 0}$. By a tiling we mean the $X_{\geq 0}^{(i)}$ cover $X_{\geq 0}$ with non-overlapping regions, so

$$X_{\geq} = \bigcup X_{\geq 0}^{(i)} , \qquad (3.3.15)$$

and the orientations of the tiles match on codimension-1 boundaries. Then the canonical form of $X_{\geq 0}$ is the sum of that of the tiles

$$\Omega(X_{\geq 0}) = \sum_i \Omega(X_{\geq 0}^{(i)}) . \qquad (3.3.16)$$

For this relation to hold true, it is fundamental that $X_{\geq}$ is a positive geometry. In fact, the union of PGs with matching orientation is not always a positive geometry. For example, consider a union of two positive geometries sharing a boundary of lower dimension, for example, two triangles touching at a vertex:

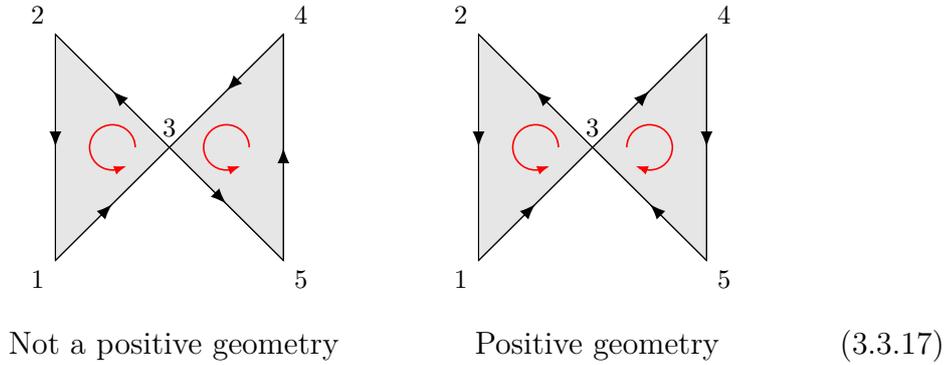

Not a positive geometry    Positive geometry    (3.3.17)

Here the case where the orientations agree is *not* a positive geometry whereas the case where they disagree is. To see this let's consider the canonical forms in the two cases. The canonical form of a triangle $\{i, j, k\}$ with standard orientation is

$$[ijk] = \frac{\langle Y d^2 Y \rangle \langle ijk \rangle^2}{\langle Yij \rangle \langle Yjk \rangle \langle Yki \rangle} , \qquad (3.3.18)$$

and thus the canonical form of the union of the two triangles, if it exists, will be given by $[123] + [345]$ or $[123] - [345]$ in the two cases respectively. In the first case the double residue corresponding to the residue at vertex 3 is[4]

$$\text{Res}_{\langle Y13 \rangle \to 0} \left( \text{Res}_{\langle Y23 \rangle \to 0} ([123] + [345]) \right) = -2 , \qquad (3.3.19)$$

which is different from $\pm 1, 0$. In the second case instead, the residue is simply zero. Double residues at the other points are equal to $\pm 1, 0$ since only one triangle at a time will contribute. Thus only the second geometry is a positive geometry. Note the difference with the previous case where the orientations had to agree for a positive

---

[4]Note that $\langle Y23 \rangle \to 0 \Leftrightarrow \langle Y35 \rangle \to 0$.

geometry, here instead they have to disagree for it to be a positive geometry!

**Push-forward**

Given two manifold X, Y, the push-forward [107] is a map from the space of differential forms on X to the space of differential forms on Y. The map $\phi$ might not be injective, but we can break the domain into $\deg(\phi)$ regions where $\phi$ is invertible and for each region we will have an inverse $\phi_1^{-1}, \cdots \phi_{\deg(\phi)}^{-1}$. The push-forward of a differential form $\omega$ on Y on a patch $(y_1, \cdots y_d)$ to a patch $(x_1, \cdots x_d)$ of X is then defined as

$$\Psi^*(\omega)(y_1, \cdots, y_d) = \sum_i^{\deg(\phi)} \omega \circ \phi_i^{-1}(y_1, \cdots, y_d) \, J_i \, dy_1 \cdots dy_d \,, \qquad (3.3.20)$$

where $J_i$ is the Jacobian of $\phi_i^{-1}$ on $(y_1, \cdots, y_d)$.

Consider two positive geometries $X_\geq \subset X$ and $Y_\geq \subset Y$ of the same dimension and a rational orientation preserving map $\phi : X \to Y$. Then, given $\Psi^*$ the push-forward from X to Y, the canonical form of $Y_\geq$ is given by

$$\Omega(Y_\geq) = \Psi^*(\Omega(X_\geq)) \,. \qquad (3.3.21)$$

As an example, consider the map $y = \phi(x) = x^2$. This is rational and has inverse for $x > 0$ equal to $x = \phi_1^{-1}(y) = \sqrt{y}$ and for $x < 0$ equal to $x = \phi_2^{-1}(y) = -\sqrt{y}$. A segment $x > 0$ will be mapped by $\phi$ to a segment $y > 0$. These two PGs have canonical form $\frac{dx}{x}$ and $\frac{dy}{y}$ respectively. We can then see that indeed the canonical form of the segment $y > 0$ is given by the push-forward of $\frac{dx}{x}$ through $\phi$

$$\frac{d\sqrt{y}}{\sqrt{y}} + \frac{d(-\sqrt{y})}{-\sqrt{y}} = \frac{dy}{y} \,. \qquad (3.3.22)$$

To our knowledge, this property has not been proved in full generality, but it has been shown to hold in a high number of non-trivial examples, in particular in the interpretation of the CHY formula [48] as the push-forward between positive geometries through the scattering equation [35, 49]. Moreover, under some technical assumptions, it was proved that the push-forward commutes with the residue operation, which is a strong indication that a proof of this property should be indeed possible.

## 3.4 The amplituhedron

The tree amplituhedron $\mathcal{A}_{n,k}$ is the subspace of $\mathrm{Gr}(k, k+4)$ defined as tree amplituhedron:   $\mathcal{A}_{n,k}(Y; Z) := \{Y = C \cdot Z \in \mathrm{Gr}(k, k+4) | \ C \in \mathrm{Gr}_>(k, n)\},$
$$\text{for } Z \in \mathrm{Gr}_>(k+4, n)\,, \tag{3.4.1}$$

where $\mathrm{Gr}_>(k, n)$ is the space of oriented k-planes for which all the maximal ordered minors are positive and is called the positive Grassmannian [108]. The positive Grassmannian is inherently real and therefore $\mathcal{A}_{n,k}$ is defined as a region in the real *oriented* Grassmannian $\widetilde{\mathrm{Gr}}(k, k+4) := \mathbb{R}^{k \times 4}/\mathrm{GL}^+(k)$, that is the space of oriented k-planes in $k+4$ dimensions. The amplituhedron is usually then viewed as being the projection of this onto the (unoriented) real Grassmannian $\mathrm{Gr}(k, k+4)$. However, we instead find it useful to remain on $\widetilde{\mathrm{Gr}}(k, k+4)$ and view the amplituhedron directly on this space. This allows for a natural universal orientation for any subset. The amplitude itself is extracted from the geometry by taking its canonical form (see section 3.3) and will therefore also initially be defined on the real Grassmannian, but can be then analytically continued to the complex numbers. This definition of Y through the matrix C is in general degenerate, that is two different C's in $\mathrm{Gr}(k, n)$ can correspond to the same $Y \in \mathrm{Gr}(k, k+4)$. We can write Y using the C matrix Plücker coordinates as

$$Y = \sum_{1 \leq i_1 < \cdots < i_k \leq n} \det(C_{i_1}, \cdots, C_{i_k}) Z_{i_1} \cdots Z_{i_k}\,, \tag{3.4.2}$$

where $C_i$ is the i-th column of the matrix C. Using (3.4.2) we can see that the brackets $\langle Y ii+1 jj+1 \rangle$ are always positive,

$$\langle Y ii + ijj + 1 \rangle = \sum_{1 \leq i_1 < \cdots < i_k \leq n} \det(C_{i_1}, \cdots, C_{i_k}) \langle Z_{i_1} \cdots Z_{i_k} Z_i Z_{i+1} Z_j Z_{j+1} \rangle > 0\,, \tag{3.4.3}$$

where we used that $Z \in \mathrm{Gr}_>(k+4, n)$ (3.4.1). Each term in the sum is positive since it is given by the product of an ordered C minor and an ordered Z minor. The $j = n$ case is special and one can check that the bracket $\langle ii+1n1 \rangle$ is positive for k odd and negative for k even. If we consider an amplituhedron for $k \neq n-4$, i.e. k not maximal. The brackets $\langle Y ii + ijj + 1 \rangle$ are the only brackets that have a fixed sign for all Y. This implies that the codimension one boundaries of the amplituhedron are a subset of the region described by the equation $\langle Y ii + 1jj + 1 \rangle = 0$.

### 3.4.1 The amplituhedron and flipping number

In [109] an equivalent, more direct definition of the amplituhedron was given as a certain subspace of the set of oriented k-planes Y in $k+4$ dimensions bounded by inequalities of the form $\langle YZ_iZ_jZ_lZ_m \rangle > 0$, together with a further topological condition to be described, but importantly with no reference to the auxiliary positive matrix C present in the original definition (3.4.1). This definition as been proved to be equivalent to the original definition first for $m=2$ in [110] and then in full generality for the tree amplituhedron in [33].

At tree-level the alternative definition of the amplituhedron (3.4.1) is as the set

$$\mathscr{A}_{n,k} := \left\{ Y \in \mathrm{Gr}(k, k+4) \;\middle|\; \begin{array}{ll} \langle Yii+1jj+1 \rangle > 0 & 1 \leq i < j-1 \leq n-2 \\ \langle Yii+11n \rangle (-1)^k > 0 & 1 \leq i < n-1 \\ \{\langle Y123i \rangle\} & \text{has k sign flips as } i = 4,..,n \end{array} \right\}$$
$$\text{for } Z \in \mathrm{Gr}_>(k+4, n)\,,$$
(3.4.4)

Here the inequalities $\langle Yii+1jj+1 \rangle > 0$ and $\langle Yii+1n1 \rangle (-1)^{k+1} > 0$ correspond to the locations of the proper poles of the amplitudes and are sometimes called proper boundaries. The second set of constraints is that the string $\{\langle Y123i \rangle\}$ as i ranges from 4 to n must change sign exactly k times, although the precise place where the sign changes is not important. This is a purely topological condition and $\langle Y123i \rangle = 0$ will not be a physical boundary, unless $i = 4, n$.

This sign flip constraint is clearly not manifestly cyclic. Cyclicity then demands that if the string $\{\langle Y123i \rangle\}$ has k sign flips, then all the strings of the form $\{\langle Yjj+1j+2i \rangle\}$ must have the same number of flips. Indeed, an even stronger statement can be proved. If the proper boundary inequalities hold, then all the strings of the form $\{\langle Yj_1j_1+1j_2i \rangle\}$ have the same number of flips as $i \neq j_1, j_1+1$ runs from $j_2+1$ to $j_2-1$ [109].

The loop amplituhedron can also be written in a similar form. The loop variables in the amplituhedron picture are represented by 2-planes $(AB)_i$ living in $Y^\perp$. The loop amplituhedron $\mathscr{A}_{n,k,l}$ is defined as the objects $\{Y, (AB)_1, .., (AB)_l\}$, with Y belonging to the tree level amplituhedron, and each $(AB)_i$ satisfying the following inequalities

$$\mathscr{A}_{n,k,l} := \left\{ Y, (AB)_1, .., (AB)_l \;\middle|\; \begin{array}{ll} Y \in \mathscr{A}_{n,k} \\ \langle Y(AB)_jii+1 \rangle > 0, & \forall j, \forall i = 1,.., n-1 \\ \langle Y(AB)_j1n \rangle (-1)^{k+1} > 0 & \forall j \\ \{\langle Y(AB)_j1i \rangle\} & \text{has k+2 flips as } i = 2,.., n, \forall j \\ \langle (AB)_i(AB)_j \rangle > 0 & \forall i \neq j \end{array} \right\}$$

$$\text{for } Z \in \text{Gr}_>(k+4, n) \,, \tag{3.4.5}$$

## 3.5  Correlahedron and squared amplituhedron

We have seen how dual superconformal invariants can be bosonized and how the superamplitude and the superamplitude squared can be associated to a differential form on the Grassmannian. The same can of course be done for superconformal invariants and in particular for the correlator. Much less is known about the correlator as a form on the Grassmannian compared to the superamplitude and its geometrical interpretation as a positive geometry is still pretty much an open question. A proposal with many compelling features for the geometry of $\mathcal{G}_{n,n-4}^{(0)}$ was made in [51] called *the correlahedron*. Here we will describe the details of the bosonization of the correlator, the correlahedron geometry and its connection to the squared amplituhedron.

### 3.5.1  Correlator potential as a differential form

As we have seen, the maximal nilpotent correlator potential $\mathcal{G}_{n,n-4}$ is a Grassmann polynomial of degree $2n-4$. All its dependence on Grassmann variables is contained in the term $\delta^{4\times(2n-4)}(X^\perp \cdot \chi)$. Following the same steps for the bosonization of super momentum-twistors, we attach $4(2n-4)$ additional Grassmann odd variables $\phi_{\alpha A}$, $\alpha = 1,..,k$, $A = 1,..,4$ to each $\chi$, thus obtaining

$$\mathcal{X}_{s,i} = \begin{pmatrix} X_i \\ \chi_{s,i} \end{pmatrix} \quad \to \quad X_{s,i} = \begin{pmatrix} X_i \\ \chi_i \cdot \phi_1 \\ \vdots \\ \chi_{s,i} \cdot \phi_k \end{pmatrix}, \tag{3.5.1}$$

where we will be using the letter X to indicate both twistors and bosonized supertwistors. Then we can write the unique maximal superconformal invariant as

$$\delta^{4\times(2n-4)}(X^\perp \cdot \chi) = \frac{1}{N(2n-4,4)} \int \prod_{i=1}^{2n-4} d^4\phi_i \langle X_1 \cdots X_n \rangle^4 \,. \tag{3.5.2}$$

Using (2.6.13), we can then write the bosonized maximal correlator as a differential form on $\text{Gr}(2n-4, 2n)$ of the form

$$G_{n,n-4} = \prod_{\alpha=1}^{2n-4} \langle Y d^4 Y_\alpha \rangle \langle X_1 \cdots X_n \rangle^4 f^{(n-4)} \,, \tag{3.5.3}$$

where we normalized by the $(N_c^2 - 1)/(4\pi^2)^n$ coefficient. We will use the letter G to indicate both the supercorrelator $G(x, y, \theta)$ and the bosonized potential correlator form $G(Y, X)$ trusting that what we mean will be clear from the context. In this thesis, we will just discuss $G_{n,n-4}$, but clearly any superconformal invariant can be bosonized as discussed in [51].

### 3.5.2 The correlahedron geometry

The correlahedron $\mathscr{G}_{n,n-4}$ is a geometrical object defined in $\text{Gr}(2n-4, 2n)$ and specified by the inequalities

$$\mathscr{G} := \langle X_1 \cdots X_n \rangle > 0, \qquad \langle YX_iX_j \rangle > 0, \quad \text{for all } i < j \leq n, \qquad (3.5.4)$$

where $X_i \in \text{Gr}(2, 2n)$ and are equivalent to points in bosonized chiral superspace. This is a very natural conjecture given the permutation symmetry of the correlator and given that its codimension one poles are given by the equations $\langle YX_iX_j \rangle = 0$.

It is not possible to interpret directly the bosonized correlator as the canonical form of the correlahedron. One reason is that the correlator form in $\text{Gr}(2n-4, 2n)$ has double poles, as we will see explicitly in equation (3.5.14) in the context of the light-like limit of the correlator. On the other hand, it is clear that this description of the correlahedron geometry is somehow redundant. To see this, we observe that $\text{Gr}(2n-4, 2n)$ is dual to $\text{Gr}(4, 2n)$ that is $Y^\perp$. So, we can give an alternative description of the maximal correlahedron as a geometrical object given by the 4-plane $X \in \text{Gr}(4, 2n)$, specified by the inequalities

$$\langle X_iX_j \rangle > 0, \quad \text{for all } i < j \leq n \qquad (3.5.5)$$

where $\langle X_iX_j \rangle$ are the minors of X. This geometry correspond to a region in the configuration space of 2n points $X_{s,i}$ in $\mathbb{R}^4$. The correlahedron as defined in (3.5.5) is clearly better described as a region in the configuration space of 2-planes, where pairs of points $X_{1,i}, X_{2,i}$ are defined up to an $\text{SL}_+(2)$ symmetry. Each $\text{SL}(2)_+$ fix 3 degrees for freedom, so the correlahedron space goes from $4(2n-4)$ to $4(n-4) + n$ degrees of freedom.

Similar ideas to reduce the dimension of the geometry have been successfully used in [51] to compute $\mathcal{G}_{5,1}$ as the canonical form of the correlahedron geometry and we believe represent a promising avenue for a better definition of correlahedron and its "canonical form".

### 3.5.3  Light-like limit in the Grassmannian

The correlator amplitude duality (2.6.16) involves two operations, one is multiplying by $\prod_i \langle X_i X_{ii+1}\rangle$ and the other is to integrate out the $\chi_{2,i}$ Grassmannian degrees of freedom. In bosonized space the procedure is the same, just we need also to integrate over $\phi_i$ for $i = n+1, \cdots 2n$ so that we get

$$\frac{1}{N(n,4)} \int \prod_i^n d^4\chi_{2,i} d^4\phi_{i+n} \langle X_1 \cdots X_n\rangle = \langle X_{1,1}\cdots X_{1,n}\rangle^4 = \langle Z_1 \cdots Z_n\rangle^4 , \quad (3.5.6)$$

as expected. The equivalent operation on the Grassmannian corresponds to taking residues on all $\langle YX_i X_{i+1}\rangle = 0$ and to make a projection.

We can compute the residue by parametrizing $Y_p$ as

$$Y_p = r_p Z_* + \sigma^s X_{s,p} + \tau^s X_{s,p+1} , \qquad \text{for } p \leq n , \quad (3.5.7)$$

and then taking the limit $r_p \to 0$. Notice that every $Y_p$ has four degrees of freedom, so one variable between $\sigma^1, \sigma^2, \tau^1, \tau^2$ must be a constant and we will choose one of the $\sigma$s. Now we have that

$$\begin{aligned}
\operatorname{Res}_{r_p=0} \frac{\langle Y d^4 Y_p\rangle}{\langle Y X_p X_{p+1}\rangle} &= \\
&= \left(\operatorname{Res}_{r_p=0} \frac{dr_p}{r_p}\right) \frac{\langle Y Z_* d^3 Y_p\rangle}{\langle Y_1 \cdots Y_{p-1} Z_* Y_{p+1} \cdots Y_{2n-4} X_p X_{p+1}\rangle} . \\
&= \langle \sigma d\sigma\rangle \langle d\tau d\tau\rangle
\end{aligned} \quad (3.5.8)$$

We can see from this computation how taking the $\langle YX_i X_{i+1}\rangle = 0$ residues effectively amount to multiply by $\langle YX_i X_{i+1}\rangle$. Notice that by taking $n$ residues we obtain a $4(2n-4)-n$ dimensional form but the superamplitude squared $A_{n,n-4}$ form is $4(n-4)$ dimensional.

The $\chi, \phi$ integration procedure corresponds to project on $(Y_1 \cdots Y_n)^\perp$ and to factor out a term purely dependent on the $Y_1, \cdots Y_n$ degrees of freedom. This operation reduce the degree of the form by $3n$ and was called in the original paper *freeze and project*, where the word freeze indicates that after the projection we fix the $(Y_1 \cdots Y_n)$ to be constant.

After projecting, the lines $X_i$ and $X_{ii+1}$ intersect in a point which will correspond in the duality to the bosonized momentum twistor $Z_i$. This means that $Z_i$ must be such that $\langle YX_i Z_i *\rangle = \langle YX_{i+1} Z_i *\rangle = 0$, where $*$ indicates an arbitrary vector $Z_*$. A solution is given by

$$Z_i = \sigma_i \cdot X_i . \quad (3.5.9)$$

In fact, $Z_i$ trivially satisfies the first equation and for the second equation we have

$$\langle YX_{i+1}Z_i * \rangle = \langle Y_1 \cdots Y_{i-1}(\sigma_i \cdot X_i) Y_{i+1} \cdots Y_{2n-4} X_{i+1}(\sigma_i \cdot X_i) * \rangle = 0 \,. \quad (3.5.10)$$

Notice that we can write the solution also as

$$Z_i = \tau_i \cdot X_{i+1} \,. \quad (3.5.11)$$

In fact, before projecting, $Z_i$ is defined up to a translation by any $Y_p$ with $p \leq n$ since such contribution is killed in the projection. Using both the equivalent ways of writing $Z_i$, we can rewrite $\langle YX_iX_j \rangle$ as a function of the $Z_i$ as

$$\langle YX_iX_j \rangle = \begin{cases} 0 \\ \frac{\langle YZ_{i-1}Z_iZ_{j-1}Z_j \rangle}{(\tau_{i-1} \cdot \sigma_i)(\tau_{j-1} \cdot \sigma_j)} \end{cases} \,. \quad (3.5.12)$$

We can also rewrite $\langle X_1 \cdots X_n \rangle$ as

$$\langle X_1 \cdots X_n \rangle = \frac{\langle Y_1 \cdots Y_n Z_1 \cdots Z_n \rangle}{\prod_i^n \tau_{i-1} \cdot \sigma_i} \,. \quad (3.5.13)$$

Therefore, observing that scaling of the $X_i$ contained in the $\langle X_iX_j \rangle$ brackets has degree 4, we will have that the light-like residue of the correlator projected on $(Y_1 \cdots Y_n)^\perp$ can be written as

$$\mathcal{A}_{n,n-4}^2 = \text{Res}_{\langle YX_iX_{i+1} \rangle = 0} \left( G_{n,n-4} \right) \Big|_{(Y_1 \cdots Y_n)^\perp} =$$
$$= \prod_{i=1}^{n} \frac{\langle \sigma d\sigma \rangle \langle d\tau d\tau \rangle}{(\tau_{i-1} \cdot \sigma_i)^2} \langle Z_1 \cdots Z_n \rangle^4 \prod_{\alpha=1}^{n-4} \langle \hat{Y} d^4 \hat{Y}_\alpha \rangle f^{(n-4)}(\langle \hat{Y} Z_i Z_{i+1} Z_j Z_{j+1} \rangle) \,, \quad (3.5.14)$$

where $\hat{Y}_i = Y_{i+n}$. Removing the $\tau, \sigma$ dependent factor we obtain the desired result.

We can also take the so-called non-maximal light-like limit which corresponds to taking light-like separated a number $p < n$ of points. In that case, we project on $(Y_1 \cdots Y_n Y_{n+1} \cdots Y_{2n-p})^\perp$, but instead of freezing all the degrees of freedom of the Y, we push-forward half of the degrees of freedom of the $Y_i$ for $n < i \leq 2n-p$ to the $X_i$ again for $n < i \leq 2n-p$ promoting them to dynamical loop variables. We refer to the correlahedron original paper for the details of this construction.

### 3.5.4 The squared amplituhedron

The Grassmannian version of the light-like limit allows making a natural guess for the geometry of the square of the superamplitude. In fact, supposing that a "canonical form like" procedure exists to extract the correlator from the correlahedron geometry, the residue $\langle YX_i X_{i+1}\rangle = 0$ should geometrically correspond to go to the boundary $\langle YX_i X_{i+1}\rangle = 0$ of the correlahedron. Moreover, using the parametrization (3.5.7) for which $Y_i \in X_i X_{i+1}$ for $i \leq n$, the freezing and project procedure also should correspond to project the correlahedron boundary on $(Y_1 \cdots Y_{2n-p})^\perp$ obtaining a geometry in $\text{Gr}(n-4, n)$.

In [51] this procedure, called the geometrical light-like limit, is explored in detail, and the geometry obtained is called the squared amplituhedron, since the authors conjectured that its canonical form should correspond to the square of the superamplitude.

We define the squared amplituhedron $\mathscr{H}_{n,n-4,l}$, following the more recent approach of [1], as the union of two regions. The latter are distinguished purely by their properties under cyclic transformations: twisted or untwisted

$$\text{Squared amplituhedron:} \quad \mathscr{H}_{n,n-4,l} := \mathscr{H}^+_{n,n-4,l} \cup \mathscr{H}^-_{n,n-4,l}, \quad (3.5.15)$$

with

$$\mathscr{H}^\pm_{n,k,l} := \left\{ Y, (AB)_1, .., (AB)_l \;\middle|\; \begin{array}{ll} \langle Y ii+1 jj+1 \rangle > 0 & 1 \leq i < j-1 \leq n-2 \\ \pm \langle Y ii+11n \rangle > 0 & 1 \leq i < n-1 \\ \langle Y(AB)_j ii+1 \rangle > 0 & \forall j, \forall i = 1, .., n-1 \\ \pm \langle Y(AB)_j 1n \rangle > 0 & \forall j \\ \langle (AB)_i (AB)_j \rangle > 0 & \forall i \neq j \end{array} \right\}$$

$$\text{for } Z \in \text{Gr}_>(k+4, n).$$

The squared amplituhedron is a much simpler geometry to study than the correlahedron. In particular, in [1] it has been proved to be a (generalized) positive geometry and that its canonical form indeed corresponds to the square of the superamplitude for $k = n - 4$.

A point that has not been stressed much in the literature regarding the derivation of the squared amplituhedron geometry from the correlahedron is that the inequalities describing it depend in general on the configuration of the plane $(Y_1 \cdots Y_{2n-p})$. One of these configurations corresponds to (3.5.15). This can be easily seen from equation (3.5.12). In principle, therefore, there are other geometries that could be associated to the square of the superamplitude. However, as we will see more in detail in section 6.6, two different geometries can have the same canonical form. It could be therefore that all these configurations have the same canonical form. A careful study of this dependence represents an interesting topic for future investigation and could provide further evidence for the correlahedron conjecture or unveil new interesting geometrical structures.

# Part II

# Loop Amplituhedron and Squared Amplituhedron as WPGs

# Chapter 4

# WPGs and Internal Boundaries

The content of this chapter has been published in the paper [2], excluding the content of section 4.4 which is only presented in this thesis.

Here, we investigate multiple residues of amplitudes and the corresponding amplituhedron boundary structure. In particular, we point out new features which have not been appreciated before [2].

First, we note that a direct consequence of amplitudes arising from positive geometries is that the amplitude should have unit maximal residues. This arises geometrically simply from the fact that maximal residues correspond to dimension 0 geometries, ie points, which can only differ by their orientation. Tree-level amplitudes indeed appear to have unit maximal residues. The tree level superamplitudes can be computed by summing a certain set of on-shell diagrams [108] arising from the BCFW recursion relation [68]. The on-shell diagrams manifestly have only logarithmic singularities and non-vanishing maximal residues equal to $\pm 1$ and have a natural geometric interpretation in amplituhedron space [15]. It was recently proven that they provide a tessellation of the amplituhedron [29] and it would be interesting to see if the details of this proof can also be used to prove that the non-vanishing maximal residues only equal $\pm 1$.[1]

What we will observe however is that, unlike at tree level, the maximal residues of the *loop* amplitude integrand take many different values in $\mathbb{Z}$. Examining the corresponding geometry, the loop amplituhedron, we find starting from 2 loops that it contains a novel feature, namely *internal boundaries*, deep within its boundary structure and find these are the geometric source of the non-unit residues. By an internal boundary we mean a codimension 1 surface separating two regions of

---

[1] It does not automatically follow since there are simple examples of geometries which can be tessellated with positive geometries but which themselves are not positive geometries, as we will see.

opposite orientation, as below:

internal boundary: 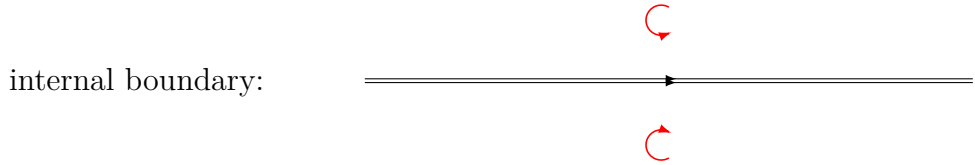

We emphasise that such internal boundaries do not appear in the amplituhedron itself, but deep within its boundary structure. That is, we claim that a certain boundary component of a boundary component of a...of the loop amplituhedron will contain an internal boundary.

There has been tremendous progress in understanding the geometry of the tree level amplituhedron and its canonical form [21–23,25–29,33]. On the loop amplituhedron and its tilings on the other hand are much less is known. A first exploration of the boundaries of the MHV loop amplituhedron was started in [16], where the 2-loop MHV amplitude was computed by tiling the amplituhedron and several cuts were discussed. Then a systematic investigation of the boundaries of the MHV loop amplituhedron was carried out up to four loops in [111] and extended to negative mutual positivity conditions in [112]. Internal boundaries, however, appear to have been missed in the construction of the stratification of the loop amplituhedron in previous works. One possible reason for this is that these boundaries cannot be labelled by the Plücker coordinates that are naturally used to describe the amplituhedron: $\langle Yijkl \rangle, \langle ABij \rangle$ and $\langle A_iB_iA_jB_j \rangle$. For example, internal boundaries arise when computing the all-in-one-point cut of [113] via consecutive single residues. By carefully looking at the boundary corresponding to three loop lines $(A_1B_1), (A_2B_2), (A_3B_3)$ all intersecting the point A, one finds that $\langle AB_1B_2B_3 \rangle = 0$ represents an internal boundary.

We conclude that we need to generalise the definition of 'positive geometry' to allow for such internal boundaries and to incorporate the loop amplituhedron. Thus, we define *generalised positive geometries* (GPGs) which include internal boundaries. Then we introduce a corresponding extension of the recursive definition of the canonical form, by adding an additional term for internal boundaries which should appear with a factor of 2. In doing so, the loop level amplitude can still be obtained as the canonical form of the amplituhedron geometry.

In practise, the most convenient way to compute canonical forms of positive geometries is via tessellation (eg via cylindrical decomposition) rather than explicit use of the recursive definition. It is important to note that tessellations work for these generalised positive geometries (GPGs) just as for positive geometries. The canonical form of a GPG can be computed simply by summing over the canonical forms of the tiles in a tessellation. Indeed, tessellation is closed for GPGs: any geometry that can be tessellated in GPGs is itself a GPG. This differs from positive geometries, which are not-closed under union. We will see examples of geometries that can be tessellated in positive geometries, but which are not themselves a positive geometry.

While the above generalisation of positive geometry to include internal boundaries is perfectly adequate to deal with the loop amplituhedron, it immediately suggests an even more general type of geometry which may be of wider use, namely 'weighted positive geometry'. Any oriented geometry is defined by specifying a region (the geometry) together with its orientation form. However, in order to compute the canonical form of a geometry containing internal boundaries, such as the loop amplituhedron, one needs some extra information encoding which points belong to an external boundary and which belong to an internal boundary. It then seems very natural to define a new object called the weighted geometry (WG). A WG is given by a pair (w, O), where w is an integer valued function we call the weight function and O is the orientation form. The value of the weight function on a point intuitively represents the number of coinciding oriented geometries at that point. For example, an ordinary oriented geometry $X_{\geq 0}$ can be described as a weighed geometry with weight function w(x) = 1 for all x ∈ $X_{\geq 0}$ and zero elsewhere, while an internal boundary will have w(x) = 2 instead (an internal boundary can be viewed as two external boundaries coinciding).

The space of weighted geometries is naturally equipped with two key operators: a sum and a projection onto boundary. The sum generalizes the union of disconnected oriented geometries by allowing for overlaps (the weights on the overlap simply sum). The projection onto boundary operator instead allows one to define the induced weight function and orientation on the boundary of a WG. As a consequence, boundaries of WGs are WGs. This construction allows for a recursive definition of the canonical form and weighted positive geometries (WPG), that treats internal and external boundaries on the same footing and makes the tiling properties of the canonical form trivial. In fact, the canonical form turns out to be a linear operator with respect to the sum of WPGs. Then, because a tiling of a geometry $X_{\geq 0}$ in this new language is nothing but a sum of WPGs, it follows trivially that the canonical form of a sum (union) of WPGs is equal to the sum of the canonical forms.

We start by presenting a simple example of a maximal residue at two loops equal to ±2 rather than ±1, and give its geometrical interpretation as an internal boundary. Then we formally define generalized positive geometries (GPGs), which allow internal boundaries, and their canonical forms. We discuss tilings of GPGs and the key property that *any* geometry that can be tiled by GPGs is a GPG, a property not shared by positive geometries. We describe the algebraic cylindrical decomposition algorithm, originally presented in [51], to compute the canonical form, and use it to identify a class of GPGs. Finally, we introduce a generalisation of GPGs which we call weighted positive geometries.

This chapter is structured as follows. In section 4.1 we present a simple example of a maximal residue at two loops equal to $\pm 2$ rather than $\pm 1$, and give its geometrical interpretation as an internal boundary. In section 4.2 we formally define generalized positive geometries (GPGs), which allow internal boundaries, and their canonical forms. We discuss tilings of GPGs and the key property that *any* geometry that can be tiled by GPGs is a GPG, a property not shared by positive geometries. Then, we introduce a generalisation of GPGs which we call weighted positive geometries. In section 4.3, we describe the algebraic cylindrical decomposition algorithm, originally presented in [51], to compute the canonical form, and use it to identify a class of GPGs. We conclude this chapter in section 4.4 discussing the internal boundary of a geometry defined by a conic and and a plain in $\mathbb{P}^2$.

## 4.1 Two loop maximal cuts and internal boundaries

The purpose of this section is to show that the loop level amplitude can have maximal residues which are not $\pm 1, 0$ and give the geometrical interpretation of this fact. Consider the four point two-loop MHV amplitude integrand written as a volume form in momentum twistor space

$$\text{MHV}(2) = \frac{\langle A_1 B_1 d^2 A_1\rangle \langle A_1 B_1 d^2 B_1\rangle \langle A_2 B_2 d^2 A_2\rangle \langle A_2 B_2 d^2 B_2\rangle \langle 1234\rangle^3}{\langle A_1 B_1 A_2 B_2\rangle \langle A_1 B_1 14\rangle \langle A_1 B_1 12\rangle \langle A_2 B_2 23\rangle \langle A_2 B_2 34\rangle} \times$$
$$\times \left[\frac{1}{\langle A_1 B_1 34\rangle \langle A_2 B_2 12\rangle} + \frac{1}{\langle A_1 B_1 23\rangle \langle A_2 B_2 14\rangle}\right] \quad + \quad A_1 B_1 \leftrightarrow A_2 B_2 \, . \quad (4.1.1)$$

Here we have external momentum twistors $Z_1, .., Z_4 \in \mathbb{C}^4$ and loop integration variables $A_i B_i \in \mathbb{C}^4$ which define a plane through the origin $aA_i + bB_i \in \mathbb{C}^4$ i.e. a line in projective twistor space. Then the bracket notation denotes the determinant of the $4 \times 4$ matrix formed by taking the four twistors inside as columns. We also surpress the Zs, so eg $\langle A_1 B_1 12\rangle := \det(A_1, B_1, Z_1, Z_2)$ etc.

Now we compute the multi-residue corresponding to taking a sequence of residues on

$$\langle A_1 B_1 12\rangle = 0, \langle A_1 B_1 34\rangle = 0, \langle A_2 B_2 12\rangle = 0 \,, \langle A_2 B_2 34\rangle = 0, \langle A_1 B_1 A_2 B_2\rangle = 0 \,,$$
(4.1.2)

followed by a residue on a hidden pole which appears at $\langle 12 B_1 B_2\rangle = 0$. To do this we first parametrise the $4 \times 4$ Z = $(Z_1 Z_2 Z_3 Z_4)$ matrix as the identity and the loops

as

$$\begin{pmatrix} A_i \\ B_i \end{pmatrix} = \begin{pmatrix} 1 & a_i & 0 & -b_i \\ 0 & c_i & 1 & d_i \end{pmatrix} . \tag{4.1.3}$$

For this choice, the brackets read

$$\langle A_i B_i 12 \rangle = b_i, \quad \langle A_i B_i 23 \rangle = d_i, \quad \langle A_i B_i 34 \rangle = c_i, \quad \langle A_i B_i 14 \rangle = a_i ,$$
$$\langle A_1 B_1 A_2 B_2 \rangle = -(b_1 - b_2)(c_1 - c_2) - (a_1 - a_2)(d_1 - d_2), \quad \langle 1234 \rangle = 1 ,$$
$$\langle A_i B_i d^2 A_i \rangle \langle A_i B_i d^2 B_i \rangle = da_i db_i dc_i dd_i . \tag{4.1.4}$$

Omitting the differential, the amplitude (4.1.1) in these coordinates reads

$$\text{MHV}(2) = -\frac{a_2 d_1 + a_1 d_2 + b_2 c_1 + b_1 c_2}{a_1 a_2 b_1 b_2 c_1 c_2 d_1 d_2 \left( (a_1 - a_2)(d_1 - d_2) + (b_1 - b_2)(c_1 - c_2) \right)} . \tag{4.1.5}$$

Now we take the first four residues in (4.1.2) namely $b_1 = 0, c_1 = 0, b_2 = 0, c_2 = 0$. We see that the complicated factor in the denominator factorises thus revealing a new pole[2] and we get

$$-\frac{a_2 d_1 + a_1 d_2}{a_1 a_2 d_1 d_2 (a_1 - a_2)(d_1 - d_2)} . \tag{4.1.6}$$

Then we take the residue in $a_1$ at $a_1 = a_2$, obtaining

$$-\frac{(d_1 + d_2)}{a_2 d_1 d_2 (d_1 - d_2)} . \tag{4.1.7}$$

We continue by taking the residue in $d_1$ at $d_1 = d_2$, obtaining

$$-\frac{2}{a_2 d_2} , \tag{4.1.8}$$

up to an overall sign due to the ordering of the differential $d^2 a_i d^2 b_i d^2 c_i d^2 d_i$. This form has clearly maximal residues equal to $\pm 2$, in apparent contradiction of the consequence of this being the canonical form of a positive geometry.

**Geometrical interpretation**

Let's try to understand how this factor of two appears geometrically from the amplituhedron. First we take boundaries corresponding to the four residues at $\{\langle A_1 B_1 12 \rangle = 0, \langle A_1 B_1 34 \rangle = 0, \langle A_2 B_2 12 \rangle = 0, \langle A_2 B_2 34 \rangle = 0\}$. The order in which these are performed is not important and the resulting geometry has each loop line $(A_i B_i)$ described by a point $A_i$ in the segment $\overline{12}$ and a point $B_i$ in $\overline{34}$, together with a further mutual positivity condition $\langle A_1 B_1 A_2 B_2 \rangle > 0$. It is natural then to parametrise $A_i, B_i$ as $A_i = Z_1 + a_i Z_2$ and $B_i = Z_3 + d_i Z_4$, so that the geometry is described by the inequalities

$$a_i > 0, \quad d_i > 0, \quad -(a_1 - a_2)(d_1 - d_2) > 0 \ . \tag{4.1.9}$$

---

[2]Such poles have been observed in the amplitude previously and been dubbed composite residues [11]. See also the example in the introduction.

Notice that the mutual positivity inequality factorizes in to the product of two terms $a_1 > a_2, d_1 < d_2$ or $a_1 < a_2, d_1 > d_2$. This is just the geometrical version of composite residues mentioned in the introduction and above (4.1.6). Because of this factorisation, the corresponding geometry is given by two regions

$$\mathcal{R}_1 := \{a_1, a_2, d_1, d_2 \mid a_1 > a_2 > 0 \wedge d_2 > d_1 > 0\} \ ,$$
$$\mathcal{R}_2 := \{a_1, a_2, d_1, d_2 \mid a_2 > a_1 > 0 \wedge d_1 > d_2 > 0\} \ . \tag{4.1.10}$$

This geometry is illustrated in the picture

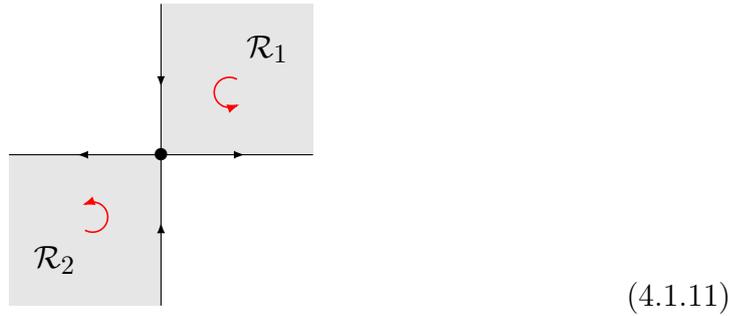

$$\tag{4.1.11}$$

where the x axis corresponds to increasing $a_1 - a_2$ and the y axis increasing $d_2 - d_1$.

These regions share only a codimension 2 boundary that is contained on the surface $(a_1 - a_2) = 0, (d_1 - d_2) = 0$. Both regions $\mathcal{R}_1, \mathcal{R}_2$ come equipped with an orientation induced by the bulk geometry, which in this case is the same for both regions. Each of these two regions is clearly a positive geometry, with canonical forms

$$\Omega(\mathcal{R}_1) = -\frac{1}{a_1 d_2 (a_1 - a_2)(d_1 - d_2)} \ ,$$
$$\Omega(\mathcal{R}_2) = -\frac{1}{a_2 d_1 (a_2 - a_1)(d_2 - d_1)} \ . \tag{4.1.12}$$

The sum of these correctly reproduces the corresponding residue of the amplitude (4.1.6). Since the two regions share a lower codimension boundary we have to see what happens on this boundary to decide if the union is or isn't a positive geometry. We can consider for example the $(d_2 - d_1) = 0$ boundary by sending $d_2 \to d_1$. This corresponds to projecting onto the x axis of (4.1.11) and thus looks as

$$\xleftarrow{\qquad\bullet\qquad\qquad}_{\mathcal{R}'_2 \qquad\qquad \mathcal{R}'_1} \tag{4.1.13}$$

We again get two regions

$$\begin{aligned}\mathcal{R'}_1 &:= \{a_1, a_2, d_1 \mid d_1 > 0 \wedge a_1 > a_2 > 0\}\,,\\ \mathcal{R'}_2 &:= \{a_1, a_2, d_1 \mid d_1 > 0 \wedge a_2 > a_1 > 0\}\,,\end{aligned} \tag{4.1.14}$$

but now, since we approach the boundary from two different directions, as we explain in detail in appendix 3.3.1, the two induced orientations are opposite. The region $\mathcal{R'}_1$ and $\mathcal{R'}_2$ share a codimension 1 boundary $(a_1 - a_2) = 0$ where the orientation changes sign. We call this an *internal boundary*.

We see that the boundary (4.1.14) is in fact not oriented (rather it flips orientation on the internal boundary $a_1 = a_2$). Part of the definition of positive geometry in [30] is that it is oriented and (by the recursive nature of the definition so are all boundaries etc.) We conclude that a generalisation of the concept of positive geometry is needed to accommodate the loop amplituhedron.

Since both $\mathcal{R'}_1$ and $\mathcal{R'}_2$ by themselves *are* positive geometries on the other hand, their respective residues at $(a_2 - a_1) = 0$ are equal to the canonical form of this. Therefore the residue on the internal boundary of $\Omega(\mathcal{R'}_1) + \Omega(\mathcal{R'}_2)$ will be equal to twice the canonical form of a positive geometry. Note that if $\mathcal{R}'_1$ and $\mathcal{R}'_2$ instead had the same orientation as each other then $(a_2 - a_1) = 0$ would be a spurious boundary and the resulting residue would vanish.

We have thus seen that the loop amplitude has non unit maximal residues and the geometrical origin of this is that the amplituhedron contains internal boundaries. In the next section we will see how to formalize what we have observed in this simple example and generalize the definition of the canonical form to geometries with internal boundaries which will then accommodate the loop amplituhedron.

## 4.2 Generalized Positive Geometries and WPGs

Two intervals of the same orientation which share a common boundary point are equivalent to the larger interval

$$[a, b] \cup [b, c] = \;\;\underset{a\quad\quad b\quad\quad c}{\bullet\!\!\to\!\!\bullet\!\!\to\!\!\bullet}\;\; = \;\;\underset{a\qquad\qquad\quad c}{\bullet\!\!\longrightarrow\!\!\bullet} \tag{4.2.1}$$

with the shared boundary point absent. This is then reflected in the addition of the corresponding canonical forms

$$\Omega([a,b]) + \Omega([b,c]) = \Omega([a,c]) \ . \tag{4.2.2}$$

In this case the point b is sometimes called a spurious boundary.

However two intervals of different orientations sharing a common boundary point

$$[a,b] \cup [c,b] = \quad \underset{a \quad\quad b \quad\quad c}{\bullet\!\!\longrightarrow\!\!\bullet\!\!\longleftarrow\!\!\bullet} \tag{4.2.3}$$

are not a positive geometry (for example it is not oriented). Nevertheless it is natural to associate to this geometry the corresponding canonical form

$$\Omega([a,b]) + \Omega([c,b]) \;=\; 2\frac{dx}{x-b} - \frac{dx}{x-a} - \frac{dx}{x-c} \ . \tag{4.2.4}$$

The point b is then special, separating two regions of opposite orientation, and we refer to this as an 'internal boundary'. It has residue twice that of each of the two individual boundaries there. This is of course exactly what we have seen occurring for the two loop amplitude in the previous section.

### 4.2.1 Generalised Positive Geometry and its Canonical Form

This then motivates a generalisation of the concept of positive geometry to incorporate such internal boundaries. Internal boundaries separate two regions of opposite orientation. So we define a generalised positive geometry as one whose internal and external boundaries are both generalised positive geometries. Both external and internal boundaries must lie inside a space defined by $f(x_i) = 0$ for some non-factorisable polynomial f. A particular subspace $f(x_i) = 0$ could contain both internal and external boundaries each of which must be (generalised) positive geometries with respective canonical forms $\omega_{\text{int}}$ and $\omega_{\text{ext}}$. Then we define the canonical form recursively as

$$\text{Res}_{f=0}\Omega = \omega_{\text{ext}} + 2\omega_{\text{int}} \tag{4.2.5}$$

Or equivalently

$$\lim_{f \to 0} f\Omega \;=\; df \wedge (\omega_{\text{ext}} + 2\omega_{\text{int}}) \ . \tag{4.2.6}$$

We see an extra term compared to the original canonical form for positive geometries (3.3.7) giving the factor of 2 associated with internal boundaries. The starting point for the recursion is the same as before, so the 0-dimensional geometries are oriented points with canonical form $\pm 1$ according to the orientation. Note that the orientation of the interior boundary is unambiguously inherited from that of the bulk just as for the exterior boundaries.

So consider the 1d example in (4.2.3). This has external boundaries (with negative orientation) at $f_a(x) = x - a = 0$ and $f_c(x) = x - c = 0$ and an internal boundary with positive orientation at $f_b(x) = x - b = 0$. Then one sees that the canonical form (4.2.4) satisfies the recursive relation (4.2.5) at all three points.

Now consider a 2d example.

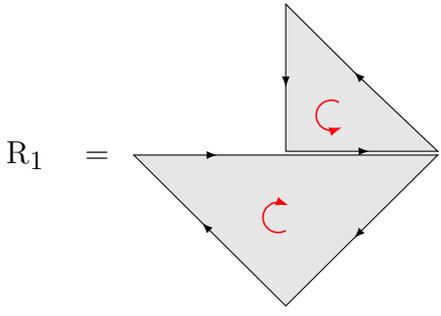

$$R_1 \quad = \qquad\qquad\qquad\qquad\qquad\qquad\qquad\qquad (4.2.7)$$

Here the x axis contains both an external boundary and an internal boundary. The canonical form can be straightforwardly obtained by simply adding together the canonical forms of the two triangles[3] giving

$$\Omega(R_1) = \frac{dx\,dy}{xy(x+y-1)} + \frac{2dx\,dy}{y(x+y+1)(x-y-1)} \ . \qquad (4.2.8)$$

Now we can see how this satisfies the recursive definition (4.2.5) along the x axis. Taking $f = y$ in (4.2.5) we have

$$\operatorname{Res}_{y=0} \Omega = dx\left(\frac{1}{x} - \frac{1}{x+1}\right) + 2dx\left(\frac{1}{x-1} - \frac{1}{x}\right) = \Omega([-1,0]) + 2\Omega([0,1]) \ , \qquad (4.2.9)$$

which is exactly as predicted by (4.2.5) since the external boundary on the x axis is the interval $[-1,0]$ and the internal boundary is $[0,1]$.

Note that internal boundaries give a contribution to the canonical form of twice that of a standard external boundary. One might think therefore that a leading singularity of any such a generalised positive geometry must be $0, \pm 1$ (as for a positive geometry) or $\pm 2$ if there is an internal boundary present. However there can be internal boundaries inside internal boundaries, leading to higher maximal residues. A very simple example of this is the region consisting of the entire plane,

---

[3]This key tessellation feature of positive geometries and the canonical forms is inherited by (and is indeed more powerful for) the generalised positive geometries as we will see.

but with the four quadrants having alternating orientations

$$R_2 \;=\; \begin{gathered}\text{[diagram]}\end{gathered} \tag{4.2.10}$$

Here each quadrant has exactly the same canonical form $dx\,dy/(xy)$ and so the geometry has non-zero canonical form

$$\Omega(R_2) = \frac{4 dx\, dy}{xy}\ . \tag{4.2.11}$$

So taking the residue on the (internal) boundary $x = 0$ gives $\lim_{x\to 0}(x\Omega) = 2dx \wedge \left(\frac{2dy}{y}\right) = 2dx \wedge \omega_{\text{int}}$ with $\omega_{\text{int}} = \frac{2dy}{y}$ satisfying (4.2.5). Here the internal boundary in $x = 0$ is $(-\infty, 0) \cup (\infty, 0)$ and this itself has an internal boundary at $y = 0$. Thus it has canonical form $\frac{2dy}{y}$ (as we get directly from (4.2.4) by taking $b \to 0$, $a, c \to \infty$ and $x \to y$). So the leading singularity at $x = y = 0$ is 4.

It is also possible to get a odd leading singularity. For example, simply take three of the four quadrants from the previous example

$$R_3 \;=\; \begin{gathered}\text{[diagram]}\end{gathered} \tag{4.2.12}$$

This geometry has canonical form

$$\Omega(R_2) = \frac{3 dx\, dy}{xy}\ . \tag{4.2.13}$$

Here taking the residue on the boundary $x = 0$ gives $\lim_{x\to 0}(x\Omega) = dx \wedge \left(\frac{dy}{y}\right) + 2dx \wedge \left(\frac{dy}{y}\right)$ in agreement with (4.2.5). This time $x = 0$ contains the external boundary $(-\infty, 0)$ as well as the internal boundary $(0, \infty)$ both of which have canonical forms $\frac{dy}{y}$. The leading singularity at $x = 0$, then $y = 0$ is 3.

## 4.2.2 Weighted Positive Geometry and its Canonical Form

Although the definition of the canonical form (4.2.5) is extremely compact, it has the downside of treating external and internal boundaries on a different footing. On the right hand side we have a weighted sum of canonical forms, so it's tempting to rewrite this as the canonical form of a weighted sum of geometries. In this section we make this intuition precise by generalising what we mean by a geometrical region to include a weight taking arbitrary integer values. This articulates the idea of having multiple coinciding geometries which we naturally have when two regions meet on an internal boundary. This concept will allow us to give an explicit formula for maximal residues.

Firstly, recall that orientation on a space can be described by a top form where we are only really interested in the sign of the top form. So orientation is the equivalence class of real top forms modulo positive rescaling, $O \in \Omega^d/\sim$ where $O \sim \lambda O$ for any $\lambda > 0$. Now we extend this to define the weighted orientation as a pair $(w, O) : X \to \left(\mathbb{Z}, \Omega^d(X)\right)/\sim$ where here the equivalence relation involves positive or negative rescaling with the negative case also flipping the weight $w$

$$(w, O) \sim (\text{sign}(\lambda)w, \lambda O') \qquad \lambda \neq 0 \;. \tag{4.2.14}$$

Thus changing the orientation is equivalent to flipping the sign of $w$. In practise we can of course always choose coset representatives of (4.2.14) such that $w > 0$ and we will mostly assume this from now on.

Weighted geometries have a natural additive structure. At any point $x \in X$ we define the sum of two weighted geometries $(w_1, O_1) \oplus (w_2, O_2)$ as

$$(w_1, O_1) \oplus (w_2, O_2) = (w_1 + \text{sign}(\lambda)w_2, O_1) \;, \tag{4.2.15}$$

where $\lambda$ is such that $O_1 = \lambda O_2$ [4]. Notice, that because of (4.2.14) this sum is symmetric and the identity element $(0, O)$ is unique.

So now we define a weighted geometry entirely by specifying its weight function and its orientation, rather than directly defining a region $X_\geq$. The region $X_\geq$ can be reconstructed by simply defining it as the set of points where $w \neq 0$. Boundaries are then places where the weighted orientation is discontinuous and they divide regions inside which the weighted orientation is continuous (and therefore $w$ is constant). We will shortly define a canonical form for weighted geometries, the existence and uniqueness of which will define weighted positive geometries (WPG). For multivariate residues to be well defined, we insist that these boundaries must be subsets of algebraic varieties – so these regions are semi-algebraic sets.

---

[4]Note that the x dependence is suppressed in the equation and that the function $\text{sign}(\lambda(x))$ is negative if the two orientations $O_1(x)$ and $O_2(x)$ are opposite and positive if they match.

A key aspect of this construction is that both weights and orientation on the boundaries are uniquely induced from those in the bulk. This happens as follows. Any boundary component, $\mathcal{C}$ can be defined through a polynomial p(x) = 0. Then one side of the boundary is p(x) > 0, with weight and orientation $(w^+, O^+)$, whereas the other side is p(x) < 0 with $(w^-, O^-)$. Now the region p(x) > 0 naturally induces the orientation $O^+|_\mathcal{C}$ on the boundary of the region p > 0 in the standard way (see appendix 3.3.1), so $O^+ = dp \wedge O^+|_\mathcal{C}$. The region p(x) < 0 on the other hand naturally induces the orientation $O^-|_\mathcal{C}$ on the boundary, where $O^- = -dp \wedge O^-|_\mathcal{C}$[5].

Given a codimension-1 variety $\mathcal{C} \in X$, then we define a projection operator $\Pi_\mathcal{C}$ that maps weighted orientations (w, O) on X to weighted orientations on $\mathcal{C}$ as

$$\Pi_\mathcal{C}(w, O) = (w^+|_\mathcal{C}, O^+|_\mathcal{C}) \oplus (w^-|_\mathcal{C}, O^-|_\mathcal{C}) . \quad (4.2.16)$$

Choosing representatives such that w > 0, we can give the following two dimensional illustration of the induced weights and orientations (denoted with arrows):

$$\text{(diagram with three cases: } w+w', \ w-w' \ (w > w'), \ w'-w \ (w < w') \text{)}$$

$$(4.2.17)$$

Note that w or w′ could have been zero in which case we have an external boundary. This definition implies that, for w(x) = 1 in $X_{>0}$ and 0 otherwise, $w|_\mathcal{C}$ will be equal to 2 on internal boundaries, to 1 on external boundaries and 0 otherwise. In this formulation internal and external boundaries are not distinguished. Furthermore note that if $w^+ = w^-$ and $\lambda < 0$ (See (4.2.14)) then there are equivalent weighted orientations on both sides (meaning the induced orientations are opposite) and thus there is no genuine boundary there (it is a spurious boundary).

An important observation now is that the projection $\Pi$ is a linear operator

$$\Pi_\mathcal{C}\Big((w_1, O_1) \oplus (w_2, O_2)\Big) = \Pi_\mathcal{C}(w_1, O_1) \oplus \Pi_\mathcal{C}(w_2, O_2) , \quad (4.2.18)$$

which can be easily proven from the definitions.

Now we can define a weighted positive geometry as a weighted geometry possessing a canonical form. The definition of the canonical form of a weighted geometry is defined recursively such that the residue of the canonical form on $\mathcal{C}$ is the canonical form of the geometry projected on $\mathcal{C}$:

$$\text{Res}_{\mathcal{C}} \Omega(w, O) = \Omega(\Pi_{\mathcal{C}}(w, O)) \, . \tag{4.2.19}$$

---

[5]The minus sign arises from the fact that the normal vector pointing inward the region p < 0 is $-\partial_p$.

The recursion starts by defining the canonical form of a zero dimensional weighted geometry (for which O is a 0 form, simply a scalar) as the product of w with the sign of O

$$\Omega(w, O) = w \, \text{sign}(O) \qquad \text{(In zero dimensions)} \, . \tag{4.2.20}$$

In zero dimensions therefore the canonical form is a linear operator

$$\Omega\Big((w_1, O_1) \oplus (w_2, O_2)\Big) = \Omega\Big(w_1 + \text{sign}(\lambda)w_2, O_1\Big) = (w_1 + \text{sign}(\lambda)w_2) \times \text{sign}(O_1)$$
$$= w_1 \text{sign}(O_1) + w_2 \text{sign}(O_2) = \Omega(w_1, O_1) + \Omega(w_2, O_2) \tag{4.2.21}$$

where recall $O_1 = \lambda O_2$. It follows by induction from the recursive definition (4.2.19) and linearity of the projection operator $\Pi$ (4.2.18) that this linearity property of $\Omega$ then holds for spaces of arbitrary dimension:

$$\Omega\Big((w_1, O_1) \oplus (w_2, O_2)\Big) = \Omega(w_1, O_1) + \Omega(w_2, O_2) \, . \tag{4.2.22}$$

Remarkably we have the feature that we can freely sum arbitrary (even overlapping) WPGs! It also follows directly from this that

$$\Omega\Big((\lambda w, O)\Big) = \lambda \Omega(w, O) \, . \tag{4.2.23}$$

Now given a sequence of boundaries $\{\mathcal{C}_1, \cdots, \mathcal{C}_n\}$, we can follow n steps of the recursion (4.2.19) and write the multi-residue of a canonical form as the canonical form of the multiply induced boundary

$$\text{Res}_{\mathcal{C}_1, \cdots, \mathcal{C}_n} \Omega(w, O) = \Omega(\Pi_{\mathcal{C}_1, \cdots, \mathcal{C}_n}(w, O)) \, , \tag{4.2.24}$$

Then taking n = d, the dimension of X, we obtain an expression for the maximal residues in terms of the canonical form at a point (4.2.20)

$$\text{Res}_{\mathcal{C}_1, \cdots, \mathcal{C}_d} \Omega(w, O) = w_{\mathcal{C}_1, \cdots, \mathcal{C}_d} \times \text{sign}\left(O_{\mathcal{C}_1, \cdots, \mathcal{C}_d}\right) \, , \tag{4.2.25}$$

where $(w_{\mathcal{C}_1, \cdots, \mathcal{C}_d}, O_{\mathcal{C}_1, \cdots, \mathcal{C}_d}) = \Pi_{\mathcal{C}_1, \cdots, \mathcal{C}_d}(w, O)$. This last equation can also be used as a direct, non-recursive, definition of the canonical form by giving all its maximal residues (the canonical form is completely determined by its maximal residues).

Note that generalised positive geometries, defined in previous subsections, should simply be WPGs for which the weight function (in the bulk) is w = ±1, 0 everywhere. Similarly positive geometries are WPGs for which the weight function w = ±1, 0 everywhere (so they are also GPGs) but also the induced weight function on all nested boundary components is also always ±1, 0.

To check this we need to show that the canonical form for the GPGs defined by (4.2.5) and that for the WPGs (4.2.19) are equivalent. By equivalent we mean that given any GPG $X_{\geq 0}$ with orientation O we associate a weighted geometry with orientation O and weight w such that w(x) = 1 for all $x \in X_{\geq 0}$ and zero otherwise, and then

$$\Omega(X_{\geq 0}) = \Omega(w, O) \, . \tag{4.2.26}$$

Now notice that the projection of (w, O) onto $\mathcal{C}$ (described in (4.2.16) and above) will have induced weight 1 or 2, depending on whether it is an external or internal boundary. So we can write $\Pi_{\mathcal{C}}(w, O) = (w_{\text{ext}}, O_{\text{ext}}) \oplus (w_{\text{int}}, O_{\text{int}})$ where $w_{\text{ext}} = 1$ on external boundaries and zero elsewhere whereas $w_{\text{int}} = 2$ on internal boundaries and zero elsewhere. Then, it follows that if we apply (4.2.19) we get

$$\text{Res}_{\mathcal{C}} \Omega(w, O) = \Omega\Big(\Pi_{\mathcal{C}}(w, O)\Big) = \Omega((w_{\text{ext}}, O_{\text{ext}}) \oplus (w_{\text{int}}, O_{\text{int}})) =$$
$$= \Omega(w_{\text{ext}}, O_{\text{ext}}) + 2\Omega\left(\frac{w_{\text{int}}}{2}, O_{\text{int}}\right) \, . \tag{4.2.27}$$

Now since $w_{\text{ext}}$ and $\frac{1}{2} w_{\text{int}}$ are both functions respectively equal to 1 on internal and external boundaries and equal to 0 otherwise, they represent with their orientations the external and internal boundaries as GPGs. This is then precisely the original defining equation of the canonical form (4.2.5). Since we showed that the recursion (4.2.19) and (4.2.5) have the same form it follows that the two definitions of the canonical form give the same result.

Finally, let us illustrate with a slightly more involved example, returning to the case considered in (4.2.12) from this new perspective

$$R_3 \quad = \quad \begin{array}{c}\text{[diagram]}\end{array} \tag{4.2.28}$$

Here we see the induced weights 2,1 on the codimension 1 boundaries x = 0, y = 0. Considering these boundaries themselves they then induce the weight 2 + 1 = 3 at the origin with positive orientation on the y axis, negative on the x axis. This is in line with (4.2.25) and the corresponding maximal residue $\text{Res}_{y=0,x=0}\Omega_{R_3} = -\text{Res}_{x=0,y=0}\Omega_{R_3} = 3$.

### 4.2.3 Tilings of WPGs

As we have seen in section 3.3.4, it can happen that a non positive geometry can be tiled by positive geometries - so the space of positive geometries is not closed under the union. This is because even if the orientation of the $X^{(i)}_{\geq 0}$ tiling $X_{\geq 0}$ matches on codimension 1 boundaries this does not imply that they will necessarily match on the boundaries of boundaries etc. This can then give rise to internal boundaries. As examples consider the following two geometries:

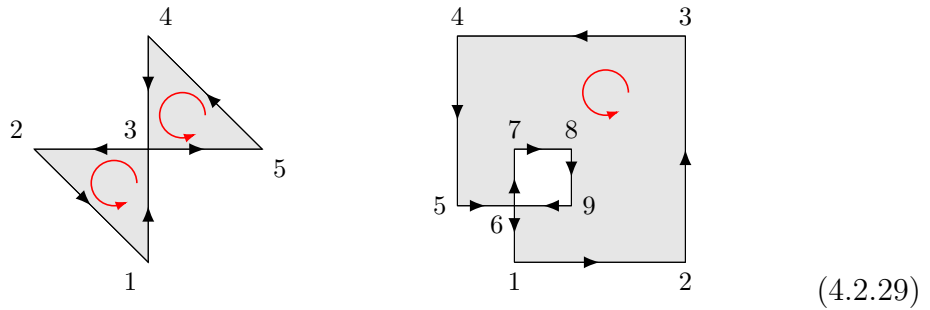

(4.2.29)

Both can be tiled by positive geometries. The first can be tiled as the oriented union of two triangles while the second as the oriented union of 4 rectangles with matching orientation. Both examples are not positive geometries themselves however. This can be seen graphically observing the orientation of the boundary, the edge 14 in the first example and 59 in the second look like (4.2.3) and have an internal boundary.

Both of these examples *are* generalised positive geometries however. And we claim more generally that if $X_{\geq 0}$ is tiled by a set of generalized positive geometries $X^{(i)}_{\geq 0}$ tiling $X_{\geq 0}$ then $X_{\geq 0}$ is a generalized positive geometry and its canonical form is given by (3.3.16). This is essentially trivial from linearity of the WPG canonical form and the definition of a GPG as a WPG with weight $0, \pm 1$ everywhere.

A beautiful consequence of the WPG formalism is that it also yields a simple proof of the tiling property of positive geometries (3.3.16). Indeed this follows trivially from the fact that the canonical form is a linear operator for weighted positive geometries (4.2.22). Translating (3.3.16) into WPG language, on the right hand side we have the canonical form of a region which is equivalently a weight function $(w, O)$ with $w = 1$ in the region and 0 outside. This region has a tesellation with tiles $(w_i, O_i)$ with

$$(w, O) = \bigoplus_i (w_i, O_i) \,. \tag{4.2.30}$$

Now linearity of WPGs give

$$\Omega(w, O) = \sum_i \Omega(w_i, O_i) \,, \qquad (4.2.31)$$

which proves (3.3.16).

## 4.3 GCD algorithm and explicit characterisation of GPGs / WPGs

In the previous subsection we gave an implicit (recursive) definition of generalised positive geometry and weighted positive geometries. Here we speculated on whether it is possible to give more explicit characterisations, so we can know in advance if a particular region is a GPG/WPG or not. The fact that GPGs/WPGs are closed under tiling as discussed in the previous subsection already suggests they ought to be more amenable to a direct characterisation than PGs. For example any characterisation of PGs would have to exclude the two examples in (4.2.29).

We first note that if we restrict ourselves to a specific class of geometry which we call *multi-linear geometries* then the characterisation is very simple. Multi-linear geometries are geometries defined by *multi-linear inequalities* in some coordinates. Note that although this may be a big restriction of the full space of positive geometries it nevertheless provides a very wide class of cases. Crucially it is straightforward to see that the amplituhedron is a multi-linear geometry. The defining inequalities of the amplituhedron are given in terms of either minors of a C matrix, or alternatively determinants of the form $\langle YL_i...\rangle$. Thus by choosing components of either the C matrix, and/or the $Y, L_i$ as coordinates, the resulting inequalities will be multi-linear in those coordinates (simply because the determinant is a multi-linear function of its components).

Note here that it is important not to confuse multi-linear geometries with linear geometries. Many of the toy examples one considers are linear geometries where defining inequalities can be given which are linear in all variables. These then have straight edges, flat planes etc. Multi-linear geometries can however be curvey. For example in 2d, boundaries of multi-linear geometries have the form $axy+bx+cy+d = 0$ (in some coordinates) which correspond to hyperbolas as well as straight lines. On the other hand circles or ellipses would include the non multi-linear terms $x^2, y^2$ and are not multi-linear. They can however be boundaries of positive geometries [30]. We will shortly return to this point.

We first claim that *any* multi-linear geometry is a (generalised) positive geometry. We can show this by explicitly and uniquely computing the canonical form for multi-linear geometries using a algorithm called Generic Cylindrical Decomposition (GCD). Given a multi-linear geometry, first use cylindrical decomposition (see also [51]) which recasts any region as a disjoint union of regions $\mathcal{R}_i$ of the form

$$\mathcal{R}_i := \{x_1, \cdots, x_d\} \quad \text{st} \quad \begin{cases} a_1 < x_1 < b_1 \\ a_2(x_1) < x_2 < b_2(x_1) \\ \cdots \\ a_d(x_1, \cdots, x_{d-1}) < x_d < b_d(x_1, \cdots, x_{d-1}) \end{cases} \quad (4.3.1)$$

for some functions $a_j, b_j$. Now changing variables to:

$$x'_j = -\frac{x_j - a_j}{x_j - b_j}, \quad (4.3.2)$$

then $\mathcal{R}_i$ becomes

$$\mathcal{R}_i := \{x'_1, \cdots, x'_d\} \quad \text{st} \quad x'_j > 0 \quad \text{for all } j. \quad (4.3.3)$$

In the new coordinates $\mathcal{R}_i$ is thus a simplex-like positive geometry with canonical form

$$\Omega(\mathcal{R}_i) = \prod_{j=1}^{d} \frac{dx'_j}{x'_j}. \quad (4.3.4)$$

An algorithm for computing such decomposition is readily available on `Mathematica` [114] under the function `GenericCylindricalDecomposition`. We discussed in sec 3.3.4 under a *rational* map the canonical forms map to each other. So as long as the change of variables (4.3.2) is rational then we have that in the original coordinates

$$\Omega(\mathcal{R}_i) = \prod_{j=1}^{d} \left( \frac{1}{x_j - a_j} - \frac{1}{x_j - b_j} \right) dx_j \quad (4.3.5)$$

and the canonical form of the full region can then be obtained by summing the contributions from all the $\mathcal{R}_i$. We see how multi-linearity is crucial here. The inequalities in (4.3.1) must arise from the defining inequalities of our region which are multi-linear. This ensures that the resulting functions $a_i$ and $b_i$ will be rational functions and thus the change of variables (4.3.2) is rational.

Let us illustrate some of these points now with a couple of examples shown in figure 4.1.

Firstly we have a region $R_1$ sandwiched between a hyperbola and a line. It is defined by the inequalities $x > 0, xy > 7, x + y < 8$. Cylindrical decomposition rewrites this as a single region written in the form of (4.3.1) as $1 < x < 7 \quad \frac{7}{x} < y <$

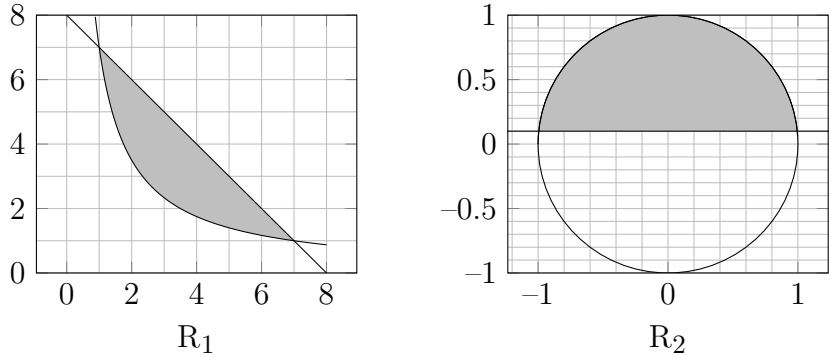

Figure 4.1: Two examples of positive geometries obtained by sandwiching a conic and a straight line. The first, involving a hyperbola, is a multi-linear geometry whereas the second is not multi-linear, but is still a positive geometry.

$8-x$. Then the simple replacement rule of (4.3.5) yields the canonical form

$$\Omega(R_1) = \left(\frac{dx}{x-1} - \frac{dx}{x-7}\right)\left(\frac{dy}{y-\frac{7}{x}} - \frac{dy}{x+y-8}\right) = -\frac{6\,dx\,dy}{(x+y-8)(xy-7)}\,. \quad (4.3.6)$$

The second example $R_2$, found in [30], is not a multi-linear geometry. This is the region between a circle and a line and is defined by the inequalities $x^2 + y^2 < 1, y > 1/10$. Let us see what happens if we attempt the same procedure to obtain its canonical form. Here cylindrical decomposition rewrites the region as

$$-\frac{3\sqrt{11}}{10} < x < \frac{3\sqrt{11}}{10}, \quad \frac{1}{10} < y < \sqrt{1-x^2}\,. \quad (4.3.7)$$

But now we encounter a problem. We see that, due to the square root $\sqrt{1-x^2}$, the change of variables needed in (4.3.2) will no longer be rational and the above procedure no longer works.

So we have seen that cylindrical decomposition gives the unique canonical form as long as all the resulting functions $a_i, b_i$ in (4.3.1) are rational. This is clearly the case for multi-linear geometries, but could also be the case for more general geometries. Further it may be possible to change coordinates so that only in the new coordinates the cylindrical decomposition map (4.3.2) is rational. So in general we can characterise generalised positive geometries to be those for which there exist coordinates and an ordering of these coordinates such that cylindrical decomposition yields a map (4.3.2) which is rational.

For example let us return to the region $R_2$ in figure 4.1 for which the cylindrical decomposition method of obtaining the canonical form didn't work as it produces an irrational map (4.3.2). Now the circle is the classic example of a rational variety.

This is a variety that has a parametrisation $t_i$ in terms of which its embedding coordinates $x_i(t_j)$ are rational functions and there is a rational inverse map $t_i(x_j)$. In this case it has a rational parametrisation given by

$$x(t) = \frac{2t}{1+t^2} \qquad y(t) = \frac{1-t^2}{1+t^2} \ . \tag{4.3.8}$$

with an inverse map from $\mathbb{R}$ onto the circle embedded in $\mathbb{R}^2$ which is also rational

$$t(x,y) = \frac{x}{1+y} \ . \tag{4.3.9}$$

Here the parameter t has the geometrical interpretation of the projection of a point on the circle, from the point $(0, -1)$ to the x axis. But this projection can clearly be extended to any point in $\mathbb{R}^2$ not just those points on the circle. So consider the change of variables from x to t, $(t, y) \to (x, y) = (t(1+y), y)$. In the new variables the region $R_2$ then has cylindrical decomposition

$$-\frac{3}{\sqrt{11}} < t < \frac{3}{\sqrt{11}}, \quad \frac{1}{10} < y < \frac{1-t^2}{t^2+1} \ , \tag{4.3.10}$$

which is now rational. The replacement rule (4.3.5), then gives the canonical form

$$\Omega(R_2) = -\frac{6\sqrt{11}\,dt\,dy}{(10y-1)\left(t^2(y+1) + y - 1\right)} = -\frac{6\sqrt{11}\,dx\,dy}{(10y-1)\left(x^2+y^2-1\right)} \ , \tag{4.3.11}$$

which is in precise agreement with the canonical form for this geometry found in [30] (see figure 1) using the recursive definition of the canonical form.

In general, if a codimension 1 boundary of a region is a rational variety, then changing coordinates from $x_i$ to $t_i, x_n$ will rationalise the final step in the cylindrical decomposition involving that boundary. In other words cylindrical decomposition in those variables will give the boundary in the form $x_n < x_n(t_i)$ which is a rational function as we saw in the above example which gave $y < y(t) = (1-t^2)/(1+t^2))$. This all suggests there should be a more intrinsic definition of a GPG/WPG in terms of rational varieties.

We would like to conclude this section highlighting that when applying the GCD algorithm it is important to check the orientation of the coordinate map. If the sign of the measure, $\prod_i \langle Y d^m Y_i \rangle$ is positive, the coordinate chart is orientation preserving, if negative it is orientation reversing and if zero it is degenerate. The reversed orientation contributes with minus sign to the canonical form. If the measure vanishes anywhere in the geometry, we need to split it into regions where the measure is everywhere non-vanishing and sum the result for the different regions with the sign contributions coming form the sign of the measure.

As a simple example consider the triangle defined by $Z_1 = (1, 0, 1)$, $Z_2 = (1, 0, -1)$,

$Z_1 = (0, 1, 0)$. On the oriented projective space (equivalent to a sphere) the coordinate chart $Y = (x, y, 1)$ does not contain this triangle ($Z_2$ clearly lies outside this coordinate chart) so we need the additional chart $Y = (x', y', -1)$. The first chart covers the northern hemi-sphere and the second the southern hemisphere. The triangle is defined by the inequalities $\langle Y i i{+}1\rangle > 0$. In the first chart this gives the region $y > 0, x > 1$, yielding canonical form $dxdy/((x-1)y)$ and in the second chart it gives the region $y' > 0, x' > 1$, yielding canonical form $-dx'dy'/((x'-1)y')$. The additional minus sign in the second case arises from the orientation of the coordinates map $\langle Y d^2 Y\rangle = 2dxdy = -2dx'dy'$, negative in the second chart. Now *after* we have obtained the canonical forms in the two charts we need to add them together. One way to do this is to covariantise the two forms above and then add them together. In order to covariantise it will be useful to introduce the additional vertex $Z_* = (1, 0, 0)$, the point where the boundary of the triangle meets the equator hence moving from the northern to the southern hemisphere. The result will just be the sum of the two triangles obtained by splitting the big triangle along the equator (see picture).

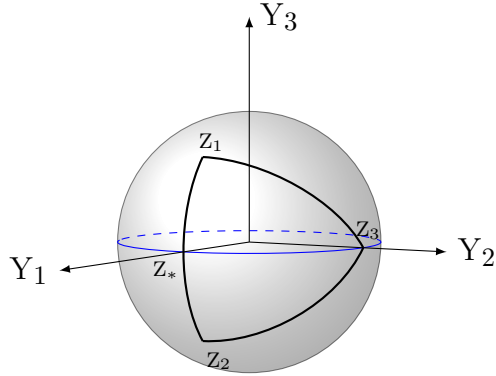

However it is also possible to add the two forms together directly at the level of coordinates. To do this we first realise that now, at the level of the form, we can safely project from the sphere to $P^2$. For the second chart we then have that $(x', y', -1) \sim (-x', -y', 1)$ and we can map safely back to $(x, y)$ coordinates as $x = -x', y = -y'$. We then have the second form directly in $x, y$ coordinates as $-dx'dy'/((x'-1)y') = -dxdy/((1+x)y)$. Now we are using the same co-ordinates for both terms, we can safely sum the two contributions together. Finally we can covariantise the final result if we like giving

$$\frac{dx\,dy}{(x-1)y} - \frac{dx\,dy}{(x+1)y} = \frac{2dx\,dy}{(x-1)(x+1)y} = \frac{\langle Y d^2 Y\rangle \langle 123\rangle^2}{\langle Y12\rangle \langle Y23\rangle \langle Y31\rangle}, \qquad (4.3.12)$$

the correct canonical form for the triangle.

## 4.4 Conics and internal boundaries

In this section we look at the geometry of a conic cut by a plane in projective space to show yet another example of the emergence of internal boundaries in very basic geometries.

The generic equation of a quadratic space can be written as

$$Q_{IJ}Y^I Y^J = 0 \,. \tag{4.4.1}$$

For $\det(Q) \neq 0$, in real projective space we can always chose coordinates such that the equation reads

$$x^2 + y^2 + z^2 = 0 \,, \tag{4.4.2}$$

or

$$x^2 + y^2 - z^2 = 0 \,. \tag{4.4.3}$$

In $\mathbb{R}^3$, the first equation has only the trivial solution $x = z = y = 0$. The second equation instead corresponds to a double cone in the z direction. A subtle but important difference to the case of a projective or Grassmannian polytope like the amplituhedron is that equation (4.4.1) is invariant under the map $Y \to -Y$. As a consequence, the projection from the oriented projective space to the projective space, that is defined by equivalence relation $Y \sim -Y$, for the region $Q_{IJ}Y^I Y^J > 0$ is two to one. We can break this symmetry for example by adding a linear constraint $W_I Y^I > 0$, which defines the region

$$Q_{IJ}Y^I Y^J > 0 \quad \wedge \quad W_I Y^I > 0 \,, \tag{4.4.4}$$

where in real projective space W is a line intersecting the conic in two points. This region for $W = (0, 1, -\frac{1}{10})$ and $QYY = 1 - x^2 - y^2$ is described in fig: 4.2. Notice that in the oriented projective space this region is given by two disconnected parts while in projective space it corresponds to the conic minus the line W. In fact, by mapping $Y \sim -Y$ we can notice that the equation $W_I Y^I > 0$ reduce to $W_I Y^I \neq 0$ in projective space. The canonical form of a conic is simply zero since it doesn't have any codimension 2 boundary, but this simple deviation from the conic allows instead for a non-trivial canonical form. By computing the orientation induced by (4.4.4) to its image in projective space, one can observe that it is discontinuous on the line $Y \cdot W = 0$, that is the line W is an internal boundary of the geometry. Using the tiling property of the canonical form, we can then write the latter as the difference of the canonical form of the two portion of the circle. These canonical forms are

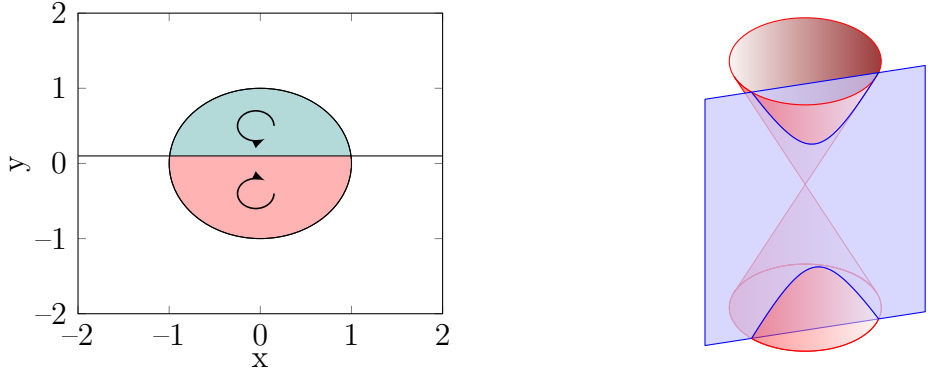

Figure 4.2: Region in between a quadric and a plane in projective space

equal but with opposite sign and given in [30] in a covariant form as

$$\frac{\sqrt{QQWW}\langle Y d^2 Y\rangle}{(QYY) Y \cdot W} \ . \tag{4.4.5}$$

Consistently with the definition of generalized positive geometries, we can observe that the residue on the line W is equal to two times the canonical form of the segment $W_1 W_2$, where $W_1$ and $W_2$ are the two intersection point given by $Q \cap W$, and that the residue on the circle, which is isomorphic to $\mathbb{P}^1$, is equal to the canonical form of the segment $W_1 W_2$ minus the canonical form of the segment $W_2 W_1$. The points $W_1$ and $W_2$ then represents codimension 2 internal boundaries and consistently the residue in both points is equal to $\pm 2$.

A particularly interesting projection is the one giving a hyperbola. In this case the quadric can be written in coordinates as

$$xy > 1 \ . \tag{4.4.6}$$

We can think to have a further constraint given by a projective line W at infinity. For this choice the projection form the oriented projective space to the projective plane is orientation preserving and therefore both regions have the same orientation. Using the GCD algorithm to compute the canonical from we obtain

$$2\frac{dxdy}{xy-1} \ , \tag{4.4.7}$$

consistently with (4.4.5).

Finally, for degenerate a quadric, that is for $\text{Det}(Q) = 0$, this equation can also be rewritten using only determinants by rewriting $Q_{IJ}$ as

$$Q_{IJ} = \epsilon_{ILM}\epsilon_{JNO} Q^L_{1,1} Q^M_{1,2} Q^N_{2,1} Q^O_{2,2} \ . \tag{4.4.8}$$

Substituting (4.4.8) in (4.4.1) we get

$$Y^I Q_{IJ} Y^J = Y^I \epsilon_{ILM} Q_{1,1}^L Q_{1,2}^M \epsilon_{JNO} Q_{2,1}^N Q_{2,2}^O Y^J = \langle Y Q_{1,1} Q_{2,2} \rangle \langle Y Q_{2,1} Q_{2,2} \rangle \, , \quad (4.4.9)$$

Which correspond of the product of two lines $Q_1$ and $Q_2$. These boundaries are given by factorizable polynomials which are the signature of codimension 2 internal boundaries. In fact, a new internal boundary emerges at $Q_1 \cap Q_2$, that is the vertex of two kissing triangles.

## 4.5 Summary

In this chapter, we highlighted some fundamental properties of the $\mathcal{N} = 4$ loop amplitudes, namely that non-vanishing maximal residues are not always $\pm 1$ as has generally been assumed, and that this fact is reflected in the presence of the loop-amplituhedron internal boundaries: codimension-1 defects separating two regions of opposite orientation. This phenomenon requires a generalisation of the concept of positive geometry and canonical form to include such internal boundaries.

We propose first a generalization of the canonical form (4.2.5) and of positive geometries, which we call generalized positive geometries (GPGs), based on distinguishing between internal and external boundaries. So, the residue of the canonical form of a boundary of a GPG is equal to the canonical form of its external boundary plus two times the canonical form of its internal boundary. Although this generalization is sufficient to describe the loop-amplituhedron and its canonical form it has the unpleasant feature that boundaries of GPGs are described by a union of two regions, that is external and internal boundaries, and the latter comes with a weight 2. In other words, the boundaries of GPGs are not always GPGs.

This suggests the utility of a further generalisation to 'weighted positive geometries'. A weighted geometry is defined by a piece-wise constant function w, the weight function, and an orientation $\mathcal{O}$. The boundaries of the geometries then correspond to the discontinuities of the pair $(w, \mathcal{O})$. Weighted geometries come equipped with two fundamental operations the sum and the projection. The sum of two weighted geometries (4.2.15) corresponds, for matching orientation, to the sum of their weight functions. The projection (4.2.19) of the pair $(w, \mathcal{O})$ on a algebraic variety instead roughly corresponds to its discontinuity on the variety, so the discontinuities of $(w, \mathcal{O})$ are codimension-1 weighed geometries . The canonical form is then simply defined on weighted geometries as the dlog form with simple poles on and only on the algebraic varieties containing the $(w, \mathcal{O})$ discontinuities and with residues equal to the canonical form of the discontinuities. The recursion terminates with the canonical form of weighted points (4.2.25).

One of the upshots of this construction is that the canonical form now acts linearly on the space of WPGs. This feature allows for a particularly simple definition of a tiling. Given a WG given by the sum of a collection of WGs, we call the latter a tiling of the former.

In section 4.3 we speculate on what properties should boundaries of WPGs have. We conclude that all WGs with multilinear-boundaries are WPGs and their canonical form can be computed using the GCD algorithm, of which we gave a review.

We concluded with section 4.4 where we presented a 2-dimensional example of codimension-1 internal boundaries emerging from the projection of a geometry from the oriented projective space to the projective space. This projection in fact, depending on the projection plane, is orientation preserving on half of the oriented projective space and orientation reversing in the other half. We explored in detail the case of a geometry defined by a linear and a quadratic inequality, showing how the internal boundary emerges in this case and how its canonical form can be computed.

# Chapter 5

# Loop-Loop Cuts

The content of this chapter has been published in the paper [2].

Another feature of the geometry of the amplituhedron which seems to have not been emphasised in the literature previously is the geometrical equivalent of the fact that multiple residues are not in general uniquely defined. One way to define multiple residues is via the residue form [115] (see also [11]) which essentially defines it via a sequence of single residues. However, taking these in different orders can give completely different results. There is a direct analogue of this fact in terms of taking boundaries of the corresponding geometry. Rather than talking about codimension 2 boundary components, instead it is the precise boundary component of the boundary component which will give the multiple residue defined by taking the corresponding simple poles in sequence. Taking these boundary components in different orders can give different results.

Consider the solid 3d geometry below, which will serve as a very simple example to illustrate the dependence on the order of taking boundaries.

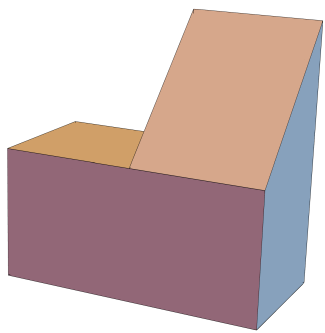

Figure 5.1: Example of a geometry where the path taken to reach the boundary is relevant.

Here we have a 3d shape which looks like a house with a flat roof on the left and an angled roof on the right. The planes in which the two roofs lie intersect along the top of the front wall which we will call the 'front eaves'. While one might wish to talk about the codimension two boundary component corresponding to the entire front eaves, taking appropriate boundaries of boundaries can give results which differ from this. One could first take the flat roof boundary component and then take the front eave boundary of that. This results in only the left half of the front eaves. If on the other hand we first take the slanted roof boundary component and then the font eaves boundary of that, we obtain the right half of the front eaves. Finally we could instead first take the boundary component on the front wall, and then the boundary component of that at the top, resulting in the entire length of the front eaves. We thus arrive at three different results from taking a sequence of boundary components.

This is completely consistent with what we get from multiple residues. Indeed, this example can be made completely precise and the corresponding canonical form and residues taken. We choose coordinates such that the flat roof lies on the plane $z = 0$, the slanted roof on $y = z$ and the front wall $y = 0$ (we also put the two side walls at $x = -1, x = 1$, the back wall at $y = 1$ and the floor at $z = -1$). The corresponding canonical form, which can easily be obtained by summing the canonical form of the living space and the roof, is

$$\Omega = \frac{1}{(x-1)x(y-1)z(y-z)} - \frac{2}{(x-1)(x+1)(y-1)yz(z+1)} \ . \tag{5.0.1}$$

The residue corresponding to the front boundary of the flat roof is $\text{Res}_{y=0}\text{Res}_{z=0}\Omega = -dx/(x(x+1)) = \Omega[-1, 0]$, which is the canonical form of the one dimensional interval $-1 \leq x \leq 0$. On the other hand, the residue corresponding to the front boundary of the slanted roof is $\text{Res}_{y=0}\text{Res}_{z=y}\Omega = -dx/(x(x-1)) = \Omega[0, 1]$, the canonical form of the one dimensional interval $0 \leq x \leq 1$. Finally, the residue corresponding to the top boundary of the front wall is $\text{Res}_{z=0}\text{Res}_{y=0}\Omega = 2dx/((x-1)(x+1)) = \Omega[1, -1]$. All cases correspond in the end to the codimension two line $z = 0, y = 0$ but the precise way we get there gives different results.

An important example of this phenomenon is the case of all l loop lines intersecting in a single point, which is a configuration closely related to the deepest cut of [113]. At 4 points, from 4 loops onward, different orderings of single residues give algebraically different results, as we will show in section 5.1.3. With this in mind, we do a detailed analysis of the all-in-one-point cut and find quite a complicated structure in general, although it seems one can always keep taking further loop loop residues to reduce to the 3 loop all-in-one-point-and-plane cut.

Here, we turn our attention to a specific boundary of the loop amplituhedron,

that is the all-in-one-point cut, and we compute its geometry and discuss its internal boundary. Finally, we take further cuts on the result of the all-in-one-point cut and obtain a new all loop formula.

## 5.1 All-in-one-point cut

We now look at a particular boundary of the loop amplituhedron related to a set of cuts on the integrand of MHV amplitudes explored in [113, 116], referred to as the *deepest cut*. This provides another example of an internal boundary as well as illustrating the other important point mentioned in the introduction, namely that the order of taking residues (or going to boundaries) can yield completely different results.

In general the deepest cut places all internal propagators on-shell

$$\left\langle (AB)_\alpha (AB)_\beta \right\rangle = 0 \quad \forall \ \alpha, \beta = 1, ..., l \,, \tag{5.1.1}$$

while leaving all external propagators $\langle (AB)_\alpha ii+1 \rangle$ generic. Geometrically there are two possible final configurations which solve (5.1.1): first, all loop lines passing through a single point A, or second, all loop lines lying on the same plane. In [113] the canonical form corresponding to these two solutions was found at any loop order. We find that this form can not be reproduced from any sequence of single residues (or any linear combination of such) acting on the amplitude and so some more complicated operation is presumably needed to reproduce it[1]. Furthermore there are many inequivalent ways of approaching this final all-in-one-point configuration via different sequences of single residues, as becomes especially apparent starting at four loops. In this section we systematically investigate all cuts ending in the all-in-one-point configuration.

We will begin by discussing the three loop all-in-one-point cut, computing its geometry and discussing the internal boundary that arises, before considering higher loops. Although we will limit the discussion to the 4-point MHV amplituhedron geometry, the derivation of the geometry is completely independent of the tree level inequalities $\langle Yijkl \rangle$ and $\langle ABij \rangle$. The results obtained in section 5.1.3 for the loop-loop inequalities of the all-in-one-point cut hold for any multiplicity and any NMHV degree by simply promoting the brackets $\langle AB_iB_jB_k \rangle$ to $\langle YAB_iB_jB_k \rangle$.

---
[1]We thank Nima Arkani-Hamed and Jaroslav Trnka for valuable discussions on this point

### 5.1.1 Three-loop all-in-one-point cut

The first case of an all-in-one-point cut is at two loops. Although we saw above that this contains a previously undetected internal boundary, since the all-in-one-point cut *is* this boundary it doesn't affect anything and the corresponding residue is simply the canonical form of two lines satisfying 1 loop inequalities as predicted in [113]. We will return to this in section 5.2.1.

We thus turn to three loops. The integrand of the three-loop MHV amplitude is given by

$$\text{MHV}(3) = \frac{\prod_{i=1}^{3}\langle A_i B_i d^2 A_i\rangle\langle A_i B_i d^2 B_i\rangle\langle 1234\rangle^3}{\langle A_1 B_1 14\rangle\langle A_1 B_1 12\rangle\langle A_1 B_1 34\rangle\langle A_2 B_2 12\rangle\langle A_2 B_2 23\rangle\langle A_3 B_3 34\rangle\langle A_1 B_1 A_3 B_3\rangle\langle A_2 B_2 A_3 B_3\rangle} \times$$
$$\times \left[\frac{1}{2}\frac{\langle 1234\rangle}{\langle A_3 B_3 12\rangle\langle A_2 B_2 34\rangle} + \frac{\langle A_1 B_1 23\rangle}{\langle A_3 B_3 23\rangle\langle A_1 B_1 A_2 B_2\rangle}\right] + \text{ symmetry}. \tag{5.1.2}$$

Here the '+ symmetry' is a sum of 23 more terms: the 3! terms generated by permutation symmetry over the loop variables (simultaneous permutation of $A_i$ and $B_i$) together with the four terms from cyclic symmetry of the external twistors, giving 24 terms in total in the sum. The first term in square brackets is the three-loop ladder integrand, and after summing it generates 12 unique terms (all with coefficient 1), while the second term is the so-called 3-loop 'tennis court' diagram generating 24 unique terms (again all will have coefficient 1).

In order to achieve an all-in-one-point final configuration – with three loop lines passing through a single point A – we must take three residues at $\langle A_i B_i A_j B_j\rangle = 0$ (5.1.1). By inspection, one can see that the first term in square brackets in (5.1.2) (the ladder integral) does not contain all three poles $\langle A_i B_i A_j B_j\rangle$ and therefore vanishes after taking these residues. From this point forward then we will only concern ourselves with the second term.

The key factor which all surviving terms contain is

$$F = \frac{\prod_{i=1}^{3}\langle A_i B_i d^2 A_i\rangle\langle A_i B_i d^2 B_i\rangle}{\langle A_1 B_1 A_2 B_2\rangle\langle A_2 B_2 A_3 B_3\rangle\langle A_1 B_1 A_3 B_3\rangle}. \tag{5.1.3}$$

We will then first consider the residue at $\langle A_1 B_1 A_2 B_2\rangle = 0$ followed by $\langle A_1 B_1 A_3 B_3\rangle = 0$. This corresponds geometrically to first intersecting the line $A_2 B_2$ with $A_1 B_1$ and then $A_3 B_3$ with $A_1 B_1$ (see 5.2). To do this, we parametrise $A_2$ and $A_3$ as

$$\begin{aligned} A_2 &= A_1 + a_2 B_1 + b_2 Z_*, \\ A_3 &= A_1 + a_3 B_1 + b_3 Z_*, \end{aligned} \tag{5.1.4}$$

where $Z_*$ is an arbitrary twistor. In this parametrisation, the limits $b_2 \to 0$ and $b_3 \to 0$ correspond to the points $A_2$ and $A_3$ moving to lie on the line $A_1 B_1$ re-spectively. Using this parametrization, we have for example that $\langle A_2 B_2 d^2 A_2\rangle = db_2 da_2 \langle A_1 B_2 Z_* B_1\rangle$ etc. and the factor (5.1.3) produces $db_2 db_3/(b_2 b_3)$. Taking the residue then gives

$$\mathop{\mathrm{Res}}_{\substack{\langle A_1 B_1 A_2 B_2\rangle = 0 \\ \langle A_1 B_1 A_3 B_3\rangle = 0}} F = \frac{da_2 da_3 \langle A_1 B_1 d^2 A_1\rangle \prod_{i=1}^{3}\langle A_i B_i d^2 B_i\rangle}{\langle A_2 B_2 A_3 B_3\rangle} \ . \tag{5.1.5}$$

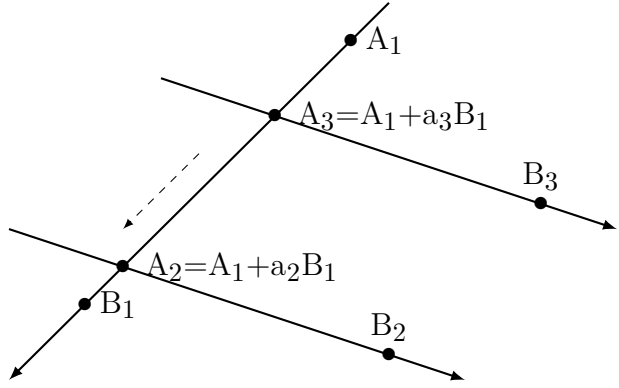

Figure 5.2: A geometrical depiction of the third residue taken when calculating the three loop all-in-one-point cut. Note the lines that pass through the points $B_2$ and $B_3$ do not in general lie on the same plane, but they do not (yet) intersect.

With the parametrisations (5.1.4), with $b_2 = b_3 = 0$ we see that the remaining singularity of F factorises into two terms $\langle A_2 B_2 A_3 B_3\rangle = (a_2 - a_3)\langle A_1 B_1 B_2 B_3\rangle$. This is another example of composite residues discussed in the introduction and section 4.1. The first factor $a_2 - a_3 = 0$ corresponds to the three loop lines intersecting in one point, while $\langle A_1 B_1 B_2 B_3\rangle = 0$ corresponds to the three lines lying in the same plane (and thus intersecting pairwise). We focus on the all-in-one-point case $(a_2 - a_3) \to 0$ (intersecting $A_3 B_3$ with $A_2 B_2$ by sliding the intersection point $A_3$ along the line $A_1 B_1$ to meet the intersection point $A_2$ see figure 5.2). We change variables from $(a_2, a_3)$ to $(a_2, \xi)$ where $\xi = (a_2 - a_3)$, so $da_2 da_3 = d\xi da_2$ and the residue at $\xi \to 0$ of (5.1.5) is

$$\mathop{\mathrm{Res}}_{\xi = 0}\left(\frac{d\xi da_2}{\xi} \frac{\langle A_1 B_1 d^2 A_1\rangle \prod_{i=1}^{3}\langle A_i B_i d^2 B_i\rangle}{\langle A_1 B_1 B_2 B_3\rangle}\right) = \frac{\langle A d^3 A\rangle \prod_{i=1}^{3}\langle A B_i d^2 B_i\rangle}{\langle A B_1 B_2 B_3\rangle} \ . \tag{5.1.6}$$

On the right-hand side we have written the expression manifestly as a function of the common intersection point of all three lines $A = A_1 + a_2 B_1$, thus $da_2 \langle A_1 B_1 d^2 A_1\rangle = \langle A d^3 A\rangle$.

Substituting (5.1.6) into (5.1.2) gives the all-in-one-point cut of the three loop

amplitude :

$$\frac{\langle Ad^3A\rangle \prod_{i=1}^{3}\langle AB_id^2B_i\rangle}{\langle AB_1B_2B_3\rangle} \times$$
$$\times \left( \frac{\langle 1234\rangle^3 \langle AB_1 23\rangle}{\langle AB_1 14\rangle\langle AB_1 12\rangle\langle AB_1 34\rangle\langle AB_2 12\rangle\langle AB_2 23\rangle\langle AB_3 23\rangle\langle AB_3 34\rangle} \quad + \quad \text{symmetry} \right) \tag{5.1.7}$$

where the sum occurs by applying cyclic symmetry of the external momenta and permutation symmetry of the $B_i$s. We see that after taking the three consecutive residues required to reach this final configuration, a new pole $\langle AB_1B_2B_3\rangle$ appears explicitly in the denominator of the integrand. This is not the same as the result given for the all-in-one-point cut in [113] which is instead the canonical form of the intersection of the hyperplane $\langle A_iB_iA_jB_j\rangle = 0$ with the amplituhedron. [2] Taking further residues of this all-in-one-point cut, starting with the pole $\langle AB_1B_2B_3\rangle$ one ends up with a maximal residue of 2 (see appendix C.1 for this computation) which, as discussed in section 4.1 suggests the existence an internal boundary.

In the next subsection we therefore look at the geometrical region corresponding to taking the above all-in-one-point cut. We will find that the pole at $\langle AB_1B_2B_3\rangle = 0$ indeed corresponds geometrically to an internal boundary of the codimension 3 boundary of the four-point three-loop amplituhedron corresponding to the all-in-one-point cut.

### 5.1.2 Geometric all-in-one-point cut

We now look to derive the geometry of the all-in-one-point cut. Following precisely the residues taken in section 5.1.1, we first intersect line $L_1$ with $L_2$, then intersect $L_3$ with $L_1$, and finally intersect $L_2$ and $L_3$ by sliding $A_3$ along $L_1$.

The four-point loop level amplituhedron is defined as the set of loop lines $L_i = (A_iB_i)$ with $i = 1,..,L$ satisfying

$$\mathcal{A}^{(L)} = \left\{ A_iB_i : \langle A_iB_i\overline{kl}\rangle > 0, \ \langle A_iB_iA_jB_j\rangle > 0, \quad 1 \le i,j \le L, \ 1 \le k < l \le 4 \right\} \tag{5.1.8}$$

Here for each loop we have the inequalities of the one loop amplituhedron

$$\begin{aligned}\langle A_iB_i 12\rangle &> 0, & \langle A_iB_i 13\rangle &< 0, & \langle A_iB_i 14\rangle &> 0, \\ \langle A_iB_i 23\rangle &> 0, & \langle A_iB_i 24\rangle &< 0, & \langle A_iB_i 34\rangle &> 0,\end{aligned} \tag{5.1.9}$$

---

[2]Note that if one instead takes an antisymmetric sum over the $B_i$ permutations in (5.1.7), the result produces a zero in $\langle AB_1B_2B_3\rangle$, cancelling the pole and reproducing the deepest cut given in [113]. We can not obtain this via an operation acting on the amplitude however but rather one would have to take a different operation on each contributing diagram.

which can all be conveniently rewritten in terms of the conjugate planes as [113]

$$\langle A_i B_i \bar{j}\bar{k} \rangle > 0 \qquad 1 \leq j < k \leq 4 \tag{5.1.10}$$

where $\bar{j} \equiv (j-1\,j\,j+1)$ and $\langle A_i B_i \bar{j}\bar{k} \rangle \equiv \langle A_i B_i (j-1\,j\,j+1) \cap (k-1\,k\,k+1) \rangle$. We then also have the loop-loop inequalities, $\langle A_i B_i A_j B_j \rangle > 0$. The all-in-one point configuration occurs when all loop lines $L_i$ pass through a single point A, so we simply set $A_i = A$. Then the loop-loop inequalities trivialise and this all-in-one-point cut geometry is

$$\mathcal{A}^{(L)}_{dc} = \mathcal{A}^{(L)}|_{A_i=A} = \left\{ A, B_i : \langle A B_i \bar{k}\bar{l} \rangle > 0, \ 1 \leq k < l \leq 4 \right\}. \tag{5.1.11}$$

This is the codimension 3 configuration of the amplituhedron corresponding to all loop lines intersecting in one point.

We now however wish to examine in detail what happens when we take a sequence of codimension 1 boundaries in order to reach such a configuration. Using the same parametrisation as (5.1.4), $A_2 = A_1 + a_2 B_1 + b_2 Z_*$, $A_3 = A_1 + a_3 B_1 + b_3 Z_*$ the loop-loop inequalities become

$$\begin{aligned}
\langle A_1 B_1 A_2 B_2 \rangle &= -b_2 \langle A_1 B_1 B_2 Z^* \rangle > 0, \\
\langle A_1 B_1 A_3 B_3 \rangle &= b_3 \langle A_1 B_1 Z^* B_3 \rangle > 0, \\
\langle A_2 B_2 A_3 B_3 \rangle &= (a_2 - a_3) \langle A_1 B_1 B_2 B_3 \rangle + (b_2 - b_3) \langle A_1 Z_* B_2 B_3 \rangle > 0.
\end{aligned} \tag{5.1.12}$$

We then consider the boundary at $a_2 = a_3$ of the boundary at $b_3 = 0$ of the boundary at $b_2 = 0$, which corresponds precisely to taking the consecutive residues of (5.1.5) and below. Here $Z_*$ is chosen arbitrarily and we can arrange it so that $\langle A_1 B_1 B_2 Z_* \rangle < 0$ and $\langle A_1 B_1 Z_* B_3 \rangle > 0$ and thus $b_2, b_3 > 0$. Notice that the third inequality factorises when $b_2, b_3 \to 0$. This is the geometric version of the factorisation discussed below (5.1.5), related to composite residues and reducible varieties. Thus the boundary at $b_2, b_3 \to 0$ is the union of two disconnected regions $\mathcal{R}_1 \cup \mathcal{R}_2$

$$\begin{aligned}
\mathcal{R}_1: &\quad a_2 > a_3, \quad \langle AB_1 B_2 B_3 \rangle > 0, \\
\mathcal{R}_2: &\quad a_2 < a_3, \quad \langle AB_1 B_2 B_3 \rangle < 0,
\end{aligned} \tag{5.1.13}$$

where $A = A_1 + a_2 B_1$. The inequalities, (5.1.13), carve out a region consisting of two almost disconnected pieces of the same orientation. This geometry is illustrated in the picture (the same as for the two-loop internal boundary case (4.1.11))

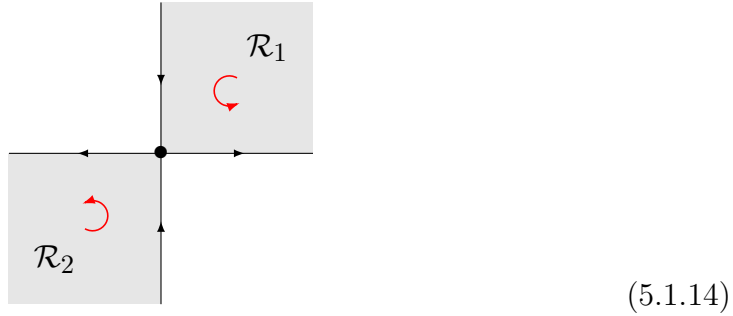

(5.1.14)

where the x axis corresponds to the region $a_2 = a_3$ and the y axis corresponds to $\langle AB_1B_2B_3 \rangle = 0$. The all-in-one-point cut corresponds to the boundary $a_2=a_3$ (so the x axis). We can then clearly see that the all-in-one-point cut consists of two regions, $\langle AB_1B_2B_3 \rangle \lessgtr 0$, with opposite orientation separated by an (internal) boundary at $\langle AB_1B_2B_3 \rangle = 0$:

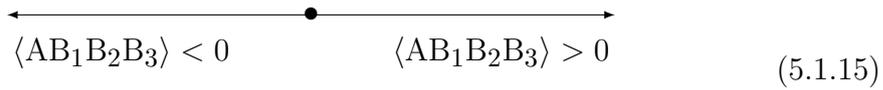

(5.1.15)

Geometrically the two regions arise from the intersection point $A_3$ approaching $A_2$ from two different directions along the line $A_1B_1$ (see figure 5.2). Importantly, after approaching the all-in-one-point cut, the mutual positivity conditions between the loops of (5.1.12) do not trivialise, but instead new inequalities emerge dictated by the sign of $\langle AB_1B_2B_3 \rangle$.

So altogether then, incorporating the inequalities resulting from (5.1.9) (rewritten as in (5.1.10)) we see that the the full geometry of the three-loop all-in-one-point cut is given by the two regions with opposite orientation

$$\mathcal{R}^{\mathrm{dc}} = \mathcal{R}_1^{\mathrm{dc}} \cup \mathcal{R}_2^{\mathrm{dc}}$$
$$\mathcal{R}_1^{\mathrm{dc}} = \mathcal{A}_{\mathrm{dc}}^{(3)} \cap \{\langle AB_1B_2B_3 \rangle > 0\} \qquad \text{positive orientation} \qquad (5.1.16)$$
$$\mathcal{R}_2^{\mathrm{dc}} = \mathcal{A}_{\mathrm{dc}}^{(3)} \cap \left\{\langle AB_1B_2B_3 \rangle < 0\right\} \qquad \text{negative orientation .}$$

Note that the deepest cut geometry $\mathcal{A}_{\mathrm{dc}}^{(3)}$ is the union of these two regions with the *same* orientation, but the actual result of taking boundaries of boundaries of boundaries requires the regions to have opposite orientation separated by an internal boundary. Also note that we made a choice of which loop lines to intersect first and which to slide etc. and one might expect different choices to give different results. This is indeed the case at higher loops. At three loops however the resulting geometry (5.1.16) is the unique geometry one obtains from approaching the all-in-one-point cut.

So to summarise we find that the all-in-one-point cut as computed as a residue corresponds to two regions of opposite orientation separated by an internal boundary. At higher loops it turns out that the all-in-one-point cut is no longer even unique but depends on the precise sequence of codimension 1 boundaries taken to reach it.

### 5.1.3  Higher loop all-in-one-point cut

We commented that at three loops the multiple residue leading to the all-in-one-point cut is unique and the corresponding geometry given by (5.1.16). For higher loops, however, there are a number of inequivalent resulting geometries depending on the sequence of single residues taken. Here, we generalise the discussion of the previous section to give the inequalities associated to any all-in-one-point cut for any loop. We show that while the final configuration is always the same – that is L lines intersecting in a point – distinct paths to reach this configuration can carve out different oriented regions.

Enforcing that a line in 3d (which has four degrees of freedom) intersects a specified point kills two degrees of freedom. Thus making all L lines go through a specified point would reduce by 2L degrees of freedom. However the intersection point itself A is not fixed and has 3 degrees of freedom, thus only 2L – 3 degrees of freedom are lost, corresponding to taking 2L – 3 single residues. We distinguish between two types of residue, each of which has a different geometrical interpretation. The first is the intersection of two loop lines which are currently not connected by any set of intersecting lines (see left-hand side of figure 5.3). Taking the maximal possible number of such intersections results in a maximal tree configuration.[3] These we will refer to simply as *intersections*. The second type occurs when we merge two separate intersection points along a line (see right-hand side of figure 5.3) which we shall call a *sliding*. An all-in-one-point cut then consists of (L–1) intersections and (L–2) slidings to make a total of 2L – 3.

To perform an intersection, for example $(A_i B_i) \cap (A_j B_j)$ depicted on the left in figure 5.3, we parameterise the point $A_j$ as $A_j = A_i + aB_i + bZ_*$ and take the residue $b = 0$. Similarly to the discussion at three loops leading to (5.1.5), any such intersection saturates one positivity condition, $\langle A_i B_i A_j B_j \rangle = 0$, and does not generate any new inequalities. The order in which these are performed is also not important. The all-in-one-point cut consists of L – 1 intersections, therefore (L – 1) of the (2L – 3) mutual positivity conditions are trivialised and no new inequalities arise.

The remaining (L – 2) mutual positivity conditions are handled by slidings. How-ever, unlike the residues corresponding to the intersections, here new inequalities *are* generated. Let us begin by determining what happens when a residue is taken corresponding to a single sliding, for example the one depicted on the right in figure 5.3. We start off with two sets of lines intersecting at two different points, with one common loop that the two intersection points lie on.

---

[3]More precisely the graph obtained by replacing each loop line with a vertex joined by edges if and only if the respective loop lines intersect should be a maximal tree on L vertices.

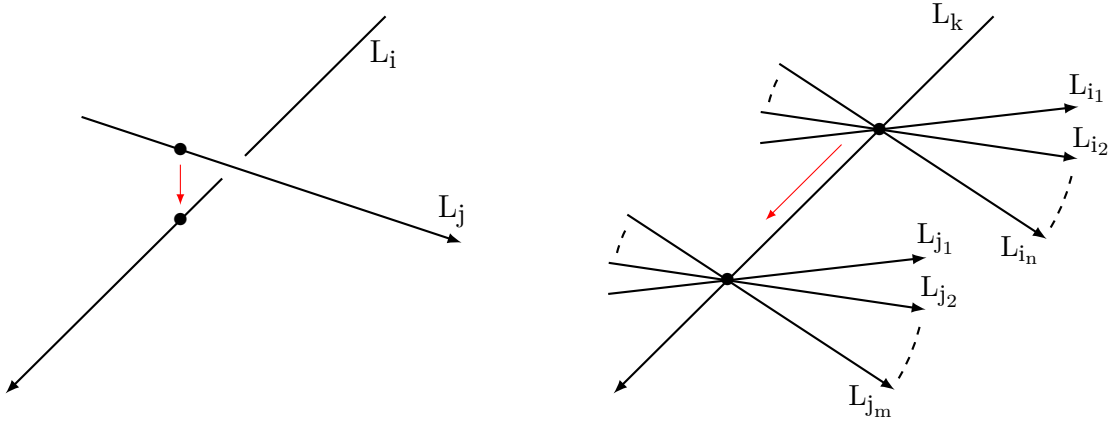

Figure 5.3: A graphical representation of the two types of residues discussed here. On the left is an intersection between lines $L_i$ and $L_j$, which we label as $(i,j)$. On the right is a sliding between the sets of lines $L_{i_1}, L_{i_2}, ..., L_{i_n}$ and $L_{j_1}, L_{j_2}, ..., L_{j_m}$, which we label as $(I,J) \equiv (i_1 i_2 ... i_n k,\ j_1 j_2 ... j_m k)$.

Let I and J be the sets of labels of the two groups of intersecting lines, $k = I \cap J$ labels the line in common, and $A, A'$ label the intersection points of the groups of lines I and J respectively, so

$$
\begin{aligned}
A &= \bigcap_{i \in I} L_i = A_k + c_1 B_k \ , \\
A' &= \bigcap_{j \in J} L_j = A_k + c_2 B_k \ .
\end{aligned}
\qquad (5.1.17)
$$

In this parametrization the mutual positivity relation between loops in I and J reads

$$\langle A_i B_i A_j B_j \rangle = (c_2 - c_1)\langle A_k B_i B_k B_j \rangle > 0 \ , \qquad \text{for all } i \in I, \ j \in J. \qquad (5.1.18)$$

As in the three loop case, the brackets factorize, giving rise to two almost disconnected regions (see (5.1.16). The geometric sliding residue is then calculated by taking the limits $(c_2 - c_1) \to 0^\pm$, leaving two regions with opposite orientation. A "positive" region for which $\langle A_k B_i B_k B_j \rangle > 0$ for all $i \in I$, $j \in J$ and a "negative" region for which $\langle A_k B_i B_k B_j \rangle < 0$ for all $i \in I$, $j \in J$.

To compute an all-in-one-point cut we must take $L - 2$ sliding residues, each of which splits the geometry of the boundary in two parts. If we label slidings by the index $a = 1, ..., L–2$ then we can identify a sub region with fixed orientation through the string $\vec{s} = \{s_1, \cdots, s_{L-2}\}$, where $s_a = \pm 1$ and keeps track of the signs of positively and negatively oriented regions. The resulting geometry is the union of these regions

$$\mathcal{R}^{dc} = \bigcup_{\vec{s}} \mathcal{R}^{dc}_{\vec{s}}$$
$$\mathcal{R}^{dc}_{\vec{s}} = \mathcal{A}^{(L)}_{dc} \cap \left\{ A, B_i : s_a \langle AB_i B_{k_a} B_j \rangle > 0, \quad a = 1, .., L–2, \; i \in I_a, \; j \in J_a \right\}$$
(5.1.19)
$$\text{orientation of } \mathcal{R}^{dc}_{\vec{s}} = \prod_a s_a$$

where we recall that $\mathcal{A}^{(L)}_{dc}$ is the deepest cut geometry, obtained by trivialising the loop-loop inequalities of the amplituhedron and sending $A_i \to A$ (5.1.11). In particular this region depends explicitly on the sequence of boundaries we took to approach the geometry through the sets $I_a, J_a$.

Note that (5.1.19) generalises directly to describe the all-in-one-point cut geometry for amplituhedrons at any number of points. One just needs to add a $Y \in \text{Gr}(k, k+4)$ into each bracket and modify $\mathcal{A}^{(L)}_{dc}$ appropriately.

**Example: Four Loop all-in-one-point Cuts**

Let us illustrate (5.1.19) by giving an explicit example at four loops. Each 4 loop all-in-one-point cut is given by 3 intersections and 2 sidings. Denoting the intersection between lines $L_i$ and $L_j$ by $(i, j)$ and a sliding between the sets of lines I and J as $(I, J)$, we explore the cut

$$\{(1,2), (1,3), (1,4) \; ; \; (12, 13), (123, 14)\},$$
(5.1.20)

represented in Figure 5.4a. From (5.1.19), the resulting geometry is given by a union of four regions:

$$\begin{aligned}
\mathcal{R}_1(\mathcal{D}_1) &= \mathcal{A}^{(L)}_{dc} \wedge \langle AB_2 B_1 B_3 \rangle > 0 \wedge \langle AB_2 B_1 B_4 \rangle > 0 \wedge \langle AB_3 B_1 B_4 \rangle > 0, \quad + \\
\mathcal{R}_2(\mathcal{D}_1) &= \mathcal{A}^{(L)}_{dc} \wedge \langle AB_2 B_1 B_3 \rangle < 0 \wedge \langle AB_2 B_1 B_4 \rangle > 0 \wedge \langle AB_3 B_1 B_4 \rangle > 0, \quad - \\
\mathcal{R}_3(\mathcal{D}_1) &= \mathcal{A}^{(L)}_{dc} \wedge \langle AB_2 B_1 B_3 \rangle > 0 \wedge \langle AB_2 B_1 B_4 \rangle < 0 \wedge \langle AB_3 B_1 B_4 \rangle < 0, \quad - \\
\mathcal{R}_4(\mathcal{D}_1) &= \mathcal{A}^{(L)}_{dc} \wedge \langle AB_2 B_1 B_3 \rangle < 0 \wedge \langle AB_2 B_1 B_4 \rangle < 0 \wedge \langle AB_3 B_1 B_4 \rangle < 0, \quad +
\end{aligned}$$
(5.1.21)

where A denotes the final point that all loops intersect, and the orientation of the regions is indicated on the right by a '+' or '–'. In particular, $\mathcal{R}_1, \mathcal{R}_4$ have the same orientation and $\mathcal{R}_2, \mathcal{R}_3$ have the same orientation but opposite to $\mathcal{R}_1, \mathcal{R}_4$.

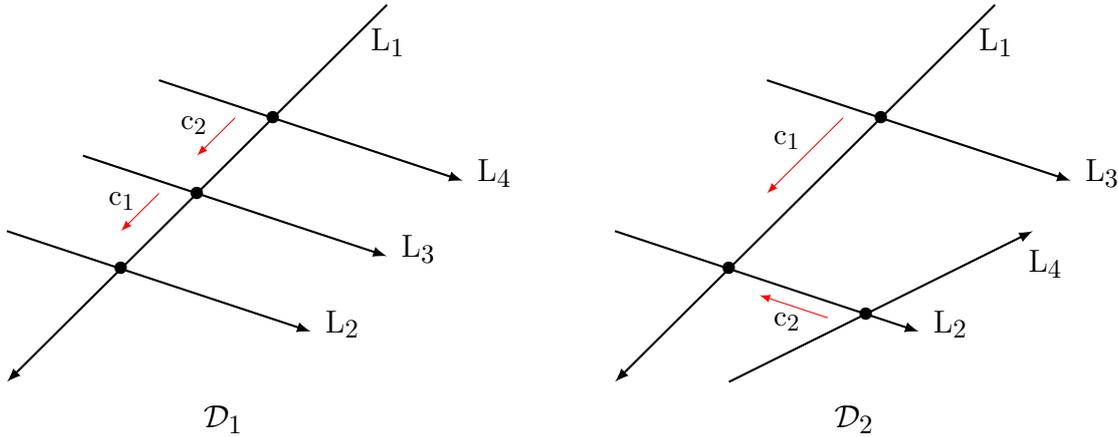

Figure 5.4: Graphical representation of the four loop all-in-one-point cut labelled in (5.1.20). The all-in-one-point cut corresponds to drawing a tree configuration, and collapsing the graph so that only one intersection point remains. The *intersections* are given by pairs of intersecting lines $L_i$, $L_j$. The *slidings* are labelled in the order they should be done, $c_1, ..., c_{L-2}$. Each slide corresponds to moving one intersection point along a line in the direction dictated by the red arrow until it meets another intersection point.

We see that this four loop all-in-one-point cut geometry has an internal boundary at $\langle AB_2B_1B_3 \rangle = 0$ and external boundaries at $\langle AB_2B_1B_4 \rangle = 0, \langle AB_3B_1B_4 \rangle = 0$. The corresponding multiple residue has poles in these positions.

Recall that at 3 loops all possible ways of reaching the all-in-one-point cut configuration result in the same geometry (5.1.16). At 4 loops on the other hand there are twelve different possible geometries. They are all equivalent to each other up to permutations. That is they are all given by (5.1.21) after permuting the $B_i$ (permuting $B_2, B_3$ in (5.1.21) gives back the same geometry up to swapping the overall orientation and so there are only 12 inequivalent permutations rather than 24).[4] Thus the corresponding action of taking residues on a permutation invariant object such as the loop integrand yields the same result for all twelve four-loop all-in-one-point cuts. From 5 loops however there are genuinely different all-in-one-point cuts giving different results when the corresponding residues are taken on a permutation invariant object.

---

[4]The choice of all-in-one-point cut $\mathcal{D}_2 = \{(1,2),(1,3),(2,4) \ ; \ (12,13),(123,24)\}$ illustrated in figure 5.4b looks like a different case at first sight but in fact results in the same geometry as (5.1.21) after permuting $B_1$ and $B_2$.

## 5.2 All in one point and plane cuts

One of the attractive features of the deepest cut defined in [113] was that its canonical form was defined by a simple formula at all loops. This was because the all-in-one-point configuration (5.1.11) consists of L independent one loop inequalities and no loop-loop inequalities and the resulting geometry thus factorises. We have seen however that any action of taking consecutive boundary components to reach the deepest cut configuration gives non-unique geometries which are more involved than (5.1.11) and in particular new loop-loop inequalities of the form $\langle AB_i B_j B_k \rangle > 0$ are generated, spoiling this factorisation. The presence of these brackets makes the computation of the canonical form much more challenging and also dependent on the particular sequence off boundaries taken to reach all-in-one-point configuration.

In this section we will show that despite this complication it can still be possible to find fairly simple all loop geometries by taking further residues after reaching the all-in-one-point cut configuration that trivialize all the new $\langle AB_i B_j B_k \rangle$ inequalities. The further cuts constrain the loop lines to all lie in the same plane as well as going through the same point. They are thus simultaneously all-in-one-point and all-in-one-plane cut configurations. We will thus refer to them as all-in-one-point-and-plane cuts or point-and-plane cuts for short. They are defined in terms of the point A which all loop lines go through together with the plane $(AP_1P_2)$ which all loop lines lie on. It is useful also to project through the point A and thus reduce the geometry to 2d, in which case we refer to the plane P instead as a line.

If we project through the common intersection point A, the geometry of the cut correspond to L points $B_i$ on an oriented line P in $\mathbb{P}^2$. Starting at 4 loops, the $\langle AB_i B_j B_k \rangle$ inequalities force some ordering between the points on P. As a practical consequence, this implies that an all-in-one-point-and-plane cut can itself also have further loop-loop type boundaries at $B_i = B_j$. Thus taking the residues / boundaries on these effectively reduces the number of free loop variables further. We call a cut for which we have exhausted all loop-loop type residues a *maximal loop-loop* cut. All the maximal loop-loop cuts that we have considered correspond – up to a permutation of the Bs and an integer factor arising from the number of internal boundaries taken in reaching there – to the three loop maximal loop-loop cut (which is also the unique all-in-one-point-and-plane cut)

$$\mathcal{A}_{\text{mll}}^{L=3} = \left\{ A, P, B_i = P_1 + b_i P_2 : \langle AB_i \bar{j}\bar{k} \rangle > 0 : \quad i = 1, 2, 3, \ 1 \leq j < k \leq 4 \right\}.$$
(5.2.1)

We conjecture this to hold in general, that is the maximal loop loop cut always reduces to the three loop one, $\mathcal{A}_{\text{mll}}^{L} = \mathcal{A}_{\text{mll}}^{L=3}$.

In this section we will show how to compute the geometry and the canonical form of the all-in-one-point-and-plane cut from the amplituhedron. We will start with the 2 and 3 loop cases, which contain all the main features of the problem. Then we will look at the geometry of the only 2 independent (up to permutations of the loop lines) all-in-one-point-and-plane cuts at 4 loops, and finally we will define a particular cut at arbitrary loops and compute its canonical form.

### 5.2.1 All-in-one-point-and-plane canonical form at 2 loops

At higher loops one can take further boundaries of the all-in-one-point configuration so that the lines all lie in a single plane. But at two loops we only have two lines intersecting in a point so they automatically lie in the same plane. Thus the all-in-one-point and the all-in-one-point-and-plane cases are identical. Nevertheless it is useful to rewrite the two loop all-in-one-point case in the same variables we will use at higher loops, namely in terms of a single line P in $\mathbb{P}^2$ (after projection through A) on which the $B_i$s lie (each now with 1 degree of freedom).

At two loops the all-in-one-point cut is obtained simply by taking the residue in $\langle A_1 B_1 A_2 B_2 \rangle$ of (5.2.9) and is thus given by

$$\mathcal{A}^{(2)}_{\rm dc} = \frac{\langle AB_1 d^2 B_1 \rangle \langle AB_2 d^2 B_2 \rangle \langle 1234 \rangle^3 \langle A d^3 A \rangle}{\langle AB_1 14 \rangle \langle AB_1 12 \rangle \langle AB_2 23 \rangle \langle AB_2 34 \rangle} \times$$
$$\times \left[ \frac{1}{\langle AB_1 34 \rangle \langle AB_2 12 \rangle} + \frac{1}{\langle AB_1 23 \rangle \langle AB_2 14 \rangle} \right] \quad + \quad B_1 \leftrightarrow B_2 \,. \quad (5.2.2)$$

The deepest cut formula of [113] is however a completely different-looking yet identical formula for $\mathcal{A}^{(2)}_{\rm dc}$ obtained by computing the canonical form of its corresponding geometry (5.1.11) (we recall that at two loops this correctly reproduces the corresponding residue but not beyond).

Following [113], the first step in computing the canonical form is to tile the A geometry into regions where the brackets $\langle Aijk \rangle$ have a well defined sign. Let's derive such a tiling for the intersection point $A = A_1 B_1 \cap A_2 B_2$. Since the intersection point A can occur at any point along a loop line, the allowed space for A can be computed as the linear combination $A = c_1 A_1 + c_2 B_1$, where $A_1 B_1$ lives in the amplituhedron. Notice, that the intersection point A is defined up to a sign, so we can fix for example $c_1 > 0$. Solving the inequalities one finds that the allowed regions for A correspond to 4 twisted cyclic permutations[5] of the solution

$$\langle A123 \rangle > 0, \quad \langle A124 \rangle > 0, \quad \langle A134 \rangle < 0, \quad \langle A234 \rangle > 0 \,. \quad (5.2.3)$$

All these cyclically related A regions are tetrahedra, and their canonical forms $\omega_i(A)$

---

[5] $Z_i \to Z_{i+1}$ for $i = 1, 2, 3$ and $Z_4 \to -Z_1$.

(where we assign the label i = 1 to region (5.2.3) and the other values to its cyclic twisted permutations) can be written as $\omega_i(A) = (-1)^i \omega(A)$, with

$$\omega(A) = \frac{\langle A d^3 A \rangle \langle 1234 \rangle^3}{\langle A123 \rangle \langle A234 \rangle \langle A134 \rangle \langle A124 \rangle} \ . \tag{5.2.4}$$

We can now project through A onto a plane not containing A and the remaining geometry is two dimensional. The configuration of $Z_i$ arising from (5.2.3) is such that $1, 2, 3$ form an anti-clockwise oriented triangle containing 4:

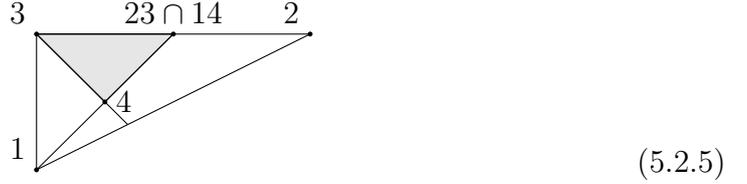

(5.2.5)

Now we can analyze the B inequalities

$$\langle AB14 \rangle > 0 \ , \quad \langle AB23 \rangle > 0 \ , \quad \langle AB34 \rangle > 0 \ , \quad \langle AB12 \rangle > 0 \ , \tag{5.2.6}$$

which one can see puts B inside the shaded triangle in (5.2.5), with vertices $\{4, 3, (23) \cap (14)\}$. For general i after cycling we have that the $B_i$ are in the triangle with edges $(i{+}2\, i{+}3)$, $(i\, i{+}4)$, $(i{+}1\, i{+}2)$ and vertices

$$W_{i1} = i{+}2 \ , W_{i2} = i{+}3 \ , W_{i3} = (i{+}1\, i{+}2) \cap (i\, i{+}3) \ . \tag{5.2.7}$$

For fixed i, our problem now simply reduces to computing the canonical form of two points $B_1$ and $B_2$ living independently inside the triangle $(W_{i1} W_{i2} W_{i3})$. Each point $B_i$ thus has the canonical form of a triangle and we obtain the two loop deepest cut form as

$$\mathcal{A}_{dc}^{(2)} = \sum_{i=1}^{4} \frac{(-1)^i \langle A d^3 A \rangle \langle 1234 \rangle^3}{\langle A123 \rangle \langle A234 \rangle \langle A134 \rangle \langle A124 \rangle} \prod_{L=1}^{2} \frac{\langle AB_L d^2 B_L \rangle \langle A W_{i1} W_{i2} W_{i3} \rangle^2}{\langle AB_L W_{i1} W_{i2} \rangle \langle AB_L W_{i2} W_{i3} \rangle \langle AB_L W_{i3} W_{i1} \rangle} \ . \tag{5.2.8}$$

Remarkably, this is indeed equal to (5.2.2).

But we now wish to rewrite this further in a way appropriate for the higher loop all-in-one-point-and-plane cut. So instead of considering the $B_i$ living in 2d, we consider first fixing a line P and then two points $B'_1, B'_2$ living on the 1d line P.

The Jacobian of the transformation from $B_1, B_2$ to $P, B'_1, B'_2$ is given by

$$\langle AB_1 d^2 B_1 \rangle \langle AB_2 d^2 B_2 \rangle = \frac{\langle AZ_* B'_1 B'_2 \rangle \langle AZ_* B'_1 dB'_1 \rangle \langle AZ_* B'_2 dB'_2 \rangle \langle APdP_1 \rangle \langle APdP_2 \rangle}{\langle AZ_* P \rangle^3} \ , \tag{5.2.9}$$

where $Z_*$ is a fixed element of $\mathbb{P}^3$ such that $\langle AB_1 B_2 Z_* \rangle \neq 0$. Notice that the 2-loop deepest cut is symmetric in $B_1$ and $B_2$, but becomes anti-symmetric in $B_1', B_2'$. This corresponds geometrically to the fact that switching $B_1, B_2$ flips the orientation of the configuration on the right. We will shortly see how this symmetry is reflected in the amplituhedron geometry in the new variables. From now on we will drop the primes on the $B_i$s.

The task is now to translate the geometry of two points in a triangle to that of a line through a triangle and two points on that line. We start by observing that the geometry of a line through a triangle can be tiled into 3 regions. These correspond to the combination of the 3 ways in which the line P can intersect the edges of the triangle. However it will turn out that we also need to consider which side $B_1$ is of $B_2$ (due to the orientation switch mentioned above) and so we in fact need to split into 6 regions.

$$\begin{array}{c} \text{(diagram: two triangles with vertices } W_{i,2} \text{ (top), } W_{i,3} \text{ (bottom-left), } W_{i,1} \text{ (bottom-right), line P passing through with points } B_1, B_2 \text{ on the left and } B_2, B_1 \text{ on the right)} \end{array} \qquad (5.2.10)$$

Let's consider one of these 6 regions, the one on the left in (5.2.10). It is described by the inequalities

$$\begin{aligned} &\langle PW_{i,1} \rangle > 0 \,, \quad \langle PW_{i,2} \rangle < 0 \,, \quad \langle PW_{i,3} \rangle > 0 \,, \\ &\langle B_l W_{i,j} W_{i,j+1} \rangle > 0 \,, \quad \text{with} \quad j = 1, 2, 3 \text{ and } l = 1, 2 \,, \\ &\langle W_{i,2} B_1 B_2 \rangle < 0 \,, \end{aligned} \qquad (5.2.11)$$

with the last inequality ensuring that $B_1$ and $B_2$ are ordered.

Then all 6 configurations can be generated by cyclic permutations of (5.2.11) together with $B_1 \leftrightarrow B_2$. We will use $p = 1, 2, 3$ to label the cyclic permutations of (5.2.11), with $p = 2$ corresponding to the case (5.2.11).

The canonical form $\lambda_{i,p}(P)$ corresponding to a line through the triangle is the same for all $p = 1, 2, 3$ and equal to

$$\lambda_{i,p}(P) = \lambda_i(P) = -\frac{\langle PdP_1 \rangle \langle PdP_2 \rangle \langle W_{i,1} W_{i,2} W_{i,3} \rangle}{\langle PW_{i,1} \rangle \langle PW_{i,2} \rangle \langle PW_{i,3} \rangle} \,, \qquad \text{with } p = 1, 2, 3 \,. \quad (5.2.12)$$

For fixed i and p, the geometry of $B_1$ and $B_2$ corresponds to two points living on the segment with vertices $I_{i,p} = (W_{i,p-1} W_{i,p}) \cap P$ and $J_{i,p} = (W_{i,p} W_{i,p+1}) \cap P$.

The canonical form of a point B on a segment (IJ) in $\mathbb{P}^1$ can be written in general as

$$[I; B; J] := \frac{\langle JdB \rangle}{\langle JB \rangle} - \frac{\langle IdB \rangle}{\langle IB \rangle} = \frac{\langle BdB \rangle \langle IJ \rangle}{\langle BI \rangle \langle JB \rangle} \,. \qquad (5.2.13)$$

For our application the 1d segment lives in $\mathbb{P}^2$ (well really in $\mathbb{P}^3$ but we already projected through A onto $P^2$). We can choose any point to project onto the segment, call this $Z_*$,[6] then all the two brackets in the above formula can be viewed as 3-brackets with an additional $Z_*$ and in turn eventually as 4-brackets with an additional A (so eg $\langle BdB \rangle = \langle Z_* BdB \rangle = \langle AZ_* BdB \rangle$ etc.)

Then, the canonical form for $B_1, B_2$ ordered on the segment (IJ), with $I < B_1 < B_2 < J$ can be written as

$$[I; B_1, B_2; J] := [I; B_1; J][B_1; B_2; J] = [I; B_1; B_2][I; B_2; J] , \qquad (5.2.14)$$

and similarly for arbitrary numbers of ordered $B_i$ on (IJ) we define inductively

$$[I; B_1, .., B_L; J] := [I; B_1, .., B_{L-1}; J][B_{L-1}; B_L; J] . \qquad (5.2.15)$$

Now we claim that the canonical form for two free points $B_1, B_2$ in a triangle translates as follows

$$\prod_{l=1}^{2} \frac{\langle AB_l d^2 B_l \rangle \langle AW_{i1} W_{i2} W_{i3} \rangle^2}{\langle AB_l W_{i1} W_{i2} \rangle \langle AB_l W_{i2} W_{i3} \rangle \langle AB_l W_{i3} W_{i1} \rangle} = \lambda_i(P) \sum_{p=1}^{3} \left( [I_{ip}; B_1, B_2; J_{ip}] - [I_{ip}; B_2, B_1; J_{ip}] \right) . \qquad (5.2.16)$$

Note in particular the minus sign between the canonical forms for the two orderings of $B_1, B_2$. This is because the orientation flips when $B_1$ passes through $B_2$ as discussed below (5.2.9).

We can finally put all the pieces together and write the canonical form $\mathcal{A}_{dc}^{(L)}$ in terms of these variables as

$$\mathcal{A}_{dc}^{(2)} = \omega(A) \sum_{i=1}^{4} (-1)^i \lambda_i(P) \sum_{p=1}^{3} \left( [I_{ip}; B_1, B_2; J_{ip}] - [I_{ip}; B_2, B_1; J_{ip}] \right) . \qquad (5.2.17)$$

An interesting aspect of this formula is that each term in the sum has a pole at $B_1 = B_2$. For i, p fixed this represent an internal boundary of the geometry. The sum of the residues over p though, as expected from (5.2.2), is equal to zero, which means that this pole is actually a spurious one. Geometrically, we have that when the only two points on P coincide the latter can rotate unconstrained on the pivotal point $B_1 = B_2$ and therefore its canonical form will be zero.

---

[6] Note that a natural point to choose for $Z_*$ is the intersection point of the two edges that P is passing through, $W_{i,p}$. Then the intersection points $I_{ip} \sim W_{i,p-1}$ and $J_{ip} \sim W_{i,p+1}$ and thus the formulae dramatically simplify.

### 5.2.2   3-loops all-in-one-point-and-plane canonical form

We can now generalize the two loop result to higher loops. To compute the canonical form of a point-plane cut, we tile the A and P geometry in the same way we did for the two loop case and then we consider the position of Bs on the line P. The general structure of the canonical form of a specific point-plane cut will depend on the details of how the cut is taken, but it will always have the general form

$$\mathcal{A}^{(L)}_{\text{point-plane}} = 2^{n_I}\omega(A) \sum_{i=1}^{4}(-1)^i \lambda_i(P) \sum_{p=1}^{3} \sum_{\sigma \in S_L} c_\sigma \times [I_{ip}; B_{\sigma_1},..,B_{\sigma_L}; J_{ip}] \,, \quad (5.2.18)$$

where $c_\sigma = \pm 1, 0$ reflecting the orientation (or absence) of a certain ordering of the $B_i$s and where $n_I$ is the number of internal boundaries approached to reach the configuration. So for example the two loop case (5.2.17) takes this form with $c_{\text{id}} = 1$ and $c_{(12)} = -1$. Turning to the three loop case then, we find, by direct computation of the residues, that the point-plane cut is given by (5.2.18) with $c_\sigma = 1$ for all 6 permutations $\sigma \in S_3$. Thus unlike the two loop case, for this case, the order of the $B_l$s on the line P is not relevant and the canonical form simplifies to that of the product of three $B_l$s

$$\mathcal{A}^{(3)}_{\text{point-plane}} = 2\omega(A) \sum_{i=1}^{4}(-1)^i \lambda_i(P) \sum_{p=1}^{3} \prod_{l=1}^{3} [I_{i,p}; B_l; J_{i,p}] \,. \quad (5.2.19)$$

Let us then see how this arises from the geometry. As we saw in section 5.1.2, the three loop all-in-one-point cut geometry is given by (5.1.16). In particular, we have a positively oriented region for $\langle AB_1B_2B_3 \rangle > 0$ and a negatively oriented region for $\langle AB_1B_2B_3 \rangle < 0$. Now consider fixing the line P (with $B_1$ and $B_2$ lying on P) and fixing $B_3$. Now consider passing $B_1$ through $B_2$ on the line P. As we saw in the two loop case the orientation for the geometry involving $B_1, B_2$ will swap, but simultaneously $\langle B_1B_2B_3 \rangle \to -\langle B_1B_2B_3 \rangle$ and so the overall orientation will also swap (see (5.1.16)). The result is no orientation change at all. We are now interested in the geometry of the internal boundary $\langle AB_1B_2B_3 \rangle = 0$ so moving $B_3$ also onto the line P. The point $B_3$ is free to go anywhere on the line P (inside the triangle). The resulting geometry is indeed just that of three free points on the line P with the canonical form (5.2.19), including the factor of 2 from taking an internal boundary.

Notice that in this case there are no remaining singularities of the form $B_i \to B_j$ so $\mathcal{A}^{(3)}_{\text{point-plane}}$ also represents what we call a maximal loop-loop cut and $2\mathcal{A}^{L=3}_{\text{mll}} := \mathcal{A}^{(3)}_{\text{point-plane}}$.

### 5.2.3 All 4-loop point-plane and maximal loop-loop cuts

We have seen that at 3-loops the all-in-one-point-and-plane cut is unique (5.2.19). At 4-loops this is not true anymore and we can have two types of geometry (modulo permutations) resulting from approaching the point-plane configuration in different ways. Each will be characterized algebraically by different coefficients $c_\sigma$ in (5.2.18) and geometrically by different ordering constraints of the Bs on the line P.

We start with the all-in-one-point cut which is unique up to permutations of the loop lines and the resulting geometry given by (5.1.21). We now consider taking further boundaries of this geometry so the loops also lie in a plane. There are 3 possible loop-loop boundaries, $\langle AB_2B_1B_3\rangle = 0, \langle AB_2B_1B_4\rangle = 0$ and $\langle AB_3B_1B_4\rangle = 0$. We start by looking at the geometry of the boundary when $B_3$ lies on the line P (on which $B_1, B_2$ lie) followed by the boundary of that geometry found when $B_4$ also approaches P (shortly we will switch the order of in which we take these boundaries).

The resulting geometry is of the four points $B_i$ on the line P inside the triangle, with the following restrictions: $B_1$ is not allowed to be between $B_2$ and $B_3$ and the orientation depends on the relative position of $B_1$ and the pair $B_2, B_3$ (with the position of $B_4$ unconstrained). This geometry is derived in appendix C.2.

The cut resulting from this geometry is then given by the general form (5.2.18) with

$$n_I = 2 \qquad c_\sigma = \begin{cases} 1 & \sigma = (2,3,1) \shuffle (4) \text{ or } (3,2,1) \shuffle (4), \\ -1 & \sigma = (1,2,3) \shuffle (4) \text{ or } (1,3,2) \shuffle (4), \\ 0 & \sigma = (2,1,3) \shuffle (4) \text{ or } (3,1,2) \shuffle (4), \end{cases} \qquad (5.2.20)$$

where $\shuffle$ is the shuffle operation (thus 4 can appear in any position). We have checked this is indeed correct by explicitly taking the corresponding residues of the 4 loop amplitude and finding perfect agreement.

Now note that even after taking the all-in-one-point *and* the all-in-one-plane configuration there are still uncancelled loop-loop poles at $B_1 = B_2$ and $B_1 = B_3$, corresponding geometrically to external boundaries.

Using the very simple residue structure of the ordered points on an interval canonical form, namely

$$\text{Res}_{B_k \to B_j}[I;..,B_i,B_j,B_k,B_l..;J] = -\text{Res}_{B_k \to B_j}[I;..,B_i,B_k,B_j,B_l..;J] = [I;..,B_i,B_j,B_l..;J],$$
$$(5.2.21)$$

one can quickly check that the residue of the four-loop point-plane cut (5.2.20) when $B_1 = B_2$ or $B_1 = B_3$ precisely reproduces the three-loop point-plane cut (5.2.19) with appropriate variables and with weight 4 instead of 2 (since this time we approached two internal boundaries, $\langle AB_2B_1B_3\rangle = 0$ and $\langle AB_2B_3B_4\rangle = 0$). This implies that the residue corresponding to this 4-loop maximal loop-loop boundary is equal to 2 times the all-in-one-point-and-plane 3-loop residue (C.1.12).

Returning to the all-in-one-point cut, we now consider the only other independent way of reaching the point-plane geometry (modulo permutation of the loop variables) by taking $\langle AB_2B_1B_4\rangle = 0$ by sending $B_4$ to the line P followed by sending $B_3$ to the line P (the other way around to what we did above). In appendix C.2 we again examine this carefully geometrically. The end result this time is the geometry of four points $B_i$ unconstrained on the line P but with the overall orientation dependent on the ordering of $B_1, B_2$. The resulting canonical form is thus given by (5.2.18) with

$$n_I = 1 \qquad c_\sigma = \begin{cases} 1 & \sigma = (2,1) \shuffle (3,4)\,, \\ -1 & \sigma = (1,2) \shuffle (3,4)\,, \end{cases} \qquad (5.2.22)$$

as we have confirmed by taking the residues explicitly and comparing.

Note that this time there is a remaining loop-loop residue apparent at $B_1 \to B_2$ (corresponding to an internal boundary). Taking the residue as above this again leads to the three loop point-plane result (5.2.19) after which no more loop loop residues are present. This final configuration corresponds to the 3-loop maximal cut $\mathcal{A}_{\mathrm{mll}}$ with weight 2.

### 5.2.4 A cut at arbitrary loop order

We have already seen from the four loop examples of the previous section that the point-plane cut depends on how you approach the configuration. However one can give specific ways of approaching the point-plane geometry at any loop order and find the resulting cut. So we conclude this section by giving precisely such an example of a cut that can be computed at arbitrary loop order. This means that we are now specifying an ordered set of residues and giving a closed formula for the result. The particular case is a generalisation of the second 4-loop case considered in the previous subsection.

We start defining what we call the simplest all-in-one-point cut. In this all loop lines first intersect the line $A_1B_1$ and then they all slide to the same intersection point in the same order as their labeling. The geometry of this boundary at L loops can be obtained from (C.3.4) and is given explicitly in appendix C.3.

After taking the above all-in-one-point cut we then constrain all loops to lie on the same line P, first $B_1, B_2$ then $B_L, B_{L-1}, .., B_3$ thus taking the ordered series of boundaries $\{\langle AB_2B_1B_L\rangle = 0, \langle AB_2B_1B_{L-1}\rangle = 0, \cdots, \langle AB_2B_1B_3\rangle = 0\}$.

Carefully examining the resulting geometry as is done explicitly in appendix C.3, we arrive at the final geometry corresponding to this point-plane cut. It is given by L points $B_i$ lying on the line P, with $B_3, B_L$ unconstrained and with $B_1$ always lying between $B_2$ and all of the points $B_4, .., B_{L-1}$. The orientation of the geometry depends on the relative order of $B_1, B_2$. The resulting canonical form at arbitrary loop order is thus

$$\mathcal{A}^{(L)}_{\text{point-plane}} = 2 \sum_{i=1}^{4} \omega(A) \sum_{p=1}^{3} \lambda_{i,p}(P)[I_{i,p}; B_3; J_{i,p}][I_{i,p}; B_L; J_{i,p}] \times$$

$$\times \left( [I_{i,p}; B_2, B_1; J_{i,p}] \prod_{l=4}^{L-1} [B_1; B_l; J_{i,p}] + (-1)^{L-1} [I_{i,p}; B_1, B_2; J_{i,p}] \prod_{l=4}^{L-1} [I_{i,p}; B_l; B_1] \right)$$
(5.2.23)

We tested (5.2.23) by computing this simplest maximal loop-loop residue up to 7 loops from the explicit from of the amplitude obtained in [98] and found complete agreement. We see that this point-plane cut has further poles when $B_l \to B_1$. We have checked up to 5 loops that taking further residues in these poles eventually leads to the 3 loop point-plane cut (5.2.19). Indeed our investigations so far indicate that after taking any all-in-one-point-and-plane geometry at any loop order, there are always L – 3 boundaries of the form $B_i \to B_j$ remaining. After further taking these boundaries we are then always lead to the three loop point-plane geometry (5.2.19). It might be possible to prove this starting from the explicit all-in-one-point geometry (5.1.19).

## 5.3 Summary

The residue of the canonical form of a WPG is equal to the canonical form of the corresponding boundary. Therefore, the knowledge of the geometry of a boundary of the amplituhedron and its canonical form gives important information on the analytic structure of the amplitude.

In this chapter, we started by describing how to compute algebraically the so-called all-in-one-point cut on any loop amplitude. In particular, we showed that the all-in-one-point cut is not unique and depends on the order in which the residues that compose it are taken. Also, we showed that, starting at three-loop, brackets of the form $\langle ABBB \rangle$ are generated and these correspond to poles of the cut.

One of the main results of [2] described in this chapter is the derivation of the geometry of any all-in-one-point cut. Given an all-in-one-point cut, its geometry is simply given by (5.1.19). The geometry of the all-in-point cut is characterized by a new set of loop-loop inequalities of the form $\langle AB_iB_jB_k \rangle > 0$. In the last section, we showed that it is possible to find fairly simple all-loop geometries by taking further residues after reaching the all-in-one-point cut configuration that trivializes all brackets of the form $\langle AB_iB_jB_k \rangle > 0$. In such a cut all loop-lines intersect at a point and lie on the same plane. We derived the canonical form of one of these cuts and give its expression in equation (5.2.23).

Even when we have trivialized all the $\langle AB_iB_jB_k \rangle$ brackets more residues involving pairs of loops are possible. These residues correspond geometrically to making two loop lines coincide. We call a cut for which we have exhausted all loop-loop type residues a maximal loop-loop cut. We conjecture that all maximal loop-loop cuts are always equal to the 3-loop maximal loop-loop cut up to an integer which can be understood geometrically as the number of internal boundaries crossed in the residue.

# Chapter 6

# Geometry of the Product of Amplitudes

The content of this chapter has been published in the paper [1].

The main purpose of this chapter is to present the results obtained in [1] in investigating the squared amplituhedron conjecture and the geometry of the product of parity conjugate superamplitudes. In particular we will present the proof that at tree-level in all cases with minimal number of points, the squared amplituhedron indeed gives the square of the amplitude.

We will focus almost entirely on the case of amplituhedron-like geometries with minimal number of points (maximal MHV degree) ie k = n–m where m = 4 in the physically interesting case but we often consider general m also. This is a big simplification and in particular means that the external data is trivialised. For the amplituhedron itself this case corresponds simply to the anti-MHV amplitude. However, for amplituhedron-like geometries there is a very rich structure even in this sector. It contains all amplitudes multiplied by their parity conjugate amplitudes, but there is evidence that the individual amplitudes themselves can be extracted from this combination [117]. Furthermore this sector corresponds to taking various light-like limits of four-point correlators about which there is a wealth of concrete information. Their integrands have a hidden permutation symmetry [90] and this has helped obtain their explicit expression up to ten loops [97, 98, 118].

We will use the following notation to distinguish between geometrical regions, the corresponding expression in bosonised superspace, and the corresponding expression in superspace:

|  | geometry | bosonised superspace | superspace |
|---|---|---|---|
| amplituhedron | $\mathscr{A}_{n,k,l}$ | $A_{n,k,l}$ | $\mathcal{A}_{n,k,l}$ |
| amplituhedron-like | $\mathscr{H}_{n,k,l}^{(f;l')}$ | $H_{n,k,l}^{(f;l')}$ | $\mathcal{H}_{n,k,l}^{(f;l')}$ |

So in particular the expression in bosonised superspace is obtained form the geometry by taking the canonical form $A_{n,k,l} = \Omega(\mathscr{A}_{n,k,l})$ and $H_{n,k,l}^{(f;l')} = \Omega(\mathscr{H}_{n,k,l}^{(f;l')})$ and the expression in superspace is obtained from the expression in bosonised superspace by integrating out the auxiliary Grassmannian variables $\phi$ appearing in (3.2.3).

The squared amplituhedron (3.5.15) is a similar geometry to the amplituhedron-like geometry, constrained just by proper boundary inequalities but with no version of the winding condition and it can thus be viewed as the union of all amplituhedron-like geometries. The square of the superamplitude with fixed MHV degree $k = n - 4$ is given by the sum over $k'$ of the product of the $N^{k'}$MHV amplitude, $\mathcal{A}_{n,k'}$ and its conjugate $\mathcal{A}_{n,n-k'-4}$. The number of terms in this sum coincides precisely with the number of inequivalent geometries tiling the squared amplituhedron. It is thus natural to propose a precise relation namely that: the amplituhedron-like geometry with flipping number f, $\mathscr{H}_{n,k}^{(f)}$, gives the product of the $N^f$MHV superamplitudes and its conjugate,

$$H_{n,n-4}^{(f)} = A_{n,f} * A_{n,n-f-4} \,. \tag{6.0.1}$$

We also make a similar proposal at loop level introducing a flipping number for loops $l'$

$$H_{n,n-4,l}^{(k',l')} = \binom{l}{l'} A_{n,k',l'} * A_{n,n-k'-4,l-l'} \,. \tag{6.0.2}$$

A proposal along these lines was previously made in [109] for the MHV case with arbitrary number of points $H_{n,0,l}^{(0;l')}$. At first sight this is a different sector to the case we consider. However, due to factorisation of anti-MHV amplitudes this in fact corresponds to $H_{n,f,l}^{(f;l')}$ and we will prove the relation for this case as well as at tree level. Here the product of amplitudes in superspace becomes a particular combination we call the star product of bosonised superamplitudes. We will give a precise definition of this star product.

It is also possible to give an alternative characterisation of amplituhedron-like geometries analogous to the original definition of the amplituhedron. Tree-level amplituhedron-like geometries with flipping number f are given in terms of a subset of the set of matrices $C \in Gr(k,n)$ projected through Z. However, rather than this subset of matrices C having positive ordered maximal minors (which would give the amplituhedron) instead it is made up by stacking two submatrices $C_1$ (an $f \times n$ matrix) and $alt(C_2)$ (a $(k-f) \times n$ matrix) where $C_1$ and $C_2$ have all positive ordered maximal minors and the matrix $alt(C_2)$ is formed from $C_2$ by flipping the sign of every odd column. A similar alternative characterisation of the amplituhedron-like geometry can also be made at loop level.

Combinations of onshell diagrams (arising from BCFW recursion) result in tilings of the amplituhedron. In a similar way we show that at tree-level pairs of onshell diagrams give a direct tiling of the amplituhedron-like geometry. This fact can then be used to prove (6.0.1) for all multiplicity and winding number. We also prove the proposal at loop level in the simplest case of maximal (or equivalently minimal) flipping number f giving MHV×anti-MHV at specified loop levels at all multiplicity and loop order.

Having understood the amplituhedron-like geometries it is interesting to return to the squared amplituhedron which is the union of amplituhedron-like geometries with different flipping number. The square of the superamplitude shares with the superamplitude the property that it has only proper poles and dlog divergences. Differently from the superamplitude however, its maximal residues are not all normalizable to $\pm 1, 0$. This implies that the squared amplituhedron is not a PG but a GPGs

Finally, all the geometries cited so far are defined by a system of inequalities depending on the kinematic data as parameters. Thus it is interesting to see if there are any other obvious further generalisations of the amplituhedron geometry for example by considering similar defining inequalities but with different choices of signs. As a modest step in this direction we examine carefully the consequence for such a geometry of demanding it has a manifest cyclic canonical form. While the canonical form, i.e. the amplitude, is invariant under the rescaling of the external data $Z_i \to \lambda Z_i$, the geometry is invariant only under positive rescaling $\lambda > 0$. Nevertheless, geometries related by such a transformation with $\lambda < 0$ have the same canonical form. We thus define geometries to be *equivalent* if they are related by a flip of some Zs. This type of observation has already been a fundamental ingredient for proving perturbative unitarity using the amplituhedron [34]. Examining all possible versions of manifest geometrical cyclicity we find that all are equivalent to either cyclic or twisted cyclic geometries, thus drastically cutting down the different geometries under consideration. As a result of this line of thinking We find an equivalence relation between amplituhedron-like geometries with complementary flipping number and new bounds for the values that they can assume. The transformation linking the two equivalent geometries corresponds to $Z_i \to (-1)^i Z_i$, a map that is closely related to parity [21]. Using similar ideas, we consider also the maximally nilpotent correlator $\mathcal{G}_{n-4,n}$ and we prove that all the geometries with the minimal requirements to be compatible with correlator pole structure are equivalent to the correlahedron.

This chapter is structured as follows. In section 6.1 we introduce the formulation of the superamplitude in dual momentum twistor variables and we review the bosonised superamplitude with some emphasis on its normalization. Then, we define the superamplitude squared and define a product between functions directly in the bosonised superamplitude space which we call * product. In section 6.2 we define the amplituhedron-like geometries and we state our conjecture for the canonical form of amplituhedron-like geometries as products of amplitudes at tree as well as loop level. Then we define the squared amplituhedron as the union of all amplituhedron-like geometries and we conjecture that its (oriented) canonical form corresponds to the square of the superamplitude. We also give an alternative definition of the amplituhedron-like geometry as a projection of the positive and the alternating positive Grassmannian which we will then use to prove our conjecture at tree level. In section 6.4 we show how any plane in the amplituhedron-like geometry can be seen as the product of two planes each belonging to one amplituhedron and we use this fact along with on-shell diagrams to prove our conjecture at tree level. We then give a proof of the conjecture at all loops for the product of MHV and anti-MHV amplitudes. We conclude the section by looking at some explicit computations for n $\leq$ 7 and to some generalized amplituhedron-like geometries for m = 2, 6, 8. In section 6.5 we formulate a refined version of our conjecture for the canonical form of regions in the amplituhedron-like geometries characterized by a precise set of inequalities called sign-flip pattern. Finally in section 6.6 we study the equivalence relations between geometries with a cyclic canonical form and we find that for each equivalence class we can always choose cyclic or twisted cyclic representatives. We then consider the maximal nilpotent correlator $\mathcal{G}_{n,n,4}$ and prove that all consistent geometries are equivalent to the correlahedron.

## 6.1 Product of amplitudes in amplituhedron space

The superamplitude squared (2.3.11) can be bosonized in the same way as the superamplitude and therefore mapped to a differential form on the Grassmannian, which we will refer as the amplituhedron space. The superamplitude squared is given by a sum of products of amplitudes and we would like therefore understand the outcome of taking the product of amplitudes directly in amplituhedron space. Note that this can not be given simply by the product of amplitudes in amplituhedron space, as these will live in different spaces. Instead we define a map we call $*$ which takes two amplitudes in amplituhedron space and produces a third amplitude in amplituhedron space which will be equivalent to the product of the two original superamplitudes under the map (3.2.2),(3.2.3):

$$\mathcal{B}_{k_1+k_2,4}\Big(\mathcal{A}_{n,k_1}(\mathcal{Z}_i)\mathcal{A}_{n,k_2}(\mathcal{Z}_i)\Big) = A_{n,k_1}(Z_i,Y_{k_1}) * A_{n,k_2}(Z_i,Y_{k_2}) \ . \qquad (6.1.1)$$

Note that the $*$ takes an object in $k_1+4$ dimensions and an object in $k_2+4$ dimensions and outputs an object in $k_1+k_2+4$ dimensions.

We now give an explicit definition of this $*$ product via its action on arbitrary dual superconformal invariants (3.2.7). So we consider the product of two dual superconformal building blocks (3.2.4) of degree $k_1$ and $k_2$ respectively. In superspace the product is clear, but what happens in the bosonised amplituhedron space when we take the product? Generalising to arbitrary m, the bosonised invariants live in dimensions, $k_1 + m$ and $k_2 + m$ dimensions respectively, and we want to write the product as an object $k_1 + k_2 + m$ dimensions. To keep track of the $\phi$ dependence we will add the subscript $k_1 + m$ to the $k_1 + m$ dimensional brackets and the subscript $k_2 + m$ to the $k_2 + m$ dimensional brackets. We label these brackets by the strings $I_a \in \binom{[n]}{k_1+m}$ and $J_b \in \binom{[n]}{k_2+m}$.

We claim that the $*$ product of bosonized brackets is given by the formula

$$\left(\prod_{a=1}^{m} \langle I_a \rangle_{k_1+m}\right) * \left(\prod_{b=1}^{m} \langle J_b \rangle_{k_2+m}\right) = \frac{(-1)^{(k_1 k_2 + k_2)m}}{m!} \sum_{\sigma \in S_m} \prod_{a=1}^{m} \langle Y(I_a \cap J_{\sigma(a)}) \rangle_{k_1+k_2+m} ,$$
(6.1.2)

where Y is in $\text{Gr}(k_1+k_2, k_1+k_2+m)$. Here $S_m$ is the set of permutations of m elements and $(I \cap J)$ represents an intersection in $k_1 + k_2 + m$ dimensions, explicitly:

$$\langle Y(I \cap J) \rangle = \sum_{i \in M(I)} \langle Yi \rangle \langle \bar{i}J \rangle \, \text{sgn}(i\bar{i}) ,$$
(6.1.3)

where $M(I) = \binom{I}{m}$, that is the set of ordered m tuples in I, and $\bar{i}$ is the ordered complement of i in I, that is $\bar{i} = I - i$.

Note that if we set Y to $Y_0$ and the Zs to $Z(\chi)$ (3.2.1) and include the normalisation factor $N(k, m)$ then the star product formula (6.1.2) must reduce to an ordinary product. (This is just from the defining equation (6.1.1) and the definition of the map $\mathcal{B}$ (3.2.2),(3.2.3)). Thus to prove the explicit form of the star product (6.1.2) we need to check that when $Z \to Z(\chi)$:

$$\frac{1}{N(k_1, m)} \left(\prod_{a=1}^{m} \langle I_a \rangle_{k_1+m}\right) \frac{1}{N(k_2, m)} \left(\prod_{b=1}^{m} \langle J_b \rangle_{k_2+m}\right) =$$

$$= \frac{1}{N(k_1 + k_2, m)} \frac{(-1)^{(k_1 k_2 + k_2)m}}{m!} \sum_{\sigma \in S_m} \prod_{a=1}^{m} \langle Y_0(I_a \cap J_{\sigma(a)}) \rangle_{k_1+k_2+m} .$$
(6.1.4)

We include the proof of this for m = 1 and some checks for m = 2 and m = 4 in the appendix.

### 6.1.1 NMHV squared example

As an example, let's look at the squared amplitude $(\mathcal{A}^2)_{6,2}$. This is given by two terms

$$(\mathcal{A}^2)_{6,2} = 2\mathcal{A}_{6,2} + (\mathcal{A}_{6,1})^2 \ . \tag{6.1.5}$$

Now we want to express (6.1.5) as a function on the bosonised amplituhedron superspace. The first term is the 6 points anti-MHV amplitude $A_{n,n-4}$ given in (3.2.9). For the second term we start with the BCFW expression for $A_{6,1}$ [73] in terms of the 5-point NMHV-invariant (3.2.6), that is

$$A_{6,1} = [12345] + [12356] + [13456] \ . \tag{6.1.6}$$

To compute the square of $A_{6,1}$ we need the (star) product of 5-brackets. Identifying a 5-bracket as $\langle \hat{i} \rangle$, where $\hat{i}$ indicates the unique twistor that is not present and a 4-bracket as $\langle \hat{i}\hat{j} \rangle$ similarly, the star product formula (6.1.2) gives

$$\langle \hat{i} \rangle^4 * \langle \hat{j} \rangle^4 = \langle Y\hat{i}\hat{j} \rangle^4 \langle 123456 \rangle^4 \ . \tag{6.1.7}$$

Indeed, as pointed out in [117], the result is completely fixed up to proportionality by matching the scaling in each Z. The square of any R-invariant will be equal to zero. We possess now all the elements to compute $(A_{6,1})^2$ and obtain

$$\begin{aligned}(A_{6,1})^{*2} &= 2\left([12345]*[12356]+[12345]*[13456]+[13456]*[12356]\right) = \\ &= 2\langle 123456\rangle^4 \frac{\langle 1245\rangle\langle 2361\rangle\langle 3456\rangle+\langle 2356\rangle\langle 3412\rangle\langle 4561\rangle+\langle 3461\rangle\langle 4523\rangle\langle 5612\rangle}{\prod_{i=1}^{3}\langle i(i+1)(i+3)(i+4)\rangle)\prod_{i=1}^{6}\langle i(i+1)(i+2)(i+3)\rangle},\end{aligned} \tag{6.1.8}$$

where the 4-brackets $\langle * \rangle$ are short-hand for $\langle Y* \rangle$. Summing this result with $A_{6,2}$ (3.2.9) we obtain $(A^2)_{6,2}$ in amplituhedron space.

### 6.1.2 Product of multiple amplitudes

The product of multiple bosonised brackets can be computed just by using the associative property of the * product. However, it's also possible to write a direct formula for the * product of multiple brackets. To do this notice that

$$\langle Y(I \cap J) \rangle = \langle I(Y \cap J) \rangle \ . \tag{6.1.9}$$

which can be checked by expanding the respective intersections on each side out over the J basis

$$\langle Y(I \cap J)\rangle = \sum_j \langle Yj\rangle \langle I\bar{j}\rangle \operatorname{sgn}(j \cup \bar{j}) ,$$
$$\langle I(Y \cap J)\rangle = \sum_j \langle I\bar{j}\rangle \langle Yj\rangle \operatorname{sgn}(\bar{j} \cup j) . \qquad (6.1.10)$$

Using this alternative expression, equation (6.1.2) for the product of 2 terms naturally generalizes to the product of t terms as

$$\left(\prod_{a=1}^m \langle I_{1,a}\rangle_{k_1+m}\right) * \cdots * \left(\prod_{a=1}^m \langle I_{t,a}\rangle_{k_t+m}\right) =$$
$$= \frac{1}{(m!)^t} \sum_{\sigma_* \in (S_m)} \prod_{a=1}^m \langle I_{1,a}(Y \cap I_{2,\sigma_2(a)}) \cdots (Y \cap I_{t,\sigma_t(a)})\rangle_{k_1+\cdots+k_t+m} , \qquad (6.1.11)$$

up to a sign which is positive for m even and depends on $k_1, \cdots, k_t$ for m odd.

## 6.2 Amplituhedron-like geometries

Having discussed the product of amplitudes in amplituhedron space we now turn to the corresponding geometries. We define a natural generalisation of the amplituhedron called "amplituhedron-like" geometries which we will prove to correspond to the product of amplitudes in the maximal k = n – 4 case.

The sign flip definition of the amplituhedron (3.4.4) has the desirable feature of treating the proper boundaries and the other constraints separately, so we can modify the second while leaving the first the same. A natural generalization of these geometries is then to relax the constraint on the number of sign flips in (3.4.4). We thus define a tree-level amplituhedron-like geometry, $\mathscr{H}_{n,k}^{(f)}$, by fixing the number of flips f. To be consistent with cyclic or twisted cyclic invariance, the proper inequalities must be tweaked accordingly. Thus concretely we define amplituhedron-like geometries

$$\mathscr{H}_{n,k}^{(f)} := \left\{ Y \in \operatorname{Gr}(k, k+4) \,\middle|\, \begin{array}{ll} \langle Yii+1jj+1\rangle > 0 & 1 \leq i < j-1 \leq n-2 \\ \langle Yii+11n\rangle (-1)^f > 0 & 1 \leq i < n-1 \\ \{\langle Y123i\rangle\} & \text{has f sign flips as } i = 4, .., n \end{array} \right\}$$
$$\text{for } Z \in \operatorname{Gr}_+(k+4, n) , \qquad (6.2.1)$$

In [109] it was proven that for a k-plane with convex Zs the maximal allowed number

of flips is exactly k so

$$0 \leq f \leq k \,. \tag{6.2.2}$$

We can see that the amplituhedron itself is then the case of an amplituhedron-like geometry with $f = k$,

$$\mathscr{A}_{n,k} = \mathscr{H}^{(k)}_{n,k} \,. \tag{6.2.3}$$

The loop amplituhedron can also be generalised in a similar fashion. Here we allow for an arbitrary flipping number, $f_j$, for each loop variable. The generalization of the loop amplituhedron is given by a Y belonging to a tree amplituhedron-like geometry and the loop variables satisfying

$$\mathscr{H}^{(f;f_1,..,f_l)}_{n,k,l} := \left\{ Y, (AB)_1, ., (AB)_l \,\middle|\, \begin{array}{ll} Y \in \mathscr{H}^{(f)}_{n,k} & \\ \langle Y(AB)_j ii{+}1 \rangle > 0, & \forall\, j,\, \forall i = 1,.,n\text{–}1 \\ \langle Y(AB)_j 1n \rangle (-1)^{f_j} > 0 & \forall\, j \\ \{\langle Y(AB)_j 1i \rangle\} & \text{has } f_j \text{ flips as } i = 2,..,n, \forall j \\ \langle Y(AB)_i (AB)_j \rangle > 0 & \forall i \neq j \end{array} \right\}$$

$$\text{for } Z \in \mathrm{Gr}_>(k+4, n)\,. \tag{6.2.4}$$

The amplituhedron itself is then the case $f = k$ and $f_j = k + 2$

$$\mathscr{A}_{n,k,l} = \mathscr{H}^{(k;k+2,k+2,..,k+2)}_{n,k,l} \,. \tag{6.2.5}$$

In the maximal k case, $k = n - 4$, Z is a square $n \times n$ matrix and thus is always in $\mathrm{Gr}_>(k+4, n)$ (or equivalently $\mathrm{Gr}_<(k+4, n)$ if the determinant is negative). Thus for much of what follows we will restrict to this case $k = n - 4$.

Now we would like to see what are the possible values for the loop flipping numbers $f_j$. If we project positive Z's through a k-plane Y with flipping number f, we obtain a configuration of Z's on $Y^\perp$ that is defined by the brackets $\langle ijkl \rangle_Y = \langle Yijkl \rangle$. The $\langle ijkl \rangle_Y$ satisfies the same inequalities as those of the $N^f$MHV amplituhedron, $\mathscr{A}_{n,f}$. In [109] it is conjectured[1] that any Z configuration $\langle ijkl \rangle$ with positive proper boundaries and flipping number equal to f can be generated as a projection of positive $\tilde{Z}$s though an f-plane $\tilde{Y} \in \mathscr{A}_{n,f}$. This conjecture implies that for any $Y \in \mathscr{H}^{(f)}_{n,k}$ there exists $\tilde{Y} \in \mathscr{A}_{n,f}$, $\tilde{Z} \in \mathrm{Gr}_>(f+m,n)$ and $\widetilde{AB} \in \tilde{Y}^\perp$ such that

$$\langle ijkl \rangle_Y = \langle \tilde{i}\tilde{j}\tilde{k}\tilde{l} \rangle_{\tilde{Y}}\,, \quad \langle ABij \rangle_Y = \langle \widetilde{AB}\tilde{i}\tilde{j} \rangle_{\tilde{Y}} \quad \forall\, i,j,k,l\,. \tag{6.2.6}$$

---

[1] The original formulation of conjecture is that given some (m»n) matrix of Zs that satisfy the winding/flip criteria, we can always add k more rows so that the resulting (k+m) × n matrix is positive.

Therefore the sign flip string $\langle YAB1i \rangle$ has the same constraints as the sign flip string $\langle \tilde{Y}\widetilde{AB}\tilde{1}\tilde{i} \rangle$. We know that the maximal flipping number for k-planes with positive Zs is k. Here $(\tilde{Y}\widetilde{AB})$ is an $(f+2)$-plane and thus has maximal flipping number $f+2$. We can then conclude that the $\langle YAB1i \rangle$ flipping number must also be less than or equal to $f+2$. Moreover, the twisted cyclicity condition for Y (second line of (6.2.1)) must be consistent with the twisted cyclicity condition for each $(AB)_j$ (third line of (6.2.4)). We thus have the following restrictions on the loop flipping numbers $f_j$ in order to obtain a sensible geometry yielding a cyclic non-trivial canonical form

$$f_j \le f+2\,, \qquad f_j = f \mod 2\,. \tag{6.2.7}$$

But there is a stronger constraint which is easiest to see by considering the following equivalence map of geometries.

If we change the sign of alternate Zs, and all loop variables, we obtain a map between amplituhedron-like spaces with different flipping numbers, $\mathscr{H}^{(f;f_j)}_{n,n-4,l} \mapsto \mathscr{H}^{(n-4-f;\,n-2-f_j)}_{n,n-4,l}$. More concretely:

$$\mathscr{H}^{(f;f_j)}_{n,n-4,l}\Big(Y,(AB)_j;Z_i\Big) = (-1)^{\lfloor \frac{n+1}{2} \rfloor} \mathscr{H}^{(n-4-f;\,n-2-f_j)}_{n,n-4,l}\Big((-1)^{\lfloor \frac{n+1}{2} \rfloor}Y, -(AB)_j; Z_i(-1)^i\Big)\,, \tag{6.2.8}$$

where an overall minus in front of $\mathscr{H}$ indicates that we also reverse all the inequalities (or equivalently send all $\langle .. \rangle \mapsto - \langle .. \rangle$). This relation can be checked by just considering the definitions on both sides. For example the sign of every second element of the string $\langle Y123i \rangle$ is swapped under $Z_i \mapsto Z_i(-1)^i$. Thus every sign flip in the original space becomes a non sign flip and vice versa, and thus the flipping number $f \mapsto n-4-f$.

The canonical forms arising from the two geometries $\mathscr{H}^{(f;f_j)}_{n,n-4,l}$ and $\mathscr{H}^{(n-4-f;\,n-2-f_j)}_{n,n-4,l}$ are identical and we thus say that the geometries are "equivalent"

$$\mathscr{H}^{(f;f_j)}_{n,n-4,l} \sim \mathscr{H}^{(n-4-f;\,n-2-f_j)}_{n,n-4,l}\,. \tag{6.2.9}$$

This equivalence then implies a much stronger bound on the allowed loop flipping numbers. We require $f_j \le f+2$ but also for the dual geometry (6.2.9) this means $n-2-f_j \le n-4-f+2$ ie $f \le f_j$. Together with (6.2.7) we then see that each loop flipping number can only take 2 possible values

$$f_j = f \quad \text{or} \quad f+2\,. \tag{6.2.10}$$

With this in mind, we only need to keep track of the relative number of $f_j$s which are equal to $f+2$ and those which are equal to f. Finally, it is also useful to define a geometry obtained by symmetrising over these variables. Thus we also define a loop amplituhedron-like geometry with just two superscripts, $f, l'$ where $l'$ is the number of loops with maximal flipping number $f+2$

$$\mathscr{H}_{n,n-4,l}^{(f;l')} := \bigcup_{\sigma \in S_l/(S_{l'} \times S_{l-l'})} \mathscr{H}_{n,n-4,l}^{(f;\sigma(\overbrace{f+2,..,f+2}^{l'},\overbrace{f,...,f}^{l-l'}))} , \qquad (6.2.11)$$

where we take the union over all inequivalent choices of taking $l'$ loop variables to have maximal flipping number $f+2$ and the remaining ones minimal flipping number f.

## 6.2.1　Conjecture: Amplituhedron-like geometries give products

Having defined a natural generalisation of the amplituhedron, the amplituhedron-like geometries, we now discuss what they correspond to physically. First at tree level, focusing on the maximal $k = n-4$, there are $k+1$ amplituhedron-like geometries $\mathscr{H}_{n,k}^{(f)}$, $f = 0,..,k$ with f equivalent to $k-f$ through (6.2.9). This perfectly mimics the possible products of two amplitudes of total Grassmann degree $k = n-4$, $\mathcal{A}_{n,k'}\mathcal{A}_{n,n-4-k'}$. We conjecture that the canonical form H of an amplituhedron-like geometry $\mathscr{H}$ gives the star product (see (6.1.1)) of superamplitudes

$$\text{main conjecture (tree-level):} \quad \boxed{H_{n,n-4}^{(f)} = A_{n,f} * A_{n,n-f-4}} . \qquad (6.2.12)$$

Note that in the case of maximal flipping number, $f = n-4$, this conjecture collapses to the standard amplituhedron conjecture (recalling that $A_{n,0} = 1$). We will define the canonical form in the next section and then in the following section describe the various proofs and checks giving evidence for this conjecture which we have performed.

The amplituhedron-like geometries at loop level depend also on the flipping number of the loop variables, $l'$ (see (6.2.11)). We thus generalize (6.2.12) to loop level and conjecture that for $k = n-4$ the canonical form of the loop amplituhedron-like geometry with $l'$ loops having maximal flipping number and $l-l'$ loops minimal is

$$H_{n,n-4,l}^{(f;\overbrace{f+2,..,f+2}^{l'},\overbrace{f,...,f}^{l-l'})} = A_{n,f,l'}(AB_1,\cdots,AB_{l'}) * A_{n,n-f-4,l-l'}(AB_{l'+1},\cdots,AB_l) , \qquad (6.2.13)$$

where the loop variables with maximal flipping number $f + 2$ belong to the first factor $A_{n,k',l'}$ and the remaining loop variables to $A_{n,n-k'-4,l-l'}$. By summing over inequivalent permutations of the loops we then obtain

$$H^{(k',l')}_{n,n-4,l} = \sum_{\sigma \in S_l/(S_{l'} \times S_{l-l'})} A_{n,k',l'}((AB)_{\sigma(i)}) * A_{n,n-k'-4,l-l'}((AB)_{\sigma(i)}) , \quad (6.2.14)$$

suppressing the explicit distribution of loop variables this can be written in the more compact form

$$\text{main conjecture (loop)-level} \quad : \quad \boxed{H^{(k',l')}_{n,n-4,l} = \binom{l}{l'} A_{n,k',l'} * A_{n,n-k'-4,l-l'} .} \quad (6.2.15)$$

One can see that (6.2.15) is consistent with the duality (6.2.9) and it's trivially true for the case $k' = n - 4, l' = l$ which collapses to the standard amplituhedron conjecture for the anti-MHV loop level amplitude

$$H^{(n-4,l)}_{n,n-4,l} = A_{n,n-4,l} * A_{n,0,0} = A_{n,n-4,l} . \quad (6.2.16)$$

Last but not least, the conjecture (6.2.15) is consistent with the squared amplituhedron conjecture. In particular the squared amplituhedron is defined in (3.5.15) as the union of two geometries defined by physical inequalities only, ie with no topological winding condition. The amplituhedron-like geometries are clearly subsets of $\mathcal{H}^{\pm}_{n,k,l}$, Furthermore the union of all even/odd flipping numbered amplituhedron-like geometries clearly gives $\mathcal{H}^{\pm}_{n,n-4,l}$ and the union of all amplituhedron-like geometries gives the squared amplituhedron $\mathcal{H}_{n,n-4,l}$

$$\mathcal{H}_{n,n-4,l} = \bigcup_{f,l'} \mathcal{H}^{(f,l')}_{n,n-4,l} . \quad (6.2.17)$$

Comparing with the expansion of the amplitude squared (2.3.15) into precisely the same products we get from the amplituhedron-like geometries, it is thus natural to conclude that the canonical form of $\mathcal{H}_{n,n-4,l}$ is the square of the amplitude. However, one has to be a bit more careful. To prove that the canonical form of the squared amplituhedron is equal to the square of the superamplitude starting from (6.2.15) we would need to specify the amplituhedron-like geometries orientation, so that the product of amplitudes sum with the right signs. We will see that such orientation choice corresponds to having all the amplituhedron like geometries with the same orientation as the oriented Grassmannian. We tested this statement by explicit computation using the GCD algorithm and we prove it by using on-shell diagrams in section 6.4.

### 6.2.2　General m amplituhedron-like geometries

As was already pointed out in the original amplituhedron paper [108], the definition of the tree amplituhedron can be generalized to arbitrary twistor dimension, m. The same can clearly be done for the tree amplituhedron-like geometries. In the generalisation we have $Y \in \text{Gr}(k, k+m)$ rather than $\text{Gr}(k, k+4)$ and the Zs live in $k+m$ dimensions rather than $k+4$. The defining inequalities of the amplituhedron-like geometry, $A_{n,k}^{(f)}$, are then similar to the m = 4 case (6.2.1)) with the following modifications. The sign flip string for generic m reads (compare with (6.2.1))

$$\{\langle Y\, 123 \cdots (m{-}1)i \rangle\} \qquad \text{has f sign flips as i} = m, m{+}1, .., n\,, \tag{6.2.18}$$

and the physical inequalities read

$$\left. \begin{array}{rcl} \langle (i_1 i_1+1) \cdots i_{\frac{m}{2}}(i_{\frac{m}{2}}+1)\rangle &>& 0 \\ \langle i_1 i_1+1 \cdots i_{\frac{m}{2}}(i_{\frac{m}{2}}+1)1n\rangle\,(-1)^f &>& 0 \end{array} \right\} \text{ for m even,}$$

$$\left. \begin{array}{rcl} \langle 1 i_1 i_1+1 \cdots i_{\frac{m-1}{2}}(i_{\frac{m-1}{2}}+1)\rangle &>& 0 \\ \langle i_1 i_1+1 \cdots i_{\frac{m-1}{2}}(i_{\frac{m-1}{2}}+1)n\rangle\,(-1)^f &>& 0 \end{array} \right\} \text{ for m odd .} \tag{6.2.19}$$

Much of the analysis that we did for m = 4 also applies to general m. In particular the duality relation (6.2.9) becomes

$$\mathscr{H}_{n,n-m}^{(f)} \sim \mathscr{H}_{n,n-m}^{(n-m-f)}\,, \tag{6.2.20}$$

and we conjecture that the canonical form of the maximal generalised amplituhedron-like geometries are products in a similar way to (6.2.12)

$$H_{n,n-m}^{(f)} = A_{n,f} * A_{n,n-m-f}\,, \tag{6.2.21}$$

where $A_{n,f} := H_{n,f}^{(f)}$, the canonical form of the standard (but generalised m) amplituhedron.

### 6.2.3　Amplituhedron-like geometries: alternative definition

The original definition of the amplituhedron was given as the projection of the positive Grassmannian $\text{Gr}_>(k, k+n)$ through positive Zs onto $\text{Gr}(k, k+m)$ (3.4.1). We have then defined amplituhedron-like geometries as generalisations of the alternative flipping number definition of the amplituhedron. It is then interesting to see if there is an alternative definition of the amplituhedron-like geometries which generalises the original definition of the amplituhedron. Here we propose precisely such an equivalent definition for the maximal case. We propose that the maximal k = n − m (generalised) amplituhedron-like geometry, $\mathscr{H}_{n,n-m}^{(f)}$, can be written as the projection of the positive Grassmannian $\text{Gr}_>(f, n)$ and the alternating positive Grassmannian $\text{alt}(\text{Gr}_>)(n - m - f, n)$ through the positive Zs onto $\text{Gr}(k, k + m)$:

$$\mathcal{H}_{n,n-m}^{(f);\text{alt}} := \begin{cases} Y = \begin{pmatrix} C_1 \\ C_2 \end{pmatrix} \cdot Z \mid C_1 \in \text{Gr}_>(f, n) \wedge C_2 \in \text{alt}(\text{Gr}_>)(n - m - f, n) & \text{for } g_{n,f} \text{ even} \\ Y = \begin{pmatrix} C_1 \\ C_2 \end{pmatrix} \cdot Z \mid C_1 \in \text{Gr}_<(f, n) \wedge C_2 \in \text{alt}(\text{Gr}_>)(n - m - f, n) & \text{for } g_{n,f} \text{ odd} \end{cases}$$
(6.2.22)

where $g_{n,f} := \lfloor \frac{n-f}{2} \rfloor + (n-f)n$. Here the alternating positive Grassmannian, $\text{alt}(\text{Gr})_>(k, n)$, is defined as the image of $\text{Gr}_>(k, n)$ under the transformation which flips the sign of the odd columns. We will give evidence for the equivalence of this definition of the amplituhedron-like geometry with the flipping number definition (6.2.1) in section 6.4.1.

Notice that for maximal $f = n - m$ this definition coincides with the original amplituhedron. However, for general $f$ the geometry splits into two copies of the amplituhedron. This product geometry manifests the conjecture that the canonical form of this geometry gives the product of the corresponding amplitudes (6.2.21).

This definition naturally extends to loops. For the amplituhedron we have that each loop can be parametrised using an auxiliary $2 \times n$ matrix $D_i$ as $AB_i = D_i \cdot Z$ with the condition $\begin{pmatrix} C \\ D_i \end{pmatrix} \in \text{Gr}_>(k + 2, n)$, which corresponds to the one loop constraints and $\begin{pmatrix} C \\ D_i \\ D_j \end{pmatrix} \in \text{Gr}_>(k+4, n)$ which corresponds to mutual positivity [15]. In analogy to the loop amplituhedron we can define the loop amplituhedron-like geometry (setting here $Z = \mathbb{1}$)

$$\mathcal{H}_{n,n-4,l}^{(f;\overbrace{f+2,..,f+2}^{l'},\overbrace{f,..,f}^{l-l'})} := \begin{cases} \begin{pmatrix} C_1 \\ D_i \end{pmatrix} \in \text{Gr}_>(f + 2, n) & \forall \quad i \leq l' \\ \begin{pmatrix} C_2 \\ D_i \end{pmatrix} \in \text{alt}(\text{Gr}_>)(n - f - 2, n) & \forall \quad i > l' \\ \det \begin{pmatrix} C \\ D_i \\ D_j \end{pmatrix} > 0 & \forall \quad i \neq j \end{cases}$$
(6.2.23)

where the tree level condition (6.2.22) is understood.

## 6.3 Maximal residues of the squared amplituhedron

Because the square of the superamplitude at tree level can be written as a sum of products of on-shell diagrams (see section 6.4.2) it only has dlog singularities just like the superamplitude itself. Differently from the tree level superamplitude however, the maximal residues of the square of the superamplitude are not all $\pm 1$. This implies that the squared amplituhedron is not a positive geometry but a GPG/WPG.

To illustrate the point about maximal residues, we give here an explicit example of two residues that have different absolute value, consider again the n = 6, k = 2 superamplitude squared, that is given by (6.1.5) lifted to amplituhedron space

$$(A^2)_{6,2} = 2A_{6,2} + A_{6,1} * A_{6,1} \,. \tag{6.3.1}$$

Note that a factor of 2 is manifest in the first term but is also present in the expression for the second term (6.1.8). These two terms then have uniform maximal residues equal to $\pm 2$ or 0. We can examine this explicitly, using the coordinates

$$Y = \begin{pmatrix} 1 & \alpha_2 + \alpha_4 + \alpha_6 + \alpha_8 & (\alpha_2 + \alpha_4 + \alpha_6)\alpha_7 & (\alpha_2 + \alpha_4)\alpha_5 & \alpha_2\alpha_3 & 0 \\ 0 & 1 & \alpha_7 & \alpha_5 & \alpha_3 & \alpha_1 \end{pmatrix}, \tag{6.3.2}$$

and setting $Z = \mathbb{1}$. These two terms then read

$$2A_{6,2} = 2\prod_{i=1}^{8} \frac{d\alpha_i}{\alpha_i}\,, \tag{6.3.3}$$

$$(A_{6,1})^2 = 2\prod_{i=1}^{8} \frac{d\alpha_i}{\alpha_i}\left(1 - \frac{\alpha_2\alpha_6 + \alpha_4\alpha_8}{(\alpha_4+\alpha_6+\alpha_8)(\alpha_2+\alpha_4+\alpha_6)}\right) \,. \tag{6.3.4}$$

From this parametrized form we can see that for example both terms contribute equally to the multi-residue corresponding to sending $\alpha_2, \alpha_4 \to 0$ (in either order) and we thus have

$$\text{Res}_{\alpha_2,\alpha_4 \to 0}(A^2)_{6,2} = 4\frac{d\alpha_1 d\alpha_3 d\alpha_5 d\alpha_6 d\alpha_7 d\alpha_8}{\alpha_1\alpha_3\alpha_5\alpha_6\alpha_7\alpha_8} \,, \tag{6.3.5}$$

and will thus yield a maximal residue of 4. On the other hand, the residue corresponding to first taking $\alpha_8 \to 0, \alpha_4 \to 0$ and then taking $\alpha_6 \to 0$ vanishes for $(A_{6,1})^2$ and so

$$\text{Res}_{\alpha_8,\alpha_4,\alpha_6 \to 0}(A^2)_{6,2} = 2\frac{d\alpha_1 d\alpha_2 d\alpha_3 d\alpha_5 d\alpha_7}{\alpha_1\alpha_2\alpha_3\alpha_5\alpha_7} \,, \tag{6.3.6}$$

yielding a maximal residue of 2. In general we will have that the maximal residues of $(A^2)_{6,2}$ are all equal to $0, \pm 2$ or $\pm 4$. Therefore, $(A^2)_{6,2}$ can not be interpreted as the canonical form of a PG.

Later in section 6.4 we will prove that on-shell diagrams provide a tessellation of amplituhedron like geometries and as a consequence of the squared amplituhedron. This allows to concluding that the squared amplituhedron is a WPG with orientation coinciding with the orientation of the positive Grassmannian and weight function equal to one inside the squared amplituhedron region described by (6.2.1).

## 6.4 Proof and checks of the conjectures

In this section we examine the equivalence of the two definitions of amplituhedron-like geometries, find a tiling of the amplituhedron-like geometry via pairs of on-shell diagrams and use this to formulate a proof of the main conjecture (6.2.12), (6.2.15) that the amplituhedron-like geometries give products of amplitudes at tree level and also at loop level in the MHV case. Note that all the proofs assume the truth of various conjectures regarding the amplituhedron itself.

### 6.4.1 Equivalence of definitions of amplituhedron-like geometries

In section 6.2.3 we proposed an alternative definition for the amplituhedron-like geometry as the image of two positive Grassmannians for $k = n-m$. This definition has the nice feature that it apparently manifests the product structure of amplituhedron-like geometries observed on taking the canonical form (6.2.12). Here we prove that this alternative definition (6.2.22) is a subset of the sign flip definition (6.2.1)

$$\mathscr{H}_{n,n-m}^{(f);\text{alt}} \subseteq \mathscr{H}_{n,n-m}^{(f)} \ . \tag{6.4.1}$$

So in other words we want to prove that any Y that can be written as

$$Y = \begin{pmatrix} C_1 \\ C_2 \end{pmatrix} \cdot Z \quad \text{where } C_1 \in \text{Gr}_{\lessgtr}(f, n) \text{ and } C_2 \in \text{alt}(\text{Gr}_{>})(n-m-f, n) \tag{6.4.2}$$

must necessarily then satisfy the defining inequalities of the sign flip definition of $\mathscr{H}_{n,n-m}^{(f)}$. To do this we first split the (n–m)-plane Y into an f-plane $Y_1$ and a (n–m–f)-plane $Y_2$

$$Y_1 = C_1 \cdot Z \,, \qquad\qquad Y_2 = C_2 \cdot Z \,, \tag{6.4.3}$$

and consider projecting the geometry onto $Y_1^\perp$. Thus we define $\langle * \rangle_{Y_1} := \langle Y_1 * \rangle$,

brackets projected onto $Y_1^\perp$. Now notice that the projected Z's satisfy

$$\langle Z_J \rangle_{Y_1} := \langle Y_1 Z_J \rangle = \Delta_{\bar{J}}(C_1) \langle Z_{\bar{J}} Z_J \rangle$$
$$= \Delta_{\bar{J}}(C_1) \langle 1 \cdots n \rangle (-1)^{\#_{\text{odd}}(J)} (-1)^{g_{n,f}}, \quad (6.4.4)$$

where J is an ordered list of n–f elements, $\bar{J}$ is the ordered complement of J in 1,..,n, $\#_{\text{odd}}(J)$ is the number of odd elements in J and $g_{n,f} := \lfloor \frac{n-f}{2} \rfloor + (n-f)n$ introduced in (6.2.3). The second equality arises simply from reordering $J, \bar{J}$: first reverse the order of J (introducing the factor $(-1)^{\lfloor \frac{n-f}{2} \rfloor}$ of $g_{n,f}$) and then permute sequentially the elements of the reversed J starting from the leftmost, into the correct position to obtain the remaining terms. Now defining $\tilde{Z}_i = (-1)^i Z_i$, since the ordered minors of $C_1$ are positive or negative according to the sign of $(-1)^{g_{n,f}}$ (6.2.3), we obtain that the projected ordered $\tilde{Z}$s are totally positive

$$\langle \tilde{Z}_J \rangle_{Y_1} > 0. \quad (6.4.5)$$

Since $C_2 \in \text{alt}(Gr)_>(n, n-f)$ then $(\tilde{C}_2)_{\alpha i} := (-1)^i (C_2)_{\alpha i} \in Gr_>(n, n-f)$ and we have that $Y_2 = C_2.Z = \tilde{C}_2.\tilde{Z}$. Thus $Y_2$ and the $\tilde{Z}$s, both projected onto $Y_1^\perp$, give a geometry equivalent to the amplituhedron $\mathscr{A}_{n,n-m-f}(Y_2; \tilde{Z})$ (3.4.1).

But then this means the projected $Y_2$ must satisfy the conditions of the equivalent sign flip definition of this amplituhedron.[2] So for example taking m = 4 for concreteness (but one can check the general case similarly) the projected brackets satisfy the sign flip definition of $\mathscr{A}_{n,n-m-f}(\tilde{Z})$:

$$\langle Y_2 \tilde{Z}_i \tilde{Z}_{i+1} \tilde{Z}_j \tilde{Z}_{j+1} \rangle_{Y_1} > 0, \quad (-1)^{n-f} \langle Y_2 \tilde{Z}_i \tilde{Z}_{i+1} \tilde{Z}_1 \tilde{Z}_n \rangle_{Y_1} > 0, \quad \{\langle Y_2 \tilde{Z}_1 \tilde{Z}_2 \tilde{Z}_3 \tilde{Z}_i \rangle_{Y_1}\} \begin{array}{l} \text{has n–4–f} \\ \text{sign flips} \end{array}$$
$$(6.4.6)$$

Therefore back in the full geometry, switching to the Zs, this becomes

$$\langle Y Z_i Z_{i+1} Z_j Z_{j+1} \rangle > 0, \quad (-1)^f \langle Y Z_i Z_{i+1} Z_1 Z_n \rangle > 0, \quad \{\langle Y Z_1 Z_2 Z_3 Z_i \rangle\} \text{ has f sign flips}$$
$$(6.4.7)$$

which are just the defining inequalities showing that $Y \in \mathscr{H}_{n,n-m}^{(f)}(Z)$.

So we have proved that the alternative definition of amplituhedron-like geometry lies inside the sign flip definition, $\mathscr{H}_{n,n-m}^{(f);\text{alt}} \subseteq \mathscr{H}_{n,n-m}^{(f)}$. To show equivalence we also therefore need to show the converse $\mathscr{H}_{n,n-m}^{(f)} \subseteq \mathscr{H}_{n,n-m}^{(f;\text{alt})}$. We have been unable to prove this in general (indeed this is similar to the situation for the two equivalent descriptions of the amplituhedron itself where only one direction has been proven) so leave it conjectural.

---

[2] It is still conjectural that the two definitions of the amplituhedron are equivalent but it has been proven that the original definition is a subset of the sign flip definition [109] which is all we need here.

Note that it is enough to prove that for any $Y \in \mathcal{H}_{n,n-m}^{(f)}$ there exists a $C_1 \in \mathrm{Gr}_>$ such that $Y = (C_1, C_2)^T.Z$. It then follows automatically that there exists a $C_2 \in \mathrm{alt}(\mathrm{Gr}_>)$ such that $Y_2 = C_2.Z$ using essentially the same logic as above. Indeed if $C_1$ is positive then $Y \in \mathcal{H}_{n,n-m}^{(f)}$ implies $Y_2 \in \mathcal{A}_{n,n-m-f}(\tilde{Z})$ (since clearly (6.4.6) $\Leftrightarrow$ (6.4.7)). Therefore, there exists a $\tilde{C}_2 \in \mathrm{Gr}_>$ such that $Y_2 = \tilde{C}_2.\tilde{Z}$ (here we are assuming that both definitions of the amplituhedron are equivalent) then letting $(C_2)_{\alpha i} := (-1)^i (\tilde{C}_2)_{\alpha i}$ gives such a $C_2$. However we have been unable to prove in general that there always exists such a positive $C_1$.

Instead then let us show this converse statement explicitly in the simplest example of $n = 6, k = 2, f = 1$. Here we initially gauge fix the C-matrix as

$$C = \begin{pmatrix} x & 1 & y & b & 0 & a \\ -c & 0 & -d & z & -1 & w \end{pmatrix}. \tag{6.4.8}$$

Imposing the physical inequalities (first two lines of (6.2.1)) gives the inequalities

$$a, b, c, d > 0,$$
$$yz + bd > 0,$$
$$xw + ac > 0,$$
$$xz + bc > 0,$$
$$yw + ad > 0.$$

We now split the space into three regions and perform the following $\mathrm{SL}(2)^+$ transformations in each region to ensure that $C_1$ (the first row of C) is strictly positive) and thus of the form (6.2.22)

$$\begin{array}{lll}
x > 0, y > 0 & C \to \begin{pmatrix} 1 & -\epsilon \\ \epsilon' & 1 \end{pmatrix} C = \begin{pmatrix} x & 1 & y & b & \epsilon & a \\ -c & \epsilon' & -d & z & -1 & w \end{pmatrix} \\[1em]
x < 0, y > \frac{dx}{c} & C \to \begin{pmatrix} 1 & \frac{x}{c}-\epsilon \\ \epsilon' & 1 \end{pmatrix} C = \begin{pmatrix} c\epsilon & 1 & y-\frac{dx}{c} & \frac{xz+bc}{c} & \frac{-x}{c} & \frac{xw+ac}{c} \\ -c & \epsilon' & -d & z & -1 & w \end{pmatrix} \\[1em]
y < 0, x > \frac{cy}{d} & C \to \begin{pmatrix} 1 & \frac{y}{d}-\epsilon \\ \epsilon' & 1 \end{pmatrix} C = \begin{pmatrix} x-\frac{cy}{d} & 1 & \epsilon d & \frac{yz+bd}{d} & \frac{-y}{d} & \frac{yw+ad}{d} \\ -c & \epsilon' & -d & z & -1 & w \end{pmatrix}
\end{array}$$
(6.4.9)

Here the variables $\epsilon, \epsilon'$ are positive but small enough so that their presence does not change the sign of any non-zero entries in the C-matrix. For simplicity we have omitted such terms in the non-zero entries of the matrix. Their job is simply to move from the boundary of the region to the interior. We now observe that in all three cases (using the inequalities (6.4.8)) the top row of C is indeed positive. Therefore, we know in advance that the second row of C must be in $\text{alt}(\text{Gr})_>(1,6)$ and indeed that is what we find. So we have shown in this example that indeed for any Y in the amplituhedron-like geometry (which gives (6.4.9)) we can find $C_1, C_2$ such that Y has the form (6.4.2).

### 6.4.2　On-shell diagrams

The superamplitude can be computed by summing a certain set of on-shell diagrams [108]. Each onshell diagram has a geometrical interpretation and the corresponding union of geometries then yields a tiling of the corresponding amplituhedron [29]. In this section we will make a similar claim for the amplituhedron-like geometries.

First we quickly review the key points we need from the standard on-shell diagram story for amplitudes. Each on-shell diagram is completely characterized by an affine (or decorated) permutation $\sigma$, which maps points $a \in \{1,..,n\}$ to $\sigma(a)$ where $a \leq \sigma(a) \leq a+n$. Each permutation, in turn, identifies a specific parametrisation of a matrix $C_\sigma(\alpha)$ in the oriented Grassmannian $\widetilde{\text{Gr}}(k,n)$, that is the set of $k \times n$ matrices modulo a $GL_+(k)$ transformation, where k is the number of a such that $\sigma(a) > n$. The evaluation of any on-shell diagram in momentum supertwistor space, labelled by an affine permutation $\sigma$, can then be written as

$$f_\sigma^{(k)} = \int \frac{d\alpha_1}{\alpha_1} \cdots \frac{d\alpha_{4k}}{\alpha_{4k}} \, \delta^{(4|4) \times k}(C_\sigma(\alpha) \cdot \mathcal{Z}) \,. \tag{6.4.10}$$

Any $C_\sigma(\alpha)$ generated from an affine permutation $\sigma$ has the property that for $\alpha_i > 0$ all its minors are $\geq 0$. The space of all elements in $\widetilde{\text{Gr}}(k,n)$ with non negative minors is called the non-negative Grassmannian and is denoted $\text{Gr}_{\geq 0}(k,n)$. So, for each affine permutation $\sigma$ we can define a region $\Pi_\sigma^> = \{C_\sigma(\alpha) : \alpha_i > 0\}$ in $\text{Gr}_{\geq 0}(k,n)$ called a positroid cell.

How is this connected with the amplituhedron and its canonical form? We know that the amplituhedron can be defined as the image of the positive Grassmannian through a map $Y = C \cdot Z$, where $Z \in \text{Gr}_>(k,k+4)$. Consider a set of on-shell diagrams labelled by the affine permutations $\sigma_i$ that give the $N^k$MHV amplitude. Then the images of the corresponding positroid cells $Z(\Pi_{\sigma_i})$, that is the regions parametrised by $Y_{\sigma_i}(\alpha) = C_{\sigma_i}(\alpha) \cdot Z$ for $\alpha_i > 0$, tile the amplituhedron. Moreover, the integrand of the onshell diagram (6.4.10) is the canonical form of the image of the positroid cell in the coordinates $Y = C_{\sigma_i}(\alpha) \cdot Z$

$$\Omega(\Pi_{\sigma_i}) = \frac{d\alpha_1}{\alpha_1} \cdots \frac{d\alpha_{4k}}{\alpha_{4k}} \,. \tag{6.4.11}$$

Thus we can compute the amplituhedron canonical form by summing the positroid canonical forms.

Now consider the product of two on shell diagrams $f_\sigma^{(k_1)}$ and $f_\tau^{(k_2)}$

$$f_\sigma^{(k_1)} f_\tau^{(k_2)} = \int \frac{d\alpha_1}{\alpha_1} \cdots \frac{d\alpha_{4k_1}}{\alpha_{4k_1}} \frac{d\beta_1}{\beta_1} \cdots \frac{d\beta_{4k_2}}{\beta_{4k_2}} \delta^{(4|4\times k)} \left( \begin{pmatrix} C_\sigma(\alpha) \\ C_\tau(\beta) \end{pmatrix} \cdot \mathcal{Z} \right) . \quad (6.4.12)$$

where $k = k_1 + k_2$. This equation makes manifest that the product of two or more on-shell diagrams has only dlog singularities and maximal residues equal to $\pm 1$, implying that the product of amplitudes has also only dlog singularities.

Now we would like to associate a corresponding geometry in the auxiliary Grassmannian $\widetilde{Gr}(k,n)$ to the product of on-shell diagrams. The naive choice would be to consider the region parametrised by $\begin{pmatrix} C_\sigma(\alpha) \\ C_\tau(\beta) \end{pmatrix}$ for $\alpha_i, \beta_i > 0$. On the other hand, for this to lie in the amplituhedron, using the alternative definition (6.2.22), we should rather have $C_\sigma$ in the positive Grassmannian,[3] $Gr_>(k_1, n)$ but $C_\tau$ in the *alternating* Grassmannian $alt(Gr_>)(k_2, n)$. From this perspective it is thus natural to associate to the product of two on-shell diagrams characterized by the auxiliary matrices $C_\sigma(\alpha) \in Gr_\geq(k_1, n)$ and $C_\tau(\beta) \in Gr_\geq(k_2, n)$, a region $\Pi_{\sigma,\tau}^>$ defined as

$$\Pi_{\sigma,\tau}^> := \left\{ C_{\sigma,\tau} = \begin{pmatrix} C_\sigma(\alpha) \\ alt(C_\tau)(\beta) \end{pmatrix} \quad \text{for} \quad \alpha_i, \beta_i > 0 \right\} , \quad (6.4.13)$$

where alt flips the sign of the odd columns of $C_2$ (which will not affect (6.4.12)).

Note that the product of on-shell diagrams can vanish ( for example the product of identical onshell diagrams must vanish). In these cases the corresponding geometry $\Pi_{\sigma,\tau}^>$ is not full dimensional.

The canonical form of this geometry, in the coordinates $Y = C_{\sigma,\tau} \cdot Z$, is then the integrand in (6.4.12)

$$\Omega(Z(\Pi_{\sigma,\tau}^>)) = \frac{d\alpha_1}{\alpha_1} \cdots \frac{d\alpha_{4k_1}}{\alpha_{4k_1}} \frac{d\beta_1}{\beta_1} \cdots \frac{d\beta_{4k_2}}{\beta_{4k_2}} . \quad (6.4.14)$$

However, although the corresponding expression in superspace is a standard product still, we therefore know that the canonical form must give the star product of the separate covariantised forms

$$\Omega(Z(\Pi_{\sigma,\tau}^>)) = \Omega(Z(\Pi_\sigma^>)) * \Omega(Z(\Pi_\tau^>)) . \quad (6.4.15)$$

---

[3]Or $Gr_<(k_1, n)$ for $g_{n,f}$ odd. When $g_{n,f}$ is odd we will need to eg flip the sign of one row of $C_\sigma(\alpha)$ so it will become an element of the negative Grassmannian. However, we will surpress this case from now on for simplicity of presentation, but it is to be understood.

### 6.4.3 Proof of the conjecture at tree-level

We now have all the ingredients needed to prove that amplituhedron-like geometries yield products of amplitudes (6.2.12). Consider two sets of on-shell diagrams $\{f^{(k_1)}_{\sigma_i}\}, \{f^{(k_2)}_{\tau_j}\}$ which each sum to separate (parity conjugate) amplitudes

$$\mathcal{A}_{n,k_1} = \sum_i f^{(k_1)}_{\sigma_i}, \qquad \mathcal{A}_{n,k_2} = \sum_j f^{(k_2)}_{\tau_j}, \tag{6.4.16}$$

with $k = k_1 + k_2 = n - m$. We would like to prove that the set of all the associated geometries $Z(\Pi_{\sigma_i,\tau_j})$ (defined in (6.4.13)) is a tiling of the corresponding amplituhedron-like geometry $\mathcal{H}^{(k_1);alt}_{n,n-m}$. That is we wish to show its elements are disjoint and their union covers the amplituhedron-like geometry. Since the canonical form of a tiling is given by the sum of the canonical forms of its elements for GPGs, this then automatically proves that this geometry yields the product of amplitudes (6.2.12).

To do this we will prove that for every $Y \in \mathcal{H}^{(k_1);alt}_{n,n-m}$, Y belongs to a unique region $Z(\Pi_{\sigma_{i^*},\tau_{j^*}})$ (defined in (6.4.13)). That is there exist unique indices $i^*, j^*$ such that Y can be written as $Y = Y_1 Y_2$ with $Y_1 = (C_{\sigma_{i^*}}(\alpha)) \cdot Z$ and $Y_2 = \text{alt}(C_{\tau_{j^*}}(\beta)) \cdot Z$ for some $\alpha, \beta > 0$, where $C_{\sigma_{i^*}}(\alpha), C_{\tau_{j^*}}(\beta)$ are the C matrices associated with the corresponding onshell diagrams $f^{(k_1)}_{\sigma_{i^*}}, f^{(k_2)}_{\tau_{j^*}}$ respectively in (6.4.16).

So we start with an arbitrary Y in the amplituhedron-like geometry, $\mathcal{H}^{(f);alt}_{n,n-m}$ (6.2.22), so

$$Y = Y_1 Y_2, \qquad \text{with} \quad \begin{pmatrix} Y_1 \\ Y_2 \end{pmatrix} = \begin{pmatrix} C_1 \\ C_2 \end{pmatrix} \cdot Z, \tag{6.4.17}$$

for some $C_1 \in \text{Gr}_>(k_1, n)$ and $C_2 \in \text{alt}(\text{Gr}_>)(n-m-k_1, n)$. We then follow the first part of the argument in section 6.4.1. Namely we project onto a $(n-k_1)$-plane orthogonal to $Y_1$, $Y_1^\perp$, and note that the resulting geometry of the projected $Y_2$ is the amplituhedron $\mathcal{A}_{n,n-m-k_1}((Y_2)_{Y_1}, (\tilde{Z})_{Y_1})$, projected on $Y_1^\perp$ and in terms of alternating $(\tilde{Z}_i := (-1)^i Z_i)$ and projected external data $(\tilde{Z})_{Y_1}$ (see the paragraph containing (6.4.5)). Here the subscript simply denotes the projection on $Y_1^\perp$. We then use the fact that we know that this amplituhedron can be described geometrically as the disjoint union of on-shell diagrams in $\text{Gr}(n-m-k_1, n-k_1)$, the space of $(n-m-k_1)$-planes in the $n-k_1$ subspace $Y_1^\perp$. Therefore, there exists a unique $j^*$ such that the projection of $Y_2$ on $Y_1^\perp$ can be written as $(Y_2)_{Y_1} = C_{\tau_{j^*}}(\beta) \cdot (\tilde{Z})_{Y_1}$ for some $\beta > 0$. Now comes the key part of the proof: we can then project back away from the hyperplane $Y_1^\perp$ by defining $\hat{Y}_2 = C_{\tau_{j^*}}(\beta) \cdot \tilde{Z} = \text{alt}(C_{\tau_{j^*}})(\beta) \cdot Z$. (In the second equality we have simply swapped the flipping of odd particles from the $\tilde{Z}$ to the C matrix). We have now that $Y = Y_1 Y_2 = Y_1 \hat{Y}_2$.

Now we can do a similar manipulation, but now projecting the geometry (both $Y_1$

and the Zs) onto the (n–$k_2$)-plane $\widehat{Y}_2^\perp$. Following similar logic to that of (6.4.4) we find that $(Y_1)_{\widehat{Y}_2}$ must live in the amplituhedron $\mathscr{A}_{n,k_1}((Y_1)_{\widehat{Y}_2}, (Z)_{\widehat{Y}_2})$ on $\widehat{Y}_2^\perp$, where $(Z)_{\widehat{Y}_2}$ lives in non negative Grassmannian, that is all its minors are either positive or zero [4]. Therefore there exists a unique $i^*$ such that $(Y_1)_{\widehat{Y}_2} = C_{\sigma_{i^*}}(\alpha) \cdot (Z)_{\widehat{Y}_2}$. This can then be projected back yielding $\widehat{Y}_1 = C_{\sigma_{i^*}}(\alpha) \cdot Z$ with $Y = \widehat{Y}_1 \widehat{Y}_2$.

We conclude that any Y satisfying (6.4.17) belongs to one and only one region associated to the product of on-shell diagrams, one in each of the sums in (6.4.16). Therefore the regions $Z(\Pi^>_{\sigma_{i^*}\tau_{j^*}})$ are disjoint and cover the corresponding amplituhedron-like geometry

$$\mathscr{H}^{(k_1)}_{n,n-m} = \bigcup_{i,j} Z(\Pi^>_{\sigma_i,\tau_j}) \,. \tag{6.4.18}$$

Finally, putting this together (6.4.15) we obtain the anticipated result. The canonical form of $\mathscr{H}^{(k_1)}_{n,n-m}$ is given by the product of amplitudes

$$H^{(k_1)}_{n,n-m} = \sum_{i,j} \Omega(Z(\Pi^>_{\sigma_i,\tau_j})) = \sum_i \Omega(Z(\Pi^>_{\sigma_i})) * \sum_j \Omega(Z(\Pi^>_{\tau_j})) = A_{k_1,n,m} * A_{k_2,n,m} \,, \tag{6.4.19}$$

which concludes our proof.

### 6.4.4 Proof of the loop level conjecture for f maximal

We can also explicitly prove the loop level conjecture (6.2.13) for maximal f flipping number. That is the loop level amplituhedron-like geometry with maximal flipping number gives the product of MHV and anti-MHV superamplitudes at all loops,

$$H^{(n-4;\overbrace{n-2,..,n-2}^{l'},\overbrace{n-4,..,n-4}^{l-l'})}_{n,n-4,l} = A_{n,n-4,l'} A_{n,0,l-l'} \,. \tag{6.4.20}$$

The first factor on the RHS, $A_{n,n-4,l'}$, is the anti-MHV $l'$-loop integrand, which itself factorizes as the tree-level anti-MHV amplitude, $A_{n,n-4,0}$, multiplied by the conjugate of the MHV amplitude $\overline{A_{n,0,l'}}$. Thus we wish to prove

$$H^{(n-4;\overbrace{n-2,..,n-2}^{l'},\overbrace{n-4,..,n-4}^{l-l'})}_{n,n-4,l} = A_{n,n-4,0} \, \overline{A_{n,0,l'}} \, A_{n,0,l-l'} \,. \tag{6.4.21}$$

---

[4] A small but important subtlety appears here in that this amplituhedron on $\widehat{Y}_2^\perp$ may be degenerate in the sense that some of the projected Z brackets $\langle Z \rangle_{\widehat{Y}_2} \geq 0$ may vanish. Nevertheless, the statement which follows in the main text is still true, the consequence of the degeneracy is simply that some on shell diagrams may vanish and the corresponding geometries be non maximally dimensional.

Nicely this factorisation can be seen straightforwardly at a purely geometric level. Firstly, we can see that the LHS, the loop level anti-MHV amplituhedron-like geometry, is the product of the tree-level anti-MHV amplituhedron, $\mathscr{A}_{n,n-4}$, (which Y lies in) and a second geometry for the loop variables lying in $Y^\perp$, a 4-plane nowhere intersecting any Y in $\mathscr{A}_{n,n-4}$. This second geometry turns out to be isomorphic to the l-loop MHV amplituhedron-like geometry, with $l - l'$ loops having maximum flipping number 2 and $l'$ loops having minimum flipping number 0. Concretely then we first have the geometric factorisation

$$\mathscr{H}_{n,n-4,l}^{(n-4;\overbrace{n-2,..,n-2}^{l'},\overbrace{n-4,...,n-4}^{l-l'})}\bigl(Y,(AB)_i;Z\bigr) = \mathscr{A}_{n,n-4}\bigl(Y;Z\bigr) \times \mathscr{H}_{n,0,l}^{(0;\overbrace{0,...,0}^{l'},\overbrace{2,...,2}^{l-l'})}\bigl(-(AB)_i;\tilde Z\bigr) ,$$
(6.4.22)

where $\tilde Z_i = (-1)^i Z_i$. This factorisation can be seen straightforwardly by simply examining the explicit definitions of the geometries involved (6.2.4). Indeed Y must lie in the tree anti-MHV amplituhedron, $Y \in \mathscr{A}_{n,n-4}$, this is just the first line of the definition of the loop amplituhedron (6.2.4). Then the 2-planes $(AB)_i$ naturally live on the 4-plane, $Y^\perp$, with effective 4-brackets defined as $\langle * \rangle_Y := \langle Y * \rangle$. The resulting effective 4-brackets involving Zs then have maximal flipping number $n-4$. Crucially the resulting inequalities are enough to fix all effective Z 4-brackets to be alternating positive:

$$(-1)^{i+j+k+l} \langle ijkl \rangle_Y > 0 , \qquad 1 \leq i < j < k < l \leq n ,$$
(6.4.23)

or equivalently $\tilde Z_j := (-1)^j Z_j$ has positive ordered effective brackets.[5] Finally, examining the inequalities $\langle Y(AB)_i jj+1 \rangle = \langle -(AB)_i \tilde j \tilde k \rangle_Y (-1)^{j+k+1}$ one can check that minimal loop flipping number $n-4$ becomes maximal loop flipping number 2 and vice-versa.

Now a second geometrical factorisation occurs for the second amplituhedron-like geometry itself, namely

$$\mathscr{H}_{n,0,l}^{(0;\overbrace{0,...,0}^{l'},\overbrace{2,...,2}^{l-l'})} = \mathscr{H}_{n,0,l'}^{(0,0)} \, \mathscr{A}_{n,0,l-l'} .$$
(6.4.24)

Examining the defining inequalities (6.2.4), only the mutual positivity $\langle AB_i AB_j \rangle > 0$ between loops with different flipping number prevents a completely factorised geometry. But a loop $(AB)_j$ with maximal flipping number 2, satisfies the same inequal-ities as the one loop MHV amplituhedron, and so we can use the original definition of the amplituhedron as the image of the positive Grassmannian to parametrise $(AB)_j$ as

$$(AB)_j = \sum_{l<m} c_{lm} Z_l Z_m , \qquad (6.4.25)$$

where $c_{lm} > 0$ for $l < m$. Then using this expression we can expand the mutual positivity condition as

$$\langle (AB)_i (AB)_j \rangle = \sum_{l<m} c_{lm} \langle (AB)_i lm \rangle . \qquad (6.4.26)$$

---

[5]Note that one might wonder why a simple factorisation of geometries like (6.4.22) does not occur for more general amplituhedron-like geometries (ie for lower values of the flipping number f). This is because in general there is no simple map $Z \to \tilde Zs$ such that all the effective $\tilde Z$-brackets are positive as there is here.

Now if $(AB)_i$ has flipping number equal to zero, all $\langle (AB)_i lm \rangle$ are positive, implying the positivity of $\langle (AB)_i (AB)_j \rangle$. As a consequence the geometry factorizes into the product of $l'$ loops with $f_{AB} = 2$ and $l - l'$ loops with $f_{AB} = 0$ implying (6.4.24).

Putting (6.4.22) and (6.4.24) together we arrive at the geometrical double factorisation

$$\mathscr{H}_{n,n-4,l}^{(n-4;\overbrace{n-2,..,n-2}^{l'},\overbrace{n-4,..,n-4}^{l-l'})} = \mathscr{A}_{n,n-4} \, \mathscr{A}_{n,0,l-l'} \, \mathscr{H}_{n,0,l'}^{(0,0)} , \qquad (6.4.27)$$

which implies (using standard the amplituhedron conjecture together with the fact that the canonical form of geometrical products gives the product of the respective canonical forms [30])

$$H_{n,n-4,l}^{(n-4;\overbrace{n-2,..,n-2}^{l'},\overbrace{n-4,..,n-4}^{l-l'})} = A_{n,n-4} \, A_{n,0,l-l'} \, H_{n,0,l'}^{(0,0)} . \qquad (6.4.28)$$

Finally to prove (6.4.21) we just need to show that $H_{n,0,l'}^{(0,0)} = \overline{A_{n,0,l'}}$, in other words that the MHV loop amplituhedron-like geometry with all loop flipping numbers *minimal* gives the conjugate of the MHV amplitude. This fact follows nicely from considering the case $l' = 0$. In this case the RHS of (6.4.28) becomes the anti-MHV loop level amplituhedron whose canonical form, the anti-MHV loop level amplitude, factorises as discussed above (6.4.21). Thus (6.4.28) with $l' = l$ reads $H_{n,n-4,l}^{(n-4,l)} = A_{n,n-4,l} = A_{n,n-4}\overline{A_{n,0,l}} = A_{n,n-4}H_{n,0,l}^{(0,0)}$ and so indeed we have shown that (as conjectured in [109] for $l = 1$)

$$H_{n,0,l}^{(0,0)} = \overline{A_{n,0,l}} . \qquad (6.4.29)$$

This then proves that amplituhedron-like geometries give products of amplitudes at loop level for maximal k and f (6.4.20).

Note that as a consequence of this derivation we have then proven an interpreta-tion for a particular sector of *non-maximal* amplituhedron-like geometries conjectured in [109]. Namely the MHV (k = 0) amplituhedron-like geometries with arbitrary n, from (6.4.24) and (6.4.29) are given by

$$H_{n,0,l}^{(0;\overbrace{0,...,0}^{l'},\overbrace{2,...,2}^{l-l'})} = \overline{A}_{n,0,l'} \, A_{n,0,l-l'} \, . \qquad (6.4.30)$$

Taking the union over all loop winding numbers to obtain the squared amplituhedron, $\mathscr{H}_{n,0,l}$, we would expect it to give the sum of these

$$H_{n,0,l} = \sum_{l'} \overline{A}_{n,0,l'} \, A_{n,0,l-l'} \, . \qquad (6.4.31)$$

Crucially all the almost disjoint amplituhedron-like geometries appearing in the union inherit consistent orientations on the oriented Grassmannian such that they indeed appear with the same sign when taking the globally canonical form and this gives the above result which is consistent with the square of the amplitude (2.3.16). In [109] it was observed for n = 5 and conjectured to hold for all n that at one loop this union of winding geometries has a (standard) canonical form corresponding to the *difference* $A_{n,0,1} - \overline{A_{n,0,1}}$ rather than the sum in (6.4.31). This therefore illustrates the importance of the carefully considering the relative orientation of the tiles in the tiling.

### 6.4.5 Checks of the tree-level general m conjecture

Explicit checks of the amplituhedron-like conjecture can and have been made for various low values of n, k, l on a computer using cylindrical decomposition (see for example [51]) but they quickly become too complicated. However, the existence of the generalised amplituhedron-like geometries nicely gives another direction in which to to perform checks.

**Explicit checks for specific values of** m, k, n

We have checked the generalised m conjecture (6.2.21) for k = 2, n = m + 2 and f = 1 for m = 2, 4, 6, 8, explicitly, that is

$$H_{m+2,2}^{(1)} = A_{m+2,1} * A_{m+2,1} \, . \qquad (6.4.32)$$

To do this we first noted that, $A_{n,1}$, is a natural generalisation of the NMHV amplitude for m even, namely

$$A_{n,1} = \sum_{i_1,\cdots,i_{m/2}} R[1 i_1 i_1 + 1 \cdots i_{m/2} i_{m/2} + 1] \, , \qquad (6.4.33)$$

where

$$R[i_1 \cdots i_{m+1}] = \frac{\langle i_1 \cdots i_{m+1} \rangle^m \langle Y d^m Y \rangle}{\langle Y i_1 \cdots i_m \rangle \cdots \langle Y i_m \cdots i_{m-1} \rangle}, \qquad (6.4.34)$$

is a generalised R-invariant. We then used this with the formula for the *-product, (6.1.2), to compute $A_{m+2,1} * A_{m+2,1}$ covariantly. On the other hand we used the GCD algorithm to compute the canonical form of $\mathcal{H}^{(1)}_{m+2,2}$ and verified that they match.

**Checks for** $m = 2$

For the case $m = 2$ the computational complexity is much lower and we have verified (6.2.21) up to $n - 2 = k = 7$. The canonical form for $k = n - 2$ reads

$$A_{n,n-2} = \frac{\langle 1, 2, \cdots, n \rangle^2}{\prod_{i=1}^{n} \langle Y i\, i+1 \rangle}. \qquad (6.4.35)$$

In [109] it was proven that for $m = 2$, the $N^k$MHV superamplitude is proportional to the product of k NMHV superamplitudes

$$\mathcal{A}_{n,k} = \frac{(\mathcal{A}_{n,1})^k}{k!}, \qquad (6.4.36)$$

and the analogous statement holds in amplituhedron space, so for example for $k = 2, m = 2$ one can verify that

$$A_{4,2} = \frac{1}{2!} A_{4,1} * A_{4,1} = \frac{\langle 1234 \rangle^2}{\langle Y12 \rangle \langle Y23 \rangle \langle Y34 \rangle \langle Y14 \rangle}. \qquad (6.4.37)$$

Thus the product of two $m = 2$ superamplitudes is

$$A_{n,k-k'} * A_{n,k'} = \frac{(A_{n,1})^{*k}}{(k-k')!k'!} = \frac{k!}{(k-k')!k'!} A_{n,k}. \qquad (6.4.38)$$

We have observed from explicit computations that in fact the geometry in the maximal case, $k = n - 2, m = 2$, with *any* valid sign flip *pattern* (ie any specific valid choice of signs for $\langle Y123i \rangle$), has a canonical form equal to (6.4.35). Since for $m = 2$, each flipping number f corresponds to $\binom{n-2}{f}$ possible flipping patterns (in $n-2$ places you either flip (f times) or don't flip ($n - 2 - f$ times)) we obtain trivially that

$$H^{(k')}_{n,n-2} = \binom{n-2}{k'} A_{n,k} = A_{n,n-2-k'} * A_{n,k'}, \qquad (6.4.39)$$

in agreement with the (generalised) amplituhedron-like conjecture (6.2.21).

## 6.5   Factorisation of sign flip patterns

In this section we note a refinement of the factorisation of amplituhedron-like geometries, noting that individual flipping pattern geometries also factorise.

The amplituhedron-like geometries can be divided into regions labelled by a specific sign-flip pattern, that is regions where all brackets $\langle 1,\ldots,m-1,i\rangle$ have a well-defined sign. We will indicate the canonical form of a region in $\widetilde{\mathrm{Gr}}(k,k+m)$ labelled by a sign flip pattern $\mathbf{p} = \{p_1,..,p_f\}$ as $h_{n,k}^{\mathbf{p}}$. Here $p_i$ denotes the position of each consecutive sign flip, so $\mathrm{sgn}(\langle 1,..,m-1,p_i-1\rangle) = -\mathrm{sgn}(\langle 1,..,m-1,p_i\rangle)$. In this notation the canonical form of an amplituhedron-like geometry $\mathscr{H}_{n,k}^{(f)}$ can be written as

$$\mathscr{H}_{n,k}^{(f)} = \bigcup_{\substack{\mathbf{p} \text{ with} \\ \text{f flips}}} h_{n,k}^{\mathbf{p}} \,. \tag{6.5.1}$$

We have observed by explicit computation for $m = 2,4$ and maximal $k = 2 = n-m$, that the canonical form of a particular sign flip pattern geometry h factorises into the following star product, mimicking the same geometrical factorisation mentioned in (6.4.6), we have

$$h_{n,n-m}^{\mathbf{p}} = h_{n,f}^{\mathbf{p}} * A_{n,n-m-f} \,, \tag{6.5.2}$$

where f is the number of sign flips in the pattern $\mathbf{p}$. Note that by taking the union over all patterns with a given flipping number, this then implies, and is therefore a refinement of, the main conjecture about amplituhedron-like geometries (6.2.12)

$$H_{n,n-m}^{(f)} = H_{n,f}^{(f)} * A_{n,n-m-f} = A_{n,f} * A_{n,n-m-f} \,. \tag{6.5.3}$$

Now geometries with complementary sign flip patterns are equivalent, yielding the same canonical form. Indeed, clearly the duality relation (6.2.8),(6.2.9) applies to the individual complementary flip pattern geometries, that is

$$h_{n,n-m}^{\mathbf{p}} = h_{n,n-m}^{\bar{\mathbf{p}}} \,, \tag{6.5.4}$$

where $\bar{\mathbf{p}}$ indicates the sign flip pattern complementary to $\mathbf{p}$. We thus also have an alternative product formula for a flip pattern geometry

$$h_{n,n-m}^{\mathbf{p}} = h_{n,n-m-f}^{\bar{\mathbf{p}}} * A_{n,f} \,, \tag{6.5.5}$$

which also implies the main conjecture (6.2.12) but this time keeping the other term in the product $A_{n,f}$ whole and reconstructing $A_{n,n-m-f}$. So one can "break apart" either of the two amplitudes appearing in the product but not both simultaneously.

It is interesting to compare this with the analogous onshell diagram story where you can indeed break apart both amplitudes.

We have also observed in all the cases that we have considered that given two flipping patterns $\mathbf{p}_1$ and $\mathbf{p}_2$ with flipping number $k'$, the following identity holds

$$h_{n,k'}^{\mathbf{p}_1} * h_{n,k-k'}^{\bar{\mathbf{p}}_2} = h_{n,k-k'}^{\bar{\mathbf{p}}_1} * h_{n,k'}^{\mathbf{p}_2} . \tag{6.5.6}$$

This relation then implies the equivalence between (6.5.2) and (6.5.5).

For $m = 2$ we can actually prove all these relations from the observation that in the maximal case all flipping patterns yield the same canonical form, $h_{n,n-2}^{\mathbf{p}} = A_{n,n-2}$, as we discussed in section 6.4.5. In fact in the non-maximal case we have observed that each individual flipping pattern contributing to the amplituhedron factorises into $k = 1$ patterns as follows

$$h_{n,k}^{\{p_1,\cdots,p_k\}} = h_{n,1}^{\{p_1\}} * \cdots * h_{n,1}^{\{p_k\}} =$$
$$= \frac{\langle (1p_1p_1+1)(Y \cap 1p_2p_2+1) \cdots \cap (Y \cap 1p_kp_k+1) \rangle^2}{\prod_{\alpha=1}^{k} \langle Y1p_\alpha \rangle \langle Yp_\alpha(p_\alpha+1) \rangle \langle Y(p_\alpha+1)1 \rangle} , \tag{6.5.7}$$

where we used (6.1.11) to compute the * product and

$$h_{n,1}^{\{p_i\}} = \frac{\langle 1p_ip_i + 1 \rangle}{\langle 1p_i \rangle \langle p_ip_i + 1 \rangle \langle (p_i + 1)1 \rangle} . \tag{6.5.8}$$

For example for the $k = 2$ amplituhedron, which has only one sign flip pattern $\{+,-,+\}$, we have

$$A_{4,2} = h_{4,2}^{\{2,3\}} = h_{4,1}^{\{2\}} * h_{4,1}^{\{3\}} = \frac{\langle 123 \rangle^2}{\langle 12 \rangle \langle 23 \rangle \langle 31 \rangle} * \frac{\langle 134 \rangle^2}{\langle 13 \rangle \langle 34 \rangle \langle 41 \rangle} . \tag{6.5.9}$$

Formulas analogous to (6.5.7) appear in [35] and in [20] and can be obtained for (6.5.7) by expanding $Y \cap (ijk)$ as

$$Y \cap (ijk) = \sum_{\alpha=1}^{k} (-1)^\alpha Y_\alpha \langle Y_1 \cdots Y_{\alpha-1} Y_{\alpha+1} \cdots Y_k ijk \rangle . \tag{6.5.10}$$

If we instead express $Y \cap (ijk)$ as a point on the 3–plane $ijk$ instead, that is

$$Y \cap (ijk) = \langle Yij \rangle Z_k - \langle Yik \rangle Z_j + \langle Yjk \rangle Z_i , \tag{6.5.11}$$

we obtain an expression for $\Omega(h_{n,k}^{\{p_1,\cdots,p_k\}})$ where only manifestly SL(2) invariant brackets, that is brackets of the form $\langle Yij \rangle$, appear. Note that expression (6.5.7) makes (6.5.6) trivial for $m = 2$.

Now consider the RHS of of (6.5.2) and expand the second term into flipping

patterns

$$h_{n,f}^{\mathbf{p}} * A_{n,n-f-2} = h_{n,f}^{\mathbf{p}} * \sum_{\substack{\mathbf{q} \text{ with} \\ \text{f flips}}} h_{n,n-f-2}^{\bar{\mathbf{q}}} \,. \tag{6.5.12}$$

Now inserting the factorisation (6.5.7), since $(h_{n,1}^{\{i\}})^{*2} = 0$, only the term $\mathbf{q} = \mathbf{p}$ in the sum will survive so that there are no repeated factors. Indeed, this surviving term will involve a product over all n – 2 available flip positions and we obtain

$$h_{n,f}^{\mathbf{p}} * A_{n,n-f-2} = h_{n,1}^{\{2\}} * \cdots * h_{n,1}^{\{n-1\}} = A_{n,n-2} \,, \tag{6.5.13}$$

proving the refinement (6.5.2) of the main conjecture (6.2.12) for m = 2.

## 6.6 Canonicalizing Cyclicity and Crossing

We have seen that the product of two parity conjugate superamplitudes is the canonical form of an amplituhedron-like geometry. One could wonder if there are more general geometries which could yield some physical object such as products of two, or more, amplitudes. In particular one could imagine tweaking the signs of the inequalities defining known geometries. At first sight this seems to give a huge choice of possibilities to investigate. An obvious property we might insist on to restrict this though is cyclic invariance. In this section we therefore consider the implications of requiring a cyclic invariant canonical form for the corresponding geometry. We saw that the amplituhedron-like geometries with even flipping number, f, are not cyclic but rather twisted cyclic, $Z_n \to -Z_1$. Nevertheless the corresponding canonical form is cyclic, simply due to the fact that the canonical form is invariant under $Z_1 \to -Z_1$. It is therefore natural to consider geometries which are cyclic up to any possible flip of the Zs. However, in this section we conclude that all such generalised cyclic geometries are equivalent to cyclic or twisted cyclic geometries. Thus one can define new generalised geometries by defining arbitrary signs for $\langle Y12ii+1 \rangle$ for each i with cyclicity giving all other physical inequalities from these. On the other hand for the correlator there is the more powerful permutation symmetry and in this case we find a unique correlahedron-like geometry.

### 6.6.1 Cyclic geometries

Recall that, as discussed below (3.4.1), it's extremely useful to consider the geometry Y as an *oriented* k-plane and the Z's as elements in *oriented* projective space $\mathbb{R}^{4k}/GL^+(1) \sim S^{4k-1}$. Then we wish to consider geometries $\mathcal{R}(Y; Z_i)$, defined as the set of $Y \in \widetilde{Gr}(k, k + m)$ satisfying a set of inequalities involving Y and $Z_i$. The inequalities will be invariant under positive rescaling of Y and Z and will be of the form $\langle YZ_{i_1}..Z_{i_m}\rangle \lessgtr 0$. Because the canonical form is a rational function, the invariance under positive rescaling of the geometry implies invariance under general rescaling of the canonical form, regardless of the sign of the scaling parameter, i.e. it will be projectively well defined.

This means two very different regions can trivially have the same canonical form: flipping the sign of Y or any Zs, the inequalities defining the geometry will change, while its canonical form will remain the same. We thus say that two geometries $\mathcal{R}_1, \mathcal{R}_2$ which are related via such sign flips are *equivalent*, $\mathcal{R}_1 \sim \mathcal{R}_2$, (and thus have the same canonical form). To this end we would first like to see if all signed cyclic symmetric geometries are equivalent to cyclic geometries and if not how many inequivalent types of flipped cyclic geometries there are.

Define $F_i$ to be the transformation which flips $Z_i$, $Z_i \to -Z_i$ and $F_I := F_{i_1} F_{i_2} ...$, where $I := \{i_1, i_2, \cdots\}$ the transformation that flips the sign of all Z's with index $i \in I$. Then in this notation the statement of equivalent geometries is that

$$\mathcal{R}' \sim \mathcal{R} \quad \Leftrightarrow \quad \mathcal{R}' = F_I \mathcal{R} \quad \text{for some I} . \tag{6.6.1}$$

These transformations clearly satisfy

$$F_I F_I = \mathbb{1} , \qquad \mathcal{C} F_I = F_{\mathcal{C}(I)} \mathcal{C} , \tag{6.6.2}$$

where $\mathcal{C}$ represents a cyclic transformation, $Z_i \to Z_{i+1}$. Now suppose we have a geometry $\mathcal{R}$ which is invariant under some flipped cyclicity $\mathcal{C} F_I$, so $\mathcal{C} F_I \mathcal{R} = \mathcal{R}$. A familiar example of this is the twisted cyclicity of the amplituhedron, $Z_i \to Z_{i+1}$ for $i = 1, .., n-1$, $Z_n \to -Z_1$ for which $I = \{n\}$, but we here imagine any possible flipped cyclic geometry.

If we now apply a further Z-flip transformation $F_J$ on our geometry $\mathcal{R}$, then using the above identities we obtain

$$\begin{aligned}
\mathcal{C} F_I \mathcal{R} = \mathcal{R} , \quad &\Rightarrow \quad F_J \mathcal{C} F_I \mathcal{R} = F_J \mathcal{R} , \\
&\Rightarrow \quad \mathcal{C} F_{\mathcal{C}^{-1}(J)} F_I F_J F_J \mathcal{R} = F_J \mathcal{R} , \\
&\Rightarrow \quad \mathcal{C} F_{\mathcal{C}^{-1}(J)} F_I F_J \mathcal{R}' = \mathcal{R}' ,
\end{aligned} \tag{6.6.3}$$

so the equivalent geometry $\mathcal{R}' := F_J \mathcal{R}$ is invariant under the flipped cyclicity $\mathcal{C} F_{\mathcal{C}^{-1}(J)} F_I F_J$.

A natural question then is whether for any list I, we can find a list J such that $F_{\mathcal{C}^{-1}(J)} F_I F_J$ is the identity, and thus obtain an equivalent geometry $\mathcal{R}'$ which is cyclic invariant (with no flips). Equivalently (by commutativity of F and using $F^2 = 1$) we ask whether for all I there exists a J such that $F_{\mathcal{C}^{-1}(J)} F_J = F_I$. Now if

$J = \{j_1, j_2, ...\}$ we have

$$F_{\mathcal{C}^{-1}(J)}F_J = (F_{j_1-1}F_{j_2-1}\cdots)(F_{j_1}F_{j_2}\cdots) = (F_{j_1-1}F_{j_1})(F_{j_2-1}F_{j_2})\cdots \quad (6.6.4)$$

since the flip operations all commute with each other. We conclude that $F_{\mathcal{C}^{-1}(J)}F_J$ can be any sign flip transformation with an even number of flips (since any such can always be constructed from sequences of adjacent flips). Thus if I contains an even number of indices, we can always find a list of indices J such that $F_{\mathcal{C}^{-1}(J)}F_I F_J = \mathbb{1}$ and so $\mathcal{R}' = F_J\mathcal{R}$ is cyclic invariant.

Instead, if the length of I is odd, but n, the total number of indices, is also odd, then the complementary set $\bar{I}$ will contain an even number of elements. Therefore, we can always choose J such that $F_{\mathcal{C}^{-1}(J)}F_J = F_{\bar{I}}$ and so $F_{\mathcal{C}^{-1}(J)}F_I F_J = F_I F_{\bar{I}} = F_{\{1,\cdots,n\}}$, so that $\mathcal{C}F_{\{1,\cdots,n\}}\mathcal{R}' = \mathcal{R}'$. The transformation $F_{\{1,\cdots,n\}}$ is simply the flipping of all Zs and thus for m even will leave all the defining inequalities $\langle Y i_1 \cdots i_m \rangle \lessgtr 0$ untouched and so we have defined an equivalent cyclic geometry $\mathcal{R}'$

$$\mathcal{C}F_{\{1,\cdots,n\}}\mathcal{R}' = \mathcal{C}\mathcal{R}' = \mathcal{R}' \,. \quad (6.6.5)$$

For m odd we will also have to flip the sign of Y.

If the length of I is odd and n is even on the other hand, the best we can do is to chose a J such that only one element is flipped. This is what is known in the literature as twisted cyclicity and one conventionally chooses the element that must be flipped to be n so $Z_i \to Z_{i+1}$, but $Z_n$ goes to $-Z_1$.

Summarizing the result of our analysis, we can say that, when n is odd (and m even), we can always map any geometry to the cyclic invariant one. When n is even instead we have two classes of geometries: cyclic and twisted cyclic.

Finally we then ask if there is a flip transformations $F_J$ mapping two geometries $\mathcal{R}_1, \mathcal{R}_2$ with the *same* type of cyclicity $\mathcal{C}F_I$. Thus we have $\mathcal{C}F_I\mathcal{R}_1 = \mathcal{R}_1$, $\mathcal{C}F_I\mathcal{R}_2 = \mathcal{R}_2$ and $\mathcal{R}_2 = F_J\mathcal{R}_1$. This implies

$$[F_J, \mathcal{C}] = F_J\mathcal{C} - F_{\mathcal{C}(J)}\mathcal{C} = 0 \quad \Rightarrow \quad F_{\mathcal{C}(J)} = F_J \,. \quad (6.6.6)$$

For a faithful representation of $F_J$ we have just one non-trivial solution, $F_J = F_{\{1,\cdots,n\}}$. However, if the representation of $F_{\{1,\cdots,n\}} = \mathbb{1}$, as is the case of m odd, then we have two further elements in the algebra that commute with $\mathcal{C}$, $F_J = F_{\text{odd}} = F_{\{1,3,5\cdots\}}$ and $F_J = F_{\text{even}} = F_{\{2,4,6\cdots\}}$. In this representation they correspond to the same operator

$$F_{\text{odd}}F_{\text{even}} = F_{1,\cdots,n} \equiv \mathbb{1} \quad \Rightarrow \quad F_{\text{odd}} = F_{\text{odd}}^{-1} = F_{\text{even}} \,. \quad (6.6.7)$$

(More generally we have that $F_I = F_{\bar{I}}$, where $\bar{s}$ is the complement of s.) This equival-ence of geometries related by $F_{odd}$ or $F_{even}$ yields the duality of amplituhedron-like geometries (6.2.8).

### 6.6.2 Crossing symmetric correlahedron geometries

In planar N = 4 SYM there is a class of fundamental observables that share many properties with amplitudes and have therefore the chance to be defined geometrically. These are the stress energy correlators. These observables can be defined on the twistor on-shell superspace [89]. A point in space time is identified by a line in twistor space, that is a pair of twistors $X_i^{IJ} = Z_{iL}^1 Z_{iM}^2 \epsilon^{LMIJ}$. In the same way, a point in the chiral super-Minkowski space is identified by a pair of super-twistors. The supercorrelator can be then organized as a sum over terms with homogeneous Grassmannian degree, usually indicated as $\mathcal{G}_{n,k}$, where n is the number of super twistors and 4(k+n) is the Grassmannian degree. In [51] the chiral super-Minkowski space is bosonised and the functions $\mathcal{G}_{n,k}$ uplifted to differential forms on the Grassmannian Gr(k+n, k+n+4). Moreover, a geometry, called the correlahedron, is defined and its canonical form is conjectured to give the bosonised supercorrelator.

The correlators exhibit a full permutation symmetry. This suggests that the correlahedron geometry be invariant under any permutation of the twistors $X_i$ up to the action of a sign flip operator $F_I$. In other terms, for each permutation $\sigma \in S_n$ there must be a flip transformation $F_\sigma$ such that $F_\sigma \sigma$ leaves the correlahedron invariant. The set of all $\tilde{\sigma} = F_\sigma \sigma$ defines a group we call the signed symmetric group or signed permutation group.

Just as for the amplitude, the correlator $\mathcal{G}_{n,k}$ is composed of two types of bracket. The n+k+4 brackets involving only Xs and the uplifted conformal invariants $\langle YX_iX_j \rangle$. We are interested now in classifying all permutation invariant geometries that are defined using these two types of brackets. The main result of this analysis will be that, for k = n – 4, there exists just one class of geometries defined using $\langle YX_iX_j \rangle$ which can be represented by the correlahedron.

The maximally nilpotent case k = n – 4 presents the advantage that there is a unique bracket involving Xs only, $\langle X_1 \cdots X_n \rangle$ and we can always fix it to be positive. Because of the permutation symmetry, we can then choose an arbitrary bracket, such as $\langle YX_1X_2 \rangle$, and use the action of the signed symmetric group to generate all the other brackets. By flipping Y → –Y if necessary, we can fix $\langle YX_1X_2 \rangle > 0$. From this moment on we will indicate $\langle YX_iX_j \rangle$ with $\langle X_iX_j \rangle$ to make the notation more compact unless there is possible ambiguity.

Since we have already studied cyclic invariance in detail to classify the inequivalent amplituhedron-like geometries, we already know we can always choose representatives invariant under cyclic or twisted cyclic symmetry. Applying powers of the cyclic permutation on $\langle X_1 X_2 \rangle > 0$ we obtain

$$\langle X_i X_{i+1} \rangle > 0, \qquad \qquad \text{for cyclic},$$
$$\langle X_i X_{i+1} \rangle > 0, \quad \langle X_n X_1 \rangle < 0 \qquad \text{for twisted cyclic}. \qquad (6.6.8)$$

Let us now consider the action under a second permutation, the transposition $(1, 2)$. This operator can come in general with a flipping sign operator $F_I$, but not all sign strings s are allowed. The transposition $(1, 2)$ leaves invariant all brackets that do not contain $X_1$ or $X_2$ and the bracket $\langle X_1 X_2 \rangle$ itself. Therefore, $F_s$ must act trivially on these brackets. The solutions for $F_s$ are

$$I = \{\}, \quad I = \{1, 2\}, \quad I = \{1, \cdots, n\}, \quad I = \{3, \cdots, n\}. \qquad (6.6.9)$$

The last two solutions are in fact equivalent to the first two and therefore there can only be two types of transposition, $(i, j, +) = (i, j)$ and $(i, j, -) = F_{\{i,j\}}(i, j)$.

We can prove that a signed cyclic geometry invariant under $(1, 2, \pm)$ is also invariant under the whole signed symmetric group. In fact, because of cyclicity, it will also be invariant under $(i, i+1, \pm)$ and the set of adjacent transpositions generates the symmetric group. This can be proven using the relation

$$(i, j)(j, k)(i, j) = (i, k), \qquad (6.6.10)$$

or more specifically

$$(i, i+1)(i+1, i+l)(i, i+1) = (i, i+l+1). \qquad (6.6.11)$$

Therefore if we start with $(i, i+1, +)$ and $l = 0$ we can then use (6.6.11) to generate all permutations. The resulting inequalities defining the geometry will read

$$\langle X_i X_j \rangle > 0, \qquad \langle X_1 \cdots X_n \rangle > 0. \qquad (6.6.12)$$

In particular we obtain that $\langle X_1 X_n \rangle > 0$, therefore the geometry generated by $(1, 2, +)$ can only be cyclic and not twisted cyclic.

If on the other hand the geometry is invariant under $(i, i+1, -)$ instead, we can see that

$$(i, i+1, -)(i+1, i+2, -)(i, i+1, -) = (i, i+2, +), \qquad (6.6.13)$$

from which we derive that

$$(i, i+1, -)(i+1, i+l, (-1)^{l-1})(i, i+1, -) = (i, i+l, (-1)^l). \qquad (6.6.14)$$

Therefore if a geometry is invariant under $(i, i+1, -)$, for all i except $i = n$, then

(6.6.14) tells us it must also be invariant under $(2, n, (-1)^n)$. If we act with $(2, n, (-1)^n)$ on $\langle X_1 X_2 \rangle > 0$ we obtain

$$(2, n, (-1)^n) \langle X_1 X_2 \rangle > 0 \quad \Rightarrow \quad (-1)^n \langle X_1 X_n \rangle > 0. \tag{6.6.15}$$

This implies that geometries generated by negative transpositions must be cyclic for n odd and twisted cyclic for n even. Therefore, for fixed n we just have two types of geometry: one invariant under positive adjacent transpositions and one invariant under negative adjacent transpositions. The geometry invariant under $(1, 2, +)$ is described by (6.6.12), while the one invariant under $\langle 1, 2, - \rangle$ is described by the following inequalities

$$\begin{aligned}
&\langle X_1 \cdots X_n \rangle > 0, \\
&(-1)^{l+1} \langle X_i X_{i+l} \rangle > 0, &&\text{for } i + l \leq n, \\
&(-1)^{n+l+1} \langle X_i X_{i+l} \rangle > 0, &&\text{for } n < i + l < 2n.
\end{aligned} \tag{6.6.16}$$

At this point we can still use $F_{\text{even}}$ or equivalently $F_{\text{odd}}$ to see if these two set of inequalities are actually equivalent. Representing the action of $F_{\text{even}}$ on the brackets we obtain

$$\begin{aligned}
&F_{\text{even}} \langle X_i X_{i+l} \rangle = \langle X_i X_{i+l} \rangle (-1)^{l+1}, &&\text{for } i + l \leq n, \\
&F_{\text{even}} \langle X_i X_{i+l} \rangle = \langle X_i X_{i+l} \rangle (-1)^{l+n+1}, &&\text{for } n < i + l < 2n.
\end{aligned} \tag{6.6.17}$$

The $F_{\text{even}}$ or $F_{\text{odd}}$ operator maps a set of inequalities invariant under $(1, 2, +)$ to one invariant under $(1, 2, -)$. Moreover, it maps cyclic to twisted cyclic for n odd. Therefore, for any n the geometry compatible with the bosonized maximally-nilpotent correlator is unique and can be described by (6.6.12).

## 6.7 Summary

In this chapter, we have used the topological characterization of the amplituhedron in terms of flipping numbers to study the geometry of the squared amplituhedron. In this new language, the amplituhedron is defined as the geometry having maximal flipping numbers and positive proper boundaries, up to the one fixed by twisted cyclicity. The squared amplituhedron corresponds instead to the union of all geometries without restriction on the flipping numbers and positive proper boundaries. We named the geometries with non-maximal flipping number amplituhedron-like geometries and proposed that these correspond to products of amplitudes (in the case of minimal number of points n) giving proofs of this at tree-level and MHV loop level. We have given an alternative non-intrinsic characterisation of the geometries (at tree and loop level) and their natural tiling as sums of pairs of on-shell diagrams (at tree level). Importantly for the verification of our results, we also proposed a formula (6.1.2) to compute the product of super amplitudes directly in the amplituhedron space.

While the superamplitude has maximal residues equal to $\pm 1$ the square of superamplitude has maximal residues in $2\mathbb{Z}$. We identified in the structure of the maximal squared amplituhedron a geometrical interpretation of this feature. In fact, we have found that the amplituhedron-like geometries that compose the squared amplituhedron are almost disconnected, which means that their interiors are disconnected but their boundaries intersect on regions of codimension smaller than 1. Each almost disconnected component is a positive geometry and therefore has a canonical form with maximal residues equal to $\pm 1$. Maximal residues with values higher than 1, correspond to points in the Grassmannian where these almost disconnected geometries touch. For this reason the union of these almost-disconnected components is not a positive geometry but a generalized positive geometry. The canonical form can act on such unions of almost disconnected positive geometries, and when acting on the squared amplituhedron gives the square of the amplitude. This square will involve a sum over pairs of equivalent geometries that is responsible for the factor of 2 in all maximal residues of the superamplitude squared.

# Chapter 7

# Outlooks

The WPGs framework, the maximal loop-loop cuts and the geometric description of the product of amplitudes raise new interesting questions on the amplituhedron geometry and open new exciting possibilities for the geometrical description of physical observables, even simply by enhancing the space of differential forms that can possibly be written as a canonical form. Here we discuss some natural continuation of the work done in [1, 2] and some more speculative ideas.

**Weighted Positive Geometries**

We began by observing that non-vanishing maximal residues of loop amplitudes are not always $\pm 1$ as has generally been assumed, but can take arbitrary values in $\mathbb{Z}$, apparently contradicting the fact that the loop amplituhedron is a positive geometry. We found the source of this apparent contradiction geometrically to be the existence of *internal boundaries* in the geometry where two regions of opposite orientation touch. This minimally requires including an extra term in the recursive definition of the canonical form to take into account these internal boundaries (4.2.5). In all the examples we have found, the internal boundaries arise from loop-loop propagators factorizing into the product of two factors. Algebraically these are examples of composite residues discussed in this context in [11] and it would be interesting to explore the relation between composite residues and internal boundaries in more detail.

As well as internal boundaries we have also stressed another under emphasised feature of the boundary structure of the amplituhedron and multiple residues, namely the simple fact that multiple-residues and corresponding multiple boundaries are non-unique. On the algebraic side a multiple residue is defined as an (ordered) sequence of simple residues. In the same way, on the geometrical side the relevant quantity one must use is 'boundaries of boundaries of...' rather than 'codimension k boundaries'. It would be interesting to revisit previous computations of the boundary structure of the loop amplituhedron, and in particular its Euler characteristic, for example [111], taking into account both internal boundaries and the above non uniqueness of codimension k boundaries. It would be natural then to try to formulate a generalization of the Euler Characteristic on WPGs. There are some challenges in coming up with a meaningful generalization but we want to point out the ordinary Euler characteristic has a well defined action over the union of two geometries. Namely, if we have two geometries $\chi(X^{(1)}), \chi(X^{(2)})$ we can write the Euler characteristic of the union as

$$\chi(X^{(1)} \cup X^{(2)}) = \chi(X^{(1)}) + \chi(X^{(2)}) - \chi(X^{(1)} \cap X^{(2)}) \,. \tag{7.0.1}$$

This formula can be used for example to compute the $\chi$ of figure 5.1 as the sum of the $\chi$ of a parallelepiped plus the $\chi$ of the roof minus the intersection of the two, that is a square. This formula could be a good starting point to address this problem together with the introduction of the concept of orientation.

We believe that the above insights will also have utility in the increasing number of wider applications of positive geometry concepts in physics beyond the amplituhedron. One closely related case is the momentum amplituhedron. For the tree-level momentum amplituhedron [45] a lot is known about its boundary stratification [119, 120] and its Euler characteristic has been proven to be equal to one [121], a strong indication that the geometry is free from internal boundaries. However, despite the very solid understanding achieved at tree level, finding the geometry of the loop momentum amplituhedron remains an open problem. In this case we expect internal boundaries to appear and the language of WPGs could give the right framework to define a loop momentum amplituhedron. Also in the search for a non-planar amplituhedron interesting ideas involving the sum of geometries over different orderings have been explored in [122, 123] and might benefit from being viewed as weighted positive geometries. Other wider applications of positive geometry which one could revisit include [36, 44, 46, 47, 49, 124–126]. Similarly weighted positive geometry may provide the right mathematical framework to deal with cosmological correlators, which contrarily to the wavefunction of the universe described by the cosmological polytopes [50, 127]s, do not currently have a geometrical description. Their maximal residues are not +/- 1 and they naively appears as a weighted sum of canonical forms of cosmological polytopes[1].

---

[1] We thank P. Benincasa for pointing this out.

**Loop-loop cuts**

The four-point planar amplitude integrand in $\mathcal{N} = 4$ SYM (and its closely related half BPS correlator) is known to ten loops using various graphical rules together with correlator insights [98]. There have also been investigations on it directly using amplituhedron insights [128–130]. A key question in this context then is whether the above all-loop cuts can be used practically to actually compute the 4-point amplitude/correlator at higher loops. Taking maximal residues (eg first the all-in-one-point cut, then further external cuts) yields a vast amount of information about the amplitude. It also has the tantalising chance of being a constructive approach: rather than using a huge basis of graphs and determining their coefficients, most of which are zero, it might be possible to use the cuts to construct only the relevant graphs and non-zero coefficients with which they appear.

Although we have focused on 4 points, the all-in-one-point cut and more general loop-loop cuts are largely independent of the number of points, and also the MHV degree, since they involve only the mutual geometry between loop lines rather than the details of the external geometry. There has been some nice recent progress in computing the amplitude for arbitrary multiplicity directly from the loop amplituhedron [23, 26, 28, 131]. Points worthy of note in the current context are that taking the maximal multiple residues involving loops at higher points yields leading singularities of amplitudes – rational coefficients – which have been extensively analysed and are given by Yangian invariant Grassmann integrals [11, 12]. Furthermore in the works [132, 133] a method for extracting a list of the physical amplitude's branch points from the amplituhedron is suggested. Here the boundaries are derived by intersecting the closure of the amplituhedron with the boundary components corresponding to vanishing brackets of the form $\langle ABij \rangle$, $\langle A_i B_i A_j B_j \rangle$. It would be extremely interesting to revisit both the above points using the insights and technology developed here.

In [16, 134] the geometry of the log of the MHV amplitude is considered and defined as a union of geometries with negative mutual positivity condition $\langle A_i B_i A_j B_j \rangle < 0$. One of these has negative mutual positivity condition for all $i, j$ and its canonical form is equal to the amplituhedron canonical form. The latter is the only term in the log of the amplitude surviving the all-in-one-point cut and therefore the all-in-one-point cut of the amplitude and the log of the amplitude are the same.

**Squared amplituhedron and amplituhedron-like geometries**

In this work we have used the topological characterization of the amplituhedron in terms of flipping numbers to study the geometry of the squared amplituhedron. In this new language the amplituhedron is defined as the geometry having maximal flipping numbers and positive proper boundaries, up to the one fixed by twisted cyclicity. The squared amplituhedron corresponds instead to the union of all geometries without restriction on the flipping numbers and positive proper boundaries. We named the geometries with non-maximal flipping number amplituhedron-like geometries and propose that these correspond to products of amplitudes (in the case of minimal number of points n) giving proofs of this at tree-level and MHV loop level. We have given an alternative non-intrinsic characterisation of the geometries (at tree and loop level) and their natural triangulation as sums of pairs of on-shell diagrams (at tree level).

Amplituhedron-like geometry generalize the topological characterization of the amplituhedron in terms of flipping numbers. One would like to investigate further generalizations of the amplituhedron geometry and what they correspond to. The most obvious thing is to consider geometries defined in a similar way to the amplituhedron-like geometries but with different signs for the inequalities. This seems to immediately lead to a vast number of cases. However restricting to non-equivalent geometries and imposing cyclicity reduces the number of possibilities. As shown in section 6.6 this reduces to examining cyclic (or twisted cyclic) geometries. So to be concrete we could imagine considering more general choices for the physical inequalities which at the moment we take to be $\langle Yii+1jj+1\rangle > 0$. We could generalise by imposing different signs for the inequalities $\langle Y12jj+1\rangle \lessgtr 0$ (with the sign of $\langle Yii+1jj+1\rangle > 0$ for other i following by (twisted) cyclicity). One needs to first examine if it makes sense to have some the analogue of flipping patterns etc in these cases. In any case it would be interesting to examine such geometries and understand what they correspond to.

Objects we might imagine appearing from more general geometries of this type are products of more than two amplitudes, the simplest example would be NMHV$^3$ at 7 points. These functions are well defined cyclic dlog forms, which can be seen by their expression as products of on-shell diagrams. As observed in [109], for m = 2 the N$^k$MHV amplituhedron is a product of NMHV amplituhedra, $\mathcal{A}_{n,k} = \frac{(\mathcal{A}_{n,1})^k}{k!}$ and as noted below (6.4.38) we find by explicit computation that in fact all flipping *patterns* of maximal amplituhedron-like geometries $\mathscr{H}^{(f)}_{n,n-2,2}$ have canonical form equal to A$_{n,k}$. For m = 4 the generalization is still unclear.

Going beyond the case of maximal k = n – m , the natural correspondence between amplituhedron-like geometries and the product of amplitudes does not seem to hold in the same way. For example we have checked by direct calculation that for k = 1, n = 6 the direct generalization of (6.2.12) is not true,

$$H^{(f)}_{n,k} \neq A_{n,f} * A_{n,1-f} \qquad \text{for n = 6, k = 1} . \tag{7.0.2}$$

Therefore two immediate questions arise:

1. What does amplituhedron-like geometries correspond to for n > k + m?

2. Is there a geometry corresponding to the product of two general (i.e. non parity conjugate) amplitudes?

We have an answer to the first question in the MHV case at loop level as the product of MHV and $\overline{\text{MHV}}$ amplitudes (6.4.31)

$$H_{n,0,l} = \sum_{l'} \overline{A}_{n,0,l'}\, A_{n,0,l-l'}\,. \tag{7.0.3}$$

However, in the above equation $\overline{A}$ is the anti-MHV amplitude divided by tree-level. Such a quantity has no analogue beyond the MHV case and so it is not clear how this formula will generalise beyond this case.

In looking at the second question note that we have checked numerically that, for k < n – m, the alternative characterisation of amplituhedron-like geometries (6.2.22) don't have a well defined flipping number, that is it is no longer equivalent to the amplituhedron-like geometries. Nevertheless, the association of a geometry to the product of on-shell diagrams described in (6.4.13) could be a starting point for a systematic derivation of the geometry corresponding to the products $A_{n,f} * A_{n,k-f}$ in a similar way as has been done in [28] for chiral pentagon integrands. In that case, the requirement that the geometry of chiral pentagons giving the 1 loop MHV amplitude must share spurious co-dimension 1 boundaries isolates a unique solution of the geometry of the chiral pentagon. It would be interesting to see if similar constraints identify a unique geometry for the product of amplitudes.

Another direction one could pursue to derive the geometry of the squared amplituhedron is to reconsider geometric light-like limit of the correlahedron. Recall the geometry of the squared amplituhedron can be derived from the correlahedron by imposing the constraints $\langle YX_iX_{i+1}\rangle = 0$ and then projecting respect to the intersection points $\tilde{Y}_i = Y \cap (X_iX_{i+1})$. The external data, Z, emerges as the points $Z_i = X_i \cap X_{i+1}$ on $\tilde{Y}^\perp$, while Y can be rewritten as $Y = \tilde{Y}\hat{Y}$, where the allowed values of $\hat{Y}$ gives the squared amplituhedron. However, the sign of the brackets $\langle Yii+1jj+1\rangle$ depends on $\tilde{Y}^\perp$ and whereas in the maximal case the Z space is unique, in the non-maximal case this is no longer true and non convex Zs could arise from the light-like limit. In particular, we suspect non-convex Z configurations on $Y^\perp$ could appear.

In conclusion, more work is needed to derive the geometry corresponding to products of amplitudes in general as well as the related non-maximal squared amplituhedron geometry.

Finally, it would also be interesting to explore if any relation between non-maximal flipping numbers and products of parity conjugate amplitudes holds in the context of the momentum amplituhedron [45], where its definition in terms of the sign flip number can be naturally generalized.

# Appendix A

# Spinor Helicity variables

At the heart of many modern progresses in amplitudes is the use of set of variables for the external kinematics that makes the on-shell condition $p^2 = m^2$ manifest. In particular, for massless particles the condition $p^2 = 0$ can be trivialized using the spinor-helicity formalism. Here we will define the spinor helicity variables and state some of their key properties.

Given a four vector $p^\mu$, we can associate a $2 \times 2$ matrix

$$p_{a\dot{b}} = p_\mu \sigma^\mu_{a\dot{b}} \,, \tag{A.0.1}$$

where $\sigma^\mu = (\mathbb{1}, \sigma^i)$ and $\sigma^i$ are the usual Pauli matrices. The matrix $p_{a\dot{b}}$, as the $\sigma$s, transforms under the $SU(2) \times \overline{SU}(2)$ representation of the Lorentz group. If $p^\mu$ is massless, then $p^2 = \det(p_{a\dot{b}}) = 0$. Therefore we can write the degenerate matrix $p_{a\dot{b}}$ as

$$p_{a\dot{b}} = |p]_a \langle p|_{\dot{b}} \,, \tag{A.0.2}$$

where $|p]_a, \langle p|_{\dot{b}}$ are two vectors and are called the angle and the square spinors. The two vectors transform respectively under $\overline{SU}(2)$ and $SU(2)$. Note that the two spinors are defined up to an overall scaling

$$\langle p| \rightarrow t \langle p| \,, \qquad |p] \rightarrow t^{-1} |p] \,, \tag{A.0.3}$$

where t is non zero. Since we want to use these variables to compute amplitudes we can chose representatives, up to a phase, expressing normalized fermionic polarization vectors in terms of spinor helicity variables. The solution of the free fermionic field equation reads

$$\Psi(x) = u(p) e^{ip \cdot x} + v(p) e^{-ip \cdot x} \,, \tag{A.0.4}$$

and we will express its polarization vectors u and v as

$$u_+ = \begin{pmatrix} |p]_a \\ 0 \end{pmatrix}, \qquad u_- = \begin{pmatrix} 0 \\ |p\rangle^{\dot{a}} \end{pmatrix}, \qquad (A.0.5)$$

for outgoing fermions, and

$$v_+ = \begin{pmatrix} |p]_a \\ 0 \end{pmatrix}, \qquad v_- = \begin{pmatrix} 0 \\ |p\rangle^{\dot{a}} \end{pmatrix}, \qquad (A.0.6)$$

for outgoing anti-fermions. Since we must have $u_\pm^* = v_\mp^*$ we obtain that

$$[p|^a = (|p\rangle^{\dot{a}})^*, \qquad \langle p|_{\dot{a}} = (|p]_a)^*. \qquad (A.0.7)$$

The over all phase of the polarization vectors is not physical but phase differences are. In fact, the polarization vector transform under little group transformations with a phase that depends on the helicity of the particle.

Raising and lowering indices is done with the Levi-Civita symbols

$$|p\rangle^{\dot{a}} = \epsilon^{\dot{a}\dot{b}} \langle p|_{\dot{a}}, \qquad [p|^a = \epsilon^{ab} |p]_b. \qquad (A.0.8)$$

The key ingredients for writing amplitudes in spinor helicity formalism are the angle and the square spinor brackets which are defined as follows

$$\langle pq \rangle = \langle p|_{\dot{a}} |q\rangle^{\dot{a}}, \qquad [pq] = [p|_a |q]^a. \qquad (A.0.9)$$

In particular we can use spinor brackets to write the scalar product of massless momenta as

$$2p \cdot q = \langle pq \rangle [pq]. \qquad (A.0.10)$$

To unlock all the technology of complex analysis, it is extremely convenient to consider external momenta as complex. Well behaved Lorenz invariant functions will then be invariant under the complexified Lorentz group. The action of the little group then correspond to (A.0.3). Consider now a generic n-point amplitude $A_n$ with momenta $(|i\rangle, |i])$ and helicities $h_i$. Since the amplitude transforms under the little group scaling in the same way as polarization vectors do we will have that in general

$$A_n(\{|1\rangle, |1], h_1\}, \cdots, \{t|i\rangle, t|i], h_i\}, \cdots) = t^{-2h_i} A_n(\{|1\rangle, |1], h_1\}, \cdots, \{|i\rangle, |i], h_i\}, \cdots). \qquad (A.0.11)$$

This scaling property constrains the amplitude and it's key in many modern amplitude computations.

# Appendix B

# Rational residue forms on rational varieties

Consider a multivariate irreducible polynomial $f(x_1, \cdots, x_n)$. We want to take the residue on $f = 0$ on the form

$$\omega = \frac{dx_1 \cdots dx_n}{f(x_1, \cdots, x_n)} g(x_1, \cdots, x_n) \,, \tag{B.0.1}$$

where g is a rational function which doesn't have a have a pole on $f = 0$. To compute the residue in $f = 0$ we have to write the differential df explicitly as

$$df = \partial_{x_1} f dx_1 + \cdots \partial_{x_n} f dx_n \,. \tag{B.0.2}$$

Then if we solve for one particular variable, let's say $x_1$ we can rewrite the full differential as

$$dx_1 dx_2 \cdots dx_n = \frac{df}{\partial_{x_1} f} dx_2 \cdots dx_n \,. \tag{B.0.3}$$

Then the residue will simply read

$$\text{Res}_{f=0}(\omega) = \frac{dx_2 \cdots dx_n}{\partial_{x_1} f} g(x_1 \cdots, x_n) \,. \tag{B.0.4}$$

This expression is defined implicitly, since $x_1$ is constrained by the equation $f = 0$. In general the equation $f = 0$ will have multiple roots in $x_1$, that is functions $q_i(x)$ for which

$$f(q_i(x_1), x_2, \ldots, x_n) = 0 \,, \tag{B.0.5}$$

each one converging part of the boundary component $f = 0$. For degree of f in $x_1$ higher then 1, $q_i$ is not a rational function and therefore also the residue will not be in general a rational function. If a rational parametrization of f is available, that is a rational functions $\tilde{x}_i$ of the variables $t = (t_1, \cdots, t_{n-1})$ covering almost all the boundary component (up to points) such that

$$f(\tilde{x}_1(t), \cdots, \tilde{x}_n(t)) = 0 , \tag{B.0.6}$$

then we can write the residue form as a rational top form in t.

As an example consider the canonical form of a unit circle cut by the line $y = a$, that is

$$\omega = \frac{2\sqrt{1-a^2} dx dy}{(1-x^2-y^2)(y-a)} . \tag{B.0.7}$$

The residue on $f = 1 - x^2 - y^2$, can be written as

$$\text{Res}_{f=0}(\omega) = -\frac{\sqrt{1-a^2} dx}{y(y-a)} , \tag{B.0.8}$$

or equivalently as

$$\text{Res}_{f=0}(\omega) = \frac{\sqrt{1-a^2} dy}{x(y-a)} . \tag{B.0.9}$$

If we write (B.0.8) explicitly in x using the map $y = \sqrt{1-x^2}$, which covers the upper half of the circle we obtain

$$\text{Res}_{f=0}(\omega) = -\frac{\sqrt{1-a^2} dx}{\sqrt{1-x^2}(\sqrt{1-x^2}-a)} , \tag{B.0.10}$$

which is not rational and has branch points in $x = \pm 1$ due to the singularity of the change of variables. If instead we use the rational parametrization of the circle $x = \frac{t^2-1}{t^2+1}, y = \frac{2t}{t^2+1}$ the form in the t variable will read

$$\text{Res}_{f=0} = \frac{2\sqrt{1-a^2} dt}{a - 2t - at^2} , \tag{B.0.11}$$

which correspond to the canonical form of the segment $\frac{1-\sqrt{1-a^2}}{a} < t < \frac{1+\sqrt{1-a^2}}{a}$, as expected.

# Appendix C

# Loop-loop cut computations

## C.1 Three loop internal Boundary and its maximal residues

Here we consider taking further residues of the all-in-one-point cut (5.1.7) to eventually arrive at the leading singularities. Then we will consider the same sequence geometrically. Indeed in [116] such maximal residues were considered leading to a final configuration in which all loop lines intersect external twistors as well as intersecting each other at a single point A.

Specifically we consider the case where loop line $L_1$ intersects $Z_1$ and $L_2, L_3$ intersect $Z_2$. We here show that the resulting residue depends on the path taken. Furthermore if the path taken involves taking a residue in $\langle AB_1B_2B_3 \rangle = 0$ first, then the resulting maximal residue has magnitude 2 suggesting that $\langle AB_1B_2B_3 \rangle = 0$ corresponds to an internal boundary.

The two routes we consider to reach the above configuration are as follows. For route 2 we first take further residues of the all-in-one-point cut (5.1.7) in the following order

$$\langle AB_1 12 \rangle = 0 \,, \ \ \langle AB_1 14 \rangle = 0 \,, \ \ \langle AB_2 12 \rangle = 0 \,, \ \ \langle AB_2 23 \rangle = 0 \,. \tag{C.1.1}$$

This corresponds to intersecting line $L_1$ with the edge $Z_1Z_4$ and then with $Z_1$ followed by intersecting $L_2$ with the edge $Z_1Z_2$ and then to $Z_2$. In the process the pole $\langle AB_1B_2B_3 \rangle \to \langle A12B_3 \rangle$ and so the final step is to take residues in this pole $\langle A12B_3 \rangle = 0$ followed by $\langle A23B_3 \rangle = 0$ corresponding to $L_3$ intersecting $Z_1Z_2$ and then sliding to $Z_2$. We take the residues explicitly by parametrizing the $B_i$ as follows

(with $Z_*$ an arbitrary twistor)

$$\begin{align}
B_1 &= c_1 Z_4 + Z_1 + d_1 Z_* \,, \\
B_2 &= c_2 Z_1 + Z_2 + d_2 Z_* \,, \\
B_3 &= c_3 Z_1 + Z_2 + d_3 Z_* \,.
\end{align} \tag{C.1.2}$$

and considering the residues at zero in $d_1, c_1, d_2, c_2, d_3, c_3$ in that order. The residues are straightforward to compute and can be done covariantly, for example:

$$\underset{d_1=0,\, c_1=0}{\mathrm{Res}} \left( \frac{\langle AB_1 d^2 B_1 \rangle}{\langle AB_1 14 \rangle \langle AB_1 12 \rangle} \right) = \frac{-1}{\langle A214 \rangle} \,. \tag{C.1.3}$$

Only the displayed term in (5.1.6) out of the 24 total terms survives this sequence of residues and it produces the final result

$$-\frac{\langle 1234 \rangle^3 \langle A d^3 A \rangle}{\langle A123 \rangle \langle A124 \rangle \langle A134 \rangle \langle A234 \rangle} = -\frac{da\, db\, dc}{abc} \,, \tag{C.1.4}$$

where on the right hand side we parametrise the point A as

$$A = Z_1 + a Z_2 + b Z_3 + c Z_4 \,. \tag{C.1.5}$$

This is the canonical form of a tetrahedron with vertices $Z_i$ and is inline with the prediction of [116]. Indeed the mutual intersection point A is now the only remaining freedom and restricting the amplituhedron geometry to this configuration results in the tetrahedron. However as we saw in the simple example in the introduction, simply restricting the geometry to a high codimension boundary will not always give the right answer and the precise order in which one takes the residues can be important.

For route 2 therefore we change the order in which we take the residues on the all-in-one-point cut (5.1.7). We first take a residue in the pole $\langle AB_1 B_2 B_3 \rangle$ making $L_3$ coplanar with L1 and L2. Then proceed taking residues as previously in (C.1.1) moving $B_1 \to Z_1$ and $B_2 \to Z_2$. Then finally we take a residue as $B_3 \to Z_2$. Explicitly then, this time we parametrize the $B_i$ as follows

$$\begin{align}
B_3 &= c_3 B_1 + B_2 + d_3 Z_* \,, \\
B_1 &= c_1 Z_4 + Z_1 + d_1 Z_* \,, \\
B_2 &= c_2 Z_1 + Z_2 + d_2 Z_* \,.
\end{align} \tag{C.1.6}$$

and take the residues at zero in the order $d_3, d_1, c_1, d_2, c_2, c_3$. The first residue is

$$\underset{\langle AB_1 B_2 B_3 \rangle = 0}{\mathrm{Res}} \left( \frac{\langle AB_3 d^2 B_3 \rangle}{\langle AB_1 B_2 B_3 \rangle} \right) = -dc_3 \,, \tag{C.1.7}$$

and then use similar results to (C.1.3) before finally taking the residue in the parameter $c_3$. This time two terms in (5.1.6) survive, the displayed term together with the term

$$\frac{\langle Ad^3A\rangle \prod_{i=1}^{3}\langle AB_i d^2 B_i\rangle \langle 1234\rangle^3 \langle 23AB_1\rangle}{\langle 12AB_1\rangle \langle 12AB_3\rangle \langle 14AB_1\rangle \langle 23AB_2\rangle \langle 23AB_3\rangle \langle 34AB_1\rangle \langle 34AB_2\rangle \langle AB_1B_2B_3\rangle} \ . \tag{C.1.8}$$

Note that this term only survives because a required pole $\langle 12AB_2\rangle$ appears from the pole in $\langle 12AB_3\rangle$ after the residues in $d_3 = 0, d_1 = 0, c_1 = 0$ have been taken. Since there are two terms now surviving, the final result turns out to be twice (C.1.4)

$$-\frac{2\langle 1234\rangle^3 \langle Ad^3 A\rangle}{\langle A123\rangle\langle A124\rangle\langle A134\rangle\langle A234\rangle} = -2\frac{\mathrm{d}a\mathrm{d}b\mathrm{d}c}{abc} \ . \tag{C.1.9}$$

Now the final configuration of these two routes is exactly the same in both cases: $B_1 \to Z_1, B_2, B_3 \to Z_2$ and yet the results differ by a factor of 2. We thus clearly see the importance of path dependence when taking residues. We will shortly see that path dependence can give different results for the all-in-one-point cut itself (rather than just when taking further residues). Furthermore, taking three further residues as $a, b, c \to 0$ clearly gives us a maximal residue of magnitude 2 indicating that there is an internal boundary present. Let us then consider this geometrically.

We can now redo the above computation geometrically by taking boundaries of the all-in-one-point cut geometry (5.1.16) and using our formula of the canonical form of a GPG (4.2.5). To do this, we parametrise A and $B_i$ just as in (C.1.5) and (C.1.6)

$$\begin{aligned} B_3 &= c_3 B_1 + B_2 + d_3 Z_* \\ B_1 &= c_1 Z_4 + Z_1 + d_1 Z_* \ , \\ B_2 &= c_2 Z_1 + Z_2 + d_2 Z_* \ , \\ A &= Z_1 + aZ_2 + bZ_3 + cZ_4 \end{aligned} \tag{C.1.10}$$

Then take boundaries in the same order with which we took residues following (C.1.9)

$$d_3 = 0, \quad d_1 = 0, \quad c_1 = 0, \quad d_2 = 0, \quad c_2 = 0, \quad c_3 = 0 \ . \tag{C.1.11}$$

The first boundary corresponds to the above internal boundary $\langle AB_1B_2B_3\rangle = 0$. Therefore, from the recursive definition of the canonical form in the presence of internal boundaries (4.2.5), the canonical form of the all-in-one-point cut geometry satisfies

$$\lim_{d_3 \to 0} d_3\big(\Omega(\mathcal{R}_1^{\mathrm{dc}}) + \Omega(\mathcal{R}_2^{\mathrm{dc}})\big) = 2\mathrm{dd}_3 \wedge \Omega(\mathcal{R}^{\mathrm{dc}}|_{\langle AB_1B_2B_3\rangle=0}). \tag{C.1.12}$$

This simply means that the residue, $d_3 = 0$, on the canonical form determined by the inequalities describing the geometry of the all-in-one-point cut (5.1.13), is equal to twice the canonical form of the interior boundary, i.e. the canonical form determined by the inequalities (5.1.9) with $B_3 = B_1 + a_3 B_2$.

Continuing with the remaining boundaries, the final geometry is described by the following set of inequalities,

$$\mathcal{R}^{\text{final}}: \quad a < 0, \quad b > 0, \quad c > 0. \tag{C.1.13}$$

The remaining residues were all on external boundaries, therefore the canonical form of the final region is $2\Omega(\mathcal{R}^{\text{final}})$, where the factor of two comes from the internal boundary residue (C.1.12). The final inequalities describe a tetrahedron with vertices $\{Z_1, -Z_2, Z_3, Z_4\}$ and therefore the final canonical form is in precise agreement with (C.1.4).

We see that the internal boundary at $\langle AB_1 B_2 B_3 \rangle = 0$ is key to obtaining the correct leading singularity from the geometry. Just as we saw algebraically above, it is also possible to reach the same final loop configuration by only going to consecutive *external* boundaries. An example of this would be to follow the residues described in (C.1.2) geometrically. Then the final canonical form would be, up to an overall sign, the canonical form of the tetrahedron without the factor of two, as predicted in [116]. We see that the precise sequence of codimension 1 boundaries taken to approach higher codimension boundaries starting from the all-in-one-point cut configuration can give different results. At higher loops this is also true for the all-in-one-point cut itself (rather than just its maximal residues as here).

## C.2 Four loop point-plane boundary geometry

We here examine the geometry of the loop-loop boundaries of the four loop all-in-one-point cut (5.1.21). We start by the boundary when $B_3$ lies on the line $P = B_1 B_2$ followed by the boundary of that geometry when $B_4$ also approaches P. We observe that the regions $\mathcal{R}_1$ and $\mathcal{R}_2$ of (5.1.21) touch on $\langle AB_2 B_1 B_3 \rangle = 0$ as do $\mathcal{R}_3$ and $\mathcal{R}_4$. Since the orientations of $\mathcal{R}_1$ and $\mathcal{R}_2$ are opposite as are $\mathcal{R}_3$ and $\mathcal{R}_4$, this is an internal boundary. Thus the geometry of the (internal) amplituhedron boundary $\langle AB_2 B_1 B_3 \rangle = 0$, with $B_3$ living on $B_2 B_1$, is given by $\mathcal{R}_{12} \cup \mathcal{R}_{34}$ where

$$\begin{aligned}\mathcal{R}_{12} &= \mathcal{A}^{(4)}_{\text{dc}} \wedge \langle AB_2 B_1 B_4 \rangle > 0 \wedge \langle AB_3 B_1 B_4 \rangle > 0 |_{B_3 \in B_2 B_1}, \; +, \\ \mathcal{R}_{34} &= \mathcal{A}^{(4)}_{\text{dc}} \wedge \langle AB_2 B_1 B_4 \rangle < 0 \wedge \langle AB_3 B_1 B_4 \rangle < 0 |_{B_3 \in B_2 B_1}, \; -.\end{aligned} \tag{C.2.1}$$

Here $\mathcal{R}_{12}$ is positively oriented (indicated by the +), while $\mathcal{R}_{34}$ is negatively oriented. Notice that both of these regions require $B_1$ to be on the same side of $B_2$ and $B_3$.

In other words $B_1$ can not lie between $B_2$ and $B_3$. Further the orientation depends which side of $B_2, B_3$, $B_1$ is on. Then after we send $B_4$ to the line P, we approach another internal boundary, with no further constraints on where $B_4$ can lie. We thus conclude that the geometry is of the four points $B_i$ on the line P inside the triangle, with $B_1$ not allowed between $B_2$ and $B_3$ and the orientation depending on the relative position of $B_1$ and $B_2, B_3$.

We can check the geometry more carefully by exploring the boundaries of (C.2.1) explicitly. To do this we make the constraint $\langle AB_1B_2B_3 \rangle = 0$ explicit by expanding $B_1$ as $c_1B_2 + c_2B_3$. Then (C.2.1) becomes

$$\mathcal{R}_{12} = \mathcal{A}_{dc}^{(4)} \wedge \left( \langle AB_2B_3B_4 \rangle > 0 \wedge c_1 < 0 \wedge c_2 > 0 \right) \vee \left( \langle AB_2B_3B_4 \rangle < 0 \wedge c_1 > 0 \wedge c_2 < 0 \right), \; + \; ,$$
$$\mathcal{R}_{34} = \mathcal{A}_{dc}^{(4)} \wedge \left( \langle AB_2B_3B_4 \rangle > 0 \wedge c_1 > 0 \wedge c_2 < 0 \right) \vee \left( \langle AB_2B_3B_4 \rangle < 0 \wedge c_1 < 0 \wedge c_2 > 0 \right), \; - \; .$$
(C.2.2)

and we see this has two external boundaries, $c_1 = 0$ and $c_2 = 0$, and one internal boundary at $\langle AB_2B_3B_4 \rangle = 0$. The two external boundaries correspond to sending $B_1 \to B_3$ and $B_1 \to B_2$ respectively. The limit internal boundary $\langle AB_2B_3B_4 \rangle \to 0^\pm$ instead corresponds to sending $B_4$ to the line P. In this case the geometry is described by the union of a positively oriented region $\mathcal{R}^+_{1234}$ and a negatively oriented region $\mathcal{R}^-_{1234}$ as

$$\begin{aligned} \mathcal{R}^+_{1234} &= \mathcal{A}_{dc}^{(4)} \wedge c_1 < 0 \wedge c_2 > 0, \; + \; , \\ \mathcal{R}^-_{1234} &= \mathcal{A}_{dc}^{(4)} \wedge c_1 > 0 \wedge c_2 < 0, \; - \; . \end{aligned}$$
(C.2.3)

We clearly see then recalling $B_1 = c_1B_2 + c_2B_3$ that $B_1$ cannot line in between the point $B_2$ and $B_3$ and the orientation depends on which side of $B2, B_3$ $B_1$ is on, with the position of $B_4$ unconstrained.

This also reveals that there are still further loop-loop type boundaries we could take even after doing the point-plane-cut. We could take a residue at $c_2 = 0$ or $c_3 = 0$ corresponding to $B_1 = B_2$ or $B_3$. The geometry we then obtain corresponds to the 3-loops maximal loop-loop cut $\mathcal{A}_{mll}^{(3)}$ but with weight 4 instead of 2 since this time we approached two internal boundaries that is $\langle AB_2B_1B_3 \rangle = 0$ and $\langle AB_2B_3B_4 \rangle = 0$. This implies that the residue corresponding to this 4-loop boundary is equal to 2 times the (C.1.12) all-in-one-point and plane 3-loop residue.

Let's now go back to the all-in-one-point cut (5.1.21) and explore the only other boundary (modulo permutation of the loop variables), at $\langle AB_2B_1B_4 \rangle = 0$. Notice that this time it is an external boundary since the 4 regions remain distinct. To be very explicit, we can approach the boundary by parametrizing $B_1$ as $c_1B_2 + c_2B_4 +$

$c_3 Z^*$ and then taking the limit $c_3 \to 0^\pm$ on (5.1.21), which becomes

$$\begin{aligned}
\mathcal{R}_1 &= \mathcal{A}_{dc}^{(4)} \wedge c_2 \langle AB_2B_3B_4 \rangle > 0 \wedge -c_1 \langle AB_2B_3B_4 \rangle > 0, \ +, \\
\mathcal{R}_2 &= \mathcal{A}_{dc}^{(4)} \wedge c_2 \langle AB_2B_3B_4 \rangle < 0 \wedge -c_1 \langle AB_2B_3B_4 \rangle > 0, \ -, \\
\mathcal{R}_3 &= \mathcal{A}_{dc}^{(4)} \wedge c_2 \langle AB_2B_3B_4 \rangle > 0 \wedge -c_1 \langle AB_2B_3B_4 \rangle < 0, \ +, \\
\mathcal{R}_4 &= \mathcal{A}_{dc}^{(4)} \wedge c_2 \langle AB_2B_3B_4 \rangle < 0 \wedge -c_1 \langle AB_2B_3B_4 \rangle < 0, \ -,
\end{aligned} \quad \text{(C.2.4)}$$

where the region 1 and 3 are positively oriented and 2 and 4 are negatively oriented. Here the geometry looks very similar to (C.2.2), but this time $c_1 = 0$ is unconstrained while $c_2$ is an external boundary. We can see that $c_1$ is free by taking the union of $\mathcal{R}_1$ and $\mathcal{R}_3$ and expanding the products into the different sign cases. The same goes for the pair $\mathcal{R}_2, \mathcal{R}_4$. Because of this we can actually rewrite the (C.2.4) as

$$\begin{aligned}
\mathcal{R}_{13} &= \mathcal{A}_{dc}^{(4)} \wedge c_2 \langle AB_2B_3B_4 \rangle > 0, \ +, \\
\mathcal{R}_{24} &= \mathcal{A}_{dc}^{(4)} \wedge c_2 \langle AB_2B_3B_4 \rangle < 0, \ -
\end{aligned} \quad \text{(C.2.5)}$$

This in turn then has two boundaries, $\langle AB_2B_1B_3 \rangle = 0$ and $c_2 = 0$. Setting $\langle AB_2B_1B_3 \rangle = 0$ obtained by sending $B_3$ to P is described by the union of a positively oriented region and a negatively oriented region as

$$\begin{aligned}
\mathcal{R}_{1234}^+ &= \mathcal{A}_{dc}^{(4)} c_2 > 0, \ +, \\
\mathcal{R}_{1234}^- &= \mathcal{A}_{dc}^{(4)} c_2 < 0, \ -.
\end{aligned} \quad \text{(C.2.6)}$$

This is the geometry of unconstrained points $B_i$ with orientation depending on the relative order of $B_1, B_2$.

This point-plane configuration has a further (internal) loop-loop boundary at $c_2 = 0$ corresponding to the limit $B_1 \to B_2$. This boundary then consists of three unconstrained points on the line $B_2, B_3, B_4$. This final configuration thus corresponds to the 3-loop maximal cut $\mathcal{A}_{mll}^{(3)}$ with weight 2.

## C.3 An all loop point-plane geometry

We here describe in detail the geometry corresponding to the specific all loop point-plane cut described in section 5.2.4.

We first take the simplest all-in-one-point cut boundary. In this all loops first intersect the line $A_1B_1$ and then they all slide to the same intersection point in the same order as their labeling. At L loops this is given by the inequalities (see (C.3.4))

$$\mathcal{R}^{\text{simplest dc}} = \bigcup_{\vec{s}} \mathcal{R}_{\vec{s}}^{\text{simplest dc}}$$

$$\mathcal{R}_{\vec{s}}^{\text{simplest dc}} = \mathcal{A}_{\text{dc}} \wedge \left( \bigwedge_{a=3}^{L} \bigwedge_{2<i<a} \{s_a \langle AB_i B_1 B_a \rangle > 0\} \right) \qquad \text{orientation} = \prod_a s_a \tag{C.3.1}$$

After taking the above all-in-one-point cut we then constrain all loops to lie in the same plane by taking the ordered series of boundaries $\{\langle AB_2 B_1 B_L \rangle = 0, \langle AB_2 B_1 B_{L-1} \rangle = 0, \cdots, \langle AB_2 B_1 B_3 \rangle = 0\}$. To do this, we will parametrize all loops, but $B_1$ and $B_2$, as $B_a = B_1 + b_a B_2 + c_a Z^*$ and we will take the limit $c_a \to 0^\pm$. We start by approaching the boundary $\langle AB_2 B_1 B_L \rangle$. What we obtain after this first limit is that for all i

$$s_L \langle AB_i B_1 B_L \rangle > 0 \to s_L b_L \langle AB_i B_1 B_2 \rangle = -s_L s_i b_L < 0 \,. \tag{C.3.2}$$

Since this inequality must hold for all i we have that all $s_i$, apart from $s_L$ must be equal and we can therefore define a single sign s as

$$s := s_{L-1} = \cdots = s_3 \,. \tag{C.3.3}$$

Moreover, we can see that $b_L$ is actually unconstrained. In fact, since for $s_L > 0$ and $s_L < 0$ we have the same orientation, (C.3.2) reduces to $b_L > 0 \vee b_L < 0$. The boundary geometry is then given by

$$\mathcal{R}^{\text{simplest dc}}|_{\langle A,B_2 B_1 B_L \rangle=0} = \mathcal{R}_{s=1}^{\text{simplest dc}}|_{\langle A,B_2 B_1 B_L \rangle=0} \cup \mathcal{R}_{s=-1}^{\text{simplest dc}}|_{\langle A,B_2 B_1 B_L \rangle=0}$$
$$\mathcal{R}_s^{\text{simplest dc}}|_{\langle A,B_2 B_1 B_L \rangle=0} = \mathcal{A}_{\text{dc}} \wedge \left( \bigwedge_{a=3}^{L-1} \bigwedge_{2<i<j} \{s \langle AB_i B_1 B_a \rangle > 0\} \right) \qquad \text{orientation} = s^{L-4} \,. \tag{C.3.4}$$

Now let's take the residue $\langle A21B_{L-1} \rangle = 0$ by taking the limit $c_L \to 0^\pm$. What we obtain is that

$$s\langle AB_i B_1 B_{L-1} \rangle > 0 = s\langle AB_i B_1 B_2 \rangle b_{L-1} > 0 = b_L < 0 \,. \tag{C.3.5}$$

This implies that $b_{L-1} \to 0^-$ corresponds to an external boundary of the geometry. Notice also that now the orientation of the two components of the geometry labeled by s = 1 and s−1 will be given by $s^{L-5}$. The series of boundaries $\langle AB_2 B_1 B_a \rangle = 0$ have a clear recursive structure such that at each step a we get an inequality of the form $b_a < 0$ and the orientation of the two components is equal to $s^{L-a-3}$. The recursion ends when we get to the last residue on $\langle AB_1 B_2 B_3 \rangle = 0$, for which no brackets of the form $\langle AB_i B_1 B_3 \rangle$ are present. At that point the two components $\langle AB_1 B_2 B_3 \rangle > 0$ and $\langle AB_1 B_2 B_3 \rangle < 0$ have the same boundary and opposite orientation and therefore this represents an internal boundary. We can conclude that the geometry of the simplest all-in-one-plane-and-point cut corresponds to

$$\mathcal{R}_{\text{s,L}}^{\text{simplest dc}}\Big|_{\bigwedge_{\text{a}}\langle A,B_2B_1B_a\rangle=0} = \mathcal{A}_{\text{dc}} \wedge \left(\bigwedge_{\text{a}=4}^{\text{L}-1} b_\text{a} < 0\right), \tag{C.3.6}$$

where the weight of geometry is equal to 2 due to the internal boundary $\langle AB_1B_2B_3\rangle = 0$ contribution.

At this point it's straight forward to compute the canonical form of this region for the 4-point MHV amplitude. The inequalities $b_\text{a} < 0$ for $\text{a} = 4, \cdots, \text{L}-1$ simply tell us that on the line P all $B_\text{a}$ must be on the same side of $B_1$ as $B_2$. So the full canonical form is as given in (5.2.23).

# Appendix D

# Checks on the bosonized product formula

## D.1 Star product proof for $m = 1$

From (6.1.4) we need to prove that when we put $Y = Y_0$ and $Z = Z(\chi)$ (3.2.1) then

$$\frac{1}{N(k_1, 1)} \langle I \rangle_{k_1} \frac{1}{N(k_2, 1)} \langle J \rangle_{k_2} = (-1)^{k_1 k_2 + k_2} \frac{1}{N(k_1 + k_2, 1)} \langle I(Y \cap J) \rangle . \quad (D.1.1)$$

Explicitly we have $N(k, 1) = (-1)^{\lfloor k/2 \rfloor} k!$, $Y = \begin{pmatrix} 0 & 0 \\ \mathbb{1}_{k_1 \times k_1} & 0 \\ 0 & \mathbb{1}_{k_2 \times k_2} \end{pmatrix}$ and each bosonised momentum twistor in k+m space is given as $Z_i = (z_i, \chi_i \phi_1, \cdots, \chi_i \phi_k)$. Defining $Y_1 = (0, \mathbb{1}_{k_1 \times k_1}, 0)$ and $Y_2 = (0, 0, \mathbb{1}_{k_2 \times k_2})$, this becomes

$$\frac{1}{N(k_1, 1)} \langle I \rangle_{k_1} \frac{1}{N(k_2, 1)} \langle J \rangle_{k_2} = \frac{(-1)^{\frac{\lfloor k_1 \rfloor}{2} + \frac{\lfloor k_2 \rfloor}{2}}}{k_1! k_2!} \langle IY_2 \rangle \left((-1)^{k_1} \langle Y_1 J \rangle \right) . \quad (D.1.2)$$

To obtain this expression we can expanded $(Y \cap J)$ on Y as

$$Y \cap J = \frac{1}{k_1! k_2!} Y_{a_1} \cdots Y_{a_{k_2}} \langle Y_{a_{k_2+1}} ... Y_{a_{k_1+k_2}} J \rangle \, \epsilon^{a_1, ..., a_{k_1+k_2}}, \quad (D.1.3)$$

to obtain

$$\langle I(Y \cap J) \rangle = \frac{1}{k_1! k_2!} \langle I Y_{a_{k_2+1}} ... Y_{a_{k_1+k_2}} \rangle \langle Y_{a_1} \cdots Y_{a_{k_2}} J \rangle \, \epsilon^{a_1, ..., a_{k_1+k_2}}, \quad (D.1.4)$$

which shows how for $m = 1$ the star product corresponds to nothing more than writing the simplest SL(k) invariant formula that combines Y, I and J and has the correct scaling. The role of the Ys as columns of the identity matrix is just to select the rows, and therefore the $\phi$s, entering the determinant. But since the $\phi$s are dummy variables that can be relabeled if we antisymmetrise respect to $Y = Y_1 Y_2$ in (D.1.2) we leave the expression unchanged. We can therefore rewrite the latter as

$$\frac{(-1)^{\frac{\lfloor k_1 \rfloor}{2} + \frac{\lfloor k_2 \rfloor}{2} - k_1}}{k!} (-1)^{k_1 k_2} \langle I(Y \cap J) \rangle, \tag{D.1.5}$$

where $(-1)^{k_1 k_2}$ come from the convention we chose for the sign of the intersection in equation (6.1.3). Now we can use that $\frac{\lfloor k_1 \rfloor}{2} + \frac{\lfloor k_2 \rfloor}{2} + \frac{\lfloor k_1 + k_2 \rfloor}{2} = k_1 + k_2 \mod 2$ and conclude that

$$\frac{1}{N(k_1, 1)} \langle I \rangle_{k_1} \frac{1}{N(k_2, 1)} \langle J \rangle_{k_2} = \frac{(-1)^{k_1 k_2 + k_2}}{N(k, 1)} \langle I(Y \cap J) \rangle \tag{D.1.6}$$

which proves the star product formula for $m = 1$.

## D.2 Bosonised product checks for $m = 2$ and $m = 4$

Here we would like to give evidence for the star product rule (6.1.2) by explicitly computing both sides of the equation for some special cases and verify that they match. We have chosen two examples that highlight how the sum over permutations in (6.1.2) is necessary to give the right result.

Consider the the following product of bosonised brackets for $m = 2$

$$(\langle 123 \rangle \langle 234 \rangle) * (\langle 123 \rangle \langle 234 \rangle) = -\frac{1}{2} \langle Y23 \rangle^2 \langle 1234 \rangle^2. \tag{D.2.1}$$

We can verify this result by using (6.1.4) and projecting both sides on a pair of on-shell Grassmannian variables $(\chi_i)^2 (\chi_j)^2$. If we project on $(\chi_1)^2 (\chi_4)^2$ for example, we obtain for the left hand side

$$\frac{1}{N(1,2)^2} \int d^2\phi_1 d^2\phi_2 \left( \langle 23 \rangle^2 \phi_1 \chi_1 \phi_1 \chi_4 \right) \left( \langle 23 \rangle^2 \phi_2 \chi_1 \phi_2 \chi_4 \right) =$$

$$= \langle 23 \rangle^4 \left( \frac{\langle \chi_1 \chi_4 \rangle}{2!} \right)^2 = -\frac{1}{8} \langle 23 \rangle^4 \langle \chi_1 \chi_1 \rangle \langle \chi_4 \chi_4 \rangle. \tag{D.2.2}$$

While for the right hand side we obtain

$$-\frac{1}{2} \langle 23 \rangle^4 \frac{1}{N(2,2)} \int d^2\phi_1 d^2\phi_2 (\phi_1 \chi_1 \phi_2 \chi_2 - \phi_1 \chi_2 \phi_2 \chi_1)^2 = -\frac{1}{8} \langle 23 \rangle^4 \langle \chi_1 \chi_1 \rangle \langle \chi_4 \chi_4 \rangle. \tag{D.2.3}$$

The two projections match as expected.

Let's now see an example of product of bosonized brackets for $m = 4$. Consider the following product of bosonized brackets

$$\left(\langle 12367\rangle^3 \langle 12357\rangle\right) * \left(\langle 134567\rangle \langle 124567\rangle^3\right) =$$
$$= \frac{1}{4}\left(\langle Y1267\rangle^2 \left(\langle Y1267\rangle \langle Y1357\rangle + 3\langle Y1257\rangle \langle Y1367\rangle\right)\langle 1234567\rangle^4\right. \quad (D.2.4)$$

To check this relation we can again use (6.1.4) and project (D.2.4) on $(\chi_3)^4(\chi_4)^4(\chi_5)^4$. Projecting on $(\chi_i)^4$ is equivalent to acting with the operator $\partial_i^{(4)} := \partial_{\chi_i}^{(4)}$. Projecting on the right and side and integrating out the $\phi$s its easy and gives

$$\frac{1}{4}\langle 1267\rangle^6 \left(\langle Y1267\rangle \langle Y1357\rangle + 3\langle Y1257\rangle \langle Y1367\rangle\right) \quad (D.2.5)$$

If we perform the same operation on the left hand side instead we obtain

$$\partial_3^{(4)}\partial_4^{(4)}\partial_5^{(4)}([45]^3[46][2][3]^3) =$$
$$= \partial_3^{(3)}[45]^3\partial_4^{(3)}\partial_5^{(3)}[3]^3\left(\partial_3^{(1)}[46]\partial_4^{(1)}\partial_5^{(1)}[2] + \partial_5^{(1)}[46]\partial_3^{(1)}\partial_4^{(1)}[2]\right) =$$
$$\langle 3333\rangle_\phi [345]^6 \left([346][245]\langle 1111\rangle_\phi \langle 2222\rangle_\phi + [456][234]\langle 1222\rangle_\phi \langle 1112\rangle_\phi\right) + (\phi_2 \leftrightarrow \phi_3),$$
$$(D.2.6)$$

where $\langle ijlk\rangle_\phi = \epsilon_{ABCD}\phi_i^A \phi_j^B \phi_l^C \phi_k^D$ and [ij] indicate a 5-bracket not containing indices i, j and analogously [ijk] indicates a 4-brackets not containing the indices i, j, k . Manipulating the expression using the identities

$$\langle a***\rangle_\phi \langle aaa*\rangle_\phi = -\binom{4}{3}^{-1}\langle aaaa\rangle_\phi \langle ****\rangle_\phi \quad (D.2.7)$$

$$\langle aa**\rangle_\phi \langle aa**\rangle_\phi = \binom{4}{2}^{-1}\langle aaaa\rangle_\phi \langle ****\rangle_\phi \quad (D.2.8)$$

and integrating out the $\phi$'s we obtain

$$[345]^6\left([346][245] - \frac{1}{4}[456][234]\right) = \langle 1267\rangle^6\left(\langle 1257\rangle \langle 1367\rangle - \frac{1}{4}\langle 1237\rangle \langle 1567\rangle\right),$$
$$(D.2.9)$$

which can be tested numerically to be equal to (D.2.5) as expected.

Bibliography                                                              169